%% file: thesis.tex
\documentclass{report}
\usepackage{amsmath, amssymb, amsthm}
\usepackage{parskip}
\usepackage{graphicx} % Required for inserting images
\usepackage{tikz}
\usetikzlibrary{quantikz2}

\usepackage{xcolor}
\usepackage{mathtools}

\usepackage[top=1in, bottom=1in, left=1.3in, right=1.3in]{geometry} 

\newif\ifblackwhite\blackwhitefalse
\ifblackwhite
    \usepackage{python_color}
    \usepackage[colorlinks=False]{hyperref}
    \hypersetup{
    allbordercolors={white}
    }
\else
    \usepackage{pythonhighlight}
    \usepackage[colorlinks]{hyperref}
    \hypersetup{
    linkcolor=[rgb]{0.3,0.3,0.6},
    citecolor=[rgb]{0.2, 0.6, 0.2},
    urlcolor=[rgb]{0.6, 0.2, 0.2},
    }
\fi
\usepackage[capitalise,noabbrev]{cleveref}

\usepackage{caption}
\usepackage[utf8]{inputenc}
\usepackage{pdfpages}

\counterwithout{footnote}{chapter}
\usepackage[maxbibnames=15, url=false, style=alphabetic, giveninits=true]{biblatex}
\DeclareSortingNamekeyTemplate{
  \keypart{
    \namepart{family}
  }
  \keypart{
    \namepart{prefix}
  }
  \keypart{
    \namepart{given}
  }
  \keypart{
    \namepart{suffix}
  }
}
\makeatletter
\AtBeginDocument{\toggletrue{blx@useprefix}}
\AtBeginBibliography{\togglefalse{blx@useprefix}}
\makeatother

\input{definitions_and_operators}

\title{\textbf{Adaptive Quantum Computers} \\
decoding and state preparation}
\author{Niels M. P. Neumann$^{a,b,c}$\footnote{This document is the online and update version of the author's PhD thesis. 
The original version is found online at~\cite{Neumann:2025}.}}
\date{$^a$ Institute for Logic, Language and Computation, University of Amsterdam, The Netherlands. \\
$^b$ Algorithms and Complexity group, National Research Institute for Mathematics and Computer Science, The Netherlands. \\
$^c$ Applied Cryptography and Quantum Applications, The Netherlands Organisation for Applied Scientific Research (TNO), The Netherlands.}

\begin{document}

\maketitle
\tableofcontents

\include{non_content/summary}
\pagestyle{headings}

% Introduction
\input{introduction/introduction}

% Constant-depth decoding
\part{Decoding}\label{part:decoding}
\input{decoding/introduction_decoding}
\input{decoding/classical_decoding}
\input{decoding/quantum_linear_codes}
\input{decoding/quadratic_codes}

% LAQCC
\part{LAQCC}\label{part:LAQCC}
\input{LAQCC/LAQCC_introduction}
\input{LAQCC/LAQCC_model}

\input{LAQCC/LAQCC_state_preparation}

\input{LAQCC/LAQCC_error_analysis}

\include{non_content/symbols}

\begingroup
\setlength{\emergencystretch}{1em}
\printbibliography
\endgroup
\end{document}

%% file: definitions_and_operators.tex
\renewcommand{\Pr}{\mbox{\rm Pr}}
\newcommand{\Exp}{{\mathbb{E}}}

\newcommand{\R}{\mathbb{R}} % reals
\newcommand{\C}{\mathbb{C}} % complex numbers
\newcommand{\D}{\mathbb{D}} % complex numbers
\newcommand{\N}{\mathbb{N}} % natural numbers
\newcommand{\Z}{\mathbb{Z}} % integers
\newcommand{\F}{\mathbb{F}} % field
\newcommand{\E}{\mathcal{E}} % event
\newcommand{\HS}{\mathcal{H}} % Hilbert space
\newcommand{\Fcal}{\mathcal{F}} % family 
\newcommand{\Acal}{\mathcal{A}} % algorithm
\newcommand{\pmset}{\{-1,1\}} % hypercube in +-1 basis
\newcommand{\bset}{\{0,1\}} % hypercube
\newcommand{\nc}{\mathsf{NC}}
\newcommand{\ac}{\mathsf{AC}}
\newcommand{\tc}{\mathsf{TC}}
\newcommand{\qnc}{\mathsf{QNC}}
\newcommand{\qac}{\mathsf{QAC}}

\DeclareMathOperator\BQP{\mathsf{BQP}}
\DeclareMathOperator\QPoly{\mathsf{QPoly}}
\newcommand{\p}{\mathsf{P}}
\newcommand{\NP}{\mathsf{NP}}
\newcommand{\Lspace}{\mathsf{L}}
\newcommand{\bigo}{\mathcal{O}} 
\DeclareMathOperator{\LAQCC}{\mathsf{LAQCC}}

\DeclareMathOperator{\im}{Im} % image

\newcommand{\one}{\mathbf{1}}

\newcommand{\st}{:\,} % "such that" to define sets

\newcommand{\eg}{{e.g.}} 
\newcommand{\eps}{\varepsilon}
\newcommand{\ip}[1]{\langle #1 \rangle}

\newcommand{\poly}{\mbox{\rm poly}}

\DeclareMathOperator{\rank}{rk}

\DeclareMathOperator{\arank}{arank}
\DeclareMathOperator{\prank}{prank}

\DeclareMathOperator{\popest}{PopEst}
\DeclareMathOperator{\degest}{DegEst}
 
\DeclareMathOperator{\bias}{bias}
\DeclareMathOperator{\Pol}{Pol}

\newcommand{\kb}[1]{\ket{#1}\bra{#1}}
\newcommand{\floor}[1]{\left\lfloor #1 \right\rfloor}
\newcommand{\ceil}[1]{\left\lceil #1 \right\rceil}

  \DeclareMathOperator{\isbal}{IsBal}
  \DeclareMathOperator{\maj}{Maj}
  \DeclareMathOperator{\LH}{LH}

\newtheorem{theorem}{Theorem}[section]
\newtheorem{lemma}[theorem]{Lemma}
\newtheorem{proposition}[theorem]{Proposition}
\newtheorem{definition}[theorem]{Definition}
\newtheorem{remark}[theorem]{Remark}
\newtheorem{corollary}[theorem]{Corollary}
\newtheorem{notation}[theorem]{Notation}
\newtheorem{claim}[theorem]{Claim}

\renewenvironment{proof}[1][]{
 	\begin{trivlist}
	\item[\hspace{\labelsep}{\em\noindent Proof#1:\/}]}
	{{\hfill$\Box$}
 	\end{trivlist}
}

\newrobustcmd*{\citeallauthors}{%
  \AtNextCite{\AtEachCitekey{\defcounter{maxnames}{999}}}%
  \fullcite}

%% file: non_content/summary.tex
\section*{Abstract}

The concept of computers dates back to the late 19th century. 
Over the years, significant progress has been made towards their physical realization. 
Today, we can hardly imagine a world without computers and we use them for a wide range of applications in our daily life.
Extensive research efforts focus on enhancing the power of computers and exploring new ways of performing computations. 
One promising approach is quantum computing.
Future quantum computers have the potential to solve specific problems significantly faster than current methods.
These quantum computers will have to interact with a standard computer to operate effectively.

Current quantum computers are still under development and have limited capabilities.
However, the interaction with a standard computer can already enhance their functionality, particularly by offloading certain computations to the standard computer. 
Quantum computers that interact with standard computers to perform computations are called \textit{adaptive quantum computers}. 
This work formalizes a model that describes these adaptive quantum computers.
As quantum computers are still under development, this work focuses on computations that terminate after a fixed number of steps, as that makes their implementation likely easier in practice. 
The work is divided into two main parts.

The first part shows that adaptive quantum computers are more powerful than standard computers.
It focuses on the practical problem of retrieving information from corrupted digital data. 
Standard computers struggle to retrieve such information within a fixed number of computation steps.
The proof uses a structure-versus-randomness approach that splits the problem in a structured and a random-like component. 
The potential of adaptive quantum computations follows from a specific example where information is retrieved from corrupted data. 
Additionally, adaptive quantum computers can even improve standard computations for this problem that are not constrained by a fixed number of computation steps. 

The second part explores how adaptive quantum computations can improve non-adaptive quantum computations, for instance by performing various quantum computations faster.
Using these faster computations, adaptive quantum computers can also prepare quantum states more efficiently than their non-adaptive counterparts. 
Quantum states describe the computational units of a quantum computer, making the task of preparing them inherently quantum. 
This work presents efficient adaptive quantum computations to prepare quantum states such as the uniform superposition state, the GHZ state, the $W$-state and the Dicke state. 
These states are often used in other quantum algorithms, so having efficient routines for preparing them also enhance the efficiency of other algorithms.
This work concludes by comparing these adaptive quantum computations with non-adaptive ones, analyzing their performance both theoretically and through quantum hardware implementations. 

%% file: introduction/introduction.tex
\chapter{Introduction}

This chapter describes the emerging technology of quantum computers, by relating them to more general computational devices and problems. 
We also discuss advances in quantum algorithms and the search for a (provable) quantum advantage.
Afterwards, the special role Fourier analysis plays in many quantum algorithms is discussed.
Next, we consider some practicalities that arise when implementing quantum algorithms on quantum hardware. 

\input{introduction/intro_computational_devices}
\input{introduction/intro_computational_problems}
\input{introduction/intro_single_quantum_solution}
\input{introduction/intro_towards_new_devices}

\input{introduction/intro_developments_quantum_algorithms}
\input{introduction/intro_quantum_advantage}
\input{introduction/intro_Fourier_analysis}
\input{introduction/intro_theory_to_practice}

%% file: introduction/intro_computational_devices.tex
\section{Computational devices}
For centuries, humans have used tools to perform computations. 
For a long time, we only had relatively simple tools: 
Our hands for counting or an abacus to perform arithmetic operations. 
With the advent of computers, we gained access to a new tool. 
These computers could carry out similar operations as we have long performed, but with the benefit of doing so automatically. 

In the 19th century, \citeauthor{Babbage:1822} was the first to propose a steam-engine powered computer-like device~\cite{Babbage:1822}.
Others later also proposed and even built devices to perform computations. 
It was not until the 1930s that \citeauthor{Turing:1937} laid the theoretical foundation for a universal computer~\cite{Turing:1937}. 
Von Neumann later introduced the von Neumann architecture, where the data and the programs are both stored in the same memory~\cite{vonNeumann:1945}. 
This architecture enables users to easily reprogram the computers and use them for different applications. 

Modern computers still heavily rely on the foundational work of \citeauthor{Turing:1937} and \citeauthor{vonNeumann:1945}. 
Today, computers work by manipulating computational units called bits in a precise order.
The speed at which bits are manipulated largely determines a computer's computational power. 
Modern day computers use transistors to manipulate bits. 
The number of transistors roughly translates to the computational power of the whole device. 
\citeauthor{Moore:1965} first noted an apparent exponential growth in the number of transistors on a chip, which he then predicted as a trend for future growth~\cite{Moore:1965}. 

Moore's law, as this predicted growth was quickly called, effectively states that the number of transistors per integrated circuit roughly doubles every two years. 
As a result, the size of the transistors decreases by roughly the same factor. 
This size decrease implies a limit to the scaling of the computational power resulting from the number of transistor per chip~\cite{Sperling:2018}. 
At some point, the transistors become so small that they no longer obey the laws of conventional physics.
Instead, quantum physics must be used to describe their behavior~\cite{Lansbergen:2012}. 

Quantum physics extends the laws of conventional physics as developed by \citeauthor{Newton:1678} and others~\cite{Newton:1678}.
With quantum physics, events can happen with certain probability, instead of with certainty. 
As an undesirable effect, electrons in transistors can `tunnel' through barriers, thereby resulting in unpredictable behavior of the transistors. 

%% file: introduction/intro_computational_problems.tex
\section{Computational problems}
As computers can be programmed to automatically perform operations and thus solve problems, it is natural to ask what problems these computers can solve. 
The field of computer science tries to group problems based on the asymptotic computational resources required to solve them. 
We call these groups complexity classes. 

The most well-known class is $\p$, the class of decision problems solvable in polynomial time\footnote{Formally, the class $\p$ is defined in terms of languages that decide a decision problem in polynomial time. 
We will instead follow the definition in terms of polynomial-time algorithms.}. 
We call problems in $\p$ efficiently solvable and we call algorithms that solve a problem in $\p$ efficient.
For an algorithm $f$ to be of polynomial time means that there exist positive constants $A$ and $c$ such that on binary input $x$ of length $n$, the algorithm computes $f(x)$ in at most $An^c$ steps. 

As both $A$ and $c$ can be any nonzero positive constant, the total number of operations required can vary widely. 
It might therefore seem counterintuitive to call \textit{all} polynomial-time algorithms efficient.
The practical applicability of algorithms significantly depends on the exact value of the constant. 
Luckily, the constants for most polynomial-time algorithms used in practice are small~\cite[Section~3]{Wigderson:2019}.

Polynomial-time algorithms have the added benefit that their sum, product, and composition remain polynomial. 
This property is desired as it allows an efficient algorithm to call another efficient algorithm as subroutine.
Their composition remains efficient. 
Subroutines are extensively used in programming languages. 

The best known algorithms for problems without known efficient solutions typically have running times that are exponential in the input length. 
As exponential functions grow significantly faster than polynomials, even for small input sizes, the running times of these inefficient algorithms become too large. 

Examples of problems that admit efficient solutions are finding the shortest route between different stops~\cite{Dijkstra:1959} and testing the primality of integers~\cite{Agrawal:2004}. 
Yet, we have not yet found efficient algorithms for the closely related problems of finding the shortest cycle between different stops (the Traveling Salesman Problem)~\cite{Robinson:1949} and finding the prime factors of an integer.
Many believe no efficient algorithm exists for these problems.
However, these two problems have efficiently verifiable solutions;
Computing the product of different primes is simple, as is comparing it with the original integer. 
This gives the complexity class $\NP$, consisting of problems that can be verified efficiently (that is, in polynomial time). 

Naturally, every problem in $\p$ is also in $\NP$, as finding a solution is more difficult than verifying one. 
However, it is still unknown whether there are any problems in $\NP$ that are not in $\p$. 
Many believe this to be the case, but it remains one of the main open questions in the field.

As one might expect, there are problems that admit no efficiently verifiable solution and are outside of $\NP$. 
A prime example is the Halting problem, which tries to decide if a computer program terminates or run indefinitely. 
\citeauthor{Turing:1937} proved that this problem is in fact undecidable~\cite{Turing:1937}, meaning that no single algorithm can decide for every program if it will ever terminate. 
\citeauthor{AaronsonKuperbergGranade:2005} provide an extensive overview of many complexity classes, including known inclusions and separations between certain classes, as well as well-known extensions~\cite{AaronsonKuperbergGranade:2005}.

\begin{figure}
    \centering
    \includegraphics[width=0.5\linewidth, trim={13cm 4.9cm 13cm 4.9cm}, clip]{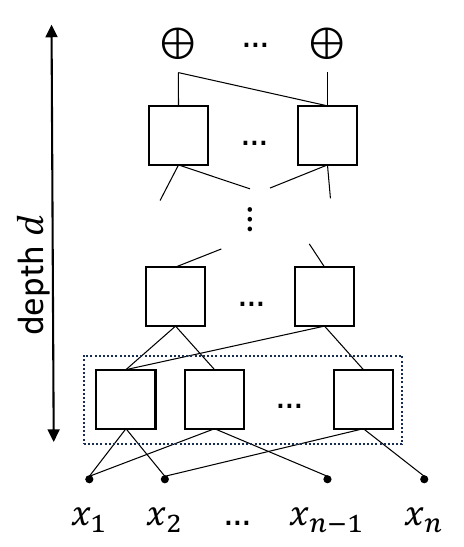}
    \caption{An example of a circuit.
    The $n$ inputs are shown at the bottom and every square box represents an elementary operation. 
    Time flows from the bottom to the top. 
    The dotted box denotes a layer and the number of numbers equals the depth of the circuit.}
    \label{fig:circuit_example}
\end{figure}

If we have a complexity class, we can often associate it with a class of circuits.
\cref{fig:circuit_example} gives a graphical example of a circuit. 
The input of length $n$ is shown at the bottom, and time flows from the bottom to the top. 
Each square represents an elementary operation, often referred to as a gate. 
We call the number of inputs to a gate the fan-in and the number of outputs the fan-out of a gate. 
Examples of these operations include the AND-gate, which outputs $1$ precisely if all inputs are $1$, and the OR-gate, which outputs $1$ precisely if at least one of the inputs is $1$. 
We can group the gates in layers, such that every gate in a layer is independent of the other gates in the layer. 
In the figure, the dotted box denotes a layer. 
The width of a single layer equals the number of operations in that layer and the width of the circuit as a whole equals the maximum width across all layers. 
The depth of a circuit is defined as the minimum number of layers required. 

In terms of circuits, the class $\p$ consists of circuits where both the width and depth of the circuit are polynomial in $n$. 
In the remainder of this work we will mainly focus on circuits.
For a complexity class $\mathsf{X}$, we refer to the corresponding circuits as $\mathsf{X}$-circuits.
In the remainder of this work, we will interpret the complexity classes in terms of these circuits. 
Differences between complexity classes translate into differences in the allowed gates, the fan-in or fan-out of gates, or the depth or width of the circuits. 

An example of a complexity class we will revisit in this work is the class $\nc^k$.
This class consists of all circuits of polynomial size in the input length $n$ and depth $\bigo((\log n)^k)$ using only bounded-fan-in AND- and OR-gates\footnote{We refer to the List of Symbols for a definition of the $\bigo$-, $o$-, $\Omega$-, $\omega$-, and $\Theta$-notation.
We can add a subscript to these notations to indicate a dependency on some fixed constant.}. 
The closely related class $\ac^k$ consists of circuits of the same depth and size, but uses unbounded-fan-in AND- and OR-gates. 
The class $\tc^k$ consists of circuits that additionally have access to unbounded-fan-in Threshold$_t$-gates. 
These gates evaluate to~$1$ precisely if the inputs sum to at least~$t$.
These three classes admit a natural hierarchy: 
\begin{equation*}
    \nc^k \subseteq \ac^k \subseteq \tc^k \subseteq \nc^{k+1},
\end{equation*}
where it is unknown if all inclusions are strict, or whether for some $k$, the inclusion is actually an equality. 
The first two inclusions follow directly from the definition.
The third inclusion follows as any gate with unbounded-fan-in can be implemented with bounded-fan-in gates in logarithmic depth. 

Another class that uses bounded-fan-in AND- and OR-gates is $\Lspace$.
This class consists of all circuits of width $\bigo(\log n)$ and has no restrictions on the circuit depth. 
\citeauthor{Johnson:1990} showed the inclusion $\Lspace \subseteq \ac^1 \subseteq \tc^1$~\cite{Johnson:1990}.

%% file: introduction/intro_single_quantum_solution.tex
\section{From two problems to a single solution}
We have now encountered two problems: 
1) The size of transistors presents a natural barrier beyond which quantum mechanical laws have to be taken into account;
2) For some problems we have no efficient algorithms. 
Where some see a problem, others see a new possibility of using quantum mechanical effects to perform computations. 

Inspired by work of others~\cite{Bennett:1973,Benioff:1980,Benioff:1982}, \citeauthor{Feynman:1982} popularized the idea of a programmable quantum device to simulate physical systems~\cite{Feynman:1982}. 
Instead of fighting against unwanted quantum effects in transistors, quantum computers embrace the quantum effects and use them to perform computations. 
Shortly after, \citeauthor{Deutsch:1985} proposed a model for a universal quantum computer~\cite{Deutsch:1985}, in line with the work for conventional computers by \citeauthor{Turing:1937}. 

\subsection{Quantum states}
Quantum computers use the properties of superposition, measurement, entanglement, and interference to perform computations on quantum bits, often called qubits. 
In the following sections, we briefly discuss these properties. 

\subsubsection{Qubits and superposition}
Quantum computers and conventional computers are closely related. 
Conventional computers operate on bits, each with $2$ (orthogonal) computational basis states, which we will denote by $\ket{0}$ and $\ket{1}$. 
With $n$ bits in total, a conventional computer is in one of the $2^n$ possible computational basis states. 

Qubits are the quantum-mechanical analogue of the conventional bit. 
A \textit{qubit}~$\ket{\psi}$ in a quantum computer additionally has the property that it can be in any \textit{superposition} of the two computational basis states
\begin{equation*}
    \ket{\psi}=\alpha_0 \ket{0} + \alpha_1 \ket{1}. 
\end{equation*}
In some sense, a superposition means that a qubit is in all computational basis states simultaneously. 
More rigorously, a superposition is a complex linear combination of the computational basis states, such that $\alpha_i$ is the complex amplitude of the state $\ket{i}$ and $|\alpha_0|^2 + |\alpha_1|^2 = 1$. 

The computational basis states of a qubit form a basis for the Hilbert space $\HS = \C^2$.
A Hilbert space is a vector space with an inner product, such that the resulting metric space is complete. 
We will use the operator norm given by
\begin{equation}\label{eq:operator_norm}
    \lVert U\rVert_{op} = \inf \{c>0 \mid \lVert U\ket{\psi} \rVert_2 \le c\lVert \ket{\psi}\rVert_2,\, \forall \ket{\psi}\in\HS\},
\end{equation}
where $\lVert \cdot \rVert_2$ denotes the Euclidean norm and gives the length of a vector. 

\subsubsection{Measurements}
Observing a quantum state, even those that are in superposition, will return precisely one computational basis state.
Before measuring, it is unknown which computational basis state is found. 
All we can say is that we observe state $\ket{i}$ with probability $|\alpha_i|^2$. 

Measurements also alter the measured quantum state:
Only the measured state remains, whereas the rest of the superposition is destroyed. 

\subsubsection{Entanglement}
The composition of two qubits is given by the tensor product:
Given two qubits $\ket{\psi}_A \in \HS_A$ and $\ket{\phi}_B \in \HS_B$, their composition is given by
\begin{equation*}
    \ket{\psi} \otimes \ket{\phi} \in \HS_A \otimes \HS_B. 
\end{equation*}
The composed Hilbert space $\HS_A \otimes \HS_B$ has as basis all possible combinations of the computational basis states of $\HS_A$ and $\HS_B$. 

If we take the tensor product of $n$ qubits, we can write the joint state as a tensor product of these $n$ single qubits: 
\begin{equation*}
    \sum_{i_1,\hdots,i_n\in \bset} \alpha_{i_1 \hdots i_n}\ket{i_1}\otimes \hdots\otimes \ket{i_n}.
\end{equation*}
We can equivalently interpret the $n$ qubits as a $2^n$-level system, described by 
\begin{equation*}
    \sum_{i=0}^{2^n-1} \alpha_i \ket{i}. 
\end{equation*}
Both notations are used interchangeably in quantum computer literature.
The used representation usually follows from the context. 

The composed Hilbert space contains more states than just the product between the states of the two individual Hilbert spaces. 
We call a quantum state in $\ket{\varphi} \in \HS_A \otimes \HS_B$ \textit{separable} if there exist $\ket{\psi}_A \in \HS_A$ and $\ket{\phi}_B \in \HS_B$ such that $\ket{\varphi} = \ket{\psi}_A \otimes \ket{\phi}_B$. 
In all other cases, we call $\ket{\varphi}$ \textit{entangled}. 
Most quantum states in the composed Hilbert space are entangled. 

A prime example of an entangled state between two parties $A$ and $B$ is 
\begin{equation}\label{eq:EPR_state}
    \frac{1}{\sqrt{2}}\left(\ket{0}_A\ket{0}_B+\ket{1}_A \ket{1}_B\right),
\end{equation}
often called the EPR-state after \citeauthor{EinsteinPodolskyRosen:1935} who first analyzed the state and the correlations between the measurement results~\cite{EinsteinPodolskyRosen:1935}.
Measuring the qubit of $A$ locally directly determines the value of the other qubit, independent of the physical distance between the two qubits. 

\citeauthor{Bell:1964} showed that the locally measured entangled variables are strictly more correlated than locally measured variables based on shared randomness. 
\citeauthor{ClauserHorneShimonyHolt:1969} introduced the CHSH game, a specific two-player game, where the players either share randomness or share an EPR-state. 
With shared randomness, the players win the game with probability at most $0.75$, whereas they do so with probability $\cos^2(\pi/8)\approx 0.85$ if they share an EPR-state~\cite{ClauserHorneShimonyHolt:1969}.
This so-called Bell inequality violation was later also realized experimentally~\cite{Bell:1964,Hanson:2015}. 

\subsubsection{Interference}
Contrary to what one might believe, quantum computing is more than simply performing computations using probability distributions. 
The complex amplitudes~$\alpha_i$ can be negative.
Different quantum states can therefore cancel or reinforce each other, an effect called \textit{interference}. 
One way to let quantum states interfere is via quantum gates. 

\subsection{Quantum gates}
Quantum gates can manipulate quantum states. 
Quantum gates are unitary operators $U:\HS\to\HS$ mapping $\ket{\phi}=\sum_{i=0}^{N-1}\alpha_i \ket{i}\mapsto\ket{\psi}=\sum_{i=0}^{N-1}\alpha_i U\ket{i}$. 
The unitary property of quantum gates implies that they are reversible. 

In this work, we will only consider quantum equivalents of the conventional circuits shown in \cref{fig:circuit_example}.
Quantum devices that perform operations represented by such circuits are often called gate-based quantum computers. 
Linear algebra provides the mathematical foundation that describes gate-based quantum computing. 
In this context, we can interpret quantum states as unit vectors, and quantum gates as unitary matrices. 
Note that unitary matrices preserve the length of vectors as desired. 

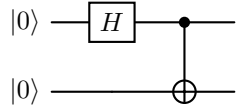
\begin{figure}
    \centering
    \begin{quantikz}
        \lstick{$\ket{0}$} & \gate{H} & \ctrl{1} & \\
        \lstick{$\ket{0}$} & &\targ{} & 
    \end{quantikz}
    \caption{An example quantum circuit. 
    Each line represents a qubit. 
    The shown quantum circuit prepares the quantum state $\tfrac{1}{\sqrt{2}}(\ket{00}+\ket{11})$.}
    \label{fig:q_circuit:example}
\end{figure}

Quantum circuits are often depicted with time running from left to right, as shown in~\cref{fig:q_circuit:example}.
This circuit prepares the EPR-state shown in \cref{eq:EPR_state}. 
Each line represents a qubit, the square denotes the single-qubit Hadamard gate, and the black circle connected to the $\oplus$-symbol represents the CNOT-gate (short for controlled-NOT-gate). 
The black circle represents a conditional control: 
The NOT-operation on the second qubit is only applied if the controlling qubit is in the $\ket{1}$-state. 
For computational basis states $\ket{x}$ and $\ket{y}$, the Hadamard gate $H$ implements the map $\ket{x}\mapsto\frac{1}{\sqrt{2}}(\ket{0}+(-1)^{x}\ket{1})$ and the CNOT-gate implements the map $\ket{x}\ket{y}\mapsto\ket{x}\ket{y\oplus x}$, where $\oplus$ denotes addition modulo~$2$.

Other common quantum gates include the three \textit{Pauli matrices},
\begin{equation}
    X:\ket{x}\mapsto\ket{1\oplus x}, \quad Y:\ket{x}\mapsto-i(-1)^x\ket{1\oplus x}, \quad Z:\ket{x}\mapsto(-1)^x\ket{x}. \label{eq:Pauli_gates}
\end{equation}
From the Pauli matrices we arrive at the Pauli group $P_n$ consisting of all $n$-qubit combinations of Pauli matrices:
\begin{equation}
    P_n = \{c p_1\otimes \hdots \otimes p_n \mid p_i\in \{I,X,Y,Z\}, c\in\{\pm 1,\pm i\}\},
\end{equation}
where $I$ is the identity matrix that leaves all states the same. 

We also have parametrized versions of the Pauli matrices:
\begin{align}
    R_X(\theta) & = e^{-i\theta X/2} = \begin{pmatrix}
        \cos\tfrac{\theta}{2} & -i\sin\tfrac{\theta}{2} \\
        -i\sin\tfrac{\theta}{2} & \cos\tfrac{\theta}{2} \\
    \end{pmatrix}, \label{eq:R_X_gate} \\
    R_Y(\theta) & = e^{-i\theta Y/2} = \begin{pmatrix}
        \cos\tfrac{\theta}{2} & -\sin\tfrac{\theta}{2} \\
        \sin\tfrac{\theta}{2} & \cos\tfrac{\theta}{2} \\
    \end{pmatrix}, \label{eq:R_Y_gate} \\
    R_Z(\theta) & := e^{i\theta/2} e^{-i\theta Z/2} = \begin{pmatrix}
        1 & 0 \\
        0 & e^{i\theta} \\
    \end{pmatrix}. \label{eq:R_Z_gate}
\end{align}
In the last line we used that global phases have no measurable effect in quantum states to simplify the expression for the $R_Z$-gate. 

Physical realizations of quantum computers often have access to a small gate set. 
Gates outside this gate set have to be constructed using the available gates. 
However, not all gate sets are equal. 
Consider for instance the \textit{Clifford group} $C_n$, which consists of all gates that normalize the Pauli group: 
\begin{equation}
    C_n = \{V\in U^{2^n\times 2^n} \mid VPV^{\dagger}\in P_n \,\forall P\in P_n\}.
\end{equation}
Quantum circuits consisting solely of gates from the Clifford group can be efficiently simulated~\cite{Gottesman:1998}, whereas it is expected that general quantum circuits do not admit efficient simulations by conventional computers. 

We call a gate set that can approximate any quantum circuit to arbitrary precision $\eps>0$ \textit{universal}.
If two quantum circuits are close with respect to the operator norm (\cref{eq:operator_norm}), the quantum states they prepare are close with respect to the Euclidean norm.
A common universal gate set is the set generated by the CNOT-gate, the Hadamard gate and the $R_Z(\pi/4)$-gate, often called the $T$-gate.
By the Solovay-Kitaev theorem, the overhead of this approximation is $\bigo(\log^c(1/\eps))$ for some small constant~$c$~\cites{Kitaev:1997}[Theorem~1]{DawsonNielsen:2005}.

\subsection{Quantum complexity theory}
Using these definitions, we can generalize the conventional complexity classes to quantum complexity classes. 
From $\nc^k$ we now obtain $\qnc^k$, the class consisting of quantum circuits of depth $\bigo((\log n)^k)$ consisting of single and two-qubit gates. 

The class $\p$ generalizes to $\BQP$ (for bounded-error polynomial time), the class of polynomial-sized quantum circuits that output the correct answer with probability at least $2/3$. 
The bounded error allows for more flexibility in the algorithm design and naturally generalizes the zero-error setting of $\p$. 
In this work, we deviate slightly from the class $\BQP$ and instead consider the class $\QPoly$ consisting of all polynomial-sized circuits that use single- and two-qubit quantum gates.

\subsection{Different computational paradigms}
Conventional computing knows two computational paradigms: digital computing and analog computing. 
Similarly, quantum computing knows different computational paradigms. 
For information purposes, we briefly discuss some of them. 

The first is adiabatic quantum computing, which works by adiabatically evolving the quantum state using a problem-specific Hamiltonian. 
Adiabatic evolution requires a slow enough evolution at low temperatures.
A measurement of the final state after evolution will give the answer to the problem. 
If the evolution is too fast or occurs at too high temperatures, an incorrect answer is found with high probability. 
Adiabatic quantum computing is proven to be equivalent to gate-based quantum computing~\cite{Aharonov:2007}. 

The second closely related paradigm is quantum annealing. 
Quantum annealing is based on simulated annealing, an optimization technique that sometimes considers a worse solution to escape local optima~\cite{Kadowaki:1998}. 
Quantum annealing algorithms evolve the system fast and the system might return the wrong answer.
As a result, the conditions imposed on the quantum hardware are less stringent compared to adiabatic quantum computers. 
To compensate the possible wrong answers returned by the quantum annealing process, the system is evolved, measured and reinitialized multiple times.
The resulting probability distribution can then be used to tackle optimization problems where often a \textit{good} solution suffices, while the \textit{best} solution might be hard to find or might not even exist. 

Measurement-based quantum computing, a universal form of quantum computing, is the third computational paradigm~\cite{GottesmanChuang:1999,RaussendorfBriegel:2001,Jozsa:2006,ClarkJozsaLinden:2007,Briegel:2009}.
In this paradigm, a quantum state is prepared and operations are performed by measuring qubits in different bases, depending on previous measurements and intermediate computations.
This method is sometimes also called one-way quantum computing because of the measurements.

Fourth, we have photonic quantum computing, which uses photons as quantum states. 
Photonic quantum computers use beam splitters, mirrors, and polarizers to manipulate the photons~\cite{Adami:1999}.
Photonic systems using two-level systems exist, as well as those using multilevel systems. 
The latter case uses the number of photons in a state as the computational unit. 
Both heuristic~\cite{AaronsonArkhipov:2013} and universal~\cite{KLM:2001} versions of photonic quantum computing have been proposed. 

%% file: introduction/intro_towards_new_devices.tex
\section{Towards new computational devices}
The actual use of these quantum effects requires physical implementations of quantum computers. 
Conventional computers operate under the laws of Newtonian mechanics and after years of development, manufacturing these devices is relatively well understood. 
Quantum computers on the other hand operate under the laws of quantum mechanics and the quantum states in a quantum computers are usually fragile and easily disturbed. 
As a result, manufacturing quantum computers raises new engineering challenges~\cite{Almudever:2017}. 

Even today, multiple hardware technologies are used to build quantum computers, some examples include trapped ions~\cite{CiracZoller:1995}, neutral atoms~\cite{Dumke:2002}, spin systems~\cite{Imamoglu:1999} and superconducting circuits~\cite{Kaminsky:2004,Houck:2009,Barends:2013,Kjaergaard:2020}.
Moreover, among the various quantum hardware technologies currently under development are postulated options such as skyrmion qubits~\cite{Psaroudaki:2021} and topological qubits based on anyons~\cite{Nayak:2008}.
One might even wonder if at some point one single hardware technology will become dominant, or that instead multiple hardware technologies will remain, each with their own specific applications in mind. 

\citeauthor{DiVincenzo:2000} laid out different requirements a quantum system should adhere to, independent of the underlying technology used~\cite{DiVincenzo:2000}. 
He describes requirements needed to perform computations such as the ability to initialize qubits in a known state and later read them out, and the access to a universal gate set. 

One of the main challenges in manufacturing quantum computers is isolating them from the environment to minimize the effect of noise and decoherence. 
Due to this difficulty, current quantum hardware still suffers greatly from the effects of noise. 
\citeauthor{Preskill:2018} saw this trend and called the current devices part of \textit{Noisy Intermediate-Scale Quantum} (NISQ) technology~\cite{Preskill:2018}. 

The NISQ devices form a stepping-stone towards general Fault-Tolerant quantum computers.  
In these fault-tolerant devices, we assume that qubits remain coherent throughout the whole execution of the algorithm and that algorithms produce correct answers. 
One step towards fault-tolerant devices is reducing error rates through the use of error-correcting codes~\cite{Shor:1995}.

Error-correcting codes impose a code on a group of qubits, such that together, the group acts as a single qubit with lower error rate. 
The individual qubits are often called physical qubits and the qubits used in quantum algorithms logical qubits. 
Simply comparing the number of qubits a quantum device has only paints a partial picture, as the error rates of these qubits and how they can interact also affect the capabilities of quantum devices in practice. 

\begin{figure}
    \centering
    \includegraphics[width=\linewidth, trim={2cm 2.3cm 1.97cm 2.1cm}, clip]{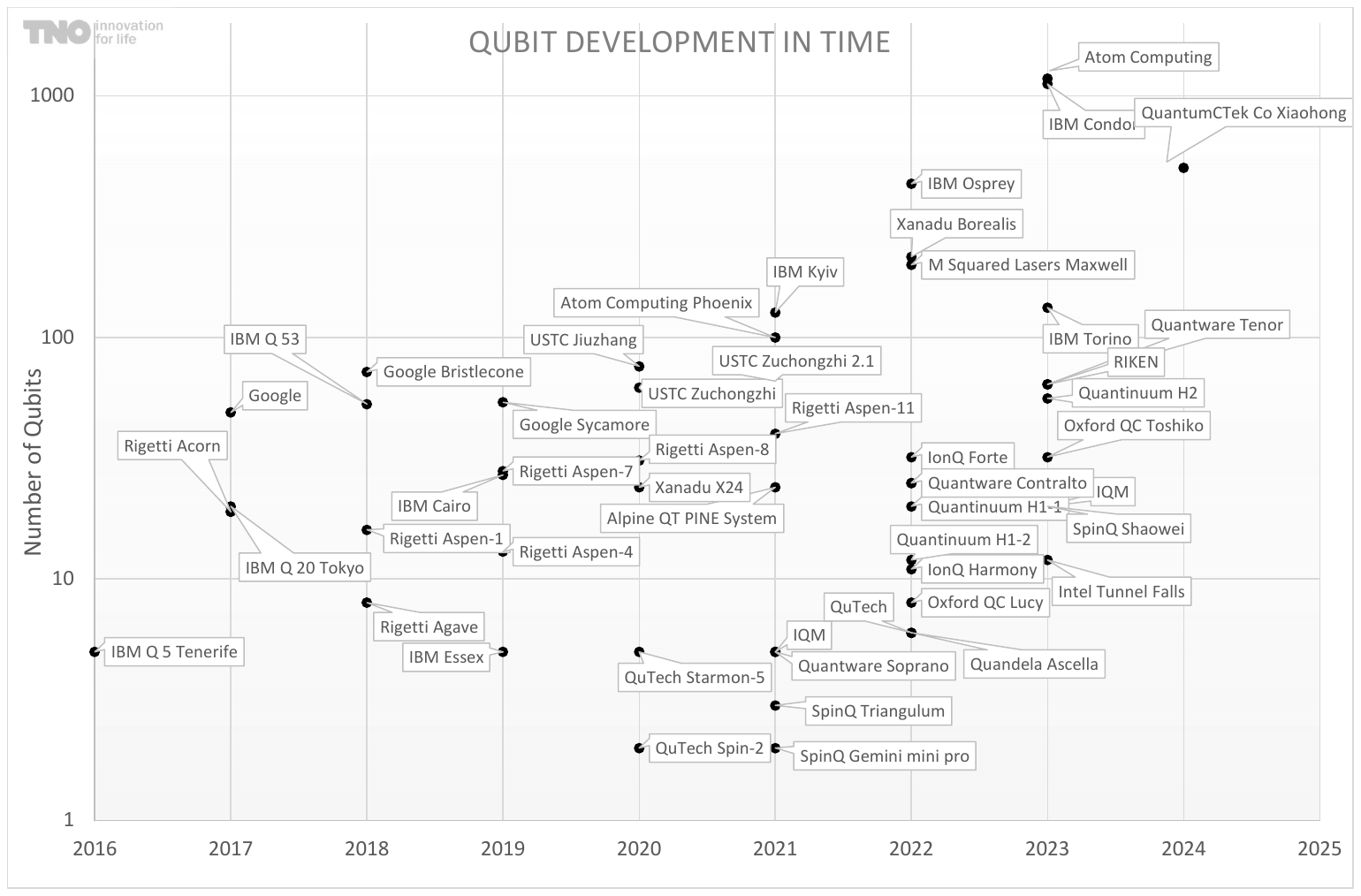}
    \caption{Overview of different quantum computers with the available number of qubits over the years.}
    \label{fig:qubit_count}
\end{figure}

The number of physical qubits in quantum hardware has been steadily rising over the last few years, as \cref{fig:qubit_count} shows. 
The error rates of the (physical) qubits in these devices remain high.
Only recently, the first results showed improved error rates by using error-correcting codes~\cite{Google:2023,DaSilva:2024,GoogleWillow:2024}.
Earlier, the error rates of physical qubits were too high for error-correcting codes to be able to work.

We can also concatenate error-correcting codes to achieve an exponential suppression of the error-rates~\cite{DevittMunroNemoto:2013}. 
Different groups of qubits are each encoded with an error-correcting code, after which the groups themselves are also encoded using a (possibly different) error-correcting code.

%% file: introduction/intro_developments_quantum_algorithms.tex
\section{Developments in quantum algorithms}
Shortly after \citeauthor{Feynman:1982} called for the development of quantum computers, the first theoretical results showed the potential power of these future devices.
\citeauthor{Deutsch:1985} and later \citeauthor{DeutschJozsa:1992} presented an oracular promise problem with provable exponential speedup~\cite{Deutsch:1985,DeutschJozsa:1992}. 
They consider the problem to decide if a Boolean function $f:\bset^n\to\bset$ is constant or balanced, provided one of the two is the case.
Their quantum algorithm answers the problem with certainty using a single query to $f$, whereas any conventional algorithm has to make an exponential number of queries to $f$ in the worst case. 

Later, \citeauthor{BernsteinVazirani:1997} presented a quantum algorithm to learn the hidden string $s\in\bset^n$ in the inner-product function $f(x)=\ip{x,s}\bmod 2$~\cite{BernsteinVazirani:1997}. 
Their algorithm again provides an exponential advantage over conventional algorithms, requiring only a single query to $f$, instead of $n$. 

\citeauthor{Grover:1996} developed an algorithm for a practical problem: 
searching an unstructured database for a specific item~\cite{Grover:1996}. 
In the worst case, you have to check every item in your database before you find the correct one. 
\citeauthor{Grover:1996}'s algorithm, achieves the same task quadratically faster. 
The difference between the quantum and conventional algorithm becomes clear in the number of queries to the database. 
Conventional algorithms can only check a single item at a time. 
Quantum algorithms can query the database ``in superposition'', which roughly means checking the entire database at once. 
The successive queries are needed for constructive interference to give the correct answer with high probability.

Each of these algorithms uses an oracle, a black-box function that queries some function and returns the results in superposition (see also \cref{sec:intro:query_based_quantum,eq:quantum_query_def}).
Oracles allow to reason about the effectiveness of algorithms. 
In practice, we have to implement the oracle. 
This implementation step might diminish, or even nullify, the speedups offered by a quantum algorithm. 

\citeauthor{Shor:1997} provided another quantum algorithm for a practical problem:
Breaking an often-used encryption protocol for which all known conventional algorithms take at least sub-exponential time.
\citeauthor{Shor:1997} presented a quantum algorithm to efficiently determine the period of a modular exponentiation function~\cite{Shor:1994,Shor:1997}. 
This computation relates to finding the prime factors of integers, which is the problem underlying the often-used RSA-encryption protocol~\cite{RivestShamirAdleman:1978}. 

Grover's algorithm offers a provable quadratic advantage over conventional algorithms.
In contrast, Shor's algorithm provides an exponential advantage over all known conventional algorithms. 
As a result, the impact of Shor's algorithm is larger: 
It provides a polynomial-time algorithm for a problem for which no efficient conventional algorithm is known. 

Note that it is still open whether the discrete logarithm problem and prime factorization, the problems solved by Shor's algorithm, are indeed hard. 
However, many believe this to be true. 

Recently, more attention has been given to variational algorithms, where a function is optimized to solve a specific problem~\cite{FarhiGoldstoneGutmann:2014,Peruzzo:2014,Hadfield:2019,Cerezo:2021}.
These variational algorithms show potential for near-term applications as they can be implemented relatively easily on NISQ hardware. 
Variational quantum algorithms use a circuit with parametrized gates. 
These gate parameters are optimized with respect to some objective function. 

Variational algorithms have the potential advantage that they can `learn' the noise present in NISQ devices. 
Specifically, unwanted bias in the hardware can be learned as systematic noise and thus be mitigated. 
Noise resulting from decoherence will be biased around the expected outcome in a noiseless situation.
Averaging repeated measurements therefore gives a correct answer with good probability. 

Research into NISQ algorithms also triggered a new trend: 
optimizing costly quantum resources. 
The simplest approach in this research is optimizing the implementation of an algorithm with respect to the number of gates or with respect to the number of $T$-gates used. 

These optimization objectives make sense, as applying quantum gates will introduce more noise than doing nothing.
Hence, if we can implement the same operation using fewer gates, noise rates will typically decrease~\cite{Kottman:2022,Perez:2023}.
Similar arguments hold for optimizing the number of $T$-gates~\cite{Reiher:2017,GidneyEkera:2021}.

The second part of this work extends this line of research by optimizing with respect to the circuit depth. 
We present state preparation protocols that have higher success probabilities than protocols using fewer gates. 
See \cref{chp:LAQCC:error_analysis} for more details.

%% file: introduction/intro_quantum_advantage.tex
\section{Achieving quantum advantage}
The advent of NISQ devices allowed for the first time to actually implement quantum algorithms on a large scale. 
The next step was to show that quantum computers can solve specific tasks faster than conventional computers in practice. 
This event is known as quantum advantage. 

The first claim for quantum advantage was made in 2019:  
a quantum computer performed a sampling task in $200$ seconds, whereas conventional computers would take $10{,}000$ years to complete the task~\cite{Arute:2019}. 
Shortly after, it was shown that conventional supercomputers could perform the task in only 2.5 days~\cite{Pednault:2019,PanChenZhang:2022}. 
Still, the result was groundbreaking and opened the way to other results showing a quantum advantage~\cite{Zhong:2020,Madsen:2022,GoogleWillow:2024}.

However, these quantum advantage experiments do not provide a \textit{provable} quantum advantage. 
The results for conventional computing are often based on extrapolation from the running times for smaller problem instances. 
To prove a quantum advantage we need to shift our focus to either low-depth circuits or to query complexity. 
The first part of this work extends this line of research in both directions.

\subsection{Constant-depth quantum advantage}\label{sec:intro:constant_depth_quantum}
Constant-depth circuits, both quantum and conventional, recently enjoyed renewed interest in the context of provable separations. 
Moreover, constant-depth circuits are likely easier to implement in practice. 
Furthermore, constant-depth circuits are one of the few settings currently amenable to provable lower bounds.
Our current techniques fail for unbounded circuit depth. 

In light of provable separations, a common extension of the classes $\nc^0$ and $\ac^0$ is the unbounded-fan-in parity gate, which computes the parity of all input bits, denoted by $\nc^0[\oplus]$ and $\ac^0[\oplus]$, respectively.  
Note that this specific example is a true extension, as parity cannot be computed by $\ac^0$ and $\nc^0$ is a proper subset of~$\ac^0$~\cite{AroraBarak:2009}.
Furthermore, the classes $\ac^0$ and~$\nc^0[\oplus]$ are incomparable since $\nc^0[\oplus]$ cannot compute the $n$-bit AND function;
indeed, $\nc^{0}[\oplus]$ circuits can compute only constant-degree polynomials over~$\F_2$ (see also \cref{sec:decoding:classical:outline}), whereas AND has degree~$n$.

\citeauthor{Moore:1999} and \citeauthor{MooreNilsson:2001} were the first to consider the quantum versions of these classes~\cite{Moore:1999,MooreNilsson:2001}. 
In contrast with their conventional analogues, the classes $\qnc^0[\oplus]$ and $\qac^0$ are known to be equivalent~\cite{GreenHomerMoorePollett:2002,HoyerSpalek:2005,Moore:1999}.
Furthermore, the works of \citeauthor{Moore:1999} and \citeauthor{GreenHomerMoorePollett:2002} showed that 
\begin{equation*}
    \qac^k[\oplus]=\qac^k[q],
\end{equation*}
where $\qac^k[q]$ extends $\qac^k$ with additional modulo-$q$ gates~\cite{Moore:1999,GreenHomerMoorePollett:2002}. 
For an integer $q\ge 2$, a modulo-$q$ gate evaluates to $1$ if the sum of its inputs equals $0\bmod q$, and evaluates to~$0$ otherwise.
In contrast, the classes $\ac^0[p]$ and $\ac^0[q]$ are incomparable if~$p$ and~$q$ are powers of distinct primes~\cite{Razborov:1987,Smolensky:1987}.

\citeauthor{HoyerSpalek:2005} proved a quantum advantage, by showing that $\qnc^0[\oplus]$ circuits can approximate with polynomially small error functions such as OR, AND, and Majority~\cite{HoyerSpalek:2005}. 
Later, \citeauthor{TakahashiTani:2013} extended these results to exact quantum circuits. 
Key in their work was the use of a quantum fanout gate
\begin{equation}
    \rm{Fanout}:\ket{x}\ket{y_1}\hdots\ket{y_n}\mapsto\ket{x}\ket{y_1\oplus x}\hdots\ket{y_n\oplus x}.
    \label{eq:quantum:fan_out}
\end{equation}
This gate is typically not supported by the native gate set of a quantum computers. 
However, using intermediate conventional processing, we can still implement this gate efficiently. 

For a long time, research on provable quantum advantages lay still. 
The breakthrough work by \citeauthor{BravyiGossetKoenig:2018} restarted this line of research into provable quantum advantages leading to multiple new results and ideas:
\begin{itemize}
    \item The 2D-Hidden Linear Function problem can be solved exactly by a $\qnc^0$-circuit, while any $\ac^0$-circuit succeeds with exponentially small probability under a certain input distribution~\cite{BeneWattsKothariSchaefferTal:2019}; this strengthened the main result of~\cite{BravyiGossetKoenig:2018} showing that this problem separates $\qnc^0$ from $\nc^0$ in the worst case. 
    \item The Relaxed Parity Halving problem can be solved exactly by a $\qnc^0$-circuit while any $\ac^0$-circuit succeeds with probability at most $\frac{1}{2} + \exp(-n^\eps)$ for some $\eps>0$ under the uniform input distribution~\cite{BeneWattsKothariSchaefferTal:2019}.
    \item The Parallel Parity Bending problem can be solved with probability $1 - \nolinebreak o(1)$ by a $\qnc^0/\mathsf{qpoly}$-circuit, while any $\ac^0[\oplus]/\mathsf{rpoly}$-circuit succeeds with probability at most $\bigo(n^{-\eps})$~\cite{BeneWattsKothariSchaefferTal:2019}.
    \item The problem of simulating correlations obtained from measuring graph states separates $\qnc^0$ and $\nc^0$, even in the average case~\cite{LeGall:2019}.
    \item The 1D Magic Square problem separates noisy $\qnc^0$-circuits from $\nc^0$-circuits, provided that the noise levels are below some threshold value independent of the input size~\cite{BravyiGKT:2020}.
\end{itemize}
Similar separations based on other relational and sampling-based problems were proven in~\cite{CoudronStarkVidick:2021, GrierSchaeffer:2020, BeneWattsParham:2023}.
A common feature of all these problems is that they were specifically designed to prove a separation between shallow quantum and conventional circuits.

\cref{chp:decoding:classical,chp:decoding:quantum} extend this line of research to a problem that occurs naturally in computer science: 
decoding heavily corrupted error-correcting codes using limited resources. 
This problem is well studied in the context of conventional complexity theory, where shallow circuits endowed with parity gates are often considered.

\subsection{Query-based quantum advantage}\label{sec:intro:query_based_quantum}
Another stylized model that admits provable separations is the query model. 
Here, black-box access is given to some function, and algorithms are compared in terms of how often the function is queried. 
The first exponential separations were shown already a few decades ago~\cite{Deutsch:1985,DeutschJozsa:1992,BernsteinVazirani:1997,Simon:1997}.

Let $f:A\to \bset$, for some set $A$. 
Conventional computers can query $f$ by inputting $x$ and obtaining $f(x)$ from the oracle. 
Quantum queries are coherent, in the sense that the input can be any superposition over elements in $A$ and the output is stored in an auxiliary bit. 
For any $x\in A$ and $b\in\bset$, a quantum query to $f$ is given by the map
\begin{equation}\label{eq:quantum_query_def}
    \ket{x}\ket{b} \mapsto \ket{x}\ket{b\oplus f(x)}.
\end{equation}
We can generalize this oracle to sets $B\supset A$ by setting $f(x)=0$ for $x\in B\setminus A$. 
We refer to this oracle as a bit-flip oracle, as it flips the state of the auxiliary bit. 

If we initialize the auxiliary qubit in the $\ket{1}$-state, we can transform the oracle query into a phase query by conjugating the oracle call with Hadamard gates on the auxiliary qubit. 
A phase oracle thus implements the map
\begin{equation*}
    \ket{x}\ket{b} \mapsto (-1)^{f(x)}\ket{x}\ket{b}.
\end{equation*}
Note that a bit-flip oracle generalizes a phase oracle, as in the latter, we cannot distinguish between $f$ and the function $1-f$, as they differ only by a global phase. 

Results based on query complexity can be translated into the circuit model. 
This translation requires a reversible implementation of the oracle as a quantum circuit. 
For some problems, this implementation of the oracle might prove difficult and thereby reduce the advantage over a conventional approach. 
The query model remains an interesting model for proving a quantum advantage.

\citeauthor{Aaronson:2010} showed the largest query-based separation known between quantum and conventional algorithms. 
He introduced a problem called Forrelation that, given two functions, asks if the first function correlates with the Fourier transform of the second~\cite{Aaronson:2010}.
A quantum algorithm only needs a single query to solve the Forrelation problem.
\citeauthor{AaronsonAmbainis:2015} showed that on input length $N$, a conventional query algorithm requires $\widetilde{\Omega}\big(\sqrt{N}\big)$ queries~\cite{AaronsonAmbainis:2015}.
They also showed that this separation is optimal for a single query and conjectured it to hold for any constant number of queries. 
\citeauthor{BansalSinha:2020} later proved this conjecture in the standard query model~\cite{BansalSinha:2020}.
\citeauthor{Sherstov:2023} independently obtained a similar result for randomized queries~\cite{Sherstov:2023}.

\citeauthor{Montanaro:2012} obtained a smaller quantum advantage in the query model, but did so for a more practical problem~\cite{Montanaro:2012}.
\citeauthor{Montanaro:2012} considered learning arbitrary polynomials of degree~$d$, a task that requires $\Omega(n^{d-1})$ quantum queries, while conventional algorithms require $\Omega(n^d)$ queries.
His algorithm generalizes the Bernstein-Vazirani algorithm, which can be interpreted as learning a polynomial of degree~$1$. 
\citeauthor{Montanaro:2012} also showed the optimality of this query lower bound in the noiseless case using Fano's inequality~\cite{FanoHawkins:1961}. 
However, the algorithm by \citeauthor{Montanaro:2012} fails if the queries can be corrupted. 

\cref{chp:decoding:quadractic} extends the work by \citeauthor{Montanaro:2012} to also work in case of corrupted query calls. 
In \cref{chp:decoding:quadractic}, we revisit a result in higher-order Fourier analysis and use a quantum subroutine to improve the query complexity. 

%% file: introduction/intro_Fourier_analysis.tex
\section{Fourier analysis: a tool for analyzing algorithms}
Many quantum algorithms, most notably that of Shor, rely on an efficient implementation of the Fourier transform on quantum computers. 
Conventional computers can implement a Fourier transform on $n$ elements in time polynomial in $n$. 
\citeauthor{Coppersmith:2002} showed how quantum computers can implement a quantum Fourier transform exponentially faster in time $\poly\log (n)$~\cite{Coppersmith:2002}. 

In the remainder of this work, we only work with finite Abelian groups. 
Two prime examples we consider are $\F_p^n$, the finite $n$-dimensional vector space over the field with $p$ elements for some prime $p$, and $\Z_n$, the cyclic group over the first~$n$ nonnegative integers. 
In this section, we present definitions only with respect to (one of) these two groups. 
They do naturally translate to other groups.
\cref{part:decoding} of this work considers applications of Fourier analysis and generalization higher-order Fourier analysis.

\subsection{The Fourier transform}
Fourier transforms prove vital to find patterns in data. 
A Fourier transform decomposes a function as a combination of character functions~\cite{Fourier:1888}.
As an example, let $f(t)$ represent some musical composition. 
The Fourier transform helps us to decompose the musical composition in individual notes (the characters) and their relative magnitude over time. 
In a sense, the Fourier transform indicates which notes are most prominent in the musical composition over time. 

Now consider $\F_p^n$ for some prime $p$ and let $\omega_p=e^{2\pi i/p}$.
We then define its character functions as $\{\chi_y | \chi_y(x) = \omega_p^{\ip{x,y}}\}$, where $\ip{x,y}=x_1y_1+\hdots x_ny_n$ denotes the inner product of $x$ and $y$. 
Character functions carry the essential information of the group elements and we often refer to them as \textit{phase functions}. 
We can then define the Fourier transform in terms of the character functions:
\begin{definition}[Fourier transform]
    Let $f:\F_p^n\to\C$ be some function, then its Fourier transform $\widehat{f}:\F_p^n\to\C$ evaluated at $y\in\F_p^n$ is given by
    \begin{equation*}
        \widehat{f}(y) = \Exp_{x\in\F_p^n}f(x)\omega_k^{-\ip{x,y}} = \frac{1}{p^n}\sum_{x\in\F_p^n} f(x)\omega_p^{-\ip{x,y}}.
    \end{equation*}
\end{definition}
The function $f$ is often said to be in physical space and the Fourier transformed function $\widehat{f}$ is said to be in frequency space. 
Both are just naming conventions. 

Both the Fourier-transformed function $\widehat{f}$ and the character functions themselves have interesting properties.
For instance, every function $f$ can be written uniquely as a sum of its Fourier coefficients, called the Fourier inversion formula:
\begin{equation}\label{eq:Fourier_inversion_formula}
    f(x) = \sum_{y\in \F_p^n} \widehat{f}(y)\omega_p^{\ip{x,y}}.
\end{equation}
This argument follows as the character functions form an orthonormal basis for~$\F_p^n$.
We have a total of $p^n$ characters, one for every $y\in\F_p^n$, and they are all orthogonal:
\begin{equation*}
    \ip{\chi_y,\chi_z} = \Exp_{x\in\F_p^n} \omega_p^{\ip{(y-z),x}} = \begin{cases}
        1 & \text{if } y=z, \\
        0 & \text{otherwise.}
    \end{cases}
\end{equation*}
The last equation uses that the expected value over the whole group of any non-identity character function vanishes, as the next lemma shows.
\begin{lemma}\label[lemma]{lem:roots_of_unity}
    Let $\chi_y$ be a non-identity character of $\F_p^n$. 
    Then 
    \begin{equation}
        \Exp_{x\in \F_p^n} \chi_y(x) = 0.
    \end{equation}
\end{lemma}
\begin{proof}
    As $\chi_y$ is not the identity element, there exists a $z\in \F_p^n$ such that $\chi_y(z)\neq 1$. 
    Then we have
    \begin{align*}
        \Exp_{x\in \F_p^n} \chi_y(x) & = \Exp_{x\in \F_p^n} \chi_y(x+z) \\
        & = \chi_y(z)\Exp_{x\in \F_p^n} \chi_y(x).
    \end{align*}
    In the first equality, we translated $x$ to $x+z$, in the second equality we used the linearity of $\chi_y$ in its phase.
\end{proof}
This proof holds in greater generality for any non-identity character of any finite Abelian group. 

An important property of Fourier transforms is that they preserve inner products, as shown by Parseval's identity from 1806~\cite{Parseval:1806}:
\begin{proposition}\label{prop:Parseval}
    For any two function $f,g:\F_p^n\to\C$ it holds that
    \begin{equation*}
        \Exp_{x\in\F_p^n} f(x)\overline{g(x)} = \sum_{y\in\F_p^n}\widehat{f}(y)\overline{\widehat{g}(y)}.
    \end{equation*}
\end{proposition}
The proof of Parseval's identity follows by applying the Fourier inversion formula twice and then use that characters are orthogonal. 
As Fourier transforms preserve inner products, they are unitary and thus valid quantum operations.

\subsection{The quantum Fourier transform}
Fourier analysis might seem like a complicated topic, yet it finds applications throughout quantum computing:
The Bernstein-Vazirani algorithm~\cite{BernsteinVazirani:1997}, Simon's problem~\cite{Simon:1997}, Shor's algorithm~\cite{Shor:1997}, the HHL algorithm~\cite{HarrowHassidimLloyd:2009} and many more use a quantum Fourier transformation over some group. 
The efficient implementation of a quantum Fourier transform by \citeauthor{Coppersmith:2002} often gives the exponential separation between quantum and conventional circuits. 

The algorithm by \citeauthor{Shor:1997} works over the cyclic group $\Z_{N}$, with $N=2^n$, instead of over $\F_2^n$. 
Luckily, a bijection between the two groups exist: 
\begin{equation*}
    x=(x_0,\hdots,x_{n-1})\in\F_2^n \leftrightarrow y=\sum_{i=0}^{n-1} 2^i x_i \in \Z_{N}.
\end{equation*}
A quantum Fourier transform over $\F_2^n$ translates to a single layer of $n$ Hadamard gates applied to every qubit. 
The quantum Fourier transform over $\Z_{2^n}$ looks slightly different and is given by
\begin{equation}\label{eq:gate_QFT}
    QFT: \ket{x}\mapsto \frac{1}{\sqrt{2^n}}\sum_{y=0}^{2^n-1}e^{2\pi i \frac{xy}{N}}\ket{y}.
\end{equation}
As character functions we now have $\omega_{N,x}(y)=e^{2\pi i \frac{xy}{N}}$. 

Implementing a conventional Fourier transform on $N$ elements requires $\bigo(nN)=\bigo(n2^n)$ operations. 
By \citeauthor{Coppersmith:2002}, we can implement an equivalent quantum Fourier transform using only $\bigo(n^2)$ quantum gates. 

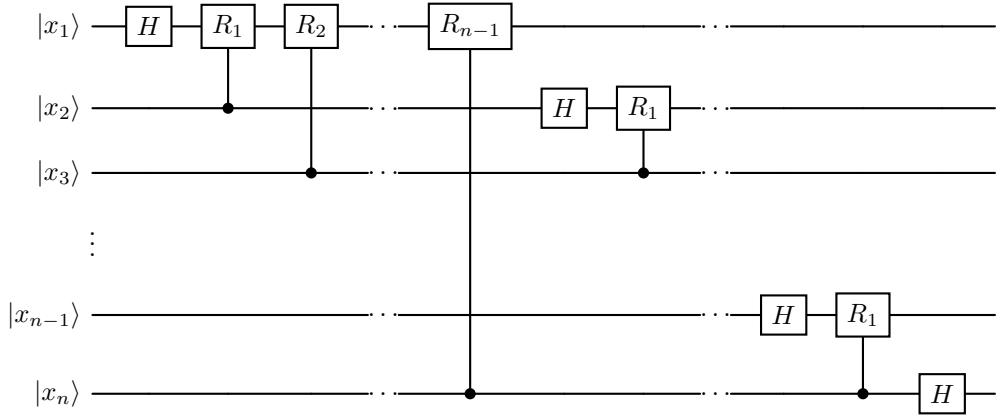
\begin{figure}
    \centering
    \begin{quantikz}[column sep=0.4cm]
        \lstick{\ket{x_1}} & \gate{H} & \gate{R_1} & \gate{R_2} & \hdots & \gate{R_{n-1}} & & & \hdots & & & & \\
        \lstick{\ket{x_2}} & & \ctrl{-1} & & \hdots & & \gate{H} & \gate{R_1} & \hdots & & & & \\
        \lstick{\ket{x_3}} & & & \ctrl{-2} & \hdots & & & \ctrl{-1} & \hdots & & & & \\
        \vdots \\
        \lstick{\ket{x_{n-1}}} & & & & \hdots & & & & \hdots & \gate{H} & \gate{R_1} & & \\
        \lstick{\ket{x_n}} & & & & \hdots & \ctrl{-5} & & & \hdots & & \ctrl{-1} & \gate{H} & 
    \end{quantikz}
    \caption{A quantum circuit that implements the quantum Fourier transform. 
    The gates $R_k$ correspond to $R_Z(\pi/2^k)$-gate.}
    \label{fig:q_circuit:QFT}
\end{figure}

\cref{fig:q_circuit:QFT} shows an implementation of a quantum Fourier transform on $n$ qubits. 
The $R_k$-gates denote an $R_Z(\pi/2^k)$-gate.
The quantum Fourier transform typically also swaps all qubits at the end of the circuit, we omitted that operation, as we can easily incorporate these operations in the rest of a quantum algorithm. 
The controlled-$R_Z$-gates have a simple decomposition into two CNOT-gates and two $R_Z$-gates, giving us a simple implementation of the quantum Fourier transform. 
As the circuit works for any computational basis state, it also works for arbitrary quantum state by linearity. 

We now define $(0.x_j\hdots x_n)$ as the binary fraction $\sum_{i=j}^n x_i/2^{i-j+1}$, to rewrite the state $\ket{x}$ after the quantum Fourier transform as
\begin{equation*}
    QFT\ket{x} = \frac{1}{\sqrt{2^n}}(\ket{0}+e^{2\pi i (0.x_1\hdots x_n)}\ket{1}) (\ket{0}+e^{2\pi i (0.x_2\hdots x_n)}\ket{1}) \hdots (\ket{0}+e^{2\pi i (0.x_n)}\ket{1}).
\end{equation*}
This representation is often used for quantum arithmetic algorithms that work with the Fourier-transformed state~\cite{RuizGarcia:2017,NeumannWezeman:2022}.

\subsection{Higher-order Fourier analysis}
Fourier analysis is an important technique for counting linear patterns in subsets of integers and other Abelian groups. 
For instance, in the field of additive combinatorics, \citeauthor{Roth:1953} used Fourier analysis to show that any sufficiently large subset of positive integers contains a triple $(a,b,c)$ such that the difference between two successive numbers is the same~\cite{Roth:1953}.
We call such triples three-term arithmetic progressions. 

Counting longer arithmetic progressions amounts to counting quadratic or higher-order patterns in subsets of integers. 
However, Fourier analysis appears insufficient to find correlations with polynomials of degree larger than one. 

An important realization is that taking the derivative of a function lowers its degree. 
For a function $f$ over a finite Abelian group, we define its multiplicative derivative in the direction $h$ as $\Delta_h f(x)=f(x+h)\overline{f(x)}$.
For a polynomial $P\in \nolinebreak \F_p[x_1,\dots,x_n]$, we often use the additive derivative instead, given by $\Delta_h P(x) = P(x+h) - P(x)$.

For a long time, a generalization of Fourier analysis was unknown, until the seminal work by~\citeauthor{Gowers:2001}~\cite{Gowers:2001}. 
He introduced the notion of higher-order Fourier analysis, which examines correlations with higher-order functions. 
Gowers also introduced his uniformity norms based on the multiplicative derivatives of a function, which we will use extensively in \cref{chp:decoding:quadractic} of this work. 
\begin{definition}\label{def:Gowers_norm}
    Let $f:\F_p^n\to\C$ be any function. 
    The Gowers $U^d$-norm is defined by 
    \begin{equation}\label{eq:Gowers_norm}
        \lVert f\rVert_{U^d} = \left(\Exp_{x,h_1,\hdots,h_d\in\F_p^n} \Delta_{h_d}\hdots\Delta_{h_1}f(x)\right)^{1/2^d}.
    \end{equation}
\end{definition}
For $d=1$, the Gowers $U^d$-norm is actually a seminorm, as 
\begin{equation*}
    \lVert f\rVert_{U^1} = \lvert \Exp_{x\in\F_p^n} f(x) \rvert
\end{equation*}
can vanish for nonzero functions, such as $f(x) = \omega_p^{x}$.
For $d\ge 2$, the Gowers $U^d$-norm is actually a norm.
For $d=2$ specifically, we can use the Fourier inversion formula to see that 
\begin{align}
    \lVert f\rVert_{U^2} & = \left(\Exp_{x,h_1,h_2\in\F_p^n} f(x)\overline{f(x+h_1)f(x+h_2)}f(x+h_1+h_2)\right)^{1/4} \nonumber \\
    & = \left(\sum\nolimits_{a\in \F_p^n}\big|\widehat{f}(a)\big|^4 \right)^{1/4} = \lVert \hat{f}\rVert_{L^4}. \label{eq:Gowers_U2_L4}
\end{align}
Hence, the Gowers $U^2$-norm of a function can be completely described in terms of the $L^4$ norm of the Fourier transform of that function. 

For larger $d$, we have no such explicit relation.
However, we can reformulate the Gowers $U^d$-norms inductively: 
\begin{equation}\label{eq:Uk_inductive}
    \|f\|_{U^{d+1}}^{2^{d+1}} = \Exp_{h\in \F_p^n}\|\Delta_{h}f\|_{U^d}^{2^d}.
\end{equation}
Repeatedly applying the Cauchy-Schwarz inequality on this inductive expressions gives the \emph{nesting property}~\cite[pp.~420]{TaoVu:2006}
\begin{equation}\label{eq:Uk_nesting}
    \|f\|_{U^1}\leq \|f\|_{U^2} \leq \cdots,
\end{equation}
a useful tool in many algorithms. 

In general, Gowers norms quantify structure in complex functions by identifying correlations of these complex functions with polynomials.
If a function correlates with a degree-$d$ function, then taking $d+1$ derivatives of that function results in an approximately zero function, which in turn correlates with an approximately constant function.
As a result, the Gowers $U^{d+1}$-norm is large for such a function. 

Interestingly, a series of works by \citeauthor{GreenTaoZiegler:2012} showed that the converse also holds~\cite{GreenTao:2008,TaoZiegler:2011,GreenTaoZiegler:2012}.
These results are known as the inverse theorem for the Gowers $U^{d}$-norm. 
Gowers norms now play a fundamental role within higher-order Fourier analysis and find many applications in theoretical computer science. 

In \cref{chp:decoding:quadractic}, we will restrict ourselves to specific functions called \textit{polynomial phase functions}:
$f:\F_p^n\to \D$, with $\D$ being the complex unit disc, such that $f=\omega_p^{P(x)}$ for $P$ some $n$-variate polynomial over $\F_p$ such that every variable has degree at most $p-1$. 
The degree of a polynomial phase function is defined as the degree of the polynomial $P$. 
Given a quantum state $\ket{x}$, we define a phase-query to $f$ as the map $\ket{x}\mapsto f(x)\ket{x}$. 

Since the $(d+1)$-fold multiplicative derivatives of a degree-$d$ polynomial phase~$f$ function equal~1, it follows from the nesting property that the correlation between~$f$ and any function $g:\F_p^n \to\D$ is bounded from above as
\begin{equation*}
    \big|\Exp_{x\in \F_p^n}g(x)\overline{f(x)}| \leq \|g\|_{U^{d+1}}.
\end{equation*}

We refer the interested reader to~\cite{HatamiHatamiLovett:2019} for an elaborate overview of higher-order Fourier analysis and its applications.

\subsection{Higher-order Fourier analysis applied to quantum computing}
The Fourier transform is widely used and also underlies multiple famous quantum algorithms. 
Higher-order Fourier analysis has found many applications in theoretical computer science. 
However, the subfield of higher-order Fourier analysis used in quantum algorithms is relatively unexplored. 

\citeauthor{Roetteler:2009} presented a quantum algorithm to learn a hidden shift in functions with a large Gowers $U^3$-norm~\cite{Roetteler:2009}. 
Later, \citeauthor{Montanaro:2012} presented a quantum algorithm to learn multilinear polynomials with improved query complexity over conventional algorithms~\cite{Montanaro:2012}.
In the context of non-local games, higher-order Fourier analysis helps to prove a relation between the bias of a quantum strategy and the bias of a conventional strategy~\cite{BanninkBrietBuhrmanLabibLee:2019}. 

Only very recently, the application of higher-order Fourier analysis in quantum computing algorithms found new applications with the work of \citeauthor{ArunachalamDutt:2024} in the context of stabilizer testing~\cite{ArunachalamDutt:2024,BaoDordrechtHelsen:2024,ArunachalamBravyiDutt:2024,ChenGongYeZhang:2024}.
These results consider the task of determining if a given unknown quantum state is close to or far from a stabilizer state. 

\cref{chp:decoding:quadractic} continues this line of research by providing a quantum analog of the quadratic Goldreich-Levin algorithm. 
With this algorithm, one can find a Reed-Muller codeword of degree at most~$2$ that correlates with the input, provided the input has large Gowers $U^3$-norm. 

A key subroutine in that chapter is the Fourier sampling routine, which generalizes sampling the Fourier spectrum of a function using standard Fourier analytic techniques. 
This Fourier sampling routine generalizes the celebrated Bernstein-Vazirani algorithm~\cite{BernsteinVazirani:1997}, which returns a string~$a\in\F_p^n$ with probability $|\widehat{f}(a)|^2$ using a single query to a polynomial phase function $f:\F_p^n\to\D$.
In particular, if~$f = \omega_{p}^{\ip{a,x} + b}$ is a linear phase function given by some vector $a\in \F_p^n$ and scalar~$b\in \F_p$, then the algorithm returns~$a$ with probability~1.

The Fourier sampling routine samples the Fourier spectrum of the multiplicative derivates of~$f$, instead of the Fourier spectrum of~$f$ itself. 
\begin{lemma}[Fourier sampling]\label[lemma]{lem:Fourier_sampling}
There is a $p$-query quantum algorithm that, given query access to a polynomial phase function $f:\F_p^n\to \D$ and input $h\in \F_p^n$, returns $a\in \F_p^n$ with probability exactly $|\widehat{\Delta_{h}f}(a)|^2$.
\end{lemma}
\begin{proof}
Define the function $T^{h}f:\F_p^n\to \D$ by $T^{h}f(x) = f(x+h)$.
Consider the unitary $S_{h}:(\C^p)^{\otimes n}\to(\C^p)^{\otimes n}$, defined by $\ket{x}\mapsto\ket{x+h}$. 
A query to~$T^{h}f$ then corresponds to applying $S_h$, querying $f$, and applying $S_{-h}$.
As one query to $f$ introduces $p$-th roots of unity to the amplitudes, $p-1$ sequential queries to $f$ are equivalent to a single query to $\overline{f}$. 

We prove the lemma by outlining the algorithmic steps taken. 
Initialize a register~$(\C^p)^{\otimes n}$ in the all-zeros state and create a uniform superposition over~$\F_p^n$ by applying a quantum Fourier transform to the initial state. 
Next, query~$T^{h}f$ and~$\overline{f}$ to obtain the state
\begin{align*}
    \frac{1}{\sqrt{p^n}}\sum_{x\in\F_p^n} f(x+h)\overline{f(x)}\ket{x}
    & = \frac{1}{\sqrt{p^n}}\sum_{x\in\F_p^n}\Delta_{h}f(x)\ket{x}.
\end{align*}
Finally, apply the inverse-quantum Fourier transform to obtain the state
\begin{align*}
    \sum_{a\in\F_p^n}\bigg(\frac{1}{p^n}\sum_{x\in \F_p^n}\Delta_{h}f(x)\,\omega_p^{-\ip{x,a}}\bigg)\ket{a}
    & = \sum_{a\in\F_p^n}\widehat{\Delta_{h}f}(a)\ket{a}.
\end{align*}
Measuring in the computational basis gives the desired distribution.
\end{proof}
Note that this lemma implicitly uses quantum gates over $p$-level qudits, instead of over $2$-level qubits.
\cref{sec:quantum:Hadamard_code_higher_fields} briefly discusses a few of these gates. 

%% file: introduction/intro_theory_to_practice.tex
\section{From theory to practice}
As quantum devices grow in computational power and more quantum algorithms are being developed, practical applications also come closer. 
With a quantum algorithm alone we however have not yet solved the computational problems encountered in practice.

Quantum computers usually solve only a small computationally hard problem in a larger sequence of algorithms, also called a workflow. 
As a result, quantum computers have to be integrated into a larger (possibly existing) workflow. 
Furthermore, with an increasing number of quantum devices becoming available it might prove hard to choose the best device to solve your problem. 
Finally, multiple steps are required to obtain an implementation on quantum hardware, given a theoretical algorithm.
This section briefly discusses these aspects relevant when performing quantum computations.

\subsection{Integration in a large workflow: Hybrid quantum computing}
Especially in the near-term, quantum computers will only solve a specific computationally hard problem. 
The outcome of the quantum algorithm is then processed by a conventional computer. 
For simple operations, conventional computers usually work significantly faster and quantum computers offer no benefit. 
Even on the long term, it is expected that quantum computers will only perform part of the computations of a larger workflow. 

The interaction between quantum and conventional computers, and their integration in the same workflow, brought forward the term hybrid quantum computing~\cite{Lanzagorta:2005}. 
However, this term is often used in varying situations, with its exact meaning differing each time~\cite{Peruzzo:2014,Li:2017,McCaskey:2020,Cerezo:2021,deLuca:2022}. 
Often, the term hybrid quantum computing means that a quantum computer interacts with a conventional computer in some unspecified way. 

Integrating conventional computers with quantum computers requires a clear definition of what is exactly meant. 
Such a definition benefits from an abstract point-of-view on the performed computations. 
\citeauthor{Weder:2021} provides such a view by considering workflows, a break-down of a computation in multiple high-level blocks~\cite{Weder:2021}.
An example workflow for clustering can for instance contain the high-level blocks \textit{Retrieve data} and \textit{Compute cluster}. 
Each block can represent a workflow itself. 

\citeauthor{Phillipson:2023} use this workflow approach to introduce new nomenclature to distinguish between various forms of hybrid quantum computing~\cite{Phillipson:2023}. 
They first make the distinction between vertical and horizontal hybrid computing. 
Vertical hybrid computing encompasses all activities required to control and operate a quantum circuit, independent of the actual algorithm.
We distinguish between three different vertical hybrid forms of hybrid: 
\begin{itemize}
    \item Decomposition hybrid: All steps to decompose a high-level quantum routine into low-level instructions that correspond to the available gate set;
    \item Implementation hybrid: All steps to map low-level instructions to the quantum hardware, taking into account the scheduling of the gates and the hardware topology;
    \item Controlling hybrid: All steps to assure the quantum computer operates correctly and stays calibrated.
\end{itemize}
For horizontal hybrid computing, we distinguish between five different forms of hybrid:
\begin{itemize}
    \item Processing hybrid: Only the pre- and post-processing steps are performed on a conventional device. 
    Shor's algorithm is a prominent example of a processing hybrid algorithm;
    \item Macro/micro hybrid split: In the whole workflow, quantum and conventional computations alternate.
    The difference between macro and micro hybrid split depends mainly on the granularity of the workflow. 
    Examples of a macro/micro hybrid splits originate typically in quantum machine learning, where data is prepared and a quantum circuit is run. 
    We also find macro/micro hybrid splits in variational circuits, where a conventional computer optimizes the gate parameters of the quantum circuit;
    \item Parallel hybrid: Computations are solved in multiple independent branches, where each branch can employ its own algorithm. 
    The workflow outputs the first (or best) solution found by the branches.
    The parallel hybrid computing is mainly used within current quantum annealing hardware~\cite{WangLuGloverHao:path_relinking:2012,WangLuGloverHao:multilevel_algorithm:2012,Booth:2017};
    \item Breakdown hybrid: A problem is decomposed in different smaller problems, each of which is solved independently, before being recombined. 
    Divide-and-conquer approaches typically fall under breakdown hybrid~\cite{Arrighi:2006,Bravyi:2016,Araujo:2021,Cojocaru:2021,Tomesh:2023,Fang:2024}.
\end{itemize}

With hybrid quantum computing, also comes forth the topic of cloud-based quantum computing. 
Quantum computers will most likely act as a secondary processor to a conventional computer, similar to current graphical processing units (GPUs).
Moreover, due to the often stringent operating conditions, quantum computers are unlikely to be hosted locally. 
Instead, we expect a few parties to host quantum computers, which others can use via remote connections. 
We call this type of computing cloud-based quantum computing. 

Cloud-based computing is common with conventional computers, where large high-performance computers (HPCs) perform complex computations. 
With quantum computers, we are offered the opportunity to redefine the way we interact with cloud-based systems, based on experience from conventional cloud-based computing. 
Moreover, for a widespread adoption of quantum computing, cloud-based solutions should offer certain user-friendly functionalities. 

Some functionalities such as appropriate device allocation, characterization of the quantum devices and automatic compilation of operations to hardware-aware instructions are similar to conventional cloud-based systems. 
Others functionalities however are completely new, first and foremost the integration of quantum computers with conventional cloud-based services. 
Additionally, communication between various quantum computers via a quantum network allows for a significantly broader range of applications.
Multiple initiatives are already underway to realize this ambition, both by companies~\cite{IBMQuantumExperience:2024}, as well as on a national and international level~\cite{EuroHPC:2024}.

Finally, these functionalities should also be tweaked based on user needs. 
Some users might want to have full control over the devices and the way operations are implemented, whereas others might only be interested in the final answer of a quantum algorithm. 
These different types of users translate into different services offered by a cloud-computing provider, as seen in conventional cloud-based computing~\cite{Medium:2018,NeumannSchootSijpesteijn:2023}.

Much work has already been done on how to set up a new cloud-based quantum platform~\cite{Singh:2014,Kaiiali:2019,NeumannSchootSijpesteijn:2023}. 
Multiple cloud-based quantum computing platforms are already available.
Most offer however only access to quantum computing hardware and do not offer integration with conventional computers or other quantum hardware.

\subsection{Choosing quantum hardware: Quantum metrics}
When using quantum devices, it is important to choose the one that best suits your goals. 
Number of qubits available is one way to choose a device, though other methods exist that might favor devices with fewer but better qubits. 
Quantum metrics provide insights in the performance of different quantum devices. 

With varying interest in using quantum computers, also come various metrics to quantify their applicability. 
Typically, quantum metrics are grouped in one of three groups~\cite{vanderSchootWezemanEendebakNeumannPhillipson:2023}:
\begin{itemize}
    \item Component-level metrics;
    \item System-level metrics;
    \item Application-level metrics.
\end{itemize}

Component-level metrics focus on individual components of a quantum computer.
This includes simple metrics such as the number of qubits available, and more elaborate metrics such as the coherence times of the qubits~\cite{Medford:2012,Youssef:2020} and the quality of single-qubit and two-qubit quantum gates~\cite{MagesanGambettaEmerson:2011,Nielsen:2021}. 
These metrics are most suited for quantum hardware developers to tune the hardware and for low-level implementations. 

System-level metrics focus on the system as a whole. 
A well-known system-level metric is the quantum volume~\cite{Cross:2019}, which quantifies how well quantum computers can run random quantum circuits of some specific form. 
The Circuit Layer Operations Per Second (CLOPS) metric uses the quantum volume metric and also introduces a time component~\cite{Wack:2021}.
Another system-level metric is based on mirror circuits, where a circuit and its inverse are applied with an intermediate layer of random Pauli gates~\cite{Proctor:2021}.
If we restrict the circuit to Clifford gates, then the inverse also consists of Clifford gates and the expected outcome and hence the device's performance is efficiently computed~\cite{Gottesman:1998}.

Application-level metrics consider the capabilities of a device to solve specific problems. 
These metrics focus on the actual use of a device in practice. 
As such, these metrics are typically agnostic to the hardware technology and, depending on the application considered, also allow to quantify the performance of heuristic methods. 

Common application-level metrics include the QED-C metric, the Q-score, and QuAS. 
The first considers various tutorial algorithms, quantum subroutines and end-user applications to measure the performance of a device~\cite{Lubinski:2021}.
The Q-score originally considered the Max-Cut problem on random Erd{\"o}s-R{\'e}nyi $(N,\frac{1}{2})$-graphs~\cite{ErdosRenyi:1959} and compared these results to a random naive approach~\cite{MartielAyralAllouche:2021}. 
The Q-score was later extended to include a time-aspect and different computational paradigms~\cite{vanderSchoot:2022,vanderSchootWezemanNeumannPhillipsonKooij:2024}.
The QuAS metric combines elements from the Q-score and the QPack metric to evaluates the performance of quantum algorithms in different dimensions~\cite{Mesman:2022,Mesman:2024}.

As mentioned, different users require different types of metrics. 
Appropriate metrics give insight in which device works best for a given situation. 
It might even be that a conventional approach performs similar compared to a quantum approach.
This comparison will however depend explicitly on the used algorithm, the used computational device and the problem instances.

\subsection{From algorithm to implementation: Quantum programming}\label{sec:quantum_programming}
If we have a quantum algorithm and chosen a quantum device, we can start implementing the algorithm. 
This requires us to program our quantum algorithm and feed it to the computer, similar to programming a conventional computer. 

Programming languages for conventional computers have been under development for decades. 
The first programming languages required the user to address the registers and program instruction close to the native hardware. 
Current conventional programming languages allow users to work with objects and classes.
Thanks to compilers, users are not concerned with the exact underlying hardware.

Quantum programming languages have not seen this development yet and many steps are still to be taken~\cite{Piattini:2021}. 
Some quantum programming languages already allow users to program somewhat higher-level instructions, such as applying a quantum Fourier transform on some qubits. 
However, most languages require users to program in a low-level hardware-agnostic manner using single- and two-qubit gates. 

Eventually, compilers and transpilers running on conventional computers will help translate hardware-agnostic instructions into hardware-aware instructions. 
These hardware-aware instructions take into account the exact way the qubits are addressed, as well as the timing between different instructions. 
Ideally, compilers would also decompose high-level instruction into hardware-agnostic instructions. 
However, such decomposition techniques currently exist only in theory~\cite{Vartiainen:2004,Zomorodi:2024}.

As qubits easily decohere, quantum computers will require error-correction methods to mitigate these effects.
There are results showing the reduction of error rates using these methods~\cite{GoogleWillow:2024}. 
However, they are not yet applied automatically on hardware devices. 
In the long term, error-correction methods should be applied automatically, for instance, by the compiler and transpiler. 

To utilize quantum solutions the soonest, developments in quantum algorithms and quantum hardware must go hand-in-hand with developments in integration of quantum solutions in larger workflows and developments in quantum programming languages.

%% file: decoding/introduction_decoding.tex
\chapter[Introduction to decoding]{Introduction to decoding error-correcting codes}\label{chp:decoding:introduction}
This chapter provides an introduction to error-correcting codes and different error models. 
We also discuss the concept of list decoding, which becomes relevant in case of many errors. 
We briefly discuss circuits and query complexity, as in the remainder of this part, we will provide results with respect to those. 
We conclude with list-decoding algorithms beyond linear codes, and a brief discussion of deviation bounds that will be used to bound the probability of events happening. 

\input{decoding/introduction_decoding/error_correcting_codes}
\input{decoding/introduction_decoding/error_models}
\input{decoding/introduction_decoding/list_decoding}
\input{decoding/introduction_decoding/circuits_and_query_complexity}

\input{decoding/introduction_decoding/list_decoding_beyond_linear}

\input{decoding/introduction_decoding/deviation_bounds}

\input{decoding/introduction_decoding/outline_intro_decoding}

%% file: decoding/introduction_decoding/error_correcting_codes.tex
\section{Error-correcting codes}
Error-correcting codes, introduced in Shannon's celebrated work~\cite{Shannon:1948}, protect digital signals from noise.
For positive integers $n\ge k$, an error-correcting code over a finite alphabet~$\Sigma$ is a map $E:\Sigma^k\to\Sigma^n$, such that any message $x\in \Sigma^k$ can be decoded from the codeword~$E(x)$ even if the codeword is partially corrupted.
The magnitude of the corruption is measured by the Hamming distance $d(x,y)$, which counts the number of entries where $x$ and $y$ differ. 

If too many errors occur, recovering the original message may become impossible.
In such cases, one can instead resort to \emph{list decoding}, an influential idea proposed in seminal works of Elias~\cite{Elias:1957} and Wozencraft~\cite{Wozencraft:1958}, which aims to give a small list of messages whose codewords are close to the received (corrupted) codeword.
Complexity considerations appear naturally in this context, as encoding and decoding ideally allow for reliable communication with limited computational resources;
they also appear because of the fundamental role played by error-correcting codes in computational complexity itself (see for instance~\cite{Trevisan:2004}).

Reed-Muller codes form a well-known class of error-correcting codes.
\citeauthor{Muller:1954} first proposed these codes and later \citeauthor{Reed:1954} found an efficient decoding algorithm for them~\cite{Reed:1954,Muller:1954}. 
The popularity of Reed-Muller codes follows from their flexibility and as they generalize many codes, including Reed-Solomon codes~\cite{ReedSolomon:1960}. 
A Reed-Muller code $RM(r,n)$ encodes a message of length $\sum_{i=0}^{r}\binom{n}{r}$ into a codeword of length $2^n$. 
The codeword corresponds to the evaluation on all inputs $x\in\bset^n$ of a polynomial whose coefficients correspond to the input message.
Polynomial interpolation helps to decode Reed-Muller codewords. 

Reed-Muller codes also generalize the Hadamard code $H$, a basic but important example of a linear error-correcting code.
The Hadamard code encodes $k$-bit messages into codewords of length $n=2^k$ by evaluating all functions of degree at most $1$: 
For message $x\in\F_2^k$, $H(x) = (\ip{x,y})_{y\in \F_2^k}$, where $\ip{x,y} = \sum_{i=0}^{k-1}x_iy_i$.

%% file: decoding/introduction_decoding/error_models.tex
\section{Error models}\label{sec:decoding:intro:error_models}
In the error model considered by Shannon~\cite{Shannon:1948}, a codeword is corrupted by some random process.
This process is described by the \emph{symmetric channel}:
for each coordinate of the codeword independently, the channel either transmits it unchanged with some probability~$\rho$, or replaces it with a uniformly random element of~$\Sigma$ with probability~$1-\rho$;
each coordinate is thus corrupted with probability $(1 - \rho)(1 - |\Sigma|^{-1})$.
We refer to~$\rho$ as the \emph{bias} of the channel.
If~$Z\in \Sigma^n$ is distributed according to the random outcome of the symmetric channel with bias $\rho$ applied to a codeword~$E(x)$, we write $Z\sim\mathcal N_\rho\big(E(x)\big)$.
In this model, the goal is to correctly decode a corrupted codeword with good probability over the noise. 

The combinatorial worst-case error model of Hamming~\cite{Hamming:1950} instead assumes that the codeword is corrupted arbitrarily on at most some $\delta$-fraction of the coordinates, for some \emph{error parameter} $\delta\in[0,1)$. 
In this setting, the number of errors that can be tolerated depends on the minimal Hamming distance between any pair of distinct codewords, or \emph{minimal distance} of the code, denoted~$d_E$.
Since the Hamming ball of diameter $d_E-1$ around any point $y\in \Sigma^n$ contains at most one codeword, a message can be retrieved if fewer than~$d_E/2$ errors occur.

Every error-correcting code has a certain capacity, the theoretical maximum error rate below which messages can always be recovered. 
An interesting question is to develop an error-correcting code that actually achieves this capacity~\cite{Shannon:1948}.

%% file: decoding/introduction_decoding/list_decoding.tex
\section{List-decoding}\label{sec:decoding:intro:list_decoding}
If more errors occur, faithful decoding is no longer possible, and list decoding enters the picture. 
For error parameter $\delta\in [0,1)$ and positive integer~$L$, a code is \emph{$(\delta,L)$-list decodable} if for any point $y\in \Sigma^n$, the Hamming ball of radius~$\delta n$ centered around~$y$ contains at most~$L$ codewords.
It is well known that any $(\delta,L)$-list decodable code satisfies $L \geq \Omega(1/\eps^2)$ when $\delta = (1-\eps) (1 - |\Sigma|^{-1})$~\cite{GuruswamiVadhan:2010}.
If fewer than a $\delta$-fraction of the codeword coordinates are corrupted, a random element from this list will give the correct message with probability at least~$1/L$.

With growing error parameter $\delta$, the list size also grows. 
However, as long as the error parameter remains below the capacity of the code, the list size remains constant. 
Yet, even list-decoding algorithms have their limits: 
If the error parameter increases too much, any list-decodable code has exponential list size. 
See Theorem 7.4.1 in Ref.~\cite{GuruswamiRudraSudan:2022} for an exact definition and a proof. 

In an influential paper in 1989, \citeauthor{GoldreichLevin:1989} gave the first efficient list-decoding algorithm~\cite{GoldreichLevin:1989}. 
They considered the Hadamard code.
This code has minimal distance~$n/2$ and is $(1/2-\eps, O(1/\eps^2))$-list decodable for any $\eps\in (0, 1/2]$, which is known to be optimal for any code~\cite{GuruswamiVadhan:2010}.

Under the symmetric channel, the Chernoff bound implies that unique decoding of the Hadamard code is possible with high probability for any constant bias $\rho > 0$.
This result even holds for any code over a large enough alphabet~\cite{RudraUurtamo:2010}. 
This is due to the fact that, with high probability, the Hamming ball of radius $(1/4 - \rho/4)n$ around a corrupted version of a codeword~$C$ contains no other codewords than~$C$ itself.

For the worst-case Hamming model, \citeauthor{GoldreichLevin:1989} famously gave an efficient list decoding algorithm for the Hadamard code that runs in time $\poly(k, 1/\eps)$, for error parameter $\delta = 1/2 - \eps$~\cite{GoldreichLevin:1989}. 
For fixed $\eps>0$, their algorithm gives a probabilistic $\ac^0$-circuit that, on input length $n$, correctly returns the original message with probability~$\Omega(1)$.

%% file: decoding/introduction_decoding/circuits_and_query_complexity.tex
\section{Constant-depth circuits}\label{sec:decoding:intro:circuits}
A well-studied problem is decoding corrupted error-correcting codes by constant-depth circuits, such as $\nc^0$ or $\ac^0$, for example in the context of black-box hardness amplification~\cite{SudanTrevisanVadhan:1999,Viola:2006,TrevisanVadhan:2007}.
\cref{sec:intro:constant_depth_quantum} showed that these circuits are amenable to provable separations. 
It is worthwhile to study whether such a separation also exists for decoding highly-corrupted error-correcting codes.

As gates in $\nc^0$-circuits have bounded-fan-in and the circuits have a constant depth, the outputs can only depend on a constant number of the inputs.
However, the output of $\nc^0[\oplus]$-circuits can depend on all inputs, due to the unbounded-fan-in parity gate.
Similarly, the output of $\qnc^0[\oplus]$-circuits can depend on the whole input.

In \cref{chp:decoding:classical} and \cref{chp:decoding:quantum} we consider the classes $\nc^0[\oplus]$ and $\qnc^0[\oplus]$ and their capabilities in decoding corrupted error-correcting codes.

%% file: decoding/introduction_decoding/list_decoding_beyond_linear.tex
\section{List-decoding beyond linear functions}
The list-decoding algorithm by \citeauthor{GoldreichLevin:1989} works in greater generality to find large Fourier coefficients for any function.

Our quantum algorithm in \cref{chp:decoding:quantum} is an extension of the original Goldreich-Levin algorithm to a quantum setting, where we again sample the Fourier spectrum to find large Fourier coefficients. 
These large Fourier coefficients correspond precisely to possible original messages. 

This quantum algorithm is inspired by the Bernstein-Vazirani algorithm~\cite{BernsteinVazirani:1997}, which in greater generality can be interpreted as a quantum algorithm to sample the Fourier spectrum of a function. 

The Hadamard code is the simplest example of a Reed-Muller code. 
The next example would be to consider Reed-Muller codewords of degree at most~$2$, for which \citeauthor{TulsianiWolf:2014} gave a conventional list-decoding algorithm~\cite{TulsianiWolf:2014}.
Their algorithm makes use of higher-order Fourier analysis. 
Furthermore, higher-order Fourier analysis helped to prove a qubic Goldreich-Levin algorithm~\cite{KimLiTidor:2023}. 

Only very few results are known that consider applications of higher-order Fourier analysis in quantum algorithms for decoding corrupted error-correcting codes.
\citeauthor{Montanaro:2012} took a first step in this direction by providing a quantum algorithm to learn uncorrupted multilinear polynomials of degree at most~$d$ using $\bigo(n^{d-1})$ quantum queries.
However, this algorithm fails in case the queries can be corrupted. 
\cref{chp:decoding:quadractic} extends this research line by considering learning functions that have large Gowers $U^3$-norm.
This includes corrupted Reed-Muller codewords of degree at most~$2$.

%% file: decoding/introduction_decoding/deviation_bounds.tex
\section{Deviation bounds}
In this part, we consider error models that corrupt codewords from some error-correcting codes. 
We interpret these codewords as strings with elements in a prime field. 
As we use probabilistic error models, we have to work with events that occur with certain probability. 
In most arguments, we will have to bound the probability of some event happening. 
Probability theory offers us the following two standard results which allow us to obtain these desired bounds. 

The first result is by Markov, which gives an upper bound on the probability that some random variable is above some threshold. 
\begin{lemma}[Markov's inequality]\label[lemma]{lem:Markov}
Let $M>0$ and let $X$ be a random variable taking values in a finite set $S\subseteq (-\infty,M]$. 
Then, for any $a\in S$, we have that
\begin{equation*}
    \Pr[X \geq a] \geq \frac{\Exp X - a}{M}.
\end{equation*}
\end{lemma}
\begin{proof}
We have
\begin{align*}
    \Exp X &= \sum_{b\in S\cap (-\infty,a)} b\Pr[X = b] + \sum_{b\in S\cap [a,M]}b \Pr[X=b]\\
    &\leq a\Pr[X\leq a] + M\Pr[X\geq a] \\
    &\leq a + M\Pr[X\geq a].
\end{align*}
\end{proof}

Additionally, Hoeffding's inequality upper bounds the probability that the sum of random variables differs from the expected value by more than a certain constant~\cite{Hoeffding:1963}. 
We will specifically use a complex version of Hoeffding's inequality. 
If all random variables are real valued, we can replace the~$4$ in the upper bound by a~$2$ in the lemma. 
\begin{lemma}[Hoeffding's inequality]\label[lemma]{lem:Hoeffding}
Let $X_1,\dots, X_n$ be independent $\C$-valued random variables such that $|X_i| \leq a_i$ for some~$a_i>0$.
Let $\overline X = n^{-1}(X_1 + \cdots + X_n)$.
Then, for any $\eps > 0$,
\begin{equation*}
    \Pr\big[\big|\overline{X} - \Exp\overline{X}\big| > \eps\big] \leq 4\exp\Bigg(-\frac{2\eps^2n^2}{\sum_{i=1}^na_i^2}\Bigg).
\end{equation*}
\end{lemma}
\begin{proof}
    Let $Y_i^0$ and $Y_i^1$ be the real and complex parts of~$X_i$, respectively.
    Then, $|Y_i^0|,|Y_i^1|\leq a_i$.
    For $b\in \bset{}$, let $\overline{Y^b} = n^{-1}(Y_1^b + \cdots + Y_n^b)$.
    By Hoeffding's inequality~\cite{Hoeffding:1963}, for $b\in \bset{}$,
    \begin{equation*}
        \Pr\big[\big|\overline{Y^b}  - \Exp\overline {Y^b}\big|>\eps\big] \leq 2\exp\Bigg(-\frac{2\eps^2n^2}{\sum_{i=1}^na_i^2}\Bigg).
    \end{equation*}
    The result now follows from the triangle inequality and the union bound.
\end{proof}

%% file: decoding/introduction_decoding/outline_intro_decoding.tex
\section{Outline}
The remainder of this Part consists of three chapters: 
\cref{chp:decoding:classical} presents a proof that polynomial maps of constant degree, and by extension also any $\nc^0[\oplus]$-circuit, cannot decode any corrupted error-correcting code with constant success probability.
\cref{chp:decoding:quantum} shows that $\qnc^0[\oplus]$ circuits can decode a corrupted Hadamard code with constant success probability for fixed error parameter. 
\cref{chp:decoding:quadractic} shows how to find a Reed-Muller codeword of degree~$2$ that correlates with a polynomial phase function given as input. 

In the next sections, $C,c>0$ will denote absolute constants whose values may be different at each occurrence.

%% file: decoding/classical_decoding.tex
\chapter{Conventional hardness of list decoding}\label{chp:decoding:classical}
In this chapter we prove our main result for conventional approaches to decoding, which holds for any error-correcting code. 
We start slowly by showing the result for linear maps, which serves as a warm-up for the more general case of polynomial maps.
We end with an improved result in the high-characteristic setting that hints towards possible improvements of our results. 

\input{decoding/classical_hardness/outline_classical_hardness}
\input{decoding/classical_hardness/warm_up_classical_linear_maps}

\input{decoding/classical_hardness/analytic_rank_polynomial_maps}

\input{decoding/classical_hardness/biased_equidistribution_high_rank_maps}

\input{decoding/classical_hardness/proof_classical_hardness}

\input{decoding/classical_hardness/high_characteristic_setting}

\input{decoding/classical_hardness/tensors_associated_to_polynomial_maps}

\input{decoding/classical_hardness/proof_high_characteristics}
\input{decoding/classical_hardness/discussion_classical_hardness}

%% file: decoding/classical_hardness/outline_classical_hardness.tex
\section{Chapter overview}\label{sec:decoding:classical:outline}
The main theorem proven in this chapter states that $\nc^0[\oplus]$-circuits cannot decode error-correcting codes with constant success probability. 
\begin{theorem}[Impossibility of decoding by {$\nc^0[\oplus]$}]\label{thm:classical_decoding_nc0plus}
For any $\rho\in [0,1)$, $d\in \N$ and $\eps\in (0,1]$, there is a $k_0 = k_0(d, \rho, \eps)\in \N$ such that the following holds.
Let $k \geq k_0$ and~$n$ be positive integers, $E:\F_2^k\to \F_2^n$ be any map and $\phi:\F_2^n\to\F_2^k$ be a map computable by an $\nc^0[\oplus]$-circuit of depth at most~$d$.
Then, for a uniformly distributed~$x\in \F_2^k$ and $Z\sim\mathcal N_\rho(0)$, we have that
\begin{equation}\label{eq:success_probability:nc0}
\Pr\big[\phi\big(E(x) + Z\big) = x] < \eps.
\end{equation}
\end{theorem}

In particular, this theorem shows that no $\nc^0[\oplus]$-circuit can correctly decode more than an $\eps$-fraction of codewords with probability higher than $\eps$ over the noise distribution, provided the messages are long enough depending on $\eps$, the error rate $(1-\rho)/2 >0$, and the depth of the circuit.
By Yao's minimax principle~\cite{Yao:1997} and the Chernoff bound, it follows that any probabilistic $\nc^0[\oplus]$-circuit also fails (with high probability) to correctly decode any binary error-correcting code in the worst-case Hamming model, for any constant error parameter~$\delta \in (0, 1/2]$.

To prove this theorem, we use the basic observation that any function $f:\F_2^n\to\nolinebreak \F_2^k$ that is computable by an $\nc^0[\oplus]$-circuit can be given by a collection of $k$ constant-degree polynomials over~$\F_2$ in~$n$ variables.
Indeed, any gate with fan-in~$d$ implements a function $g:\F_2^d\to\F_2$ and any such function can be represented by a $d$-variable polynomial of total degree at most~$d$.
Degree is multiplicative under composition and composition occurs only between different layers of the circuit.
Since the parities amount to addition in~$\F_2$ and $\nc^0$-circuits have constant depth, the total degree of the output is bounded.

We will therefore study the distribution of polynomial maps under biased input distributions.
We will do so in a slightly more general setting over arbitrary finite fields of prime order\footnote{The restriction to prime order is done for notational reasons and for ease of exposition. 
Our arguments can be readily adapted to the case of non-prime finite fields.}.
For $\rho\in [0,1]$, an $\F_p$-valued random variable~$Z$ is \emph{$\rho$-biased} if with probability~$\rho$ it equals~0 and with probability $1 - \rho$ it is uniformly distributed over~$\F_p$.
Note that this corresponds to the noise $\mathcal{N}_\rho(0)$ added by the symmetric channel when the alphabet is~$\F_p$.

A mapping $\phi:\F_p^n\to\F_p^k$ is called a \emph{polynomial map} if there exist polynomials $f_1,\dots,f_k\in \F_p[x_1,\dots,x_n]$ such that $\phi = (f_1,\dots,f_k)$.
The degree of~$\phi$ is the maximal degree among the~$f_i$.
To prove \cref{thm:classical_decoding_nc0plus}, it thus suffices to prove the following result.
\begin{theorem}[Impossibility of decoding by polynomial maps]\label{thm:classical_decoding_polynomial_maps}
For every integer $d$ and any $\rho,\eps\in (0,1)$, there exists an integer $k_0 = k_0(p,d,\rho,\eps)$ such that the following holds.
Let $k \geq k_0$ and $n$ be integers, $\phi:\F_p^n\to\F_p^k$ be a polynomial map of degree at most~$d$ and $E:\F_p^k\to\F_p^n$ be an arbitrary function.
Then
\begin{equation}\label{eq:success_probability:polynomial_maps}
\Pr_{x\in \F_p^k, Z\sim \mathcal N_\rho(0)}\big[\phi\big(E(x) + Z\big) = x\big] \leq \eps.
\end{equation}
\end{theorem}

Studying the distribution of polynomial maps in many variables over a finite field falls within the purview of additive combinatorics.
In the ``unbiased'' situation where~$Z$ is uniformly distributed, higher-order Fourier analysis provides powerful tools to study the distribution of~$\phi(Z)$.
In particular, \citeauthor{GreenTao:2009}~\cite{GreenTao:2009} proved that if~$\phi$ is ``regular'' (random-like), then~$\phi(Z)$ is approximately uniformly distributed over~$\F_p^k$.
This implies that, for every~$x\in\F_p^n$, the probability of the event $\{\phi(E(x) + Z) = x\}$ is small.
A ``regularity-type'' lemma proved in~\cite{GreenTao:2009} shows that one can ``force'' $\phi$ to be regular by restricting it to a partition defined by sufficiently many polynomial equations of degree less than the degree of~$\phi$.
However, these techniques cause the size of the polynomial map $\phi$ considered to blow up considerably, and are only effective if~$k$ is an extremely slowly growing function of~$n$.

To address this issue and to adapt these results to the case where~$Z$ is no longer uniform but biased, we employ a dichotomy often used in additive combinatorics.
This approach separately studies the ``pseudorandom'' case of regular maps and the ``structured'' case of maps that carry a certain algebraic structure.
This is done in \cref{sec:decoding:classical:analytic_rank} by defining and studying a new notion of rank for (high-dimensional) polynomial maps, which we call the \emph{analytic rank}\footnote{A very similar notion of rank was defined for multilinear forms by \citeauthor{GowersWolf:2011}~\cite{GowersWolf:2011}, who coined the term analytic rank when studying linear systems of equations over $\F_p^n$.
We use the same name to highlight the similarity between our two notions, which are relevant for distinct types of mathematical objects.}, 
and which measures how equidistributed the values taken by the considered map are.

We consider the pseudorandom case in \cref{sec:decoding:classical:biased_equidistribution}.
The main tool we use in this case is a random restriction result for high-rank polynomial maps proved by \citeauthor{BrietCastroSilva:2022}~\cite{BrietCastroSilva:2022}.
We use this result to show that the distribution of values taken by a high-rank polynomial map will be close to uniform even under a biased input distribution.
This implies that the event considered in the theorem has very low probability for any fixed~$x$, in which case we can conclude by averaging.

In the structured case we deal instead with polynomial maps of low rank, whose values are in a sense poorly distributed.
Results from higher-order Fourier analysis then imply that they can be determined by ``few'' lower-degree polynomial maps (plus a few extra polynomials);
by a simple Fourier-analytic argument we can reduce the analysis of a low-rank polynomial map to those lower-degree maps that specify it, making it amenable to an inductive argument.
We use this idea in \cref{sec:decoding:classical:main_result} where we prove \cref{thm:classical_decoding_polynomial_maps}.

The decay found on the probability in \cref{eq:success_probability:nc0} of correct message retrieval as a function of the message length is extremely slow, making \cref{thm:classical_decoding_nc0plus} a qualitative result rather than quantitative.
Nevertheless, we conjecture that the true decay of this probability is exponential in the message length~$k$;
this would clearly be optimal, as can be seen by taking a constant map~$\phi$ which always returns some fixed message.
In \cref{sec:classical:high_char,sec:decoding:classical:proof_high_char,sec:decoding:classical:tensors} we will provide some evidence to support this conjecture.

We start off gradually by first proving in \cref{sec:warm_up:classical} that linear maps have exponentially small success probability in decoding corrupted error-correcting codes. 

%% file: decoding/classical_hardness/warm_up_classical_linear_maps.tex
\section{Impossibility of decoding linear maps}\label{sec:warm_up:classical}
As a warm-up to our later arguments, we here present a proof of the first nontrivial case of \cref{thm:classical_decoding_polynomial_maps}, namely that of maps~$\phi: \F_p^n \to \F_p^k$ of degree~$1$.
In this case, there is a matrix~$U\in \F_p^{k\times n}$ and a vector~$v\in \F_p^k$ such that
\begin{equation*}
\phi(y) = Uy + v \quad \text{for all } y\in \F_p^n.
\end{equation*}

Let~$x$ be a uniformly distributed random variable over~$\F_p^k$ and let~$Z$ be an $\F_p$-valued \emph{$\rho$-biased} random variable, meaning that with probability $\rho$ it equals $0$, and with probability $1-\rho$ it is uniformly distributed over $\F_p$. 
We denote this distribution by~$\mathcal{N}_{\rho}(0)$.
Our goal is then to bound the probability of the event
\begin{equation}\label{eq:warm_up:base_classical}
    U(E(x) + Z) + v = x.
\end{equation}

We follow a structure-versus-randomness approach, commonly used in additive combinatorics. 
For that we distinguish between $U$ having low rank, the \textit{structure} case, and $U$ having high rank, the \textit{randomness} case. 
We bound the probability of event~$\E$ for both cases separately.
Let~$r=k/2$ be an integer.

\subsection{Structured low-rank case} % Structure
If~$U$ has rank at most~$r$, then its image~$\im(U)$ is a subspace of size at most~$p^r$.
If event \eqref{eq:warm_up:base_classical} holds, then~$x$ is contained in the coset~$v + \im(U)$ of this subspace, which (for~$x$ uniform over~$\F_p^k$) happens with probability at most~$p^r/p^k$.
Hence, event \eqref{eq:warm_up:base_classical} holds with probability at most~$p^{-(k-r)}$ in this case.

\subsection{Pseudorandom high-rank case} % Randomness
For the ``pseudorandom case'' of high-rank matrices, we make the following simple but important observation:
one can sample~$Z \sim \mathcal{N}_{\rho}(0)$ by first sampling the set~$I \subseteq [n]$ of ``corrupted coordinates'', then sampling the ``noise'' $y$ uniformly at random from~$\F_p^I$ and setting\footnote{Given $x\in \F_p^n$ and $I\subseteq [n]$, we denote by $x_{|I}\in \F_p^I$ the restriction of $x$ to the coordinates indexed by $I$.}
$Z_{|I} = y$, $Z_{|[n]\setminus I} = 0$.
Each index~$i\in [n]$ has probability~$1-\rho$ of belonging to the random set~$I$, with these events being mutually independent;
we denote this sampling scheme by~$I\sim [n]_{1-\rho}$.

If we denote by~$U_I\in \F_p^{k\times I}$ the restriction of~$U$ to the columns indexed by~$I\subseteq [n]$, it follows that the random variable~$UZ$ has the same distribution as the random variable~$U_I y$, where $I\sim [n]_{1-\rho}$ and $y$ is uniformly distributed over $\F_p^I$.
Thus, for any given~$x\in \F_p^k$, we have
\begin{equation*}
\Pr_{Z\sim \mathcal{N}_{\rho}(0)}\big[ U(E(x)+Z) + v = x \big] = \Exp_{I\sim [n]_{1-\rho}} \Pr_{y\in \F_p^I}\big[ U_I y = x - UE(x) - v \big].
\end{equation*}
Now, if~$I \subseteq [n]$ is fixed and~$y$ is uniformly distributed over~$\F_p^I$, then the random variable~$U_I y$ is uniformly distributed over~$\im(U_I)$;
hence
\begin{equation*}
\max_{w\in \F_p^k} \Pr_{y\in \F_p^I}\big[ U_I y = w \big] =  \frac{1}{|\im(U_I)|} =  \frac{1}{p^{\rank(U_I)}}.
\end{equation*}
Taking the expectation over~$I\sim [n]_{1-\rho}$ and~$x\in \F_p^k$, we conclude that event \eqref{eq:warm_up:base_classical} holds with probability at most~$\Exp_{I\sim [n]_{1-\rho}} p^{-\rank(U_I)}$.

Suppose now that~$U$ has rank at least~$r$, and let~$J\subseteq [n]$ be a set of~$r$ linearly independent columns of~$U$.
By the Chernoff bound (see \eg~\cite{HagerupRub:1990}), we have that
\begin{equation*}
\Pr_{I\sim[n]_{1-\rho}}\bigg[|I\cap J| \leq  \frac{(1-\rho) r}{2}\bigg] \leq e^{-(1-\rho) r/8}.
\end{equation*}
Thus, the matrix $U_I$ will contain more than~$(1-\rho)r/2$ linearly independent columns with probability at least~$1 - e^{-(1-\rho) r/8}$;
whenever this happens we have $\rank(U_I) \geq (1-\rho)r/2$.
It follows that
\begin{align*}
    \Pr_{x\in \F_p^k, Z\sim \mathcal{N}_{\rho}(0)}\big[ U(E(x) + Z) + v = x \big]
    &\leq \Exp_{I\sim [n]_{1-\rho}} p^{-\rank(U_I)} \\
    &\leq e^{-(1-\rho) r/8} + p^{-(1-\rho) r/2}.
\end{align*}

The choice of $r = k/2$ implies an exponential decay of the probability that event~\eqref{eq:warm_up:base_classical} holds for both cases, concluding the analysis.

%% file: decoding/classical_hardness/analytic_rank_polynomial_maps.tex
\section{The analytic rank of polynomial maps}\label{sec:decoding:classical:analytic_rank}
\citeauthor{GreenTao:2009} introduced the notion of rank for polynomials~$P\in \F_p[x_1,\dots,x_n]$, defined as the smallest number of lower-degree polynomials needed to compute~$P$.
It is related to the \emph{bias} of the polynomial~$P$, or more specifically to the bias of the symmetric~$\deg(P)$-multilinear form associated to~$P$.
The bias of a function $f: \F_p^n\to \F_p$ is an analytic measure of how well-equidistributed the values of~$f$ are when evaluated on a uniformly random input;
formally,
\begin{equation}\label{def:bias}
\bias(f) = \left|\Exp_{x\in \F_p^n} \omega_p^{f(x)}\right|.
\end{equation}

When dealing with a polynomial~$P$ of some bounded degree~$d$, having non-negligible bias implies that it has a significant amount of internal structure.
Such a result was first proven by \citeauthor{GreenTao:2009}~\cite{GreenTao:2009} in the case of polynomials with degree~$d$ smaller than the characteristic~$p$ of the field considered, and motivated the introduction of both their notion of rank and \citeauthor{GowersWolf:2011}'s notion of analytic rank~\cite{GowersWolf:2011}.
We will need a similar result, proven by \citeauthor{KaufmanLovett:2008}~\cite{KaufmanLovett:2008}, that generalizes this theorem to higher characteristics~$p \leq d$ and also gives more precise information on the structure of the polynomial.

The following result of \citeauthor{KaufmanLovett:2008} shows that polynomials with large bias must be highly structured:
\begin{theorem}[Bias implies low rank; Theorem~4~\cite{KaufmanLovett:2008}]\label{thm:KaufmanLovett}
For every~$d\in \N$ and~$\eps>0$, there is an~$r = r(p,d,\eps)\in \N$ such that the following holds.
If~$P\in \F_p[x_1,\dots,x_n]$ is a polynomial of degree at most~$d$ with~$\bias(P) \geq \eps$, then there exist~$h_1,\dots,h_r\in \F_p^n$ and a map~$\Gamma:\F_p^r\to\F_p$ such that
\begin{equation*}
P(x) \equiv \Gamma\big(\Delta_{h_1}P(x),\dots, \Delta_{h_r}P(x)\big).
\end{equation*}
\end{theorem}

For integers $d, n, k \geq 1$, we denote by $\Pol_{\leq d}(\F_p^n, \F_p^k)$ the space of all polynomial maps $\phi: \F_p^n \to \F_p^k$ of degree at most~$d$.
\begin{definition}[Analytic rank]\label{def:analytic_rank}
Given a polynomial map $\phi\in \Pol_{\leq d}(\F_p^n, \F_p^k)$, we define its analytic rank $\arank_d(\phi)$ by
\begin{equation*}
\arank_d(\phi) = -\log_p \bigg( \max_{\psi:\F_p^n\to\F_p^k,\, \deg(\psi)<d} \Pr_{x\in \F_p^n} \big[\phi(x) = \psi(x)\big] \bigg).
\end{equation*}
\end{definition}

Note that, for affine-linear maps $\phi \in \Pol_{\leq 1}(\F_p^n, \F_p^k)$, this definition coincides with the usual notion of rank for the matrix~$U \in \F_p^{k\times n}$ encoding its linear part.
Indeed, suppose~$\phi(x) = Ux + v$ for some~$v \in \F_p^k$.
Since~$Ux$ is uniformly distributed over~$\im(U) \simeq \F_p^{\rank(U)}$ when~$x$ is uniformly distributed over~$\F_p^k$, we have that
\begin{equation*}
\Pr_{x\in \F_p^n} \big[Ux+v = w\big] =
\begin{cases}
 p^{-\rank(U)} &\text{if } w-v \in \im(U), \\
 0 &\text{if } w-v \notin \im(U).
\end{cases}
\end{equation*}
This example might help explain the reason for the~$-\log_p$ in the definition of analytic rank, as well as the need to maximize the probability of agreement over all lower-degree maps.

Another useful way of viewing the analytic rank of a polynomial map~$\phi$ is as a measure of how well-equidistributed its values are in~$\F_p^k$, up to lower-degree perturbations.
Indeed, we can equivalently write
\begin{equation*}
\arank_d(\phi) = \min_{\psi:\F_p^n\to\F_p^k,\, \deg(\psi)<d}-\log_p \big(\Exp_{v\in \F_p^k, x\in \F_p^n}\omega_p^{\langle v,\, \phi(x) - \psi(x)\rangle}\big).
\end{equation*}
The expectation inside the logarithm above is analogous to the notion of bias given in \cref{def:bias}, and can be seen as an analytic measure of how close to uniformly distributed over~$\F_p^k$ the values taken by~$\phi-\psi$ are.

It is clear from the definition that the function $\arank_d$ is nonnegative (since probabilities are bounded by~$1$), and that $\arank_d(\phi) = 0$ if and only if $\deg(\phi) \leq d-1$.
It also shares several useful properties with the rank of matrices;
in order to state them we will need some notation for considering coordinate restrictions:

\begin{definition}[Restriction]
For a polynomial map~$\phi:\F_p^n\to\F_p^k$ and subset~$I\subseteq [n]$, we define the restriction~$\phi_{|I}:\F_p^I\to\F_p^k$ to be the map given by
$\phi_{|I}(y) = \phi(y_I)$,
where~$y_I\in \F_p^n$ agrees with~$y$ on the coordinates in~$I$ and is zero elsewhere.
\end{definition}

The properties of analytic rank that will be important to us are summarized in the next lemma.
\begin{lemma}[Properties of analytic rank] \label[lemma]{lem:natural_rank}
For all integers $d, n, k \geq 1$, the analytic rank function $\arank_d$ satisfies:
\begin{enumerate}
    \item Symmetry: \\
    $\arank_d(\phi) = \arank_d(-\phi)$ for all~$\phi\in \Pol_{\leq d}(\F^n, \F^k)$.
    \item Sub-additivity: \label{it:sub_additivity} \\
    $\arank_d(\phi + \gamma) \leq \arank_d(\phi) + \arank_d(\gamma)$ for all~$\phi, \gamma\in \Pol_{\leq d}(\F^n, \F^k)$.
    \item Monotonicity under restrictions: \label{it:monotinicity} \\
    $\arank_d(\phi_{|I}) \leq \arank_d(\phi)$ for all~$\phi\in \Pol_{\leq d}(\F^n, \F^k)$ and all sets~$I \subseteq [n]$.
    \item Restriction Lipschitz property: \label{it:Lipschitz} \\
    $\arank_d(\phi_{|I \cup J}) \leq \arank_d(\phi_{|I}) + |J|$ for all polynomial maps~$\phi\in \Pol_{\leq d}(\F^n, \F^k)$ and all sets~$I, J \subseteq [n]$.
\end{enumerate}
\end{lemma}

\begin{proof}
The first property is trivial.
To prove property \ref{it:sub_additivity}, let~$\psi, \chi: \F_p^n\to \F_p^k$ be polynomial maps of degree at most~$d-1$ such that
\begin{align*}
    &\arank_d(\phi) = -\log_p \Pr_{x\in \F_p^n} \big[\phi(x) = \psi(x)\big], \\
    &\arank_d(\gamma) = -\log_p \Pr_{x\in \F_p^n} \big[\gamma(x) = \chi(x)\big].
\end{align*}
Then~$p^{-\arank_d(\phi) -\arank_d(\gamma)}$ can be expressed as
\begin{align*}
    \Pr_{x, y\in \F_p^n} &\big[\phi(x) = \psi(x) \wedge \gamma(y) = \chi(y)\big] \\
    &= \Pr_{x, y} \big[\phi(x) = \psi(x) \wedge \phi(x) + \gamma(x+y) = \psi(x) + \chi(x+y)\big],
\end{align*}
where we performed the change of variables~$(x, y)\mapsto (x, x+y)$.
Now, as 
\begin{equation*}
\gamma(x+y) = \gamma(x) + \Delta_y \gamma(x),
\end{equation*}
this equals
\begin{align*}
\Pr_{x, y} \big[\phi(x) = \psi(x) &\wedge \phi(x) + \gamma(x) = \psi(x) + \chi(x+y) - \Delta_y \gamma(x)\big] \\
&\leq \Pr_{x, y} \big[\phi(x) + \gamma(x) = \psi(x) + \chi(x+y) - \Delta_y \gamma(x)\big].
\end{align*}

Note that, for any fixed~$y\in \F_p^n$, the function
\begin{equation*}
x \mapsto \psi(x) + \chi(x+y) - \Delta_y \gamma(x)
\end{equation*}
is a polynomial map of degree at most~$d-1$, as every term has degree at most $d-1$.
The last probability above is then bounded by
\begin{align*}
\max_y \Pr_x \big[\phi(x) + \gamma(x) = \psi(x) &+ \chi(x+y) - \Delta_y \gamma(x)\big] \\
&\leq \max_{\xi: \F_p^n \to \F_p^k,\, \deg(\xi) < d} \Pr_x \big[\phi(x) + \gamma(x) = \xi(x)\big] \\
&= p^{-\arank_d(\phi + \gamma)}.
\end{align*}
Sub-additivity now follows by taking logarithms.

To prove property \ref{it:monotinicity} it suffices to show that $\arank_d(\phi_{|[n]\setminus\{i\}})\leq \arank_d(\phi)$ for any $i\in[n]$, which can then be applied iteratively.
Assume for notational convenience that~$i=n$, and let $\psi:\F_p^n\to\F_p^k$ be a polynomial map of degree at most~$d-1$ which satisfies
\begin{equation*}
p^{-\arank_d(\phi)} = \Pr_{x\in \F_p^n}[\phi(x) = \psi(x)].
\end{equation*}
Factoring out the variable~$x_n$ allows us to write the probability on the right-hand side as
\begin{align*}
\Exp_{x_n\in \F_p} \Pr_{y\in \F_p^{n-1}}\big[\phi_{|[n-1]}(y) + \phi'(y,x_n)x_n = \psi_{|[n-1]}(y) + \psi'(y,x_n)x_n\big],
\end{align*}
where~$\phi'$ and~$\psi'$ are some polynomial maps of degree at most~$d-1$.
By the averaging principle, this is at most
\begin{align*}
\max_{x_n\in \F_p}\Pr_{y\in \F_p^{n-1}} &\big[\phi_{|[n-1]}(y) = \psi_{|[n-1]}(y) + \psi'(y,x_n)x_n - \phi'(y,x_n)x_n \big] \\
&\leq \max_{\xi: \F_p^{n-1} \to \F_p^k,\, \deg(\xi) < d} \Pr_{y\in \F_p^{n-1}} \big[\phi_{|[n-1]}(y) = \psi_{|[n-1]}(y) + \xi(y)\big] \\
&= p^{-\arank_d(\phi_{|[n-1]})},
\end{align*}
showing that $\arank_d(\phi_{|[n-1]})\leq \arank_d(\phi)$ as wished.

Finally, for the Lipschitz property \ref{it:Lipschitz}, let~$\psi: \F_p^I \to \F_p^k$ be a map with $\deg(\psi)<d$ maximizing the agreement probability $\Pr_{x\in \F_p^I} \big[\phi_I(x) = \psi(x)\big]$, and suppose without loss of generality that $J\cap I = \emptyset$.
Then
\begin{align*}
    p^{-\arank_d(\phi_{|I \cup J})}
    &\geq \Pr_{x\in \F_p^I,\, y\in \F_p^J} \big[\phi_{|I \cup J}(x,y) = \psi(x)\big] \\
    &\geq \Pr_{x\in \F_p^I,\, y\in \F_p^J} \big[\phi_{|I \cup J}(x,0) = \psi(x) \wedge y=0\big] \\
    &= p^{-|J|}\, \Pr_{x\in \F_p^I} \big[\phi_I(x) = \psi(x)\big] \\
    &= p^{-\arank_d(\phi_I) - |J|},
\end{align*}
and the restriction Lipschitz property follows.
\end{proof}

%% file: decoding/classical_hardness/biased_equidistribution_high_rank_maps.tex
\section{Biased equidistribution of high-rank maps}\label{sec:decoding:classical:biased_equidistribution}
As in the degree-$1$ case considered in \cref{sec:warm_up:classical}, we will need to
study the distribution of values~$\phi(Z)$ taken by a polynomial map~$\phi$ when the input is a $\rho$-biased random variable~$Z\sim \mathcal{N}_{\rho}(y)$.
This can be done by considering restrictions of~$\phi$ to random subsets of variables, which model the coordinates ``corrupted'' by the random process.

As the analytic rank satisfies all properties of \cref{lem:natural_rank}, Theorem~1.8 of Ref.~\cite{BrietCastroSilva:2022} immediately gives that the random restrictions of a high-rank polynomial map will also have high rank with high probability, summarized in the next theorem.
\begin{theorem}[Random-restriction theorem]\label{thm:restriction}
For every integer $d$ and every $\sigma,\eps\in (0,1]$, there exist~$\kappa = \kappa(d, \sigma)>0$ and~$R = R(d,\sigma,\eps)\in\N$ such that the following holds.
For every map~$\phi\in \Pol_{\leq d}(\F^n, \F^k)$ with~$\arank_d(\phi) \geq R$, we have that
\begin{equation*}
\Pr_{I\sim [n]_{\sigma}}\big[ \arank_d(\phi_{|I}) \geq \kappa\cdot \arank_d(\phi)\big] \geq 1 - \eps.
\end{equation*}
\end{theorem}

With the help of this theorem, it is easy to show that high-rank polynomial maps are approximately equidistributed even under biased inputs:
\begin{lemma}[Biased equidistribution lemma]\label[lemma]{lem:biased_equi_distribution}
For every integer $d$ and every $\rho, \eps \in (0, 1)$ there exists a constant $R_0 = R_0(d, \rho, \eps)>0$ such that the following holds.
If $\phi\in \Pol_{\leq d}(\F^n, \F^k)$ satisfies~$\arank_d(\phi) \geq R_0$, then
\begin{equation*}
\Pr_{Z \sim \mathcal N_\rho(0)}\big[ \phi(y+Z) = w \big] \leq \eps \quad \text{for all $y\in \F_p^n, w\in \F_p^k$.}
\end{equation*}
\end{lemma}
Recall that~$I\sim [n]_{\sigma}$ denotes the random process of sampling a subset~$I\subseteq [n]$ where each~$i\in [n]$ belongs to~$I$ with probability~$\sigma$, all events being mutually independent.
\begin{proof}
It suffices to prove the special case where both $y$ and $w$ are zero, that is
\begin{equation*}
\Pr_{Z \sim \mathcal N_\rho(0)}\big[ \phi(Z) = 0 \big] \leq \eps.
\end{equation*}
Indeed, for fixed~$y\in \F_p^n$ and~$w\in \F_p^k$, the map~$\tilde{\phi}: x \mapsto \phi(y+x)-w$ has the same degree and analytic rank as~$\phi$, and satisfies~$\tilde{\phi}(x) = 0$ if and only if~$\phi(y+x)=w$.

We can sample~$Z \sim \mathcal N_\rho(0)$ by first sampling~$I \sim [n]_{1-\rho}$ (the ``corrupted coordinates''), then sampling~$z$ uniformly from~$\F_p^I$ (the ``noise'') and setting~$Z_{|I} = z$, $Z_{|[n]\setminus I} = 0$;
thus
\begin{align*}
\Pr_{Z \sim \mathcal N_\rho(0)}\big[\phi(Z) = 0\big]
&= \Exp_{I\sim [n]_{1-\rho}} \Pr_{z\in \F_p^I} \big[\phi_{|I}(z) = 0 \big] \\
&\leq \Exp_{I\sim [n]_{1-\rho}} p^{-\arank_d(\phi_{|I})}.
\end{align*}
Let~$R = R(d, 1-\rho, \eps/2)$ and~$\kappa = \kappa(d, 1-\rho)$ be the constants guaranteed by Theorem~\ref{thm:restriction}.
If~$\arank_d(\phi) \geq R$, from that result we obtain
\begin{equation*}
\Exp_{I\sim [n]_{1-\rho}} p^{-\arank_d(\phi_{|I})} \leq \eps/2 + p^{-\kappa\cdot \arank_d(\phi)}.
\end{equation*}
Taking\footnote{Note that this bound is non-increasing on the value of~$p$, so we can obtain a field-independent bound by considering the smallest case~$p=2$.}
$R_0 = \max\big\{ R,\, \log_p(2/\eps)/\kappa \big\}$ we conclude that
\begin{equation*}
\Pr_{Z \sim \mathcal N_\rho(0)}\big[\phi(Z) = 0\big] \leq \Exp_{I\sim [n]_{1-\rho}} p^{-\arank_d(\phi_{|I})} \leq \varepsilon
\end{equation*}
whenever~$\arank_d(\phi) \geq R_0$, as wished.
\end{proof}

%% file: decoding/classical_hardness/proof_classical_hardness.tex
\section{The proof of \texorpdfstring{\cref{thm:classical_decoding_polynomial_maps}}{Theorem 3.1.2}}\label{sec:decoding:classical:main_result}
We are now ready to present the proof of \cref{thm:classical_decoding_polynomial_maps}, which proceeds by induction on the degree~$d$.
For degree-$1$ maps the result was already proven in the warm-up section,\footnote{It would also be possible to start the induction from the trivial base case~$d=0$ of constant maps, but we thought it is more instructive to first present the argument for degree-$1$ maps in order to gain some intuition.}
so let~$d \geq 2$ and assume the result holds for maps of degree at most~$d-1$.

Similar to the base case, we will divide the argument into two parts, corresponding to whether the analytic rank of~$\phi$ is ``high'' (the pseudorandom case) or ``low'' (the structured case).
The pseudorandom case immediately follows from \cref{lem:biased_equi_distribution}, the biased equidistribution lemma:
let~$R_0 = R_0(d, \rho, \eps)$ be the constant guaranteed by that lemma, and suppose that~$\arank_d(\phi) > R_0$.
Then for every~$x \in \F_p^k$ we have that
\begin{equation*}
\Pr_{Z \sim \mathcal N_\rho(0)}\big[ \phi \big(E(x)+Z\big) = x \big] \leq \eps,
\end{equation*}
and we conclude by averaging over all such~$x$.

For the structured case, suppose that~$\arank_d(\phi) \leq R_0$, and let~$\psi: \F_p^n \to \F_p^k$ be a map of degree at most~$d-1$ such that~$\Pr_{x\in \F_p^n} \big[\phi(x) = \psi(x)\big] \geq p^{-R_0}$.
For convenience, denote $\tilde{\phi} = \phi-\psi$, and let~$P\in \F_p[y_1,\dots,y_n,v_1,\dots,v_k]$ be the polynomial given by 
\begin{equation*}
P(y,v) = \ip{v, \tilde{\phi}(y)}.
\end{equation*}
This polynomial has nonnegligible bias:
\begin{equation*}
\bias(P) = \Exp_{y\in \F_p^n} \Exp_{v\in \F_p^k} \omega_p^{\ip{v, \tilde{\phi}(y)}} = \Exp_{y\in \F_p^n} \one \big[\tilde{\phi}(y)=0\big] \geq p^{-R_0}.
\end{equation*}
By \cref{thm:KaufmanLovett}, there exist~$s = s(p, d, R_0)\in \N$, a map~$\Gamma:\F_p^s\to\F_p$ and pairs $(h_1,w_1), \dots, (h_s,w_s)\in \F_p^n\times \F_p^k$ such that
\begin{equation*}
P(y,v) = \Gamma\big(\Delta_{(h_1,w_1)}P(y,v),\dots, \Delta_{(h_s,w_s)}P(y,v)\big).
\end{equation*}
Let~$f:\F_p^s\to \C$ be the map given by~$f(t) = \omega_p^{\Gamma(t)}$ and let~$\widehat f:\F_p^s\to \C$ be its Fourier transform,
\begin{equation*}
\widehat f(\alpha) = \Exp_{t\in \F_p^s}f(t)\omega_p^{-\ip{\alpha,t}}.
\end{equation*}
Since~$P$ is linear in $v$, the last~$k$ coordinates, it follows that
\begin{align*}
\Delta_{(h,w)}P(y,v) & = P(y+h, v+w) - P(y+h,v) + P(y+h,v) - P(y,v) \\
& = \ip{w, \tilde{\phi}(y+h)} + \ip{v, \Delta_h\tilde{\phi}(y)}.
\end{align*}
By the Fourier inversion formula (\cref{eq:Fourier_inversion_formula}), we conclude that
\begin{align*}
\omega_p^{P(y,v)} & = f\big(\Delta_{(h_1,w_1)}P(y,v),\dots, \Delta_{(h_s,w_s)}P(y,v)\big) \\
& = \sum_{\alpha\in \F_p^s}\widehat f(\alpha) \omega_p^{Q_\alpha(y) + \ip{v, \gamma_\alpha(y)}},
\end{align*}
where for~$\alpha\in \F_p^s$ we denote
\begin{align*}
Q_\alpha(y) & = \sum_{i=1}^s \ip{\alpha_i w_i, \tilde{\phi}(y+h_i)},\\
\gamma_\alpha(y) & = \sum_{i=1}^s\alpha_i \Delta_{h_i}\tilde{\phi}(y).
\end{align*}
Note crucially, that~$\deg(\gamma_\alpha) \leq d-1$ for all~$\alpha \in \F_p^s$, which is what will eventually allow us to apply the induction hypothesis.

It follows from our expression for~$\omega_p^{P(y, v)}$ that
\begin{align*}
\one[\phi(y) = x] & = \Exp_{v\in \F_p^k}\omega_p^{\ip{v, \phi(y) - x}} \\
& = \Exp_{v\in \F_p^k}\omega_p^{P(y,v) +\ip{v, \psi(y)-x}} \\
& = \sum_{\alpha\in \F_p^s}\widehat f(\alpha) \omega_p^{Q_\alpha(y)}\, \Exp_{v\in \F_p^k}\omega_p^{\ip{v,\, (\gamma_\alpha + \psi)(y) - x}}.
\end{align*}
Taking~$y = E(x) + Z$, we then obtain
\begin{align*}
\Pr\big[\phi\big(E(x) + Z\big) = x\big]
&=
\Exp_{x,Z}\one\big[\phi(E(x) + Z\big) = x\big]\\
&\leq
\sum_{\alpha\in \F_p^s}|\widehat f(\alpha)|\, \Exp_{x,Z}\big| \Exp_{v\in \F_p^k}\omega_p^{\ip{v,\, (\gamma_\alpha + \psi) (E(x) + Z) - x}}\big|\\
&\leq
\bigg( \sum_{\alpha\in \F_p^s}|\widehat f(\alpha)| \bigg) \max_{\alpha\in \F_p^s} \Exp_{x,Z} \one\big[(\gamma_\alpha + \psi)\big(E(x) + Z\big) = x\big]\\
&\leq
p^{s/2} \max_{\alpha\in \F_p^s} \Pr\big[(\gamma_\alpha + \psi)\big(E(x) + Z\big) = x\big],
\end{align*}
where we have used the Cauchy-Schwarz inequality and Parseval's identity in the last line.
Since~$\deg(\gamma_{\alpha} + \psi) \leq d-1$ and~$s$ ultimately depends only on~$p$, $d$, $\rho$ and~$\varepsilon$, by taking
\begin{equation*}
k \geq k_0(p,d,\rho,\eps) := k_0(p, d-1, \rho, \eps\, p^{-s/2})
\end{equation*}
we conclude from the induction hypothesis that
\begin{equation*}
\Pr\big[\phi\big(E(x) + Z\big) = x\big] \leq \varepsilon
\end{equation*}
in this structured case as well.
The theorem follows.

%% file: decoding/classical_hardness/high_characteristic_setting.tex
\section{The high-characteristic setting}\label{sec:classical:high_char}
The extremely slow decay found in \cref{thm:classical_decoding_nc0plus} makes the result qualitative rather than quantitative.
Yet, results from high characteristics seem to indicate room for improvements up to an exponential decay in the message length $k$.  
This is done by proving a ``high-characteristic'' analogue of \cref{thm:classical_decoding_polynomial_maps} with much better bounds, presented below;
as we see no reason to believe a result of this kind has a strong dependence on the characteristic of the finite field considered\footnote{While the result is stated in the setting of prime fields~$\F_p$, it easily generalizes to the case of non-prime finite fields $\F_q$, with only minor modifications in the proof.},
we believe that a theorem also holds for low-characteristic fields such as~$\F_2$.
\begin{theorem}[Exponential decay in high characteristic]\label{thm:classical:high_char}
For every $d\in \N$ and $\rho\in [0,1)$ there exist constants $C=C(\rho,d)$ and $c=c(\rho,d) > 0$ such that the following holds.
Let $p > d$ be a prime, and let $n$, $k$ be integers with $k\geq p$.
Then for every polynomial map $\phi:\F_p^n\to\F_p^k$ of degree at most~$d$ and every function $E:\F_p^k\to\F_p^n$ we have
\begin{equation}\label{eq:exponential_decay_high_char}
\Pr_{x\in \F_p^k, Z\sim \mathcal N_\rho(0)}\big[\phi\big(E(x) + Z\big) = x\big] \leq C e^{-c k/(\log k)^{d^2}}.
\end{equation}
\end{theorem}

\begin{remark}
The presence of the polylogarithmic term in the exponential in \cref{eq:exponential_decay_high_char} is due to a polylogarithmic loss when passing between two distinct notions of tensor rank in our proof of \cref{thm:classical:high_char}.
It is a widely-believed conjecture in additive combinatorics that these two notions of rank (see \cref{sec:decoding:classical:tensors}) are within a constant multiplicative factor of one another, in which case our proof would give an upper bound of the form $C e^{-c k}$ for the probability of correct message retrieval, which is optimal.
\end{remark}

We use induction on the degree~$d$ to prove \cref{thm:classical:high_char} and use the degree-$1$ case shown in \cref{sec:warm_up:classical} as the base case.
The inductive argument again relies on a structure-versus-randomness dichotomy based on a notion of rank associated with the polynomial map~$\phi$.
The better bounds we obtain stem from the fact that, in the high-characteristic case, one can work with tensors (i.e., multilinear forms) rather than with general polynomial maps.
In the quasirandom case of our argument, we can then use a stronger version of the random restriction theorem for the analytic rank of tensors, while in the structured case we use a connection between analytic rank and partition rank of tensors~\cite{MoshkovitzZhu:2022}.

%% file: decoding/classical_hardness/tensors_associated_to_polynomial_maps.tex
\section{Tensors associated to polynomial maps}\label{sec:decoding:classical:tensors}
Given a polynomial map $\phi: \F_p^n \to \F_p^k$ of degree at most $d$, we can define a $(d+1)$-tensor $T: (\F_p^n)^d \times \F_p^k \to \F_p$ associated to it by
\begin{equation*}
T(y_1, \dots, y_d, v) = \ip{v,\, \Delta_{y_1} \cdots \Delta_{y_d} \phi(0)}.
\end{equation*}
While not immediately obvious, the formula above is indeed linear in each variable separately and thus defines a tensor.
This follows from the fact that $\Delta_{y_1} \cdots \Delta_{y_d} \phi$ does not depend on the order of the derivatives, and that the polynomial map $\Delta_{y_1} \cdots \Delta_{y_{d-1}} \phi$ has degree at most~1 (since~$\phi$ has degree at most~$d$);
note that, if~$\psi$ is a linear map, then~$h\mapsto\Delta_h\psi$ is linear in~$h$.

If the characteristic~$p$ of the field in strictly higher than the degree~$d$, then we also have the \emph{integration formula}
\begin{equation*}
\phi(y) = \frac{1}{d!} T(y, \dots, y, \cdot) + \psi(y) \quad \text{for all $y\in \F_p^n$,}
\end{equation*}
where $y$ is repeated $d$ times inside $T$ and $\psi$ is a polynomial map of degree at most~$d-1$.
This follows from the (discrete) Taylor expansion theorem, and allows us to pass back and forth between tensors and polynomial maps.

We will use the following two notions of rank for tensors, originally introduced by \citeauthor{GowersWolf:2011}~\cite{GowersWolf:2011} and by \citeauthor{Naslund:2020}~\cite{Naslund:2020}, respectively.
\begin{definition}[Tensor analytic rank]\label{def:tensor_analytic_rank}
Let $X_1,\dots,X_r$ be positive integers and~$T:\F_p^{X_1}\times\cdots\times\F_p^{X_r}\to \F_p$ be an $r$-tensor.
The bias of~$T$ is defined as
\begin{equation*}
\bias(T) = \Exp_{x_1\in \F_p^{X_1},\dots,x_r\in \F_p^{X_r}} \omega_p^{T(x_1,\dots,x_r)}.
\end{equation*}
The bias is always real and positive\footnote{It is not hard to show that $\bias(T) = \Pr_{x_1\in \F_p^{X_1},\dots,x_{r-1}\in \F_p^{X_{r-1}}} \one\big[T(x_1,\dots, x_{r-1}, \cdot) \equiv 0\big]$.},
and the analytic rank of~$T$ is defined by
\begin{equation*}
\arank(T) = -\log_{p} \bias(T).
\end{equation*}
\end{definition}

\begin{definition}[Partition rank]\label{def:partition_rank}
For positive integers $X_1,\dots,X_r$, a nonzero $r$-tensor $T:\F_p^{X_1}\times\cdots\times\F_p^{X_r}\to \F_p$ is said to have partition rank~1 if there exists a nonempty strict subset $I\subset [r]$ and tensors $S: \prod_{i\in I}\F^{X_i}\to \F$ and $R: \prod_{i\in [r]\setminus I}\F^{X_i}\to \F$ such that~$T$ can be factored as $T = SR$.
The partition rank of~$T$, denoted $\prank(T)$, is defined as the least $m\in \N$ such that there is a decomposition $T = T_1 + \cdots + T_m$ where each $T_i$ has partition rank~1.
\end{definition}

While these two notions of rank are defined in very different ways, it turns out that they are intimately related to each other.
\citeauthor{Lovett:2019} showed that for all tensors~$T$, it holds that $\arank(T) \leq \prank(T)$~\cite{Lovett:2019}.
It is a well-known open problem to determine whether a similar inequality holds in the converse direction, up to an absolute multiplicative factor.
\citeauthor{MoshkovitzZhu:2022} proved that the relation between these two rank functions is at worst quasilinear~\cite{MoshkovitzZhu:2022}.
\begin{theorem}[Moshkovitz--Zhu]\label{thm:MoshkovitzZhu}
For every $r\geq 2$ there exists $L_r>0$ such that for every $r$-tensor~$T$ over any finite field, we have
\begin{equation}\label{eq:MoshkovitzZhu}
\arank(T) \leq \prank(T) \leq L_r\arank(T)\,\log^{r-1}\big(1 + \arank(T)\big).
\end{equation}
\end{theorem}

This result will be an important ingredient in our proof of \cref{thm:classical:high_char};
we note that the decay obtained could be improved to $C e^{-ck}$ if \cref{thm:MoshkovitzZhu} were proven without the polylogarithmic factor on the right-hand side of~\cref{eq:MoshkovitzZhu}.
Another important ingredient is the following random restriction theorem for tensors~\cite{BrietCastroSilva:2022}, stated here for the special case of the analytic rank.

\begin{theorem}[Tensor random restriction theorem]\label{thm:tensor_restriction}
For every integer $d$ and every $\sigma \in (0,1]$, there exist constants $C,\kappa>0$ such that for any order-$d$ tensor~$T$ over any field, we have that
\begin{equation*}
\Pr_{I\sim [n]_\sigma}\big[\arank(T_{|I}) \geq \kappa\cdot \arank(T)\big]
\geq
1 - Ce^{-\kappa\,\arank(T)}.
\end{equation*}
\end{theorem}

%% file: decoding/classical_hardness/proof_high_characteristics.tex
\section{The proof of Theorem~\ref{thm:classical:high_char}}\label{sec:decoding:classical:proof_high_char}
The proof will proceed by induction on the degree of the polynomial map.
Recall that \cref{sec:warm_up:classical} provides the proof for the base case of degree-1 maps.

Let $\phi: \F_p^n \to \F_p^k$ be a polynomial map of degree at most $d$, with $2 \leq d < p$, and suppose the theorem holds for polynomial maps of degree at most $d-1$.
Define the $(d+1)$-tensor $T: (\F_p^n)^d \times \F_p^k \to \F_p$ by
\begin{equation*}
T(y_1, \dots, y_d, v) = \ip{v,\, \Delta_{y_1} \dots \Delta_{y_d} \phi(0)}.
\end{equation*}
We split the analysis into two cases, depending on whether the analytic rank of~$T$ is above or below some cutoff value $r = \Theta\big(k/(\log k)^{d^2}\big)$.

\subsection{Pseudorandom case}
Assume that $\arank(T) \geq r$.
We will show that the probability
\begin{equation} \label{eq:pseudo_case}
\Pr_{Z\sim \mathcal N_\rho(0)}\big[\phi\big(E(x) + Z\big) = x\big]
\end{equation}
decays exponentially in $\arank(T)$ for every $x\in\F_p^n$;
we then conclude the pseudorandom case by averaging over all $x$.
Now fix some $x\in \F_p^n$.
As before, we write
\begin{align*}
    \Pr_{Z\sim \mathcal N_\rho(0)}\big[\phi\big(E(x) + Z\big) = x\big]
    & = \Exp_{I\sim [n]_{1-\rho}} \Pr_{y\in \F_p^I}\big[\phi\big(E(x) + y\big) = x\big] \\
    & = \Exp_{I\sim [n]_{1-\rho}} \Exp_{y\in \F_p^I, v\in \F_p^k} \omega_p^{\ip{v,\, \phi(E(x)+y) - x}}.
\end{align*}
Note that we can write $\phi(E(x)+y) - x = \phi(y) + \psi(y)$, where
\begin{equation*}
\psi(y) := \Delta_{E(x)}\phi(y) - x
\end{equation*}
has degree at most $d-1$.
Using this identity and the triangle inequality, it follows that
\begin{align*}
    \Pr_{Z\sim \mathcal N_\rho(0)}\big[\phi\big(E(x) + Z\big) = x\big]
    & = \Exp_{I\sim [n]_{1-\rho}} \Exp_{y\in \F_p^I, v\in \F_p^k} \omega_p^{\ip{v,\, \phi(y) + \psi(y)}} \\
    & \leq \Exp_{I\sim [n]_{1-\rho}}\Exp_{v\in \F_p^k} \big| \Exp_{y\in \F_p^I} \omega_p^{\ip{v,\, (\phi + \psi)(y)}} \big|.
\end{align*}

Repeated applications of the Cauchy-Schwarz inequality (or equivalently, the monotonicity property of the Gowers uniformity norms (\cref{eq:Uk_nesting}) shows that, for any fixed $v\in \F_p^k$, $I\subseteq [n]$, we have
\begin{align*}
    \big| \Exp_{y\in \F_p^I} \, \omega_p^{\ip{v,\, (\phi + \psi)(y)}} \big|
    \leq \big( \Exp_{y_0, y_1, \dots, y_d\in \F_p^I} \, \omega_p^{\ip{v,\, \Delta_{y_1} \dots \Delta_{y_d}(\phi + \psi)(y_0)}} \big)^{1/2^d}.
\end{align*}
We then conclude that
\begin{align*}
    \Pr_{Z\sim \mathcal N_\rho(0)} & \big[\phi\big(E(x) + Z\big) = x\big] \\
    & \leq \Exp_{I\sim [n]_{1-\rho}}\Exp_{v\in \F_p^k} \left( \Exp_{y_0, y_1, \dots, y_d\in \F_p^I} \, \omega_p^{\ip{v,\, \Delta_{y_1} \dots \Delta_{y_d}(\phi + \psi)(y_0)}} \right)^{1/2^d} \\
    & \leq \Exp_{I\sim [n]_{1-\rho}} \left(\Exp_{v\in \F_p^k} \Exp_{y_0, y_1, \dots, y_d\in \F_p^I} \, \omega_p^{\ip{v,\, \Delta_{y_1} \dots \Delta_{y_d}(\phi + \psi)(y_0)}} \right)^{1/2^d},
\end{align*}
where we applied H\"{o}lder's inequality once (or, alternatively, Cauchy-Schwarz $d$ further times).

We now need to relate this last expression to the analytic rank of~$T$.
Derivating~$d$ times a polynomial map of degree at most $d-1$ gives the zero map, and so
$\Delta_{y_1} \dots \Delta_{y_d}\psi(y_0) \equiv 0$.
Moreover, since $\deg(\phi) \leq d$, the $d$-th derivative $\Delta_{y_1} \dots \Delta_{y_d}\phi$ is a constant map.
We conclude that
\begin{equation*}
\Exp_{v\in \F_p^k} \Exp_{y_0, y_1, \dots, y_d\in \F_p^I} \omega^{\ip{v,\, \Delta_{y_1} \dots \Delta_{y_d}(\phi + \psi)(y_0)}}
= \Exp_{v\in \F_p^k} \Exp_{y_1, \dots, y_d\in \F_p^I} \omega^{\ip{v,\, \Delta_{y_1} \dots \Delta_{y_d}\phi(0)}}.
\end{equation*}
For each $v\in \F_p^k$, let $S(v)$ be the $d$-tensor given by $T(\cdot,\dots,\cdot,v)$.
Then the above is precisely the bias of the restricted tensor $S(v)_{|I^d}$, averaged over~$v$, which (by definition) equals the average of $p^{-\arank(S(v)_{|I^d})}$.
The probability in \cref{eq:pseudo_case} is then bounded from above by
\begin{equation*}
\Exp_{v\in \F_p^k}\Exp_{I\sim [n]_{1-\rho}} p^{-\arank(S(v)_{|I^d})/2^d}.
\end{equation*}
Theorem~\ref{thm:tensor_restriction} now implies that for some absolute constant $C = C(d,\rho)>0$, the last quantity is bounded from above by $Cp^{-\arank(T)/C}$.

\subsection{Structured case}
Now we assume that $\arank(T) < r$.

Denote the partition rank of $T$ by $s := \prank(T)$.
\cref{thm:MoshkovitzZhu} shows that $s \leq L_{d+1} r (\log r)^d$, where $L_{d+1}$ is a universal constant.
We can then write
\begin{equation*}
T(y_{[d]}, v) = \sum_{i=1}^s R_i(y_{I_i}) S_i(y_{I_i^c}, v)
\end{equation*}
for some nonempty sets $I_i \subseteq [d]$, $|I_i|$-tensors $R_i$ and $(d-|I_i|+1)$-tensors $S_i$.
Since $d<p$, by Taylor's expansion theorem we have that
\begin{equation*}
\phi(y) = \frac{1}{d!} \Delta_y \dots \Delta_y \phi(0) + \psi_0(y), \quad \deg(\psi_0) < d.
\end{equation*}
Define $q_i: \F_p^n \to \F_p$, $\psi_i: \F_p^n \to \F_p^k$, for $i\in [s]$, by
\begin{equation*}
q_i(y) = \frac{1}{d!} R_i(y_{I_i}), \quad \ip{v, \psi_i(y)} = S_i(y_{I_i^c}, v),
\end{equation*}
and note that $\deg(\psi_i) < d$ for all $i\in [s]$.
By the definition of $T$ we conclude that
\begin{equation*}
\phi(y) = \psi_0(y) + \sum_{i=1}^s q_i(y) \psi_i(y), \quad \text{with $\deg(\psi_i) < d$ for $0\leq i\leq s$.}
\end{equation*}

Let $\mathcal{A} = \{A_1,\dots,A_m\}$ be the partition of $\F_p^n$ given by the level sets of the polynomial map $(q_1,\dots,q_s):\F_p^n\to\F_p^s$;
note that $m \leq p^s \leq p^{L_{d+1} r (\log r)^d}$.
For each $j\in [m]$, $\phi$ will coincide on $A_j$ with a polynomial map $\psi_{A_j}: \F_p^n \to \F_p^k$ of degree at most $d-1$;
simply substitute the $q_i(y)$ in the formula above by their value on $A_j \in \mathcal{A}$.
Define the random events
\begin{equation*}
\mathcal{E}_j = \big\{ E(x)+Z \in A_j:\, x\sim \mathcal{U}(\F_p^k),\, Z\sim \mathcal{N}_\rho(0)\big\}, \quad j\in [m].
\end{equation*}
Since these events partition the probability space, it follows that
\begin{align*}
    \Pr_{x\in \F_p^k, Z\sim \mathcal N_\rho(0)} &\big[\phi\big(E(x) + Z\big) = x\big] \\
    &= \sum_{i=1}^m \Pr_{x, Z}\big[\phi\big(E(x) + Z\big) = x \wedge \mathcal{E}_j\big] \\
    &= \sum_{i=1}^m \Pr_{x, Z}\big[\psi_{A_j}\big(E(x) + Z\big) = x \wedge \mathcal{E}_j\big] \\
    &\leq m \cdot \max_{1\leq j\leq m} \Pr_{x, Z}\big[\psi_{A_j}\big(E(x) + Z\big) = x\big] \\
    &\leq p^{L_{d+1} r (\log r)^d} \cdot \max_{\deg(\psi)<d} \Pr_{x, Z}\big[\psi\big(E(x) + Z\big) = x\big],
\end{align*}
where the last maximum is over all polynomial maps $\psi: \F_p^n \to \F_p^k$ of degree at most $d-1$.
By the induction hypothesis we have that this maximum probability is at most $C' e^{-c' k/(\log k)^{(d-1)^2}}$, where $C' = C(d-1, \rho)$ and $c' = c(d-1, \rho)$;
we conclude that
\begin{align*}
    \Pr_{x\in \F_p^k, Z\sim \mathcal N_\rho(0)} &\big[\phi\big(E(x) + Z\big) = x\big] \\
    &\leq C'\exp\left((\log p) L_{d+1} r (\log r)^d - \frac{c'k}{(\log k)^{(d-1)^2}}\right).
\end{align*}

Taking
\begin{equation*}
r = \frac{c'}{2L_{d+1}} \frac{k}{(\log k)^{d^2}},
\end{equation*}
and using our assumptions $k \geq p$ and $d\geq 2$, we have that
\begin{align*}
    (\log p) L_{d+1} r (\log r)^d
    &\leq (\log k) L_{d+1} \frac{c'}{2L_{d+1}} \frac{k}{(\log k)^{d^2}} (\log k)^d \\
    &= \frac{c'}{2} \frac{k}{(\log k)^{d^2-d-1}} \\
    &\leq \frac{c'}{2} \frac{k}{(\log k)^{(d-1)^2}}.
\end{align*}
We conclude that
\begin{equation*}
    \Pr_{x\in \F_p^k, Z\sim \mathcal N_\rho(0)} \big[\phi\big(E(x) + Z\big) = x\big]
    \leq C'\exp\bigg(-\frac{c'k}{2(\log k)^{(d-1)^2}}\bigg)
\end{equation*}
in this case, and the theorem follows.

%% file: decoding/classical_hardness/discussion_classical_hardness.tex
\section{Reflections and outlook}
This chapter proved that polynomial maps of constant-degree cannot decode any error-correcting code for any positive error rate with constant success probability (\cref{thm:classical_decoding_polynomial_maps}).
As a result, any $\nc^0[\oplus]$ circuit could also not do so (\cref{thm:classical_decoding_nc0plus}). 
However, the decay in the theorems is slow, making the results qualitative rather than quantitative. 
We conjecture that this decay results from the used techniques, and that in general an exponential decay holds (\cref{thm:classical:high_char}). 
We provided evidence for this conjecture by studying the high-characteristic case. 

The main strength of \cref{thm:classical_decoding_polynomial_maps} is that it holds for any code and for any positive error rate.
Complementary results are known for restricted classes of codes, and also for when the error rate tends to~$1/2$.

A code $E:\F_2^k\to\F_2^n$ is \emph{$t$-wise independent} if, for any size-$t$ subset of coordinates~$S\subseteq [n]$ and a uniformly random $X\in \F_2^k$, the restriction $E(X)_{|S}$ is uniformly distributed over~$\F_2^{|S|}$.
Many codes have this property;
for instance, the dual code of a linear code of distance~$d$ is $(d-1)$-wise independent.

Under the same noise model considered in this chapter, \citeauthor{LeeViola:2017}~\cite{LeeViola:2017}, using earlier work of \citeauthor{Viola:2009}~\cite{Viola:2009}, showed that~$\nc^0[\oplus]$ circuits cannot distinguish a corrupted uniformly random codeword of an $\omega(1)$-wise independent linear code from a uniformly random element of~$\F_2^n$.
Note that this problem is formally easier than (list) decoding.

However, the result of \citeauthor{LeeViola:2017} does not cover the Hadamard code, as it is not even 3-wise independent.
Indeed, the Hadamard code is also easy to distinguish, as it contains the sub-code $(x_1,x_2, x_1+x_2)$.
Since the parity of these three bits is always zero, the parity under noise is biased towards zero and is therefore easily distinguished from the parity of a random string.

Future work can explore the use of the techniques in this chapter to the situation sketched by \citeauthor{LeeViola:2017} to see how their result extends to a broader class of codes. 
Additionally, it is interesting to see how our quantitative bounds can be improved, for instance, using the techniques by \citeauthor{LeeViola:2017}. 

The techniques used in this chapter use novel ideas from additive combinatorics. 
These techniques are of greater interest, and it would therefore be interesting to see where else they can be applied. 

Finally, an interesting direction for future work is to see if we can prove similar results for other types of circuits. 
An example is $\nc^0[\oplus]$-circuits that additionally have access to a fixed number of unbounded-fan-in AND- or OR-gates. 
These circuits are strictly stronger than $\nc^0[\oplus]$, yet do not admit the power that $\ac^0[\oplus]$-circuits have. 
Another potential class of circuits would be those using bounded-fan-in gates and having non-constant depth $\bigo(\log\log n)$. 
In these circuits, before the parity functions, each output bit can only depend on $\bigo(\log n)$ input bits, instead of on a constant number of bits. 
The polynomials associated with these classes can have non-constant degree $\bigo(\log n)$. 

%% file: decoding/quantum_linear_codes.tex
\chapter[Decoding the Hadamard code]{Decoding the Hadamard code with quantum circuits}\label{chp:decoding:quantum}

In this chapter, we present a constant-depth quantum algorithm that can list-decode a heavily corrupted Hadamard codeword. 
To provide intuition for the algorithm, we start by formulating a non-local game that can list-decode a corrupted Hadamard code. 
We then turn this non-local game into a quantum circuit. 
With this quantum circuit, we arrive at a separation between quantum and conventional constant-depth circuits. 
We furthermore revisit our quantum circuit and show how it can be used to implement majority gates. 
We conclude by providing directions for quantum circuits for different error-correcting codes. 

\input{decoding/quantum_decoding/outline_quantum_decoding}
\input{decoding/quantum_decoding/warm_up_decoding_hadamard}

\input{decoding/quantum_decoding/details_quantum_decoding}
\input{decoding/quantum_decoding/complexity_quantum_decoding}

\input{decoding/quantum_decoding/separation_decoding}

\input{decoding/quantum_decoding/list_decoding_to_majority}

\input{decoding/quantum_decoding/circuit_hadamard_higher_characteristic}
\input{decoding/quantum_decoding/discussion_quantum_decoding}

%% file: decoding/quantum_decoding/outline_quantum_decoding.tex
\section{Chapter overview}
The main result of this chapter is a family of quantum circuits that can retrieve a codeword such that, with high probability, its encoding is close (in Hamming distance) to the corrupted message. 
\begin{theorem}[Decoding Hadamard code with {$\qnc^0[\oplus]$}]\label{thm:quantum_decoding_Hadamard_circuit}
There is a family of $\qnc^0[\oplus]$-circuits $(\mathcal C_n)_{n\in \N}$ such that the following holds.
Let $k\in \N$, $n = 2^k$ and $\eps \in (0,1/2]$.
Then, for any $y\in \F_2^n$ and $x\in \F_2^k$ satisfying $d\left(y, H(x)\right)\leq (\frac{1}{2} - \eps)n$, on input~$y$ the circuit~$\mathcal C_n$ returns~$x$ with probability~$\Omega(\eps^2)$.
\end{theorem}
We note that the bound~$\Omega(\eps^2)$ obtained in the theorem is optimal, since in general there can be $\Theta(\eps^{-2})$ messages~$x\in \F_2^k$ satisfying $d\big(y, H(x)\big)\leq (\frac{1}{2} - \eps)n$.
Note that this bound is only nontrivial if $\eps = \Omega(1/\sqrt{n})$, as there are~$n$ possible messages.

The family of quantum circuits of \cref{thm:quantum_decoding_Hadamard_circuit} originates directly from an $n$-player non-local game presented as warm-up in \cref{sec:warm_up:quantum:Hadamard}.
The players employing a quantum strategy share $k$ GHZ states and are given as input coordinates of the corrupted Hadamard codeword. 

In \cref{sec:quantum:details_quantum_algorithm}, we turn the optimal quantum strategy for this non-local game into a quantum circuit. 
For the GHZ states, we use a technique by \citeauthor{BeneWattsKothariSchaefferTal:2019}~\cite{BeneWattsKothariSchaefferTal:2019}. 
We first prepare a \emph{poor man's cat state}: $\tfrac{1}{\sqrt{2}}(\ket{z}+\ket{\bar{z}})$ for some binary vector $z$, and then correct it to a standard GHZ state. 

We input the coordinates of the corrupted Hadamard codeword in the circuit using controlled phase flips. 
By carefully choosing the controls, we can link the $i$-th (possibly corrupted) bit with the $\ket{i}$-state. 
Standard techniques using stacked Toffoli-gates and auxillary qubits require depth $\bigo(\log\log n)$. 
Instead, we use an unbounded-fan-in AND-gate 
\begin{equation*}
\ket{x_1}\hdots \ket{x_k}\ket{b}\quad\mapsto\quad \ket{x_1}\hdots \ket{x_k}\ket{\text{AND}(x_1,\hdots,x_k)\oplus b}
\end{equation*}
to apply the phase flips. 
Conjugating the inputs of the AND-gate with $X$-gates ensures we target only the $\ket{i}$-state. 
We implement this gate using a constant-depth exact OR-routine by \citeauthor{TakahashiTani:2013}~\cite{TakahashiTani:2013}.

We consider the size and depth of the quantum circuit in \cref{sec:decoding:quantum:complexity}, and show that the circuits have size $\bigo(n^2\log n)$ and depth $65$.

\cref{sec:separation_quantum_classical} discusses a separation between conventional and quantum circuits based on decoding corrupted error-correcting codes. 
Note that running the circuit in parallel outputs the desired list of possible messages. 
We then introduce the \textit{List-Hadamard problem} to prove a separation between $\nc^0[\oplus]$ and $\qnc^0[\oplus]$ for decoding the Hadamard code for any error rate $\delta>0$, see also \cref{thm:quantum_classical_separation}. 

In the high-error regime where the parameter $\delta$ approaches the information-theoretic limit of $1/2$ (which is relevant for hardness amplification), a stronger separation follows by combining \cref{thm:quantum_decoding_Hadamard_circuit} with a result of Sudan showing the hardness of noisy decoding by $\ac^0[\oplus]$-circuits, implying a separation between $\ac^0[\oplus]$ and $\qnc^0[\oplus]$. 
We prove this separation in \cref{sec:quantum:list_decoding_majority} and also give a quantum circuit for Majority, based on a quantum circuit for list decoding. 

Next, \cref{sec:quantum:Hadamard_code_higher_fields} discusses the extension of the Hadamard circuit to higher field characteristics, similar to how \cref{thm:classical_decoding_polynomial_maps} extends \cref{thm:classical_decoding_nc0plus}. 

%% file: decoding/quantum_decoding/warm_up_decoding_hadamard.tex
\section{Quantum decoding the Hadamard code in a non-local game}\label{sec:warm_up:quantum:Hadamard}
Our quantum algorithm is inspired by the analysis of a particular non-local game.
In a non-local game, a referee randomly sends questions to a set of players, according to a probability distribution known to the players in advance.
Then, without communicating with each other, the players individually answer the referee.
Finally, the referee determines whether the players win or lose based solely on the questions and answers.
The rule used by the referee is known to the players in advance as well.
With a (deterministic) conventional strategy, the players decide before the game starts what to answer to each possible question.
With a quantum strategy, the players base their answers on the outcomes of local measurements of their respective parts of a shared entangled state.
We refer to~\cite{CHTW:2004} for further background on non-local games.

Let $H:\F_2^k\to \F_2^n$ be the Hadamard code and let $\eps \in (0,1/2)$ be some constant independent of~$n$.
We identify the codewords~$H(x)$ with functions $\F_2^k\to\F_2$ given by~$H(x)(y) = \ip{x,y}$.
We consider the following $n$-player non-local game, which we shall refer to as the \emph{Hadamard game}:
Each player is labeled uniquely with an element in~$\F_2^k$.
The referee picks a uniformly chosen message~$x\in \F_2^k$ and randomly corrupts the codeword~$H(x)$ using the binary symmetric channel with error rate $1/2 - \eps$, resulting in a function~$c:\F_2^k\to \F_2$.
He then sends player~$y$ the value~$c(y)$.
The players each return a string in~$\F_2^k$ and they win the game if the sum of their answers equals~$x$.

Let the $n$ players share an $n$-partite GHZ state of local dimension~$2^k$:
\begin{equation*}
\frac{1}{\sqrt{n}}\sum_{y\in \F_2^k} \ket{y}\otimes\hdots\otimes\ket{y}.
\end{equation*}
Consider the following strategy inspired by the famous Bernstein-Vazirani algorithm~\cite{BernsteinVazirani:1997}: 
\begin{enumerate}
    \item Upon receiving input~$c(y)$, player~$y$ applies a conditional phase flip on their part of the shared state:
    \begin{equation}\label{eq:Hadamard_game:conditional_phase}
    \ket{z} \mapsto \begin{cases}
        (-1)^{c(y)}\ket{z} & \text{if } z=y, \\ 
        \ket{z} & \text{otherwise.}
    \end{cases}
    \end{equation}
    \item Each player applies a $k$-qubit Hadamard gate to their local register;
    \item Each player measures their local register in the computational basis;
    \item Each player returns the measurement results. 
\end{enumerate}

\begin{theorem}\label{thm:quantum:strategy_proof}
    The players win the Hadamard game with probability $\Omega(\eps^2)$, if they follow the strategy outlined above. 
\end{theorem}
\begin{proof}
Once all players have applied their conditional phase flips, they share the state
\begin{equation}\label{eq:quantum:hadamard:final}
\frac{1}{\sqrt{n}}\sum_y (-1)^{c(y)}\ket{y}\otimes\hdots\otimes\ket{y}. 
\end{equation}
The $k$-qubit Hadamard gates map this state to
\begin{equation*}
n^{-(n+1)/2}\sum_{y\in \F_2^k} \sum_{b_1,\hdots,b_n\in\F^k_2}(-1)^{c(y)}(-1)^{\langle y,b_1+\cdots +b_n\rangle}\ket{b_1}\otimes\hdots\otimes\ket{b_n}. 
\end{equation*}

The probability that the measurement results sum to some string $z\in\F^k_2$ is now given by
\begin{align}
\Pr\bigg[\sum_{i=1}^nb_i= z\bigg] & = \frac{1}{n^{(n+1)}}\sum_{b_1+\cdots+b_n=z}\bigg| \sum_{y\in\F_2^k} (-1)^{c(y)+\langle y,z\rangle}\bigg|^2 \nonumber \\
& = \left( 1 - 2\frac{d\big(c, H(z)\big)}{n}\right)^2. \label{eq:quantum:hadamard:succes_probability}
\end{align}
It follows from the Chernoff bound~\cite{HagerupRub:1990} that for fixed~$x\in \F_2^k$ and the random~$c$ obtained by corrupting the codeword~$H(x)$,
\begin{equation*}
\Pr\left[\frac{d\big(c, H(x)\big)}{n} \geq \frac{1-\eps}{2}\right] \leq \exp(-C\eps^2 n).
\end{equation*}
Hence, by the union bound, for fixed $\eps \in (0,1/2)$, the players win with probability at least $C\eps^2$, where the probability is taken over the message~$x$, the noise corrupting the codeword~$H(x)$ and the measurements done by the players.
\end{proof}

Note that this strategy succeeds with probability~$C\eps^2$ for \emph{every}~$x$ and whenever at most any $(1/2 - \eps)$-fraction of the coordinates of~$H(x)$ are flipped, independent of the error-model.

%% file: decoding/quantum_decoding/details_quantum_decoding.tex
\section{Details of quantum algorithm}\label{sec:quantum:details_quantum_algorithm}
Below we give more details on how to generate the GHZ states, as well as on how to implement the controlled phase gates.

\subsection{Generating GHZ states}
Preparing the GHZ state
\begin{equation*}
    \ket{GHZ}=\frac{1}{\sqrt{2}}\left(\ket{0}^{\otimes n}+\ket{1}^{\otimes n}\right)
\end{equation*}
requires depth $\Omega(\log n)$ given an all-to-all connectivity. 
This exceeds the constant-depth requirement of our circuits. 
Instead, we generate an intermediate state and then use the conventional parity gates to implement correction terms to correct the intermediate state in a GHZ state. 

\begin{figure}[ht]
	\centering
\begin{quantikz}
\lstick{$\ket{0}_1$} & \gate{H} & \ctrl{1} & \qw & \qw & \qw & \qw & \qw \\
\lstick{$\ket{0}_2$} & \qw & \targ{} & \targ{} & \meter{} & \gate[style={white}][1.5cm]{d_1}\setwiretype{c} & \cwbend{1} \\
\lstick{$\ket{0}_3$} & \gate{H} & \ctrl{1} & \ctrl{-1} & \qw & \qw & \targ{} & \qw \\
\lstick{$\ket{0}_4$} & \qw & \targ{} & \targ{} & \meter{} & \gate[style={white}][1.5cm]{d_1\oplus d_2}\setwiretype{c}& \cwbend{1} \\
\lstick{$\ket{0}_5$} & \gate{H} & \qw & \ctrl{-1} & \qw & \qw & \targ{} & \qw  
\end{quantikz}
	\caption{The quantum circuit to generate a 3-qubit GHZ state. 
    First, we prepare a poor man's cat state $\frac{1}{\sqrt{2}}(\ket{z}+\ket{\bar{z}})$ with each $z\in\F_2^3$ equally likely to be found. 
    The parity gates compute a prefix sum on the measurement results $d_1$ and $d_2$ and determine if a qubit has to be flipped to obtain the GHZ state.}
	\label{fig:q_circuit:poor_man_cat_state}
\end{figure}
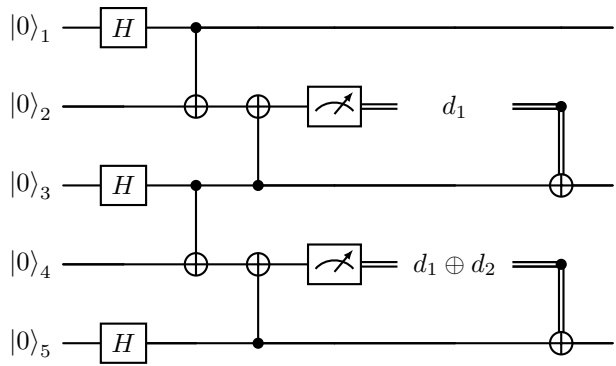

Starting with $2n-1$ qubits, we apply Hadamard gates to all $2i-1$ qubits for $i\in[n]$. 
Next, we apply two layers of parallel CNOT-gates: 
In the first layer, from qubit $2i-1$ to qubit $2i$, for $i\in[n-1]$;
In the second layer, from qubit $2i+1$ to qubit $2i$, for $i\in[n-1]$. 
Next, all even-numbered qubits are measured, giving measurement results $d_i$, $i\in[n-1]$. 
The resulting poor man's cat state is then corrected to a GHZ state by applying an $X$-gate to qubit $i$ based on a prefix sum computation, that is, an $X$-gate is applied to qubit $2i-1$ if and only if $\sum_{j=1}^{i-1} d_j\bmod 2\equiv 1$ for $i\in[n]$. 
\cref{fig:q_circuit:poor_man_cat_state} shows the quantum circuit to generate a 3-qubit GHZ state. 

With this construction, the depth of the circuit remains constant, independent of~$n$. 
Also note that this circuit corresponds to a Clifford ladder, which is discussed later in \cref{sec:LAQCC:Clifford_circuits}.

\subsection{Quantum fanout gate}
\begin{figure}
    \centering
    \begin{quantikz}
    \lstick{$\ket{\phi}$}\slice[style=gray]{t=0} & \ctrl{3}\slice[style=gray]{t=1} & \slice[style=gray]{t=2} & \slice[style=gray]{t=3} & & \slice[style=gray]{t=4} & \slice[style=gray]{t=5} & \slice[style=gray]{t=6} & \slice[style=gray]{t=7} & \gate{Z}\slice[style=gray]{t=8} & \\
    \lstick{$\ket{x_1}$} & & & & \targ{} & & & & & & \\
    \lstick{$\ket{x_2}$} & & & & & \targ{} & & & & & \\
    \lstick[wires=3]{$\ket{GHZ_3}$} & \targ{} & \meter{} & \ctrl[vertical wire=c]{1}\setwiretype{c} & \setwiretype{n} &&&&&& \\
     & & & \targ{} & \ctrl{-3} & & \gate{H} & \meter{} & \gate[2,disable auto height]{\begin{array}{c}\text{Parity}\setwiretype{c} \\ d_1\oplus d_2\end{array}} & \ctrl[vertical wire=c]{-4} \\
     & & & \targ{}\vcw{-1} & & \ctrl{-3} &  \gate{H} & \meter{} & \setwiretype{c} & \setwiretype{n} & 
    \end{quantikz}
    \caption{Implementation of a quantum fanout gate with one control qubit $\ket{\phi}$ and two target qubits $\ket{x_1}$ and $\ket{x_2}$. 
    Only single- and two-qubit gates and conventional parity gates are used. 
    The bottom three qubits are in the GHZ$_3$ state.
    The dotted lines denote time steps and which gates can be applied in parallel.}
    \label{fig:q_circuit:non_local_cnot}
\end{figure}
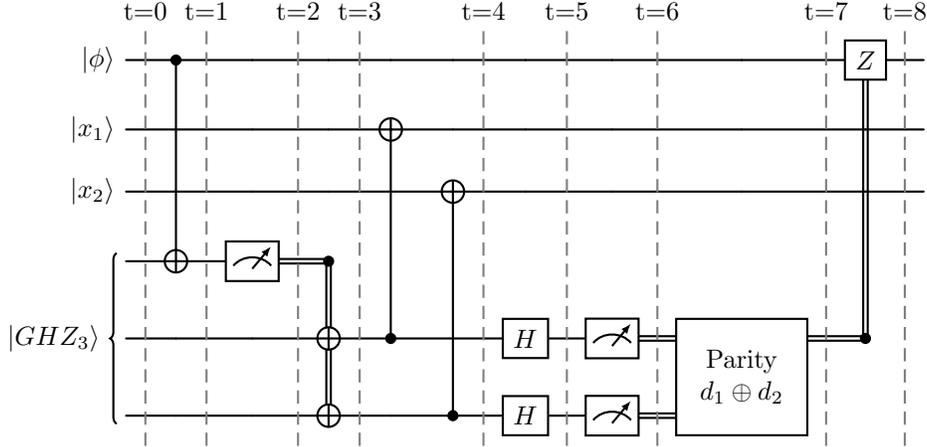

We will use a quantum fanout gate to implement the controlled phase flips.
For our construction, we will use ideas from distributed quantum computing. 
\citeauthor{Eisert:2000} and \citeauthor{YimsiriwattanaLomonaco:2004} introduced a non-local CNOT-gate, using single-qubit gates, CNOT-gates and shared GHZ states~\cite{Eisert:2000,YimsiriwattanaLomonaco:2004}. 
We extend their construction to a quantum fanout gate, by using GHZ states shared by more parties. 
A circuit for the quantum fanout gate for $n=3$ is given in \cref{fig:q_circuit:non_local_cnot}.
The last $Z$-gate is applied only if the parity of all measurement results equals $1$. 
The time steps denote which gates can be applied in parallel. 

Later, \citeauthor{PhamSvore:2013} introduced another way to implement a quantum fanout gate~\cite[Figure~4]{PhamSvore:2013}.
At first sight, their circuit seems to have a shallower depth.
Yet, their circuit uses Bell-basis measurements that decompose into a CNOT-gate, a Hadamard gate and two standard basis measurements.
Furthermore, the state produced corresponds to a GHZ state up to Pauli corrections, which depend on the measurement outcomes. 

Even though both approaches implement a quantum fanout gate, we use the approach inspired by \citeauthor{Eisert:2000} and \citeauthor{YimsiriwattanaLomonaco:2004} for its simplicity and its broader applicability in, for instance, distributed quantum computing. 

\begin{lemma}
The circuit of \cref{fig:q_circuit:non_local_cnot}, extended to arbitrary $n$, implements a quantum fanout gate. 
\end{lemma}
\begin{proof}
Let $\ket{x}$ be an $n$-qubit computational basis state and $\ket{\phi}=\alpha\ket{0}+\beta\ket{1}$ be any single qubit quantum state. 
We will prove that the circuit implements the quantum fanout gate on the state $\ket{\phi}\ket{x}$. 
The lemma then follows by linearity of the operations. 

The action of the quantum fanout gate on the quantum state is given by 
\begin{equation}
    \ket{\phi}\ket{x} %= \alpha \ket{0} \ket{x}  + \beta \ket{1} \ket{x} 
    \overset{\text{Fanout}}{\mapsto} \alpha \ket{0} \ket{x} + \beta \ket{1} X^{\otimes n} \ket{x} = \alpha \ket{0} \ket{x} + \beta \ket{1} \ket{\bar{x}},
    \label{eq:action_fanout}
\end{equation}
where $\ket{\bar{x}}$ is the computational basis state $\ket{x}$ with all qubits flipped. 

The circuit indeed implements this map (when generalized to arbitrary $n$). 
Assume we have a GHZ$_{n+1}$ state, up to a normalization factor of $1/\sqrt{2}$, we have:
\begin{align*}
     \big[\alpha & \ket{0} + \beta \ket{1}\big] \ket{x} \otimes\big[\ket{00 \cdots 0} + \ket{11 \cdots 1}\big]  \\
    & \xmapsto{(1)} \alpha \ket{0}\ket{x}\otimes\big[\ket{00 \cdots 0} + \ket{11 \cdots 1}\big]   + \beta \ket{1} \ket{x}\otimes\big[\ket{10 \cdots 0} + \ket{01 \cdots 1}\big]  \\
    & \xmapsto{(2)} \alpha \ket{0}\ket{x}\ket{d_0 0 \cdots 0}  + \beta \ket{1} \ket{x}\ket{d_01 \cdots 1} \\
    & \xmapsto{(3)} \alpha \ket{0}\ket{x}\ket{d_0 0 \cdots 0}  + \beta \ket{1} X^{\otimes n}\ket{x}\ket{d_01 \cdots 1} \\
    & \xmapsto{(4)} \frac{1}{2^{n-1}} \sum_{d\in\F_2^n} \big[\alpha\ket{0}\ket{x} + \beta (-1)^{d_1 + \hdots + d_n}\ket{1} X^{\otimes n}\ket{x}\big]\ket{d_0 d_1\hdots d_n} \\
    & \xmapsto{(5)} \alpha\ket{0}\ket{x}\ket{d_0 d_1\hdots d_n}  + (-1)^{d_1 + \hdots + d_n}\beta \ket{1} X^{\otimes n}\ket{x}\ket{d_0 d_1\hdots d_n} \\
    & \xmapsto{(6)} \big[\alpha\ket{0}\ket{x} + \beta\ket{1}\ket{\bar{x}}\big]\ket{d_0d_1\hdots d_n}.
\end{align*}
Step (1): Perform a CNOT-gate from the control qubit to the first qubit of the GHZ state;
Step (2): Measure that qubit, with outcome $d_0$, and apply an $X$-gate to the remaining $n$ qubits of the GHZ state if $d_0=1$;
Step (3): Perform $n$ parallel CNOT-gates between the $i+1$-th qubit of the GHZ state and the $i$-th target qubit;
Steps (4) and (5): Apply Hadamard gates to each unmeasured qubit of the GHZ state and subsequently measure it;
and, Step (6): Compute the parity $d_1\oplus\hdots\oplus d_n$ and apply a $Z$-gate to the control qubit if this parity equals one. 
The circuit thus implements the quantum fanout gate as desired. 
\end{proof}

\subsection{Applying the phase flip}
We apply the conditional phase flip by first conjugation the input qubits corresponding to the zeros in the binary representation of the index with $X$-gates and then computing the AND of these qubits.

As we have the identity AND$(z_1,\hdots,z_k)=\neg \text{OR}(\neg z_1,\hdots,\neg z_k)$, we can use the exact OR-gate by \citeauthor{TakahashiTani:2013}~\cite[Lemma~1]{TakahashiTani:2013} to implement the phase flip.
They used the Fourier inversion formula of \cref{eq:Fourier_inversion_formula}, to rewrite the OR as the sum of the parities of all nonempty subsets of inputs. 
Their method uses single- and two-qubit gates, as well as quantum fanout gates.

We can use quantum fanout gates to compute the parity of all subsets in parallel. 
Via a phase kickback trick, implemented via a fanout gate combined with a Hadamard gate, we have computed the OR in an auxiliary qubit. 
We use that auxiliary qubit to apply the phase flip in the GHZ state. 

%% file: decoding/quantum_decoding/complexity_quantum_decoding.tex
\section{Circuit complexity}\label{sec:decoding:quantum:complexity}
We now count all single-qubit gates, CNOT-gates and unbounded-fan-in parity gates used in the circuit to determine its size and width.
In the circuit, we will also use controlled-$R_Z$-gates, which we can implement in depth four, using two CNOT-gates and three single-qubits gates~\cite[Corollary~4.2]{NielsenChuang:2010}. 
\begin{lemma}
    The circuit for decoding the Hadamard code has size $\bigo(n^2\log n)$ and depth $65$.
\end{lemma}
\begin{proof}
We break down the steps of the previous section and count the size and depth. 
Generating $n$-qubit GHZ states requires $2n-1$ qubits and depth $6$. 
Each of the GHZ states can be prepared in parallel. 

The quantum fanout gates used require depth $8$. 
The GHZ states used to implement the quantum fanout gate can be initialized at the same time with the GHZ states used for the initial superposition. 

The conditional phase gates can be applied in parallel. 
Hence, for the depth, we consider only a single instance for index $i$, which requires the operations:
\begin{enumerate}
    \item Apply $X$-gates corresponding to the zeros in the binary representation of index $i$ and at the same time apply an $X$-gate to an auxillary qubit;
    \item Compute the OR of the input in this auxilliary qubit;
    \item Apply a $Z$-gate to the auxillary qubit, conditioned on the input bit $c(i)$;
    \item Uncompute the OR of the input;
    \item Apply $X$-gates corresponding to the zeros in the binary representation of index $i$ and at the same time apply an $X$-gate to an auxillary qubit.
\end{enumerate}
The odd steps correspond to single qubit gates, and hence add $1$ to the circuit depth. 
The second and fourth step apply the circuit for the OR-gate, which has depth~$28$~\cite[Lemma~2]{TakahashiTani:2013}. 
In the circuit, auxiliary GHZ states are used, which can be prepared in parallel with the initial uniform superposition. 
\begin{enumerate}
    \item For every of the $k$ local qubits part of the GHZ state, apply a fanout gate from that qubit to $n-1$ auxiliary qubits;
    \item Compute the parity of all possible nonempty subsets of the $k$ indices in parallel using the quantum fanout gate, conjugated by Hadamard gates.
    The conjugation by Hadamard gates can be integrated in the circuit for the quantum fanout gates and therefore does not increase the circuit depth;
    \item Apply controlled-$R_Z$-gates from the parity computation circuit to a GHZ state;
    \item Apply a fanout gate on the auxiliary GHZ state to reduce it to a single bit;
    \item Apply a Hadamard gate for a phase kick-back into the target qubit. 
\end{enumerate}
As the depth of the fanout gate is $8$, the depth of the OR-gate is $3\cdot 8+4=28$.
This gives a total circuit depth of $65$.

For input size $k$, the size of circuit for applying a conditional phase gate is $\bigo(k2^k)=\bigo(n\log n)$. 
As we have to apply $n$ conditional phase gates in parallel, the overall circuit size is $\bigo(n^2\log n)$. 
\end{proof}

We can use an OR-reduction by \citeauthor{HoyerSpalek:2005} to further reduce the circuit size~\cite[Lemma~5.1]{HoyerSpalek:2005}.
This OR-reduction uses a $\bigo(k\log k)$-sized constant-depth quantum circuit to prepare a quantum state on $\ceil{\log(k+1)}$ qubits, such that the OR on these $\ceil{\log(k+1)}$ qubits evaluates to the same value as the OR on the original $k$ qubits. 
\begin{corollary}
    There exists a quantum circuit for decoding the Hadamard code of size $\bigo(n\log n\log\log n)$ and depth $103$.
\end{corollary}
\begin{proof}
    The result follows by noting that the OR-reduction has depth $19$ and counting the gates of the exact OR-routine by \citeauthor{TakahashiTani:2013}.
\end{proof}

Note that we can use a fanout gate to simplify the proof of \cref{thm:quantum:strategy_proof} slightly:
First, apply a fanout gate to map the state of \cref{eq:quantum:hadamard:final} to obtain
\begin{equation}\label{eq:quantu:hadamard:final_simplified}
    \frac{1}{\sqrt{n}}\sum_y (-1)^{c(y)}\ket{y}\ket{0}^{\otimes n-1},
\end{equation}
Next, apply Hadamard gates only to the first register. 
A final measurement of the first register then directly gives a bitstring according to the probability distribution.
For the Hadamard code, this approach only simplifies the proof; for other codes, this step is necessary for correct decoding.

%% file: decoding/quantum_decoding/separation_decoding.tex
\section{Separating \texorpdfstring{$\nc^0[\oplus]$}{NC0[+]} from \texorpdfstring{$\qnc^0[\oplus]$}{QNC0[+]}}\label{sec:separation_quantum_classical}
In this section we combine our conventional and quantum results, \cref{thm:classical_decoding_polynomial_maps} and \cref{thm:quantum_decoding_Hadamard_circuit} respectively, to show a separation between $\nc^0[\oplus]$, $\ac^0[\oplus]$, and $\qnc^0[\oplus]$. 
Note that our quantum circuit outputs a single possible message, whereas the conventional circuit returns a list. 
This requires minor changes and gives the following theorem.
\begin{corollary}\label[corollary]{cor:quantum:hadamard_to_list}
There is a family of $\qnc^0[\oplus]$-circuits $(\mathcal C_n)_{n\in \N}$ such that the following holds.
Let $k\in \N$, $n = 2^k$ and $\eps \in [1/\sqrt n,1/2]$.
Then, on any input~$y\in \F_2^n$, with probability $1 - \bigo(\eps)$ the circuit~$\mathcal C_n$ returns a list~$L(y)$ of size $\bigo(\eps^{-2}\log (1/\eps))$ which contains every $x\in \F_2^k$ with $d\big(y, H(x)\big)\leq (\frac{1}{2} - \eps)n$.
\end{corollary}
\begin{proof}
For a large enough constant~$C>0$, consider $C\eps^{-2}\log (1/\eps)$ parallel instances of the circuit from \cref{thm:quantum_decoding_Hadamard_circuit}.
This gives a list~$L(y)$ of the claimed size such that any message $x\in \F_2^k$ satisfying $d\big(y, H(x)\big)\leq (\frac{1}{2} - \eps)n$ appears in~$L(y)$ with probability at least $1 - \eps^3$.
Since there are at most $O(1/\eps^2)$ such messages, it follows from the union bound that with probability at least~$1-\bigo(\eps)$ every such message appears in~$L(y)$.
\end{proof}
\begin{remark}
Note that the circuits obtained in this corollary also output several messages whose codewords differ from the input $y$ in more than $(\frac{1}{2} - \eps)n$ coordinates;
this differs from the usual notion of the list decoding problem, which aims to output a list of all messages $x\in \F_2^k$ with $d\big(y, H(x)\big)\leq (\frac{1}{2} - \eps)n$ and none other.
One can also solve the usual list-decoding problem for the Hadamard code using $\qnc^0[\oplus]$-circuits, by making use of Majority gates (and more general threshold gates) to prune the obtained list (see also \cref{sec:quantum:list_decoding_majority}).
\end{remark}

As a consequence we arrive at the main result of this chapter: 
List-decoding the Hadamard code separates the classes $\nc^0[\oplus]$ and $\qnc^0[\oplus]$. 
This result thus gives a quantum advantage for a problem appearing naturally. 

In the high-error setting, where the error parameter $\delta$ approaches the information-theoretic limit of $1/2$ (which is relevant for hardness amplification), a stronger separation follows by combining \cref{thm:quantum_decoding_Hadamard_circuit} with a result of Sudan showing hardness of noisy decoding by $\ac^0[\oplus]$-circuits.
Stating this problem requires us to consider a slightly different problem: 

\medskip
\noindent
\textbf{List-Hadamard problem:}
Let $\eps: \N\to (0,1]$ be a function.
For each dyadic number $n=2^k$ we define the problem $\LH_n(\eps)$ as follows:
given $y\in \F_2^n$, output a list of at most $n/4$ elements in $\F_2^k$ that contains every $x\in \F_2^k$ satisfying $d\big(y, H(x)\big)\leq \big(\frac{1}{2} - \eps(n)\big) n$.
\medskip

In the next section, we discuss the hardness of the list-Hadamard problem and its implications further. 
For now, it gives us the tools to state the most general form of our quantum advantage result:
\begin{theorem}[Quantum-vs-conventional separation]\label{thm:quantum_classical_separation}
For every constant $\delta \in (0,\frac{1}{2})$, list decoding the Hadamard code with error parameter $\delta$ separates $\qnc^0[\oplus]$ from $\nc^0[\oplus]$.
Moreover, for any $(\log n)/\sqrt{n} \leq \eps(n) \leq 1/(\log n)^{\omega(1)}$, the list-Hadamard problem $\LH_n(\eps)$ separates $\qnc^0[\oplus]$ from $\ac^0[\oplus]$.
\end{theorem}

%% file: decoding/quantum_decoding/list_decoding_to_majority.tex
\section{A quantum circuit for Majority}\label{sec:quantum:list_decoding_majority}
This section shows how to obtain a $\qnc^0[\oplus]$-circuit that computes Majority, which uses our $\qnc^0[\oplus]$-circuit for list decoding the Hadamard code. 
As a corollary, we prove the second statement of \cref{thm:quantum_classical_separation}.

We first show the reduction from list decoding to Majority by introducing a new problem called $\isbal$ and then transform that into a $\qnc^0[\oplus]$-circuit for Majority. 

\subsection{From list decoding to Majority}
Sudan (see~\cite[Section~6.2]{Viola:2006}) showed that list decoding with error parameter $1/2 - \eps$ requires probabilistic~$\ac^0[\oplus]$-circuits to have size $\exp(\poly(1/\eps))$.
The hardness of the list-Hadamard problem will follow as a corollary. 
For concreteness, we state his result restricted to the Hadamard code;
as can be easily seen from its proof, one could instead consider any other error-correcting code.
\begin{theorem}[List-Hadamard implies Majority]
\label{thm:Sudan:majority_from_list_Hadamard}
Let~$\mathcal{C}$ be a probabilistic circuit that solves the list-Hadamard problem $\LH_n(\eps)$ with probability at least $3/4$.
Then there exists a (deterministic) oracular $\ac^0$-circuit $\mathcal{D}$ of size $\poly(n, 1/\eps)$ that, when given oracle access to $\mathcal{C}$ and the ability to fix its random bits, computes Majority on $\Omega(1/\eps)$ bits.
\end{theorem}
As a corollary, the circuit lower bound for Majority due to Razborov~\cite{Razborov:1987} and Smolensky~\cite{Smolensky:1987} gives the following (known) hardness result for list decoding the Hadamard code.
\begin{corollary}[Hardness of list-Hadamard] \label{cor:quantum:hardness_list_Hadamard}
If $\eps(n) \leq 1/(\log n)^{\omega(1)}$, then the list-Hadamard problem $\LH_n(\eps)$ cannot be solved by a probabilistic $\ac^0[\oplus]$-circuit with probability~$\Omega(1)$.
\end{corollary}

Combining this corollary with our~$\qnc^0[\oplus]$-circuits for list-Hadamard given in \cref{cor:quantum:hadamard_to_list}, we obtain the second separation of complexity classes stated in \cref{thm:quantum_classical_separation}.

To prove \cref{thm:Sudan:majority_from_list_Hadamard}, let $\maj_t$ denote the Majority function on $t$ bits.
We first introduce a promise problem called $\isbal_t$, which asks to determine whether a given binary string is balanced.
We then show that a (probabilistic) circuit solving $\isbal_t$ can be turned into a deterministic circuit computing $\maj_t$. 
Finally, we show how a circuit for $\LH_n(\eps)$ can be used to solve $\isbal_t$ for $t =\Omega(1/\eps)$.

\begin{definition}[The $\isbal_t$ problem]\label{def:isbal}
For an even positive integer~$t$, define $\isbal_t:\{x\in \F_2^t :\, |x|\leq t/2\}\to \F_2$ by
\begin{equation*}
\isbal_t(x) = \left\{
\begin{array}{ll}
	1 &\text{if $|x| = t/2$,} \\
	0 &\text{otherwise.}
\end{array}
\right.
\end{equation*}
Given an arbitrary~$x\in \F_2^t$, define the $\isbal_t$ problem as returning $\isbal_t(x)$ if $|x| \leq t/2$, and an arbitrary bit otherwise.
\end{definition}

A probabilistic circuit for the $\isbal_t$ problem takes in a string $x$ and random coin tosses $u$. 
We say that a probabilistic circuit solves the $\isbal_t$ problem if the probability it correctly outputs $1$ on balanced inputs is at least $2/3$, and the probability it correctly outputs $0$ on unbalanced inputs is at least $1/2$, both taken over the random coin tosses. 
Next, we have the following derandomization lemma to obtain a deterministic circuit from a probabilistic one. 
\begin{lemma}[Derandomization lemma]\label[lemma]{lem:probabilistic_IsBal}
Let~$\mathcal{C}$ be a probabilistic circuit that solves $\isbal_t$ with probability at least~$2/3$ for every input.
There exists a deterministic oracle $\ac^0$-circuit~$\mathcal{C'}$ that, when given oracle access to~$\mathcal{C}$ and the ability to fix its random bits, solves~$\isbal_t$.
\end{lemma}
\begin{proof}
For some large enough constant~$c\in \N$, consider~$ct$ parallel instances of~$\mathcal{C}$.
It follows from the Chernoff bound that, for any fixed $x\in \F_2^t$ given to all of these instances, with probability $1- \exp(-10\, t)$ at least 55\% of the instances solves the $\isbal_t$ problem on input~$x$.

By the union bound, one can fix the randomness in the instances of~$\mathcal{C}$ in order to get a deterministic conventional circuit that, for every input~$x\in\F_2^t$ with $|x| \leq t/2$, returns a $ct$-bitstring whose Hamming weight is at least $0.55 t$ if $\isbal_t(x) = 1$ and at most $0.45 t$ if $\isbal_t(x) = 0$.
Distinguishing these two types of strings is known as the approximate majority problem, for which there is an $\ac^0$-circuit~\cite{Ajtai:1983}.
Combining these circuits gives the result.
\end{proof}

We now show how a deterministic circuit that solves $\isbal_t$ can be used to compute $\maj_t$. 

\begin{lemma}\label[lemma]{lem:deterministic_IsBal}
Let~$\mathcal{C}$ be a deterministic circuit for $\isbal_t$.
There exists an oracle $\ac^0$-circuit~$\mathcal{D}$ that, given oracle access to~$\mathcal{C}$, computes $\maj_t$.
\end{lemma}
\begin{proof}
For $x\in\F_2^t$ and $i\in\{0,\hdots,t\}$, define $x_i$ as the string $x$ with the first~$i$ bits set to zero and the rest of the bits equal to those of~$x$. 
So, for instance, $x_0 = x$ and $x_t$ is the all-zeros string. 
Let $\mathcal{D}$ be the circuit that runs $t+1$ parallel instances of $\mathcal{C}$ with inputs $x_0,x_1,\dots,x_t$, respectively, and returns the OR of the $t+1$ outputs.

We claim that~$\mathcal{D}$ computes $\maj_t$.
Indeed, if $x$ has fewer than $t/2$ ones then~$\mathcal{C}$ returns $0$ for each input $x_i$, as the number of ones only decreases with $i$. 
If $x$ has at least~$t/2$ ones, then $\mathcal{C}$ returns $1$ for at least one $i$, since $x_0$ has at least $t/2$ ones, whereas $x_t$ is the all-zeros string. 
This completes the proof.
\end{proof}

Towards turning a circuit $\mathcal{C}$ for $\LH_n(\eps)$ into a circuit for $\isbal_t$, we associate with each input $x\in \F_2^t$ to $\isbal_t$ a random error vector~$N_x$ over~$\F_2^n$ as follows:
independently, each coordinate of~$N_x$ is a uniformly random entry of~$x$.
In particular, for balanced $x$, the error vector $N_x$ will correspond to an error rate of $1/2$ and we refer to it as $N_{1/2}$. 
The next lemma shows that there exists some message $m\in\F_2^k$ that has small probability of recovery by $\mathcal{C}$ under the error vector $N_{1/2}$.

\begin{lemma}\label[lemma]{lem:decoding_random_message}
Let $\mathcal{C}$ be a probabilistic circuit that, on input $y\in \F_2^n$, returns a (random) list $L(y)\subseteq \F_2^k$ of at most~$2^k/4$ elements.
Then there exists $m\in \F_2^k$ such that
\begin{equation}\label{eq:upper_bound_random_decoding}
\Pr[m\in L(H(m) + N_{1/2})] \le 1/4,
\end{equation}
where the probability is taken over~$L$ and~$N_{1/2}$.
\end{lemma}

\begin{proof}
Note that, for any $y\in \F_2^n$, the vector $y+N_{1/2}$ is uniformly distributed over~$\F_2^n$;
in particular, it has the same distribution as~$N_{1/2}$.
Let $M\in \F_2^k$ be a uniformly distributed random element.
Then, by independence of $M$, $L(y)$ and~$N_{1/2}$, get that
\begin{align*}
\Pr_{M,L,N_{1/2}}[M\in L(H(M) + N_{1/2})] & = \Pr_{M, L, N_{1/2}}[M\in L(N_{1/2})] \\
& \leq \frac{1}{2^k}\, \Exp_{L,N_{1/2}}\, |L(N_{1/2})|  \\
& \leq 1/4.
\end{align*}
Hence, there exists a value~$m$ of~$M$ such that~\cref{eq:upper_bound_random_decoding} holds. 
\end{proof}

Finally, we prove that the circuit $\mathcal{C}$ in \cref{thm:Sudan:majority_from_list_Hadamard} can solve~$\isbal_t$. 

\begin{lemma}\label[lemma]{lem:list_decoding_implies_IsBal}
Let $\mathcal{C}$ be a probabilistic circuit as in \cref{thm:Sudan:majority_from_list_Hadamard}. 
There exists a probabilistic oracle $\ac^0$-circuit $\mathcal{D}$ of size $\poly(n,1/\eps)$ that, when given oracle access to $\mathcal{C}$, solves $\isbal_t$ with probability at least~3/4 for $t = \Omega(1/\eps)$. 
\end{lemma}

\begin{proof}
We may assume without loss of generality that $\eps \leq 1/4$.
Let $\delta \in [\eps, 1/4]$ be minimized such that $t = 1/(2\delta)$ is an even integer;
note that, since $\delta \geq \eps$, the circuit~$\mathcal{C}$ also solves $\LH_n(\delta)$ with probability at least~$3/4$.
Fix a message $m$ as in \cref{lem:decoding_random_message}, and let $x\in \F_2^t$ be some string to serve as input to~$\mathcal{D}$.

The circuit~$\mathcal{D}$ has three layers.
The first layer has the string~$H(m)$ hardwired into it and uses~$n$ independent uniform samples to the coordinates of~$x$ to compute the random string $H(m) + N_x$.
This layer is a probabilistic circuit using~$n$ parallel two-bit XOR-gates.
The second layer consists of the circuit~$\mathcal{C}$, which produces a random list $L(H(m)+ N_x)$ of size at most~$n/4$.
The third layer consists of an~$\ac^0$-circuit of size $\poly(n)$ that returns $0$ if and only if $m \in L(H(m)+ N_x)$. 
This can be done by checking equality between~$m$ and the~$\bigo(n)$ elements of the list.
We claim that this solves $\isbal_t$.

If~$x$ is balanced then it follows from \cref{lem:decoding_random_message} that~$\mathcal{D}$ correctly returns~$1$ with probability at least~$3/4$. 
If~$x$ has Hamming weight strictly less than~$t/2$, then each coordinate of~$N_x$ is~1 with probability at most $1/2 - 1/t = 1/2 - 2\delta$.
%$(1/2\cdot 1/(2\eps') - 1)/(1/(2\eps')) = 1/2 - 2\eps'$. 
By the Chernoff bound,~$N_x$ has Hamming weight at most $(1/2 - \delta)n$ with probability $1- \exp(-\Omega(\delta^2 n))$.
%then bounds the number of ones in $N_x$ to be at most $1/2 - 2\eps' + \eps' = 1/2-\eps' \le 1/2 -\eps$ with high probability. 
Hence, the properties of the circuit~$\mathcal{C}$ imply that in this case~$\mathcal{D}$ correctly outputs $0$ with probability at least $3/4$. 
\end{proof}

\cref{thm:Sudan:majority_from_list_Hadamard} now follows directly by combining \cref{lem:probabilistic_IsBal,lem:deterministic_IsBal,lem:list_decoding_implies_IsBal}.

\subsection{Obtaining a quantum circuit}
In the previous section we discussed conventional circuits for Majority using list-decoding circuits. 
Now we sketch how the above proof can be used to turn our $\qnc^0[\oplus]$-circuit for decoding the Hadamard code into one that computes Majority with polynomially small error.

Let $\eps = n^{-1/4}$ and let~$\mathcal{C}$ be the circuit from \cref{cor:quantum:hadamard_to_list}.
Since~$\mathcal{C}$ returns lists of size at most~$n^{3/4}$, a stronger version of \cref{lem:decoding_random_message} holds, where the probability of~\cref{eq:upper_bound_random_decoding} -- taken additionally over the measurement outcomes of~$\mathcal{C}$ -- is bounded from above by~$n^{3/4}/2^k = n^{-1/4}$.

The proof of \cref{lem:list_decoding_implies_IsBal} then gives an oracle $\qnc^0[\oplus]$-circuit~$\mathcal{D}$ of size~$\poly(n)$ that, given oracle access to~$\mathcal{C}$, solves $\isbal_t$ with probability $1 - \bigo(n^{-1/4})$ for $t = \Omega(n^{1/4})$.
Here, the $\ac^0$-circuit used to check membership of~$m$ can be replaced with our $\qnc^0[\oplus]$-circuit for the OR-function (see above) applied to the entrywise sum of~$m$ with each element in the list.

Now let $t' = \floor{n^{1/8}}$, and note that the same circuit $\mathcal{D}$ above can be used to solve $\isbal_{t'}$ with probability $1 - \bigo(n^{-1/4})$:
it suffices to pad the input with zeros and ones in the same number until we have a string of the correct size.
Finally, with the proof of \cref{lem:deterministic_IsBal} and the union bound we obtain a $\qnc^0[\oplus]$-circuit~$\mathcal{D'}$ that, given oracle access to~$\mathcal{D}$, solves $\maj_{t'}$ with probability $1 - \bigo(n^{-1/8})$.
We again use our $\qnc^0[\oplus]$-circuit for the OR-function as discussed in \cref{sec:quantum:details_quantum_algorithm}.

%% file: decoding/quantum_decoding/circuit_hadamard_higher_characteristic.tex
\section{Hadamard code for higher field characteristics}\label{sec:quantum:Hadamard_code_higher_fields}
In this section, we extend the $\qnc^0[\oplus]$-circuit for decoding the Hadamard code to higher field characteristics. 
Before doing so, we first briefly discuss quantum gates for higher characteristics.

\subsection{Quantum gates for higher field characteristics}
Most definitions for qubits generalize directly to multilevel systems. 
The $p$-level generalization of a qubit, a qudit, has computational basis states $\ket{0},\hdots,\ket{p}$.
These form the basis for the Hilbert space $\HS=\C^p$.
The quantum gates for qubits also generalize to qudits (see also, for example, \cite{Wang:2020}). 
Let $+_p$ denote addition modulo~$p$, then we have the following quantum gates:
\begin{align*}
    X_p: & \ket{i}\mapsto\ket{i+_p 1} & Z_p: & \ket{i}\mapsto\omega_p^i\ket{i} \\
    R_{Z,p}(\theta): & \ket{i}\mapsto\omega_p^{i\theta}\ket{i} & H_p: & \ket{i}\mapsto\frac{1}{\sqrt{p}}\sum_{j\in\F_p} \omega_p^{ij}\ket{j} \\
    CNOT_p: & \ket{x}\ket{y}\mapsto \ket{x}\ket{x+_py} & QFT_p: & \ket{x} \mapsto \frac{1}{\sqrt{p^n}}\sum_{y\in \F_p^n}\omega_p^{\ip{x,y}}\ket{y}
\end{align*}

\subsection{Hadamard code for higher field characteristics}
The Hadamard code over larger prime fields is similar to the definition used in previous sections over a binary field.
Let $n=p^k$ for some prime $p$. 
The Hadamard code over $\F_p^k$ is then given by
\begin{equation*}
    H:\F_p^k\to\F_p^n, \qquad x\mapsto \langle i,x\rangle_{i\in\F_p^k},
\end{equation*}
where the inner product is now taken modulo~$p$. 

A quantum circuit for list decoding the Hadamard code over $\F_p^n$ looks similar to the circuit exposed in \cref{sec:quantum:details_quantum_algorithm}, with the key difference that we use quantum gates that operate on qudits instead of qubits. 
In the noiseless case we have:
\begin{align*}
    \ket{0} & \xrightarrow{(1)} \frac{1}{\sqrt{p^k}}\sum_{i\in\F_p^k}\ket{i} \\
    & \xrightarrow{(2)} \frac{1}{\sqrt{p^k}}\sum_{i\in\F_p^k}\omega_p^{\langle i,x\rangle}\ket{i} \\
    & \xrightarrow{(3)} \frac{1}{p^k}\sum_{j\in\F_p^k}\sum_{j\in\F_p^k}\omega_p^{\langle i,x+j\rangle}\ket{j} \\
    & = \ket{x}
\end{align*}
Step $(1)$: Create a uniform superposition of $n$ states using $k$ $H_p$-gates;
Step~$(2)$: Apply for every index in parallel a conditional phase flip $Z_p$-gate controlled by the corrupted input and the specific index;
and, Step $(3)$: Apply an inverse generalized Hadamard gate $H_p^{-1}$. 
The last equality follows directly from the orthogonality of the $\omega_p$, see also \cref{lem:roots_of_unity}.

In case of errors, some of the inputs $c(i)$ are corrupted. 
The probability to find output $z$ is then given by:
\begin{equation}
\Pr[z] = \frac{1}{p^{2k}}\Bigg| \sum_{y\in\F_p^k} \omega_p^{c(y)+\ip{y,z}}\Bigg|^2.
\end{equation}
In the binary case, every corrupted term cancels precisely one uncorrupted term. 
For higher characteristics, this cancellation depends on the exact value of the terms, and hence on the corruption. 
Considering the worst-case corruption, we obtain a lower bound on the probability of correct decoding: 
\begin{align}
\Pr[z] & = \frac{1}{p^{2k}}\bigg| \sum_{y\in\F_p^k} \omega_p^{c(y)+\langle y,z\rangle}\bigg|^2 \nonumber \\
& \ge \bigg(1 - 2\frac{d(c,H(z))}{p^k}\bigg)^2,
\end{align}
where $d$ denotes the Hamming distance. 
By comparing this expression with \cref{eq:quantum:hadamard:succes_probability} we see that the success probability increases with higher field characteristics. 
By the same reasoning as before, the circuit for higher characteristics remains of constant depth. 

%% file: decoding/quantum_decoding/discussion_quantum_decoding.tex
\section{Reflections and outlook}
This chapter presented a $\qnc^0[\oplus]$-circuit that can list decode the Hadamard code and that successfully returns the message with probability $\Omega(\eps)$ for error rate $\frac{1}{2}-\eps$.

A similar circuit can also be constructed using the Goldreich-Levin algorithm and the implementation of the Majority gate using a $\qnc^0[\oplus]$-circuit~\cite{HoyerSpalek:2005,TakahashiTani:2013}. 
However, the circuit based on the Goldreich-Levin algorithm depends on the error parameter~$\eps$, and as a result, the size of the Majority gates will depend on~$\eps$. 
In contrast, the quantum circuits presented in \cref{thm:quantum_decoding_Hadamard_circuit} and \cref{cor:quantum:hardness_list_Hadamard} have size independent of~$\eps$.

\citeauthor{AdcockCleve:2002} considered a quantum version of the Goldreich-Levin algorithm~\cite{AdcockCleve:2002}. 
However, their focus was on the cryptographic applications instead of decoding error-correcting codes, leading to more complex proofs.
The circuit resulting from their approach also depends on the error parameter $\eps$. 

\citeauthor{KawachiYamakami:2010} considered quantum list-decoding algorithms for conventional error-correcting codes~\cite{KawachiYamakami:2010}.
They introduce shuffled codeword states, similar to \cref{eq:quantu:hadamard:final_simplified}, and a way to retrieve a list of possible messages from them.
However, preparing shuffled codeword states is nontrivial for most codes, especially with restricted computational resources.
This chapter provides an explicit construction of the shuffled codeword state for the Hadamard code. 
In a follow-up work, \citeauthor{Yamakami:2016} extend this line of research by considering faulty quantum circuits that implement the encoding~\cite{Yamakami:2016}.

None of these approaches focused on the circuit depth.
Thus, our result of separation (\cref{thm:quantum_classical_separation}) is the first separation with respect to decoding error-correcting codes. 
Future work can extend our result in multiple directions:
\begin{enumerate}
    \item Provide a $\qnc^0[\oplus]$-circuit for practical error-correcting codes other than the Hadamard code considered in this chapter. 
    In \cref{chp:decoding:quadractic} we take a first step in this direction by considering Reed-Muller codewords of degree at most $2$;
    \item Prove a general statement on the existence of $\qnc^0[\oplus]$-circuits for any error-correcting code. 
    Such a result would be an important next step, as the result for $\nc^0[\oplus]$-circuits holds for any error-correcting code, whereas the quantum result focused on the Hadamard code;
    \item Extend the result to noisy quantum circuits. 
    Current quantum devices are noisy, limiting implementations of quantum algorithms. 
    We expect that specific noise models can correspond to corruptions in the codeword. 
    If error rates are sufficiently low, we can use the $\qnc^0[\oplus]$-circuits presented in this chapter to retrieve the correct message with high probability.
\end{enumerate}

%% file: decoding/quadratic_codes.tex
\chapter{Decoding quadratic codes}\label{chp:decoding:quadractic}
In this chapter we extend the results from the previous two chapters to retrieve a Reed-Muller codeword of degree at most~$2$, given as input a polynomial phase function with large Gowers $U^3$-norm. 
We revisit earlier work of \citeauthor{TulsianiWolf:2014} and show an improved query complexity by employing a quantum Fourier sampling routine, omitting a Fourier estimation subroutine, and by using a different algorithmic version of the Balog-Szemer\'edi-Gowers theorem with a simpler proof and better quantitative bounds~\cite{BalogSzemeredi:1994,Gowers:1998,Schoen:2014}. 
This algorithmic version might be of independent interest.

\input{decoding/quadratic_decoding/outline_quadratic_decoding}

\input{decoding/quadratic_decoding/warm_up_quadratic_decoding}
\input{decoding/quadratic_decoding/non_uniformity_to_weak_linearity}
\input{decoding/quadratic_decoding/variant_BSG}
\input{decoding/quadratic_decoding/sampling_high_degree_elements}

\input{decoding/quadratic_decoding/constructing_approximating_matrix}
\input{decoding/quadratic_decoding/query_lower_bound}
\input{decoding/quadratic_decoding/discussion_quadratic_decoding}

%% file: decoding/quadratic_decoding/outline_quadratic_decoding.tex
\section{Chapter overview}
The previous two chapters considered list decoding corrupted error-correcting codes and a quantum approach for list decoding the corrupted Hadamard code. 
In this chapter, we extend this line of research by considering Reed-Muller codewords of degree at most~$2$. 
For corrupted linear Reed-Muller codewords, i.e., the Hadamard code, Fourier analysis provides the tools to retrieve the codeword.
Fourier analysis is closely related to the Gowers $U^2$-norm, see also \cref{eq:Gowers_U2_L4}. 
Higher-order Fourier analysis provides the tools to analyse the results for corrupted Reed-Muller codewords of degree at most~$2$. 
These corrupted codewords have large Gowers $U^3$-norm. 

The algorithm described in this chapter can find a Reed-Muller codeword of degree at most~$2$, such that it correlates with the corrupted input with large Gowers $U^3$-norm.
We let $\delta(f,g)$ denote the normalized Hamming distance between two functions $f$ and $g$, where both functions are evaluated on all possible inputs. 
We let $\delta_{\mathrm{RM}_2}(f)$ denote the minimum Hamming distance between the function $f$ and any Reed-Muller codeword of degree at most~$2$. 
The next theorem will be the main theorem of this chapter.
\begin{theorem}\label{thm:quantum_decoding_quadratic}
    For any $\eta>0$, there is an $\eps>0$ such that the following holds:
    There exists a quantum algorithm that, given a function $f:\F_2^n\to\F_2$ satisfying $\delta_{\mathrm{RM}_2}(f) \leq \frac{1}{2} - \eps$, makes at most $\bigo_\eps(n\log n)$ queries to~$f$ and returns a polynomial~$g:\F_2^n\to\F_2$ of degree at most~$2$ such that $\delta(f,g) \leq \frac{1}{2} - \eta$.
\end{theorem}
The dependency of $\eta$ on $\eps$ in the theorem is exponential: $\eta=\exp(-1/\eps^C)$ for some constant $C>0$. 
The bulk of the work in proving \cref{thm:quantum_decoding_quadratic} goes into finding a quadratic phase function that correlates with $f$ sufficiently well, as summarized in the next lemma.
Note that this lemma holds for any fixed prime~$p$.

\begin{lemma}\label[lemma]{lem:quadratic_decoding:main_lemma}
    There is a quantum algorithm that, given a polynomial phase function $f:\F_p^n\to \D$ satisfying $\|f\|_{U^3} \geq \gamma$, makes $\poly(1/\gamma)n\log n$ queries to~$f$ and with probability $\poly(\gamma)$, returns a matrix $M\in\F_p^{n\times n}$ such that
    \begin{equation*}
        \Exp_{h\in \F_p^n}\big|\widehat{\Delta_hf}(Mh)\big|^2 \ge \exp(-\poly(1/\gamma)).
    \end{equation*}
\end{lemma}

Once we have such a matrix $M$, we can use the next two lemmas, together with the Bernstein-Vazirani algorithm, to obtain (with good probability) a quadratic phase that correlates with~$f$. 
We distinguish between odd primes and $p=2$. 
\begin{lemma}[Green--Tao \cite{GreenTao:2008}]\label[lemma]{lem:symmetrization}
Let~$p$ be an odd prime, $M\in \F_p^{n\times n}$ and $f:\nolinebreak\F_p^n\to \D$ be such that
\begin{equation*}
    \Exp_{h\in \F_p^n}\big|\widehat{\Delta_hf}(Mh)\big|^2 \geq \delta.
\end{equation*}
Then, there is a $b\in \F_p^n$ such that the symmetric matrix $M' = (M+M^\mathsf{T})/2$ satisfies
 \begin{equation*}
    \big|\Exp_{x\in\F_p^n} f(x)\omega_p^{\ip{x,M'x}/2 + \ip{x,b}}\big| \geq \Omega_{p,\delta}(1).
\end{equation*}
\end{lemma}

For the case $p = 2$, we have a similar result, due to \citeauthor{Samorodnitsky:2007}~\cite{Samorodnitsky:2007} (see also~\cite[Section~4.3]{HatamiHatamiLovett:2019}).
\begin{lemma}[\citeauthor{Samorodnitsky:2007}]\label[lemma]{lem:symmetrization_F2}
Let $M\in \F_2^{n\times n}$ and $f:\F_2^n\to \pmset{}$ be such that
\begin{equation*}
    \Exp_{h\in \F_2^n}\big|\widehat{\Delta_hf}(Mh)\big|^2 \geq \delta.
\end{equation*}
Let~$W = \ker(M+M^{\mathsf{T}})$ and $W^\perp\subseteq\F_2^n$ be its orthogonal complement.
Let~$M'\in \F_2^n$ be the matrix satisfying $M'x = Mx$ for all $x\in W$ and $\ker(M') = W^\perp$.
Let~$v$ be the diagonal of the matrix~$M'$ and let $M'' = M' + vv^\mathsf{T}$.
Then, there is a $b\in \F_2^n$ such that 
\begin{equation*}
    \big|\Exp_{x\in\F_2^n} f(x)\omega_p^{\ip{x,M''x} + \ip{x,b}}\big| \geq \Omega_{\delta}(1).
\end{equation*}
\end{lemma}

\cref{lem:symmetrization} and \cref{lem:symmetrization_F2} show how, given a matrix~$M$ as in \cref{lem:quadratic_decoding:main_lemma}, to obtain a matrix~$M'$ such that the function $g(x) = f(x)\omega_p^{-\ip{x,M'x}}$ satisfies $|\widehat{g}(b)|\geq \nolinebreak\Omega_{p,\gamma}(1)$ for some~$b\in \F_p^n$.
Since we can query~$g$ by sequentially querying~$f$ and then the quadratic phase~$\omega_p^{-\ip{x,M'x}}$, we can use the Bernstein-Vazirani algorithm to find such a~$b$ with positive probability.
Finally, we can find a $c$ such that 
\begin{equation*}
\Re\left(\Exp_{x} f(x)\omega_p^{-\ip{x,Mx + b} - c}\right) \ge \Omega_{\gamma}(1).
\end{equation*}
Specifically, as $p$ is a fixed constant, we can evaluate this expression for all~$p$ possible values of $c$ and choose the one among them that maximizes the expression. 

The next two subsections give an outline of the proof of \cref{lem:quadratic_decoding:main_lemma}. 
Again, we start of gradually with a warm-up.
\cref{sec:quadratic:warm_up} shows how the Fourier sampling algorithm introduced in \cref{lem:Fourier_sampling} can be used to decoded corrupted quadratic Reed-Muller codewords within the unique decoding radius. 
With this warm-up, we can get accustomed to the ideas underlying the later proofs. 
The remainder of this chapter then provides the algorithmic proofs of the results in the next two subsections. 
Specifically, \cref{sec:quadratic:non_uniformity_to_weak_linearity} provides an outline for \cref{sec:quadratic:quantum_algorithmic_weak_linearity};
\cref{sec:quadratic:weak_linearity_to_true_linearity} provides an outline for both \cref{sec:quadratic:variant_BSG} and \cref{sec:quadratic:sampling_high_degree_elements};
and, \cref{sec:quadratic:constructing_approximating_matrix} will discuss how to obtain a matrix $M$ that satisfies the constraints outlined in \cref{lem:symmetrization} and \cref{lem:symmetrization_F2}, thereby completing the proof of \cref{lem:quadratic_decoding:main_lemma} and hence of \cref{thm:quantum_decoding_quadratic}.
Finally, \cref{sec:quadratic:query_lower_bound} proves a lower bound on the query complexity, that matches the query complexity of our algorithm up to a logarithmic factor.

\subsection{From non-uniformity to weak linearity}\label{sec:quadratic:non_uniformity_to_weak_linearity}
The starting point for proving \cref{lem:quadratic_decoding:main_lemma} is the following basic result.
\begin{proposition}\label[proposition]{prop:U3_to_weak_linearity}
Suppose that $f:\F_p^n\to \D$ satisfies $\|f\|_{U^3} \geq \gamma$. 
Then, there is a set $S\subseteq \F_p^n$ of size at least $\gamma^8p^n/2$ and a map $\phi:S\to\F_p^n$ such that for all $h\in S$, we have
\begin{equation*}
    \big|\widehat{\Delta_{h}f}\big(\phi(h)\big)\big|^2 \geq \frac{\gamma^8}{2}.
\end{equation*}
\end{proposition}
\begin{proof}
    It follows by the nesting property of the Gowers norms (\cref{eq:Uk_nesting}) and Parseval's identity (\cref{prop:Parseval}) that
    \begin{align*}
        \gamma^8 &\leq \|f\|_{U^3}^8\\
        &= \Exp_{h\in \F_p^n}\|\Delta_hf\|_{U^2}^4\\
        &\leq \Exp_{h\in \F_p^n} \max_{a\in \F_p^n}|\widehat{\Delta_hf}(a)|^2.
    \end{align*}
Letting $\phi:\F_p^n\to\F_p^n$ be any map such that $|\widehat{\Delta_hf}(\phi(h))|$ is maximal for each $h\in \F_p^n$ and applying Markov's inequality (\cref{lem:Markov}) then gives the result.
\end{proof}

The next step establishes that a map~$\phi$ as in \cref{prop:U3_to_weak_linearity} satisfies a weak form a linearity.
For a finite Abelian group~$G$, an \emph{additive quadruple} is a four-tuple $(a,b,c,d)\in G^4$ satisfying $a+b=c+d$.
The \emph{energy} of a set~$A\subseteq G$ is then defined as the number of additive quadruples contained in~$A$.
A set $A\subseteq G$ can be shown to have energy~$|A|^3$ if and only if~$A$ is the coset of a subgroup.
For two sets $A,B\subseteq{G}$, we define $A+B$ as the set $\{a+b\mid a\in A, b\in B\}$. 
Then, if $\phi:\F_p^n\to\F_p^n$ is some map and its graph
\begin{equation*}
    A = \big\{(h,\phi(h)\big)\mid h\in \F_p^n\big\} \subseteq \F_p^n\times \F_p^n
\end{equation*}
has energy $p^{3n}$, it turns out that~$\phi$ must be an affine linear map (and vice versa).
\begin{lemma}[Weak linearity]\label[lemma]{lem:weak_linearity}
    For a map $f:\F_p^n\to \D$, a set $S\subseteq \F_p^n$ and a map $\phi:S\to \F_p^n$, suppose that 
    \begin{equation*}
        \sum_{h\in S} \big|\widehat{\Delta_hf}\big(\phi(h)\big)\big|^2 \geq \eps p^n.
    \end{equation*}
    Then, the set $\{(h,\phi(h))\mid h\in S\}$
    has energy at least $\eps^4 p^{3n}$.
\end{lemma}

\subsection{From weak linearity to true linearity}\label{sec:quadratic:weak_linearity_to_true_linearity}
Given the set $A$ from \cref{lem:weak_linearity}, we now use results from additive combinatorics to lift the weak form of linearity of~$\phi$ to true linearity on a large affine subspace. 
Let $A\subseteq G$, we call $A+A$ the doubling set of $A$, and the relative size $|A+A|/|A|$ the doubling constant. 
In practice, determining the doubling constant exactly is hard. 
Yet, lower bounds can often be given. 
The next two results show that the set $A$ from \cref{lem:weak_linearity} has small doubling and that this implies that it is contained in a slightly larger affine subspace. 
On this affine subspace, $\phi$ behaves linearly. 
\begin{theorem}[Balog--Szemer\'edi--Gowers]\label{thm:BSG:Schoen}
  There is an absolute positive constant~$C$ such that the following holds.
  Let~$G$ be a finite Abelian group and suppose that $A\subseteq G$ has energy at least $\delta|A|^3$.  
  Then, there is a set $A'\subseteq A$ of size $|A'| \geq \delta^C|A|$ such that $|A'+A'|\leq \delta^{-C}|A'|$.
\end{theorem}
\begin{theorem}[Fre\v{\i}man--Ruzsa]\label{thm:FreimanRuzsa}
For any $K\geq 1$ and prime number~$p$, there is a $K'>0$ such that the following holds.
Suppose $A\subseteq \F_p^n$ satisfies $|A+A|\leq K|A|$.
Then, the linear span of~$A$ satisfies $|\ip{A}| \leq K'|A|$.
\end{theorem}
\begin{remark}\label[remark]{rm:poly_FR}
    Recently, \citeauthor{GowersGreenMannersTao:2025} proved the longstanding Marton conjecture, also known as the polynomial Fre\v{\i}man-Ruzsa conjecture~\cite{GowersGreenMannersTao:2025}, improving the dependency of $K'$ on $K$ in the previous theorem. 
    The setting considered by \citeauthor{GowersGreenMannersTao:2025} differs however from the situation considered in our work. 
    Additionally, their result is non-algorithmic. 
    A future algorithmic version of their result might improve the bounds found in this chapter. 
\end{remark}

It then follows that there exists a set
\begin{equation*}
A'\subseteq \big\{\big(h,\phi(h)\big) \mid h\in \F_p^n\big\}    
\end{equation*}
of large size $|A'| \geq c p^n$ such that $|\ip{A'}| \leq C |A'|$, for some~$c,C>0$ depending on~$\eps$ and~$p$ only.
Moreover, each pair $(h,\phi(h))\in A'$ satisfies
\begin{equation*}
    \big|\widehat{\Delta_{h}f}\big(\phi(h)\big)\big| \geq c.
\end{equation*}
Let $(h_1,\phi(h_1)),\dots,(h_m,\phi(h_m))$ be a maximal set of linearly independent elements in~$A'$.
Let~$M\in\F_p^{n\times n}$ be a matrix satisfying the set of linearly independent equations
\begin{equation*}
    Mh_i = \phi(h_i)
\end{equation*}
for all $i\in[m]$.
The map~$\phi$ then behaves linearly on all of~$A'$, in the sense that $\phi(h) = Mh$ for all $(h,\phi(h))\in A'$.
Hence, due to the relative size of~$A'$ in its linear span, we get that
\begin{align}
\Exp_{h\in \F_p^n}\big| \Exp_{x\in \F_p^n} \Delta_hf(x) \omega_p^{\ip{x,Mh}}\big|^2&=
    \Exp_{h\in \F_p^n} \big|\widehat{\Delta_hf}(Mh)\big|^2\label{eq:strong_linearity}\\
    &\geq
    \Exp_{h\in \F_p^n}\one\big[\big(h,\phi(h)\big)\in A'\big] \big|\widehat{\Delta_hf}(Mh)\big|^2\nonumber\\
    &\geq c.\nonumber
\end{align}

%% file: decoding/quadratic_decoding/warm_up_quadratic_decoding.tex
\section{Quantum decoding of quadratic codes in a noiseless case}\label{sec:quadratic:warm_up}
Below we discuss how to learn quadratic functions in a noiseless setting and how we can modify this algorithm to learn the function with high probability in a low-error setting, in both cases using the Fourier sampling subroutine (\cref{lem:Fourier_sampling}).

\subsection{Noiseless case}\label{sec:noiseless_case}
\citeauthor{Montanaro:2012} considered a problem of learning an unknown multilinear polynomial of degree at most~$d$ and presented a $\bigo(n^{d-1})$-query quantum algorithm that solves it~\cite{Montanaro:2012}.
\begin{theorem}\label{thm:quadratic:noiseless}
    Given a quadratic phase function $f:\F_p^n\to\D$, we can learn $f$ using $pn+2$ quantum queries to $f$.
\end{theorem}
The algorithm uses Fourier sampling to learn the quadratic phase function. 
In the noiseless case, we have the following lemma on the Fourier coefficients of multiplicative derivatives found using Fourier sampling. 
\begin{lemma}\label[lemma]{lem:Fourier_sampling:delta_factor_noiseless}
    Let $f:\F_p^n\to \D$ be a quadratic phase given by $f(x) = \omega_p^{\ip{x,Mx} + \ip{x,b} + c}$, for some matrix $M\in \F_p^{n\times n}$, vector $b\in\F_p^n$ and scalar $c\in\F_p$.
    Then, 
    \begin{equation*}
    \widehat{\Delta_h f}(y)= 
        \begin{cases}
            1 & \text{if $y=(M+M^T)h$}\\
            0 & \text{else.}
        \end{cases}
    \end{equation*}
\end{lemma}
\begin{proof}
    We expand the derivative and see which terms cancel:
    \begin{align*}
        \widehat{\Delta_h f}(y) & = \frac{1}{p^n}\sum_{x\in\F_p^n} \overline{f(x)}f(x+h)\omega_p^{\ip{x,y}} \\
        & = \frac{1}{p^n}\sum_{x\in\F_p^n} \omega_p^{\ip{x,(M+M^T)h} + \ip{b, h}}\omega_p^{\ip{x,y}} \\
        & = \frac{1}{p^n}\omega_p^{\ip{b,h}}\sum_{x\in\F_p^n} \omega_p^{\ip{x,(M+M^T)h + y}} \\
        & = \begin{cases}
            1 & \text{if } y=(M+M^T)h \\
            0 & \text{otherwise.}
        \end{cases}
    \end{align*}
\end{proof}
\begin{proof}[ of \cref{thm:quadratic:noiseless}]
\cref{lem:Fourier_sampling,lem:Fourier_sampling:delta_factor_noiseless} show that for fixed $h\in\F_2^n$ and using $p$ queries to $f$, we obtain the output $(M+M^T)h$ with certainty. 

Choosing $n$ linearly independent $h$'s (for instance the standard basis vectors) gives $n$ linearly independent pairs of the form $(h,(M+M^T)h)$. 
We now learn~$M$ using Gaussian elimination combined with the fact that we can set $M$ to be upper diagonal. 

Next, we learn $b$ by running the Bernstein-Vazirani algorithm, where $f$ is queried and multiplied by the phase $\omega_p^{\ip{x,Mx}}$.
We learn $c$ by querying $f(0)$.  
\end{proof}

\subsection{Decoding within the unique-decoding radius}
In the noisy case, the Fourier coefficients no longer behave as delta functions, but instead as ``approximate delta functions''.
By this, we mean that in the Fourier spectrum, almost all terms have small magnitude and only a bounded number have larger amplitude. 
The number of terms with larger magnitude depends on the actual error rate. 

Below we illustrate what happens in case of few errors and how our algorithm changes. 
We restrict ourselves to $p=2$ for simplicity, yet the results extend to odd primes~$p$. 
We know that within the unique-decoding radius, a unique codeword exists that is closest to the corrupted codeword. 
In this case, essentially the same algorithm works as in the noiseless case, with the main difference that the Fourier transforms of the multiplicative derivatives are no longer delta functions. 
Instead, they satisfy that there exists a unique Fourier coefficient whose absolute valued squared is strictly greater than~$1/2$. 
The existence of this Fourier coefficient implies that $\bigo(\log n)$ repetitions of the Fourier sampling algorithm can locate the large Fourier coefficient with high probability via Hoeffding's inequality. 
\begin{theorem}\label{thm:quadratic:low_error}
    Let $\eps>0$. 
    Let $f'$ have relative distance at most $\frac{1}{4} - \frac{1}{4}\sqrt{\frac{1}{2} + \eps}$ from a degree-$2$ Reed-Muller codeword $f$.
    Then, with success probability at least $1-\tfrac{1}{n}$, we can learn~$f$ using $\bigo_\eps(n\log n)$ quantum queries. 
\end{theorem}
Note that the allowed relative distance in the theorem is slightly smaller than the unique decoding radius of $\frac{1}{8}$ for Reed-Muller codewords of degree at most~$2$. 
\begin{proof}
    With probability $\frac{3}{4} + \frac{1}{4}\sqrt{\frac{1}{2} + \eps}$, $f'(x)=f(x)$ for a random $x$. 
    By the union bound, it holds for uniformly random $x$ and $h$ that $f'(x+h)\overline{f'(x)}=f(x+h)\overline{f(x)}$ with probability at least $\tfrac{1}{2}+\frac{1}{2}\sqrt{\frac{1}{2}+\eps}$. 
    
    Let $c(x)$ be the indicator variable that evaluates to~$1$ if $f'(x)\neq f(x)$. 
    Then, revisiting the proof of \cref{lem:Fourier_sampling:delta_factor_noiseless}, we find that the probability for the correct measurement outcome $y$ is given by: 
    \begin{align*}
        \Pr[y] & = |\widehat{\Delta_h f}(y)|^2 \\
        & = \frac{1}{2^{2n}}\Big|\sum_{x\in\F_2^n} (-1)^{c(x)+c(x+h) + \ip{x,(M+M^T)h+y}}\Big|^2
    \end{align*}
    For $y=(M+M^T)h$, this thus evaluates to 
    \begin{align}
        \Pr[y=(M+M^T)h] & = \frac{1}{2^{2n}}\Big|\sum_{x\in\F_2^n} (-1)^{c(x)+c(x+h)}\Big|^2 \nonumber \\
        & \ge \frac{1}{2}+\eps. \label{eq:warm_up:success_probability_quadratic}
    \end{align}

    We can now amplify the probability of correctly finding $y$ by taking the majority of multiple independent runs of the algorithm. 
    By Hoeffding's inequality, the majority over the measurement outcomes of $m=\tfrac{1}{\eps^2}\log n$ independent runs for the same $h$ gives a valid pair $(h,(M+M^T)h)$ with probability $1-\tfrac{1}{n^2}$. 
    
    Now run the above procedure for $n$ linearly independent $h$-values (for instance the $n$ standard basis vectors) to obtain $n$ pairs $(h_i,(M+M^T)h_i)$. 
    By the union bound, all pairs are correct with probability $1-\tfrac{1}{n}$. 

    Similar to \cref{thm:quadratic:noiseless}, we can learn $M$ from these $n$ linearly independent pairs. 
    With two additional queries (one in superposition and one to a single index) we learn $L$ and $c$ with good probability, and thereby learn $f$. 
\end{proof}

Conventional queries only return a single $f(h)$ value.
Polynomial interpolation is therefore needed to learn a pair $(h,(M+M^T)h)$. 
Additionally, every query can be corrupted, requiring more queries to obtain a valid pair with high probability. 

As the error rate increases beyond the unique-decoding radius, the Fourier coefficients of the multiplicative derivatives are no longer peaked at a single point.
In fact, multiple Fourier coefficients can have the same absolute value  squared.
As a result, the outcome for various directions $h$ and $h'$ may result in pairs originating from different quadratic phases, that is, we might obtain two pairs $(h,(M+M^T)h)$ and $(h', (M'+M'^T)h')$ for distinct matrices $M\neq M'$. 

As a result, combining the obtained pairs might result in a combined matrix with a distance to the correct quadratic Reed-Muller codeword larger than expected. 

%% file: decoding/quadratic_decoding/non_uniformity_to_weak_linearity.tex
\section{Quantum-algorithmic weak linearity}\label{sec:quadratic:quantum_algorithmic_weak_linearity}
In this section, we show how to sample from a set with high energy provided we have a polynomial phase function $f:\F_p^n\to \D$ satisfying $\|f\|_{U^3} \geq \gamma$.

Given some $\delta \in (0, 1)$, define the \emph{spectral set} $S_{\delta}$ to be the set of all $(h,a)$ pairs with high Fourier coefficient:
\begin{equation}\label{eq:spectral_set}
    S_{\delta} =  \left\{(h,a)\in \F_p^{n}\times\F_p^n \mid |\widehat{\Delta_hf}(a)| \geq \delta\right\}.
\end{equation}
Let $\delta_1 = \poly(\gamma)$ be some fixed parameter.

Let $\phi:\F_p^n\to\F_p^n$ be the random map where each coordinate is independently sampled according to the Fourier sampling algorithm from \cref{lem:Fourier_sampling}, namely, for each $h\in \F_p^n$,
\begin{equation} %\label{eq:prob_phi}
    \Pr\big[\phi(h) = a\big] = |\widehat{\Delta_{h}f}(a)|^2.
\end{equation}

Based on the map~$\phi$, define the random set
\begin{equation} \label{eq:setA0}
    A_0 = \big\{\big(h,\phi(h)\big) \mid h\in \F_p^n\big\}.
\end{equation}

Note that we can sample a uniformly random element from~$A_0$ by first sampling a uniformly random $h\in\F_p^n$ and then sampling~$\phi(h)$ using Fourier sampling.
This way, we sample sequentially from the joint probability distribution imposed on~$A_0$. 
Later, we will also have to check if a given pair $(h,a)$ belongs to~$A_0$. 
We do this by applying the Fourier sampling routine on $h$ and comparing the outcome with~$a$. 

We will keep track of the spectral elements in this set
\begin{equation}\label{eq:setA1}
    A_1 = A_0\cap S_{\delta_1}.
\end{equation}
\begin{proposition}\label{prop:size_A1}
    With probability~$\poly(\gamma)$, we have that $|A_1| \geq \poly(\gamma)p^n$.
\end{proposition}
\begin{proof}
    \cref{prop:U3_to_weak_linearity} and Markov's inequality (\cref{lem:Markov}) show that there are at least $\frac{\gamma^8}{2}p^n$ many $h\in \F_p^n$ such that $\max_a|\widehat{\Delta_{h}f}(a)|^2 \geq \frac{\gamma^8}{2}$.
    Define the indicator random variables
    \begin{equation*}
        X_h = \one\left[\phi(h) = \arg\max_a|\widehat{\Delta_{h}f}(a)|\right]
    \end{equation*}
    and let $\overline{X} = \sum_{h\in \F_p^n}X_h$.
    Note that if~$X_h$ holds, then $\big(h,\phi(h)\big)\in S_{\delta_1}$.
    Then,
    \begin{equation*}
        \Exp\big[\overline{X}\big] \geq \frac{\gamma^{16}}{4}p^n.
    \end{equation*}
    By Markov's inequality, we have that $\overline{X} \geq \frac{\gamma^{16}}{8}p^n$ with probability at least~$\frac{\gamma^{16}}{8}$.
\end{proof}

It follows from \cref{lem:weak_linearity} that if the event from \cref{prop:size_A1} holds, then the (random) set~$A_1$ has energy at least~$\poly(\gamma)|A_1|^3$.
As the size of~$A_1$ is large, a sample from $A_0$ is in $A_1$ with constant probability. 
As a result, contrary to the approach of \citeauthor{TulsianiWolf:2014}, we do not have to sample from~$A_1$ directly, thereby reducing the query complexity. 
The next step is to construct an algorithmic version of the Balog-Szemer\'edi-Gowers theorem.

%% file: decoding/quadratic_decoding/variant_BSG.tex
\section{A variant of the Balog-Szemer\'edi-Gowers theorem}\label{sec:quadratic:variant_BSG}
In this section, we prove a variant of the Balog--Szemer\'edi--Gowers theorem that is inspired by a proof of this theorem due to \citeauthor{Schoen:2014}~\cite{Schoen:2014}.
The variant outlined here is tailored for the purpose of turning it into an algorithm that we give in the following section.

Let $G$ be a finite Abelian group and let $A\subseteq G$, let~$E(A)$ denote the energy of~$A$. 
For an element $x\in G$, define the \emph{popularity} of~$x$ relative to~$A$ by
\begin{equation*}
    r_A(x) = |\{(a,b)\in A^2 \st a-b = x\}|.
\end{equation*}

The next lemma defines two sets $H_1\subseteq H_0$ such that $H_0$ has small doubling and~$H_1$ is large. 
We can interpret the lemma as an algorithmic way of preparing a bipartite graph. 
Two nodes in the bipartite graph are connected if their difference is popular. 
It turns out that a large fraction of the vertices in the graph has high degree, which we can use to construct the set of small doubling. 
\begin{lemma}\label[lemma]{lem:BSG_graph}
Let $\delta,\eta>0$ and suppose that $A_1\subseteq A_0\subseteq G$ are sets such that $|A_1| = \eta|A_0|$ and~$A_1$ has energy at least $\delta|A_1|^3$.
Let~$X,Y$ be independent uniformly distributed random variables over~$A_1$ and define the random sets
\begin{equation}
    B_i = A_i\cap (A_i - X+Y). \label{eq:set:B_i}
\end{equation}
Define subsets~$H_1\subseteq H_0$ by
\begin{align}
    H_0 & = \Big\{a\in B_0 \mid \sum_{b\in B_0} \one\big[r_{A_0}(a-b) \geq \tfrac{\delta^2}{32}|A_0|\big]  \geq \tfrac{5}{8}|B_0|\Big\} \label{eq:set:H_0}, \\
    H_1 & = \Big\{a\in B_1 \mid \sum_{b\in B_0} \one\big[r_{A_0}(a-b) \geq \tfrac{\delta^2}{16}|A_0|\big]  \geq \tfrac{3}{4}|B_0|\Big\}. \label{eq:set:H_1}
\end{align}
Then, with probability at least~$\tfrac{\delta^2\eta^2}{4}$ over the choice of~$X$ and~$Y$, we have that
\begin{equation*}
    |H_1| \geq \tfrac{\delta}{C}|A_1| \quad\text{and}\quad|H_0 + H_0| \leq \frac{C}{\delta^{11}\eta^{11}}|A_0|,
\end{equation*}
where~$C>0$ is an absolute constant.
\end{lemma}
By this lemma, the set $H_0$ has small doubling and the set $H_1$ is large.
As also $A_1\subseteq A_0$, we see that there exists a large set of small doubling interspersed between $A_1$ and $A_0$.
The next lemma states that most vertices in the bipartite graph induced by the sets $H_0$ and $H_1$ have high degree. 
\begin{lemma}\label{lem:BSG_Schoen}
Let $A_1\subseteq A_0$ be finite additive sets such that $|A_1| = \eta|A_0|$ and $E(A_1) =\delta|A_1|^3$.
Let $X,Y$ be independent uniformly distributed $A_1$-valued random variables.
For $i \in \bset$, let
\begin{align*}
    B_i = A_i\cap (A_i- X+Y).
\end{align*}
Then, for any $\eps > 0$, with probability at least $\frac{1}{4}\delta^2\eta^2$, we have that $|B_1| \geq \frac{1}{4}\delta |A_1|$ and $B_1\times B_0$ has at least $\big(1 - \frac{2\eps}{\delta^2\eta^2}\big)|B_1||B_0|$ pairs $(a,b)$ such that $r_{A_0}(a-b) \geq \eps|A_0|$.
\end{lemma}
\begin{proof}
    Let $\one_{A}[a]$ denote the indicator value of the event $a\in A$. 
    Since $E(A_1)$ equals the number of triples $a,x,y$ such that $a+x-y\in A_1$, we have
    \begin{equation*}
        \delta|A_1| = \frac{E(A_1)}{|A_1|^2} = \Exp_{x,y\in A_1}\sum_{a\in A_1}\one_{A_1}[a+x-y] = \Exp|B_1|.
    \end{equation*}

    Define the set of unpopular pairs
    \begin{equation*}
        S = \{(a,b)\in A_1\times A_0 \mid r_{A_0}(a-b) < \eps |A_0|\}.
    \end{equation*}
    Let $T = B_1\times B_0\cap S$.
    Note that $\Pr[X-Y = z] = r_{A_1}(z)/|A_1|$.
    Then,
    \begin{align*}
        \Exp|T| & = \Exp_{x,y\in A_1}\Big[\sum_{(a,b)\in S}\one_{A_1 - x + y}[a]\one_{A_0 - x+y}[b]\Big] \\
        & = \frac{1}{|A_1|}\sum_zr_{A_1}(z)\sum_{(a,b)\in S}\one_{A_1 -z}[a]\one_{A_0 -z}[b] \\
        & \leq \frac{1}{|A_1|}\sum_{(a,b)\in S}\sum_z\one_{A_1}[a+z]\one_{A_0}[b+z] \\
        & \leq \frac{1}{|A_1|}\sum_{(a,b)\in S}\sum_z\one_{A_0}[z]\one_{A_0}[z-a+b].
    \end{align*}
    We see that this last expression is equal to 
    \begin{equation*}
        \frac{1}{|A_1|}\sum_{(a,b)\in S}r_{A_0}(a-b),
    \end{equation*}
    which we can upper bound by $\eps |A_0|^2$.
    Using the inequalities for $\Exp|T|$ and $\Exp|B_1|$, it follows from the Cauchy-Schwarz inequality that for $\mu = \frac{\delta^2\eta^2}{2\eps}$, we have
    \begin{equation*}
        \Exp\big(|B_1|^2 - \mu|T|\big) \geq \big(\Exp|B_1|\big)^2 - \mu\Exp|T| \geq \frac{\delta^2\eta^2}{2}|A_0|^2.
    \end{equation*}
    Hence, by Markov's inequality, with probability at least $\frac{\delta^2\eta^2}{4}$, we have 
    \begin{equation*}
        |B_1|^2 - \mu|T| \geq \frac{\delta^2\eta^2}{4}|A_0|^2.
    \end{equation*}
    This gives the desired lower bound on~$|B_1|$.
    Since $B_1\subseteq B_0$, it also implies that
    \begin{equation*}
        |B_1||B_0| \geq \mu|T|.
    \end{equation*}
    Rearranging, this last expression shows that there are at most $\frac{2\eps}{\delta^2\eta^2}|B_1||B_0|$ pairs $(a,b)\in B_1\times B_0$ such that $r_{A_0}(a-b) < \eps|A_0|$.
\end{proof}

In proving \cref{lem:BSG_graph}, we will use a triangle-inequality-like result~\cite[Corollary~7.3.6]{Zhao:2023}.
\begin{lemma}\label[lemma]{lem:triangle_like_inequality}
    For any three sets $A,B,C\subseteq G$, we have that
    \begin{equation*}
        |A||B+C| \leq |A+B||A+C|.
    \end{equation*}
\end{lemma}
\begin{proof}[ of \cref{lem:BSG_graph}]
Let~$B_1,B_0$ be as in \cref{lem:BSG_graph} and set $\eps = \frac{\delta^2\eta^2}{16}$. Define the bipartite graphs
\begin{align*}
    \Gamma_1 &= \big\{(a,b)\in B_0\times B_0 \mid r_{A_0}(a-b) \geq \tfrac{\delta^2\eta^2}{32}|A_0|\big\}\\
    \Gamma_2 &= \big\{(a,b)\in B_1\times B_0 \mid r_{A_0}(a-b) \geq \tfrac{\delta^2\eta^2}{16}|A_0|\big\}.
\end{align*}
Then, $|\Gamma_2| \geq \frac{7}{8}|B_1||B_0|$.
Define the sets
\begin{align*}
    H_0 &= \big\{a\in B_0 \mid \deg_{\Gamma_1}(a) \geq \tfrac{5}{8}|B_0|\big\}\\
    H_1 &= \big\{a\in B_1 \mid \deg_{\Gamma_2}(a) \geq \tfrac{3}{4}|B_0|\big\}.
\end{align*}

Due to the size of~$\Gamma_2$, it follows that $|H_1| \geq \frac{1}{8}|B_1|$.

By the inclusion-exclusion principle, since any two distinct left vertices $u,v\in H_0$ have such a high degree in~$\Gamma_1$, they must have at least $\tfrac{1}{4}|B_0|$ common (right) neighbors.
If~$x\in B_0$ is a right neighbor of~$u$, then $r_{A_0}(u-x) \geq \tfrac{\delta^2\eta^2}{32}|A_0|$ and similarly for~$v$.
It follows that
\begin{equation*}
    \sum_{x\in B_0}r_{A_0}(u-x)r_{A_0}(v-x) \geq \big(\tfrac{1}{4}|B_0|\big)\big(\tfrac{\delta^2\eta^2}{32}|A_0|\big)^2 \geq 2^{-14}\delta^5\eta^5|A_0|^3.
\end{equation*}
Observe that the left-hand side of the equation is a lower bound on the number of four-tuples $(a,b,c,d)\in A_0^4$ such that $(a-b) - (c-d) = u-v$.
For each $w\in H_0 - H_0$, let $u_w,v_w\in H_0$ be an arbitrary pair such that $u_w - v_w= w$.
Then,
\begin{equation*}
    |A_0|^4 \geq \sum_{w\in H_0 - H_0}\sum_{x\in B_0}r_A(u_w-x)r_{A_0}(v_w-x) \geq 2^{-14}\delta^5\eta^5|H_0 -H_0||A_0|^3.
\end{equation*}
Rearranging then gives
\begin{equation}\label{eq:diffset}
    |H_0 - H_0| \leq \frac{2^{14}}{\delta^5\eta^5}|A_0|.
\end{equation}
Since we have $H_0\subseteq H_1$, combining~\cref{eq:diffset} with \cref{lem:triangle_like_inequality} applied with $A = -B = -C = H_0$ then gives the desired bound on the doubling of~$H_0$:
\begin{equation*}
    |H_0 + H_0| \leq \frac{|H_0 - H_0|^2}{|H_0|} \leq \frac{2^{31}}{\delta^{11}\eta^{11}}|A_0|.
\end{equation*}
\end{proof}

%% file: decoding/quadratic_decoding/sampling_high_degree_elements.tex
\section{Sampling high-degree elements}\label{sec:quadratic:sampling_high_degree_elements}
This section gives an algorithmic version of \cref{lem:BSG_graph}.
Let $X,Y$ be uniformly random elements from the (random) set~$A_0$ as given in~\cref{eq:setA0}. 
Based on~$A_0$, and the spectral set~$A_1$ as in~\cref{eq:setA1}, define the sets~$B_0$ and~$B_1$ as in \cref{eq:set:B_i} of \cref{lem:BSG_graph}.
Define the event
\begin{equation}\label{eq:event_E1}
    \E_1 = \big\{|A_1| \geq \poly(\gamma)p^n \quad \wedge\quad X,Y\in A_1\big\}.
\end{equation} 
By \cref{prop:size_A1}, this event holds with probability~$\poly(\gamma)$.
Conditioned on this event, both~$X$ and~$Y$ are uniformly distributed over~$A_1$.

Let $\delta_2 = E(A_1)/|A_1|^3$ and $\eta = |A_1|/p^n$.
Define subsets~$H_0\supseteq H_1$ as in~\cref{eq:set:H_0,eq:set:H_1} of \cref{lem:BSG_graph} using $\delta=\delta_2$. 
Let~$C$ be the constant from \cref{lem:BSG_graph} and define the event
\begin{equation} \label{eq:event_E2}
    \mathcal{E}_2 = \big\{ |H_1| \geq \tfrac{\delta_2}{C}|A_1|\text{ and } |H_0 + H_0| \leq \tfrac{C}{\delta_2^{11}\eta^{11}}|A_0|\big\}.
\end{equation}
Conditioned on~$\E_1$, event $\E_2$ holds with probability~$\poly(\gamma)$.

\subsection{Approximating sets of small doubling}
By event~$\E_1$, we learn that $A_1$ is large.
By event~$\E_2$ know that $H_1$ has size at least a constant fraction of the size of $A_1$, hence $H_1$ is also large. 
The same event also tells us that $H_0$, which contains $H_1$, has doubling bounded by a constant times the size of $A_0$. 
As $A_1 \subseteq A_0$, we see that there should exist a set, called~$B_2$, approximately between $H_1$ and $H_0$ that is large and has small doubling. 
Below, we give an algorithm that allows us to sample from~$B_2$.
\begin{definition}[$m$-Popularity Estimator]
    Given input $u\in \F_p^{2n}$, independently and uniformly sample $v_1,\dots,v_m$ from~$A_0$.
    Return
    \begin{equation*}
        \popest_m(u) := \frac{1}{m}\sum_{i=1}^m \one[u+v_i\in A_0].
    \end{equation*}
\end{definition}

Note that the indicator function requires membership queries to~$A_0$.
\begin{claim}\label{claim:upperbound_popest}
    Let $\delta \in (0, 1)$ and $L = \lceil 12/\delta^2 \rceil$.
    Conditioned on $A_0$, the random events
    \begin{equation*}
        \big\{\big|\popest_L(u) - \tfrac{r_{A_0}(u)}{|A_0|}\big| > \delta/2 \big\}, \quad u \in \F_p^{2n}
    \end{equation*}
    are jointly independent, and each occurs with probability at most $1/100$.
\end{claim}
\begin{proof}
    The independence follows as the samples in the popularity estimator are taken independent of the input $u$. 
    The second claim follows from Hoeffding's inequality (\cref{lem:Hoeffding}) and the fact that $\Exp_u \popest_L(u)=|A_0|^{-1}r_{A_0}(u)$. 
\end{proof}

\begin{definition}[$(m,\delta)$-Degree Estimator]
    Given input $w\in B_0$, independently and uniformly sample $u_1,\dots,u_m$ from~$B_0$.
    Return
    \begin{equation*}
        \degest_{m, \delta}(w) := \frac{1}{m} \sum_{i=1}^m \one \big[\popest_{L}(w-u_i) \geq \delta\big],
    \end{equation*}
    where $L = \lceil 12/\delta^2 \rceil$.
\end{definition}

Let $m_2 = \Omega(\log n)$ and define the random set 
\begin{equation}\label{eq:setB1}
    B_2 = \big\{ w\in B_0 \mid \degest_{m_2, \delta_2}(w) \geq 3/4\big\}.
\end{equation}
We show that with high probability -- conditioned on the events~$\E_1$ and~$\E_2$ -- a negligible fraction of~$B_2$ lies outside of~$H_0$, while it contains nearly all of~$H_1$.
To this end, define the random event
\begin{equation}\label{eq:event_E3}
    \mathcal{E}_3 = \big\{ |B_2 \cap H_0| \geq \big(1 - \tfrac{1}{n^2}\big) |B_2| \text{ and } |B_2 \cap H_1| \geq \big(1 - \tfrac{1}{n^2}\big) |H_1|\big\},
\end{equation}
which indicates that $B_2$ is very nearly sandwiched between $H_0$ and $H_1$.
We show that $\E_3$ holds with high probability:

\begin{lemma}\label{lem:lower_bound_e3_event}
    Conditioned on $\E_1$ and $\E_2$, the probability of event $\E_3$, taken over the random choices of the Popularity and Degree Estimators, is at least $1-1/n^2$.
\end{lemma}
\begin{proof}
    We prove that for all $w\not\in H_0$ we have $\Pr[w\in B_2]\le 1/n^2$.
    The case where if $w\in H_1$, then $\Pr[w\not\in B_2]\le 1/n^2$ follows via similar arguments.

    Assume $w\in B_0\setminus H_0$, then 
    \begin{equation*}
        \left|\left\{v\in B_0\mid \tfrac{r_{A_0}(w-v)}{|A_0|}<\frac{\delta^2_2}{32}\right\}\right|\ge \frac{3|B_0|}{8}.
    \end{equation*}
    Let $I\subseteq [m_2]$ denote the random set 
    \begin{equation*}
        \left\{i\in[m_2]\mid \frac{r_{A_0}(w-u_i)}{|A_0|}< \frac{\delta^2_2}{32}\right\},
    \end{equation*}
    where $u_1, \hdots, u_{m_2}\in B_0$ are sampled independent and uniformly at random and are accepted by the $(m_2,\delta_2)$-degree estimator. 
    By Hoeffding's inequality, we have
    \begin{equation}\label{eq:Ibound}
        \Pr\Big(|I|< \tfrac{7m_2}{24}\Big) \le 2e^{-2m_2/9}.
    \end{equation}
    Note that for any $\xi$ and~$L$, we have that
    \begin{equation*}
        \sum_{i=1}^{m_2} \one[\popest_L(w-u_i)\ge \xi] \le \sum_{i\in I} \one[\popest_L(w-u_i)\ge \xi] + m_2 - |I|.
    \end{equation*}
    It follows that
    \begin{align*}
        & \Pr\Bigg(\sum_{i=1}^{m_2} \one[\popest_{L}(w-u_i)\ge \xi] \ge \frac{3m_2}{4}\Bigg) \\
        & \quad \le \Pr\Bigg( \sum_{i\in I} \one[\popest_{L}(w-u_i)\ge \xi] \ge |I| - \frac{m_2}{4}\Bigg) \\
        & \quad \le \Pr\bigg(|I| < \frac{7m_2}{24}\bigg) + \Pr\Bigg( \sum_{i\in I} \one[\popest_{L}(w-u_i)\ge \xi] \ge \frac{m_2}{24} \,\Big|\, |I|\ge\frac{7m_2}{24}\Bigg).
    \end{align*}
    The first term is small as \cref{eq:Ibound} shows. 
    For the second term, we use that for every $i\in I$,  \cref{claim:upperbound_popest} combined with $\xi=\delta_2^2/16$ and $L=\ceil{12/\xi^2}$ gives
    \begin{align*}
        \Pr\big(\popest_{L}(w-u_i)\ge \delta_2^2/16\big) & \le \Pr\left(\left|\popest_{L}(w-u_i) - \tfrac{r_{A_0}(w-u_i)}{|A_0|}\right|> \tfrac{\delta_2^2}{32}\right) \\
        & \le 1/100.
    \end{align*}
    Hoeffding's inequality then gives
    \begin{equation*}
        \Pr\left(\sum_{i\in I} \one\Big[\Big|\popest_{L}(w-u_i) - \tfrac{r_{A_0}(w-u_i)}{|A_0|}\Big|> \tfrac{\delta_2^2}{32}\Big] \ge \frac{|I|}{24}\right) \le 2e^{-|I|/288}.
    \end{equation*}
    From this and~\cref{eq:Ibound}, we can conclude that 
    \begin{align*}
        \Pr & \left(\sum_{i\in I} \one\left[\left|\popest_{L}(w-u_i)\right| \ge \tfrac{\delta_2^2}{16}\right] \ge \frac{m_2}{24} \,\Big|\, |I|\ge \frac{7m_2}{24}\right) \\
        & \le \Pr\left(\sum_{i\in I} \one\left[\left|\popest_{L}(w-u_i) - \tfrac{r_{A_0}(w-u_i)}{|A_0|}\right| \ge \tfrac{\delta_2^2}{32}\right] \ge \frac{m_2}{24} \,\Big|\, |I|\ge \frac{7m_2}{24}\right) \\
        & = \frac{\Pr\Bigg(\sum_{i\in I} \one\left[\left|\popest_{L}(w-u_i) - \tfrac{r_{A_0}(w-u_i)}{|A_0|}\right| \ge \tfrac{\delta_2^2}{32}\right] \ge \frac{m_2}{24} \wedge |I|\ge \frac{7m_2}{24}\Bigg)}{\Pr\left(|I|\ge \frac{7m_2}{24}\right)} \\
        & \le \frac{\Pr\left(\sum_{i\in I} \one\left[\left|\popest_{L}(w-u_i) - \tfrac{r_{A_0}(w-u_i)}{|A_0|}\right| \ge \tfrac{\delta_2^2}{32}\right] \ge \frac{|I|}{24} \wedge |I|\ge \frac{7m_2}{24}\right)}{\Pr\left(|I|\ge \frac{7m_2}{24}\right)} \\
        & \le \frac{\Pr\left(\sum_{i\in I} \one\left[\left|\popest_{L}(w-u_i) - \tfrac{r_{A_0}(w-u_i)}{|A_0|}\right| \ge \tfrac{\delta_2^2}{32}\right] \ge \frac{|I|}{24}\right)}{\Pr\left(|I|\ge \frac{7m_2}{24}\right)} \\
        & \quad \le 4e^{-|I|/288},
    \end{align*}
    where in the last line we upper bounded \cref{eq:Ibound} by $1/2$. 

    Combined, this gives $\Pr(w\in B_2)\le 2e^{-2m_2/9} + 4e^{-m_2/288} \le 1/n^2$ for $w\not\in H_0$.
\end{proof}

\subsection{Finding a small spanning set}
Next, we work towards obtaining a basis for the set~$B_2$ of small doubling.
Let $\mu = |B_2\cap H_1|/|B_2|$ and let 
\begin{equation*}
    B_3 = \{v_1,\dots,v_{20n/\mu}\}\subseteq B_2
\end{equation*} 
be a set of independent uniformly random elements from~$B_2$.
Define the event
\begin{equation*}
    \E_4 =  \{B_3\subseteq H_0 \text{ and } |\langle B_3\rangle\cap H_1| \geq |H_1|/2\}.
\end{equation*}

\begin{lemma}\label{lem:prob_schoen}
    We have that $\Pr[\E_4 \mid \E_1 \wedge \E_2 \wedge \E_3]\ge 1/3$.
\end{lemma}
\begin{proof}
    Assume events~$\E_1, \E_2, \E_3$ all hold true.
    Since $|B_2\cap H_0|\ge (1-\frac{1}{n^2})|B_2|$, a uniform sample from $B_2$ lies in $H_0$ with probability at least $1-\frac{1}{n^2}$ and hence by the union bound, $B_3\subseteq H_0$ holds with probability $1-\bigo(\frac{1}{n})$. 
    
    Let $H_1' = B_2\cap H_1$ and let $I = \{i \mid v_i\in H_1'\}$.
    For each $i\in I$, the random variable~$v_i$ is uniformly distributed over~$H_1'$.
    Since $|H_1'| = \mu|B_2|$, by Hoeffding's inequality, $|I|\geq 10n$ with probability $1-\bigo(\frac{1}{n})$.
    Assume this is the case and that $B_3\subseteq H_0$; note that both events hold simultaneously with probability $1-\bigo(\frac{1}{n})$.
    
    Write $V_{\emptyset} = \{0\}\subseteq \F_p^{2n}$ and $V_J = \mathrm{span}\{ v_i\mid i\in J\}\subseteq\F_2^{2n}$ for $J\subseteq I$. 
    
    We show by contradiction that
    \begin{equation*}
        \Pr_{\{v_i\mid i\in I\}}\big(|V_I\cap H_1'|\ge |H_1'|/2\big) \ge 1/2.
    \end{equation*}
    Suppose that this is false. Then for every $J\subseteq I$, 
    \begin{equation*}
        \Pr_{\{v_i\mid i\in J\}}\big(|V_J\cap H_1'|\le |H_1'|/2\big) \ge 1/2.
    \end{equation*}

    For any strict subset $J\subset I$ and $j\in I\setminus J$, we can then conclude that
    \begin{align*}
        & \Pr\big(v_{i}\not\in V_J\big) \\
        & \quad \ge \Pr\big(|V_J\cap H_1'|\le |H_1'|/2\big)\, \Pr\big(v_j\not\in V_J\mid |V_J\cap H_1'|\le |H_1'|/2 \big) \\
        & \quad \ge \frac{1}{2}\Pr\big(v_j\in H_1'\setminus V_J\mid |H_1'\setminus V_J|\ge |H_1'|/2 \big) \\
        & \quad \ge \frac{1}{4}.
    \end{align*}
    For simplicity, assume that $I = \{1,\dots,|I|\}$. As a result of the above
    \begin{equation*}
        \Exp\dim(V_I) = 1 + \sum_{i=1}^{|I|-1}\Pr\big(v_{i+1}\not\in V_{\{v_1,\dots,v_i\}}\big) \ge 10n/4.
    \end{equation*}
    This is a contradiction since $\dim(V_I) \le \dim(\langle H_1'\rangle)\leq \dim(H_0)\le 2n$.
    Taking the union bound over the three events then completes the proof. 
\end{proof}

%% file: decoding/quadratic_decoding/constructing_approximating_matrix.tex
\section{Constructing an approximating matrix for~\texorpdfstring{$\phi$}{phi}}\label{sec:quadratic:constructing_approximating_matrix}
Recall that from event~$\E_4$, we have that $B_3\subseteq H_0$ and $|\langle B_3\rangle \cap H_1| \geq |H_1|/2$.
Moreover, by event~$\E_2$, we have $|H_0 + H_0| \leq \poly(1/\gamma)|H_0|$ and from event~$\E_1$ we have that $|H_1| \geq \poly(\gamma)|H_1|$.
It thus follows from \cref{thm:FreimanRuzsa} that
\begin{equation}\label{eq:quadratic:upper_bound_FR}
    |\langle B_3\rangle| \leq |\langle H_0\rangle| \leq \exp(-\poly(1/\gamma))|H_0|.
\end{equation}
Note that in this upper bound an algorithmic polynomial version of \cref{thm:FreimanRuzsa} gives improved dependencies on $\gamma$; see also \cref{rm:poly_FR} for a discussion. 

Denote $V = \langle B_3\rangle$ and let $\pi:V\to \F_p^n$ be the projection to the first~$n$ coordinates.
Let $T\subseteq V$ be the complement of the subspace $\ker(\pi)$ such that $V = T\oplus \ker(\pi)$.
Combing two results of \citeauthor{GreenTao:2008}~\cite{GreenTao:2008} (see also~\cite[Proposition~2.6]{Green:2007}) and \citeauthor{Samorodnitsky:2007}~\cite[Lemma~6.10]{Samorodnitsky:2007}\footnote{The lemma is proved only for~$\F_2^n$ but the same proof works for larger prime fields.} shows that 
\begin{equation}\label{eq:GTS}
    |H_1\cap T| \geq \frac{|H_1|}{2|\ker(\pi)|} \geq \exp(-\poly(1/\gamma))|H_0|,
\end{equation}
where we used that $|\pi(H_1)| = |H_1|$.
Since~$T$ is a linear subspace of~$V$ on which~$\pi$ acts injectively, it follows that there exists a matrix~$M\in \F_p^{n\times n}$ such that
\begin{equation*}
    T = \big\{(h,Mh) \mid h\in \pi(T)\big\}.
\end{equation*}
By standard linear algebra techniques, we can find a basis for~$\pi(T)$ which can be used to get an explicit description of such a matrix~$M$.

By event~$\E_2$, we have that $|H_0| \geq |H_1| \geq \poly(\gamma)|A_1|$.
Since $H_1\subseteq A_1\subseteq A_0$, it follows from \cref{eq:GTS} and event~$\E_1$ that
\begin{equation*}
    |A_1\cap T| \geq \exp(-\poly(1/\gamma))|A_1|.
\end{equation*}
It follows that
\begin{align*}
    \Exp_{h\in \F_p^n}\big|\widehat{\Delta_hf}(Mh)\big|^2 &\geq
    \Exp_{h\in \F_p^n}\one_{S_{\delta_1}\cap A_0\cap T}[(h,Mh)] \, \big|\widehat{\Delta_hf}(Mh)\big|^2\\
    &\geq \exp(-\poly(1/\gamma)).
\end{align*}

%% file: decoding/quadratic_decoding/query_lower_bound.tex
\section{Lower bound on query complexity}\label{sec:quadratic:query_lower_bound}
\citeauthor{Montanaro:2012}~\cite{Montanaro:2012} obtained tight bounds on the query complexity of learning multilinear polynomials over a finite field $\F_p$, both in the conventional setting and in the quantum setting.
His arguments for proving the lower bounds were based on elementary information theory, and easily generalize to the following result:
\begin{lemma} \label[lemma]{lem:Montanaro_query_lower_bound}
    Let $\Fcal \subseteq \{f: \F_p^n \to \F_p\}$ be a family of functions, and let~$f$ be a uniformly chosen element of $\Fcal$.
    Then:
    \begin{enumerate}
        \item Any conventional query algorithm that learns $f$ with bounded error must make $\Omega \big(\log |\Fcal|/(\log p)\big)$ queries to $f$. \label{item:classical_query}
        \item Any quantum query algorithm that learns $f$ with bounded error must make $\Omega \big(\log |\Fcal|/(n\log p)\big)$ queries to $f$. \label{item:quantum_query}
    \end{enumerate}
\end{lemma}
\begin{proof}
    We view~$f$ as a random variable uniformly distributed over $\Fcal$, which thus has entropy $H(f) = \log |\Fcal|$.
    We will use the intuitive fact that one cannot learn~$f$ from much fewer than~$H(f)$ bits of information.
    
    \emph{Fano's inequality} formalizes this fact: 
    for any random variable $X$ we have\footnote{Here $H(f \mid X)$ is the entropy of $f$ conditional on knowing $X$, defined as $H(f \mid X) = H(f) - I(f:X)$ where $I(f:X)$ is the mutual information.}
    \begin{equation*}
        H(f \mid X) \leq H\big(\mathbf{1}[f\neq X]\big) + \Pr[f \neq X] (\log |\Fcal|-1).
    \end{equation*}
    See for example~\cite[Chapter 12]{NielsenChuang:2010} for a proof of this inequality.
    One can think of $X$ as being our best guess for the function $f$ based on some partial information.
    Using that $H(f \mid X) = \log |\Fcal| - I(f:X)$ and that $H\big(\mathbf{1}[f\neq X]\big) \leq 1$, we obtain
    \begin{equation}\label{eq:Fano_inequality}
    \Pr[f \neq X] \geq 1 - \frac{I(f:X)+1}{\log |\Fcal|}.
    \end{equation}

    To prove \cref{item:classical_query}, suppose that a conventional algorithm $\Acal$ makes $r$ queries to $f$ and let~$X$ be the output of $\Acal$ after making those queries.
    Every query returns an element from $\F_p$ and thus provides at most $\log p$ bits of information about $f$.
    After $r$ queries we then must have $I(f:X) \leq r\log p$, and so by \cref{eq:Fano_inequality}
    \begin{equation*}
        \Pr[f \neq X] \geq 1 - \frac{r\log p + 1}{\log |\Fcal|}.
    \end{equation*}
    In order for this error probability to be bounded away from $1$, $\Acal$ must make $\Omega\big(\log |\Fcal|/(\log p)\big)$ queries as wished.

    To prove \cref{item:quantum_query}, we interpret a quantum query algorithm~$\Acal$ as a communication protocol between two players:
    Alice, who runs the algorithm, and Bob, who knows the function $f$.
    Each query can be regarded as an exchange of messages where Alice sends Bob $n+1$ \emph{qudits} (say $\ket{x}\ket{b}$ for some $x\in \F_p^n$, $b\in \F_p$) and Bob sends~$n+1$ qudits back to Alice (the query oracle output $\ket{x}\ket{b+f(x)}$).
    It follows as a consequence of Holevo's theorem~\cite[Chapter~12]{NielsenChuang:2010} that, after $r$ such rounds of communication, Alice obtains at most $2r(n+1)\log p$ bits of information (see~\cite[Theorem~2]{CleveDamNielsenTapp:1998} for a derivation of this result).
    We conclude that
    \begin{equation*}
        \Pr[f \neq X] \geq 1 - \frac{2r(n+1)\log p + 1}{\log |\Fcal|}.
    \end{equation*}
    This is bounded away from $1$ when $r = \Omega\big(\log |\Fcal|/(n\log p)\big)$, as wished.
\end{proof}

Using this lemma with $\Fcal = \Pol_d(\F_p^n)$, the set of all polynomials $f:\F_p^n\to\F_p$ of degree at most $d$, we immediately obtain Montanaro's lower bounds on the conventional and quantum query complexities of learning degree-$d$ multilinear polynomials, as we have $|\Fcal| = n^d$.

We will next prove that the \emph{same} lower bound holds even if we only require the algorithm to learn \emph{any} polynomial~$Q$ which is at a nontrivial distance from the original polynomial~$P$.
\begin{theorem}
    Let $P\in \Pol_d(\F_p^n)$ be a polynomial of degree $d\geq 2$ and let $\eps>0$.
    Suppose $\Acal$ is a query algorithm which, with nonnegligible probability, outputs a polynomial $Q\in \Pol_d(\F_p^n)$ such that $\delta(P, Q) \leq 1 - 1/p - \eps$.
    Then:
    \begin{enumerate}
        \item If $\Acal$ is a conventional algorithm, it must make $\Omega(n^d)$ queries to $P$.
        \item If $\Acal$ is a quantum algorithm, it must make $\Omega(n^{d-1})$ queries to $P$.
    \end{enumerate}
\end{theorem}

This result is a simple consequence of the next theorem, which concerns (a special case of) the Gowers inverse theorem.
\begin{theorem} \label{thm:Uk_lower_bound}
    Let $f: \F_p^n \to \C$ be a polynomial phase function of degree $d\geq 2$ and let $\eps>0$.
    Suppose $\Acal$ is a (conventional or quantum) query algorithm that, with nonnegligible probability, outputs a polynomial $Q\in \Pol_d(\F_p^n)$ such that
    \begin{equation*}
        \big|\Exp_{x\in \F_p^n} f(x) \omega_p^{Q(x)}\big| \geq \eps.
    \end{equation*}
    Then:
    \begin{enumerate}
        \item If $\Acal$ is a conventional algorithm, it must make $\Omega(n^d)$ queries to $f$.
        \item If $\Acal$ is a quantum algorithm, it must make $\Omega(n^{d-1})$ queries to $f$.
    \end{enumerate}
\end{theorem}
Proving this theorem requires the following lemma. 
\begin{lemma} \label[lemma]{lem:high_prank_family}
    Let $d\geq 2$ be an integer.
    For every prime $p$ and every sufficiently large integer $n$, there exists a family $\Fcal \subseteq (\F_p^n)^{\otimes d}$ of symmetric zero-diagonal tensors with size $|\Fcal| \geq p^{n^d/(10d!)}$ for which
    \begin{equation*}
        \prank(S-T) \geq \frac{n}{10 d!} \quad \text{for all distinct $S, T\in \Fcal$.}
    \end{equation*}
\end{lemma}
\begin{proof}
    We will show that a randomly chosen family of $p^{n^d/(10d!)}$ symmetric zero-diagonal tensors will satisfy the desired property with high probability.

    We first determine the number of symmetric zero-diagonal tensors in $(\F_p^n)^{\otimes d}$.
    Note that the upper-diagonal indices $\Delta_+ = \big\{(i_1, \dots, i_d) \in [n]^d:\: i_1 < i_2 < \dots < i_d\big\}$ of any such tensor $T$,
    fully determine the tensor:
    all other entries of $T$ follow either by symmetry from the upper diagonal entries or are zero if two indices coincide. 
    For any index $(i_1, \dots, i_d) \in \Delta_+$ it holds that $T(i_1, \dots, i_d) \in \F_p$. 
    As $|\Delta_+| = \binom{n}{d}$, it follows that the number of such tensors $T$ is precisely $p^{\binom{n}{d}}$.
    
    Next, we bound the number of tensors in $(\F_p^n)^{\otimes d}$ having partition rank at most~$r$ (for any given $r\in \N$). 
    Per the definition of the partition rank, each such tensor admits a decomposition
    \begin{equation*}
        \sum_{i=1}^{r} S_{I_i} \otimes T_{[d]\setminus I_i},
    \end{equation*}
    where $I_i \neq \emptyset$ is a proper subset of~$[d]$ for each $1\leq i\leq r$, $S_{I_i} \in (\F_p^n)^{\otimes I_i}$ and $T_{[d]\setminus I_i} \in (\F_p^n)^{\otimes [d]\setminus I_i}$.
    Suppose the size of the $i$-th set $I_i$ is~$j$ for some $1\leq j\leq d-1$, then there are $\binom{d}{j}$ ways of choosing the set $I_i\subset [d]$, $p^{n^j}$ ways of choosing $S_{I_i}$ and $p^{n^{d-j}}$ ways of choosing $T_{[d]\setminus I_i}$.
    It follows that the total number of such $r$-term decompositions is at most
    \begin{equation*}
        \left(\sum_{j=1}^{d-1} \binom{d}{j} p^{n^j} p^{n^{d-j}}\right)^r \leq \big(2^d p^{2n^{d-1}})^r.
    \end{equation*}
    We conclude that the number of tensors in $(\F_p^n)^{\otimes d}$ having partition rank at most~$r$ is at most $2^{dr} p^{2rn^{d-1}}$.

    Now take $m := p^{n^d/(10d!)}$ symmetric zero-diagonal tensors $T_1, \dots, T_m \in (\F_p^n)^{\otimes d}$ uniformly at random.
    As for any given $i\neq j$, the tensor $T_i - T_j$ is also uniformly distributed over the set of symmetric zero-diagonal tensors, it follows that
    \begin{equation*}
        \Pr\big[\prank(T_i-T_j) \leq r\big] \leq \frac{2^{dr} p^{2rn^{d-1}}}{p^{\binom{n}{d}}} \quad \text{for all $1\leq i < j \leq m$.}
    \end{equation*}
    By the union bound, the probability that there exist two indices $i, j\in [m]$ such that $\prank(T_i-T_j) \leq r$ is at most
    \begin{equation*}
        \sum_{1\leq i<j \leq m} \Pr\big[\prank(T_i-T_j) \leq r\big] \leq \binom{m}{2} \frac{2^{dr} p^{2rn^{d-1}}}{p^{\binom{n}{d}}}.
    \end{equation*}
    If we take $r = n/(10d!)$ then this last quantity will tend to zero as $n$ grows, and thus the random collection of tensors will satisfy the property in the statement with high probability.
\end{proof}

Using this lemma we can now prove \cref{thm:Uk_lower_bound}, which gives the desired lower bound on the query complexity. 
\begin{proof}[ of \cref{thm:Uk_lower_bound}]
    Let~$\Fcal$ be the family of tensors promised by \cref{lem:high_prank_family}.
    For each tensor $T\in \Fcal$, define the multilinear polynomial $P_T: \F_p^n \to \F_p$ by
    \begin{equation*}
        P_T(x_1, \dots, x_n) = \sum_{i_1<\dots<i_d} T(i_1, \dots, i_d) \prod_{j=1}^d x_{i_j};
    \end{equation*}
    in other words, $P_T$ denotes the homogeneous degree-$d$ form on $\F_p^n$ associated with the upper-diagonal part of the tensor~$T$.
    Now define $\Fcal' = \{P_T: T\in \Fcal\}$.

    We now apply \cref{lem:Montanaro_query_lower_bound} to the family~$\Fcal'$, which by construction has size at least $p^{n^d/(10d!)}$.
    Hence, any query algorithm that learns a uniformly random element from~$\Fcal'$ must make either $\Omega(n^d)$ conventional queries or $\Omega(n^{d-1})$ quantum queries.
    In order to complete the proof, it suffices to show that one can infer any element $P_T\in \Fcal'$ from any polynomial $Q\in \Pol_{d}(\F_p^n)$ such that
    \begin{equation*}
        \big|\Exp_{x\in \F_p^n} \omega_p^{P_T(x)} \omega_p^{-Q(x)}\big| \geq \eps.
    \end{equation*}

    Fix a function $P_T\in \Fcal'$ and a degree-$d$ polynomial~$Q$ such that the above inequality holds.
    Let $S$ denote the partial derivative of $Q$, that is, let $S := \partial Q$.
    We can check that $\partial P_T = T$.
    Using the nesting property of the Gowers norms (\cref{eq:Uk_nesting}), we now have
    \begin{equation*}
        \left|\Exp_{y_1, \dots, y_d\in \F_p^n} \omega_p^{T(y_1, \dots, y_d) - S(y_1, \dots, y_d)}\right| = \left\|\omega_p^{P_T - Q}\right\|{U^d}^{2^d} \geq \left|\Exp_{x\in \F_p^n} \omega_p^{P_T(x) -Q(x)}\right|^{2^d} \geq \eps^{2^d}.
    \end{equation*}
    From this lower bound, we conclude that $\arank(T-S) \leq 2^d \log_p(1/\eps)$.
    By the qualitative equivalence between analytic rank and partition rank (\cref{thm:MoshkovitzZhu}), it follows that (say) $\prank(T-S) \leq \nolinebreak n/(30d!)$ if~$n$ is large enough.

    Now we are essentially done:
    by assumption we have $\prank(T'-T) \geq n/(10d!)$ for all $T'\in \Fcal \setminus\{T\}$, and by subadditivity we have
    $$\prank(T'-T) \leq \prank(T'-S) + \prank(T-S).$$
    We then conclude that $\prank(T'-S) \geq 2n/(30d!)$ for all $T'\neq T$ in~$\Fcal$ while $\prank(T-S) \leq n/(30d!)$, so learning $S=\partial Q$ suffices to infer $T\in \Fcal$.
    Thus, learning $Q\in \Pol_{d}(\F_p^n)$ suffices to infer $P_T\in \Fcal'$, concluding the proof.
\end{proof}

%% file: decoding/quadratic_decoding/discussion_quadratic_decoding.tex
\section{Reflections and outlook}
This chapter revisited the work of \citeauthor{TulsianiWolf:2014}~\cite{TulsianiWolf:2014} on list decoding corrupted Reed-Muller codewords of degree at most two. 
We provided a novel proof of their work using new techniques from higher-order Fourier analysis discovered since. 
We specifically used an improved algorithmic version of the result by Balog--Szemer\'edi--Gowers (see also \cref{thm:BSG:Schoen}), with a simpler proof and better quantitative bounds. 
This algorithmic Balog--Szemer\'edi--Gowers theorem can also be of independent interest in non-quantum applications. 
Additionally, our new proof makes no use of Fourier estimation as a subroutine, giving a simpler prove and reducing the query complexity. 
Finally, by employing the Fourier sampling subroutine we managed to reduce the query complexity by a factor~$n$ by using quantum queries.

Multiple paths remain open to consider in follow-up work:
First, the Fre\v{\i}man-Ruzsa theorem gives an exponential upper bound in the doubling constant for the size of the set of small doubling, as shown in \cref{eq:quadratic:upper_bound_FR}. 
A recent result by \citeauthor{GowersGreenMannersTao:2025} considers a related, but slightly different, situation, where they improve the upper bound to a polynomial dependence on the doubling constant~\cite{GowersGreenMannersTao:2025}. 
\citeauthor{Arunachalam:2025} turned this non-algorithmic result on the Fre\v{\i}man-Ruzsa theorem into an algorithmic result~\cite{Arunachalam:2025}.
As a first step, we can therefore look into applying this algorithmic version in the approach outlined in this chapter. 
This should in turn provide an improved dependency on $\gamma$ in \cref{eq:quadratic:upper_bound_FR}.

Second, in this chapter, we restricted our attention to polynomial phase functions. 
It is interesting to see how our results extend to general functions $f:\F_p^n \to \mathbb{D}$ such that $\lVert f\rVert_{U^3}\ge \eps$, for some $\eps>0$. 
The main reason for the restriction in our work is that the Fourier sampling subroutine requires a query to $\overline{f}$, the complex conjugate of $f$. 
For polynomial phase functions, $n-1$ queries to $f$ correspond to a single query to $\overline{f}$.
For general functions $f$ however, we have no such equivalence. 
Instead, only probabilistic methods exist that query $\overline{f}$~\cite{Ebler:2023}. 
Using such a probabilistic method affects both the success probability of the algorithm and the query complexity. 

Third, a recent result showed an algorithmic approach to the inverse theorem for the $U^4$-norm, with which corrupted Reed-Muller codes of degree at most $3$ can be decoded~\cite{KimLiTidor:2023}.
Follow-up work might look at combining the techniques used by \citeauthor{KimLiTidor:2023} and the techniques used in this chapter. 
This way, we might improve the techniques in this chapter further, and we might improve the qubic Goldreich-Levin algorithm by \citeauthor{KimLiTidor:2023}.
Additionally, we believe that the quantum Fourier sampling routine directly gives a factor $n$ improvement in the query complexity for the qubic Goldreich-Levin algorithm. 

Fourth, it is interesting to see at what other places quantum algorithms can provide improvements within higher-order Fourier analysis. 
In this chapter, we only use a quantum subroutine to efficiently sample from the Fourier spectrum. 
This direction for future research relates to the more general question of the relation of higher-order Fourier analysis and quantum algorithms. 

Finally, inverse theorems of the $U^k$-norm are only known algorithmically for $k\le 4$.
For $k\ge 5$, quantitative bounds are only known for large field characteristics. 
Providing quantitative bounds for low field characteristics for any $k\ge 5$ or finding an algorithmic approach for $k\ge 4$ will be a significant breakthrough. 
Ideas from quantum computing might help finding such a breakthrough, this research line recently received new attention~\cite{ArunachalamDutt:2024}.

%% file: LAQCC/LAQCC_introduction.tex
\chapter{Introduction to shallow-depth computing}
This chapter provides an introduction to shallow-depth quantum computing.
We provide a rationale for considering such circuits and the added benefit of intermediate conventional computing. 
We also discuss multiple quantum states for which we later provide shallow-depth quantum circuits that prepare them. 

\input{LAQCC/introduction_LAQCC/near_term_quantum_computers}
\input{LAQCC/introduction_LAQCC/quantum_conventional_computations}
\input{LAQCC/introduction_LAQCC/computations_to_complexity}
\input{LAQCC/introduction_LAQCC/state_prep_intro}
\input{LAQCC/introduction_LAQCC/introducing_noise}
\input{LAQCC/introduction_LAQCC/outline_intro_LAQCC}

%% file: LAQCC/introduction_LAQCC/near_term_quantum_computers.tex
\section{Near-term quantum computers}
Current quantum hardware is unable to carry out universal quantum computations due to the buildup of errors that occur during the computation. 
The magnitude of the individual error is currently above the value that the Threshold Theorem requires in order to kick-start quantum error correction and fault-tolerant quantum computation~\cites{AharonovBenOr:1997,KnillLaflammeZurek:1998,Kitaev:1997}[Section 10.6]{NielsenChuang:2010}. 
Although the experimentally achieved fidelity rates are promising and the error bounds are inching closer to the required threshold, we will have to work for the foreseeable future with quantum hardware with errors that build-up during the computation.
This implies that we can only do a limited number of steps before the output of the computation has become completely uncorrelated with the intended one.

For fault-tolerant quantum computing, we repeat four steps: 
1) Apply a number of single- and two-qubit quantum gates, in parallel whenever possible; 
2) Perform a syndrome measurement on a subset of the qubits; 
3) Perform fast conventional computations to determine which errors occurred and how to correct them; 
and, 4) Apply correction terms.
We then repeat these four steps with the next sequence of gates. 
These four steps are essential for fault-tolerant quantum computing. 

The starting point of this work is to use the four steps outlined above, not to carry out error correction and fault-tolerant computation, but to enhance short, constant-depth, {\em uncorrected} quantum circuits that perform single-qubit gates and {\em nearest-neighbor} two-qubit gates. 
Since in the long run we will have to implement error-correction and fault-tolerant computation anyhow, and this is done by such a four-step process, why not make other use of this architecture? 
Moreover, on some of the quantum hardware platforms, these operations are already in place.
Embracing this idea we naturally arrive at the question: what is the computational power of \textit{low-depth} quantum-conventional circuits organized as in the four steps outlined above? 
We thus investigate circuits that execute a small, ideally constant, number of stages, where at each stage we may apply, in parallel, single-qubit gates and {\em nearest-neighbor} two-qubit gates, followed by measurements, followed by low-depth conventional computations of which the outcome can control quantum gates in later stages. 
It is not clear, at first, whether such circuits, especially with constant depth, can do anything remotely useful. 
But we will see that this is indeed the case: many quantum computations can be done by such circuits in constant depth. 
By parallelizing quantum computations in this way, we improve the overall computational capabilities of these circuits, as we do not incur errors on qubits that are idle, simply because qubits are not idle for a very long time. 
Furthermore, reducing the depth of quantum circuits, at the cost of increasing width, allows the circuit to be run faster even if errors occur.

%% file: LAQCC/introduction_LAQCC/quantum_conventional_computations.tex
\section{Quantum-conventional computations}
The first usage of such a four-step process, not to do error correction, but to perform computations, can be found in measurement-based quantum computing.

\citeauthor{PhamSvore:2013} were the first to formalize the four-step process for performing computations~\cite{PhamSvore:2013}. 
They included specific hardware topologies by considering two-dimensional graphs for imposing constraints on qubit interactions. 
In their model, they develop circuits for particularly useful multi-qubit gates and presented the cost of these circuits in terms of the width, number of qubits, depth, number of consecutive time steps, size, and total number of non-identity operations.
They use these gates to construct an algorithm that factors integers in polylogarithmic depth.
\citeauthor{Browne:2011} showed that the main tool in the work by \citeauthor{PhamSvore:2013}, the fanout gate, can also be replaced by additional log-depth conventional computations in the measurement-based quantum computing setting~\cite{Browne:2011}.

More recently, \citeauthor{PiroliStyliarisCirac:2021} introduced a scheme to implement unitary operations involving quantum circuits combined with Local Operations and Classical Communication ($\mathsf{LOCC}$) channels: $\mathsf{LOCC}$-assisted quantum circuits~\cite{PiroliStyliarisCirac:2021}. 
Similarly to the four-step process we just described, they allow for a short-depth quantum circuit, followed by one round of $\mathsf{LOCC}$, in which auxiliary qubits are measured and local unitaries are applied based on the measurement outcomes. 
They show that in this model any 1D transitionally-invariant matrix-product state (MPS) with fixed bond dimension is in the same phase of matter as the trivial state. 
Similar ideas on efficiently preparing matrix product states and topological states using intermediate circuit measurements and feedforward can be found in~\cite{Tantivasadakarn:2023,Smith:2023,Tantivasadakarn:2024,Smith:2024}.

We propose a new model called \textit{Local Alternating Quantum-Classical Computations} ($\LAQCC$), where we bound the power of both the quantum circuits (in computational power and locality) and the conventional computations following intermediate quantum circuit measurements. 
The outcome of the conventional computations is used to control operations in future quantum circuits.
We allow for flexibility in this model by imposing different constraints on the power of both the quantum circuits and the conventional circuits as well as the number of alternations between them. 
Most attention will be given to $\LAQCC$ containing quantum circuits of constant depth, conventional circuits of logarithmic depth and at most a constant number of alternations between them. 
Any circuit constructed in this model is considered to be of constant depth. 
We restrict the conventional computations to be of logarithmic depth, as this is the first natural and nontrivial extension beyond constant-depth conventional computations. 

The definition of $\LAQCC$ sharpens the original definition of \citeauthor{PhamSvore:2013} by adding constraints to the intermediate conventional computations. 
The bound on the conventional computations allows for a bound on the power of these circuits as a whole. 
The $\mathsf{LOCC}$-assisted circuits of \citeauthor{PiroliStyliarisCirac:2021} are not low-depth, as they allow for long sequential measurement-based corrections of the auxiliary qubits needed for their calculations.
These measurement-based operations are considered as sequential alternations between the quantum and conventional circuits in $\LAQCC$, resulting in increasing the total depth.

%% file: LAQCC/introduction_LAQCC/computations_to_complexity.tex
\section{From computations to complexity}
The considered $\LAQCC$-circuits are bounded in power. 
Specifically, we show that $\LAQCC \subseteq \qnc^1$ and that $\LAQCC$-circuits are unlikely to be efficiently simulatable as they contain instantaneous quantum polynomial-time (IQP)-circuits~\cite{BremnerJozsaShepherd:2010,ShepherdBremner:2009}.
An efficient conventional simulation of IQP-circuits implies the collapse of the polynomial hierarchy to the third level, which is believed to be false~\cite{BremnerMontanaroShepherd:2017}. 

It is interesting to see how the computational power of $\LAQCC$ changes with more quantum resources, more conventional resources, or more alternations between the two. 
This new view asks for a complexity theoretical analysis beyond standard decision problems. 
\citeauthor{Aaronson:2004} considered a relative complexity, where the complexity is measured between two given states and corresponds to the number of gates required from a given gate set to transform one state into the other state~\cite{Aaronson:2004}. 
\citeauthor{RosenthalYuen:2022}, and \citeauthor{MetgerYuen:2023} instead consider classes based on sequences of quantum states preparable by polynomial-sized quantum circuits, where the circuits are uniformly generated by a computational class.
For instance, the class $\mathsf{PSPACE}$ implies the class $\mathsf{StatePSPACE}$~\cite{RosenthalYuen:2022,MetgerYuen:2023}.

We combine the ideas from these notions.
We omit the uniformity constraint from~\cite{RosenthalYuen:2022,MetgerYuen:2023} and define a class $\mathsf{StateX}$ as a states preparable by $\mathsf{X}$-circuits. 
This notion is similar to the relative complexity from~\cite{Aaronson:2004}, where one state is the $\ket{0}$-state and instead of counting the number of gates, we consider the set of states preparable by a bounded number of gates. 
We use this notion of state complexity to show that any state preparable by a $\LAQCC^*$-circuit --an instance of $\LAQCC$ with enhanced conventional and quantum computational power-- is also preparable by a $\mathsf{PostQPoly}$-circuit, the class of circuits of polynomial depth with an additional post-selection gate, see also \cref{sec:LAQCC:complexity_results_powerful}. 

%% file: LAQCC/introduction_LAQCC/state_prep_intro.tex
\section{State preparation}
Even though the power of $\LAQCC$ is bounded for solving computational problems, $\LAQCC$-circuits can still prove useful as a subroutine in other quantum algorithms. 
In light of the new notion of complexity, it is important to know how this new $\LAQCC$-model performs in preparing often-used quantum states. 
We show that $\LAQCC$-circuits with constant quantum depth and logarithmic conventional depth can prepare states with long-range entanglement.

\cref{chp:LAQCC:state_preparation} introduces new efficient state-preparation routines for three types of non-stabilizer states. 
Efficient circuits for preparing stabilizer states are already known from measurement-based quantum computing, as discussed in \cref{sec:LAQCC:Clifford_circuits}.

\cref{sec:LAQCC:uniform_superposition_modulo_q} presents a $\LAQCC$-circuit to prepare the first non-stabilizer state: a uniform superposition over an arbitrary number of states.
This circuit uses an exact version of Grover search, which returns a marked item with certainty~\cite{Long:2001}. 

\cref{sec:LAQCC:W_state} gives a $\LAQCC$-circuit for the second state: the $W$-state.
This state corresponds to the uniform superposition over all computational basis states of Hamming weight~$1$ and is a natural long-range entangled state that displays entanglement fundamentally different from the Greenberger–Horne–Zeilinger (GHZ) state~\cite{DurVidalCirac:2000}.
$\LAQCC$-type constant-depth circuits for the GHZ state were already known~\cite{PhamSvore:2013, PiroliStyliarisCirac:2021}. 
The $W$-state is often used as benchmark for new quantum hardware~\cite{Haffner:2005,Neeley:2010,GarciaPerez:2021}. 
A circuit for preparing the $W$-state was given in~\cite{PiroliStyliarisCirac:2021}, but this implementation requires sequentially alternating measurements followed by local unitaries, which is not considered constant depth in the $\LAQCC$-model. 
We improve their protocol by giving a $\LAQCC$-implementation of the $W$-state, based on an uncompress-compress method that links the one-hot representation and binary representation of integers.

The third state considered is the Dicke state, which generalizes the $W$-state to a superposition over all computational basis states with Hamming weight~$k$~\cite{Dicke:1954}. 
Dicke states prove useful for various applications, including quantum game theory~\cite{OzdemirShimamuraImoto:2007}, quantum storage~\cite{BaconChuangHarrow:2006,PleschBuzek:2010}, quantum error correction~\cite{Ouyang:2014}, quantum metrology~\cite{Toth:2012}, and quantum networking~\cite{Prevedel:2009}. 
Dicke states have been used as a starting state for variational optimization algorithms, most notably in Quantum Alternating Operator Ansatz~\cite{Hadfield:2019}, to find solutions to problems such as Maximum $k$-vertex Cover~\cite{Brandhofer:2022,CookEidenbenzBartschi:2020}.

Dicke states also arise naturally in physics.
The ground states of physical Hamiltonians describing one-dimensional chains, such as those states resulting from the Bethe ansatz, tend to show resemblance to Dicke states making them an ideal starting state when investigating the ground state behavior of these Hamiltonians~\cite{TsyplyatyevDelftLoss:2010,Brandes:2013,Bezvershenko:2021}. 
For instance, the algorithm by \citeauthor{Dyke:2021}, which prepares the Bethe ansatz eigenstates of the spin-1/2 XXZ spin chain, starts by first preparing a Dicke state~\cite{Dyke:2021}. 

Efficient deterministic circuits for preparing Dicke states have been proposed by \citeauthor{BartschiEidenbenz:2019}~\cite{BartschiEidenbenz:2019, BartschiEidenbenz:2022}. 
They provide a quantum circuit of depth $\bigo(k \log(\frac{n}{k}))$, allowing arbitrary connectivity, to prepare a Dicke state, which they conjecture to be optimal for constant~$k$.
We present a $\LAQCC$-circuit that prepares the Dicke state with better depth than their conjecture already for constant~$k$.
However, this does not directly disprove their conjecture, as we allow for intermediate measurements and conventional computations. 
More significantly, in \cref{sec:LAQCC:Dicke_small_k} we construct constant-depth $\LAQCC$-circuits for $k = \bigo(\sqrt{n})$ greatly improving their conjectured lower bound.
This construction extends the compress-uncompress method for the $W$-state combined with additional subroutines. 
In \cref{sec:LAQCC:Dicke_in_LAQCC_LOG}, we provide a log-depth circuit that prepares the Dicke state for arbitrary~$k$ by mapping efficiently between the factoradic representation and the combinatorial number representation of positive integers.

%% file: LAQCC/introduction_LAQCC/introducing_noise.tex
\section{Introducing noise}
The main idea behind $\LAQCC$ is to parallelize operations, such that qubits are idle only briefly. 
Near-term hardware will remain noisy and it is unclear when error-rates will decrease sufficiently far for long elaborate algorithms to be run faithfully. 
Some even doubt whether quantum computers can ever overcome the error threshold imposed by the Threshold Theorem~\cite{Kalai:2016,Kalai:2020}, claiming that the error rate will scale linearly with the number of qubits, whereas they hypothesize that the effort to suppress these errors scales exponentially. 
As long as error rates remain above the error threshold, error-correcting techniques fail to scale quantum computers to the fault-tolerant setting.

In addition to improved quantum hardware and error-correcting codes, conventional mitigation techniques can also help further reduce error rates~\cite{Kandala:2019,Cai:2023}. 
Examples include zero-noise extrapolation~\cite{TemmeBravyiGambetta:2017}, dynamic decoupling~\cite{ViolaLloyd:1998,ViolaKnillLloyd:1999,Pokharel:2018} and improved compilation and transpilation routines~\cite{WallmanEmerson:2016}.
Our $\LAQCC$-model also opens the way to improved quantum circuits by off-loading part of the computations to a conventional controller. 

However, even with shallow-depth quantum circuits, errors cannot be ruled out. 
A careful analysis of the success probability of different quantum circuits is needed. 
This careful analysis also helps compare different quantum circuits and determine which is best for specific situations, given a problem instance and quantum hardware. 

Such an analysis naturally requires a noise model that describes the decohering behavior of the qubits. 
Common noise models include single- and two-qubit gate errors, read-out errors, depolarizing noise, and dephasing noise. 
These noise models typically assume some underlying physical behavior. 
Naturally, a trade-off exists between the complexity of the noise model and the correspondence with the actual behavior. 

%% file: LAQCC/introduction_LAQCC/outline_intro_LAQCC.tex
\section{Outline}
The remainder of this part consists of three chapters: 
\cref{chp:LAQCC:model_definition} formally introduces the class of Local Alternating Quantum-Classical Computations ($\LAQCC$).
In that chapter, we also provide some complexity theoretic results on $\LAQCC$ and show that all Clifford circuits have an equivalent $\LAQCC$-circuit and we discuss some complex gates from literature that admit a $\LAQCC$-implementation. 
\cref{chp:LAQCC:state_preparation} presents $\LAQCC$-circuits for preparing three types of non-stabilizer quantum states: the uniform superposition over an arbitrary number of states, the $W$-state and the Dicke state for $k=\bigo(\sqrt{n})$. 
We then provide a protocol that prepares the Dicke state for any $k$, at the cost of a logarithmic number of alternations of quantum and conventional circuits. 
\cref{chp:LAQCC:error_analysis} presents an error analysis for preparing a GHZ state and preparing a $W$-state. 
We compare $\LAQCC$-circuits for both states with standard methods and determine which protocol should work best. 
We also implement the protocols and compare the results found on quantum hardware with our theoretical estimates. 

%% file: LAQCC/LAQCC_model.tex
\chapter{The \texorpdfstring{$\LAQCC$}{LAQCC}-model}\label{chp:LAQCC:model_definition}
This chapter formally defines the $\LAQCC$-model.
We show that Clifford circuits are in a specific instance of $\LAQCC$, as well as multiple more complex gates. 
We conclude this chapter with some complexity theoretical results for $\LAQCC$ and a version of $\LAQCC$ with enhanced conventional and quantum computational capabilities, a version we call $\LAQCC^*$.

\input{LAQCC/model_definition/outline_model_definition}
\input{LAQCC/model_definition/model_definition}

\input{LAQCC/model_definition/Clifford_in_LAQCC}
\input{LAQCC/model_definition/useful_gates_in_LAQCC}

\input{LAQCC/model_definition/complexity_for_LAQCC}
\input{LAQCC/model_definition/Powerful_LAQCC}

\input{LAQCC/model_definition/discussion_LAQCC_model}

%% file: LAQCC/model_definition/outline_model_definition.tex
\section{Chapter overview}\label{sec:LAQCC:model_definition:outline}
This chapter introduces a new computational model to bring structure to hybrid quantum-conventional circuits where quantum and conventional computations are alternated. 
We define a new class of \emph{Local Alternating Quantum-Classical Computations} ($\LAQCC$). 
This class alternates $d$ times between quantum computations $\mathcal{Q}$ and conventional computations $\mathcal{C}$ and extends other commonly used classes, as those in general do not impose bounds on the conventional computations. 
\cref{def:LAQCC:model_definition} gives the formal definition. 

The remainder of the chapter mainly focuses on a specific instance of this new model, where constant-depth quantum circuits are alternated with logarithmic-depth conventional computations. 
Unless stated otherwise, $\LAQCC$ will refer to this specific instance of $\LAQCC(\mathcal{Q},\mathcal{C}, d)$. 
\cref{sec:LAQCC:Clifford_circuits} will show that every Clifford circuit has an equivalent $\LAQCC$-circuit. 
\begin{lemma}
    Every $n$-qubit Clifford unitary has an equivalent $\LAQCC$-circuit. 
\end{lemma}
We prove this lemma by first noting that every Clifford unitary has an equivalent linear-depth implementation on a linear nearest-neighbor architecture and then for these so-called Clifford-grid circuits show the existence of an equivalent $\LAQCC$-circuit. 
This $\LAQCC$-circuit requires a precomputation to account for the correction terms to be applied. 
However, this precomputation only depends on the intermediate measurement outcomes and is independent of the circuit itself.

We continue in \cref{sec:LAQCC:useful_gates_and_subroutines} by discussing different quantum gates implementable with $\LAQCC$-circuits. 
Most of these gates critically depend on the fanout gate, which generalizes the CNOT-gate to multiple targets. 
In this same section, \cref{lem:grover_constant_fraction} will give a version of exact Grover search not to search a database, but instead to prepare quantum states. 
This circuit corresponding to this lemma can prepare a uniform superposition based on a unitary that identifies target states. 
The next chapter will use this lemma to prepare different quantum states. 

This chapter finishes with a complexity theoretical analysis of the $\LAQCC$-model. 
First, \cref{sec:LAQCC:complexity_results_standard} will bound the computational power of $\LAQCC$:
\cref{lem:LAQCC_QNC1} shows how every $\LAQCC$-circuit has an equivalent $\QPoly$-circuit.
However, despite this upper bound, we still expect $\LAQCC$-circuits to be hard to simulate in general.
We support this intuition in \cref{lem:IQP_in_LAQCC} by linking $\LAQCC$-circuits to instantaneous quantum polynomial-time (IQP)-circuits. 

\cref{sec:LAQCC:complexity_results_powerful} considers an instance of $\LAQCC(\mathcal{Q},\mathcal{C}, d)$ with unbounded conventional computations. 
We will show that, even for this powerful instance, its computational power with respect to state preparation is upper bounded by circuits equipped with post-selection gates, where the quantum state is considered conditional on an auxiliary qubit being in the one state. 

%% file: LAQCC/model_definition/model_definition.tex
\section{Model definition}
We define the computational model \emph{Local Alternating Quantum-Classical Computations} ($\LAQCC$) as follows:
\begin{definition}\label{def:LAQCC:model_definition}
Let $\LAQCC(\mathcal{Q}, \mathcal{C}, d)$ be the class of circuits such that
\begin{itemize}
\item every quantum layer implements a quantum circuit $Q\in\mathcal{Q}$ constrained to a grid topology;
\item every conventional layer implements a conventional circuit $C\in\mathcal{C}$;
\item there are $d$ alternating layers of quantum and conventional circuits;
\item after every quantum circuit $Q$, a subset of the qubits is measured;
\item the conventional circuit receives the measurement outcomes as input;
\item the conventional circuit can control quantum operations in future layers.
\end{itemize}
A circuit in $\LAQCC(\mathcal{Q}, \mathcal{C}, d)$ is required to deterministically prepare a pure state starting from the $\ket{0}^{\otimes n}$-state.
\end{definition}

The grid topology imposed on the quantum operations implies that qubits can only interact with their direct neighbors.
A circuit in $\LAQCC(\mathcal{Q}, \mathcal{C}, d)$ can use the output of conventional intermediate layers to control quantum operations in future layers. 
Thus, in some sense, information is fed forward in the circuit. 
Note, as conventional computations are in general significantly faster than the quantum operations, we only count the quantum operations towards the depth of the circuit, unless specified otherwise.

Even with this definition, a trade-off between the power of the quantum and conventional layers exists, as the following example shows: 
\begin{remark}
We have the equivalence
\begin{equation*}
\LAQCC(\QPoly(n), \p, \bigo(1)) = \LAQCC(\qnc^0, \p, \bigo(\mathrm{poly}(n))).
\end{equation*}
We can prove this equivalence by noting that any $\QPoly(n)$-circuit can be written as $\mathrm{poly}(n)$ concatenated $\qnc^0$-circuits, with no intermediate measurements and trivial conventional computations. 
\end{remark}

Our main focus will be on a specific instance of $\LAQCC(\mathcal{Q},\mathcal{C}, d)$.
\begin{notation}
Unless specified otherwise, in the remainder, $\LAQCC$ refers to the instance $\LAQCC(\qnc^0, \nc^1, \bigo(1))$.
The quantum layers can use all single-qubit gates and the two-qubit CNOT-gate. 
\end{notation}
The class $\nc^1$ is a natural nontrivial class beyond constant-depth complexity classes.
$\LAQCC$ contains many useful gates and subroutines already, as outlined in the next section, specifically the fanout gate.
Note that $\LAQCC$ equals $\qnc^0[\oplus]$ if the conventional computations are restricted to parity computations only. 

The current definition of $\LAQCC(\mathcal{Q},\mathcal{C}, d)$ concerns circuits. 
Many quantum applications consider the potential of circuits in approximating quantum states. 
Approximating quantum states requires a different notion of complexity classes:
\begin{definition}
Let $\HS_n$ be a Hilbert space on $n$ qubits, then define 
\begin{equation*}
    \mathsf{StateX}_{n,\eps}=\{\ket{\psi}\in\HS_n\mid \exists\, \mathsf{X}\text{-circuit $A$ such that } |\bra{\psi}A\ket{0}^{\otimes n}| \ge \eps\},
\end{equation*}
as the set of $n$-qubit quantum states that are $\eps$-close to a state preparable by an $\mathsf{X}$-circuit. 
Define $\mathsf{StateX}_{\eps} =\bigcup_{n\in\N}\mathsf{StateX}_{n,\eps}$.
\end{definition}
This definition extends already existing ideas and definitions of state complexity~\cite{Susskind:2018,Aaronson:2020,RosenthalYuen:2022}.
Our definition shows similarities to state complexity defined in~\cite{MetgerYuen:2023}, but with the uniformity requirement dropped.

%% file: LAQCC/model_definition/Clifford_in_LAQCC.tex
\section{Clifford circuits in \texorpdfstring{$\LAQCC$}{LAQCC}}\label{sec:LAQCC:Clifford_circuits}
The idea of intermediate measurements with subsequent computations is closely related to measurement-based quantum computing. 
A famous result from this field shows that all Clifford circuits can be parallelized using measurements. 
We now use this result to show that any Clifford circuit has a $\LAQCC$-implementation. 

This result is best understood in the teleportation-based quantum computing model~\cite{Jozsa:2006}, a specific instance of measurement-based quantum computing that applies quantum operations using Bell measurements. 
In teleportation, qubits are measured in the Bell basis, which projects the measured qubits onto an entangled two-qubit Bell state, up to local Pauli corrections.
This projection combined with an entangled Bell state teleports a quantum state between qubits.
After teleportation, one needs to correct the local Pauli gate introduced by the Bell measurement. 
A similar process also allows to implement quantum gates. 

Most gates do not commute with Pauli gates, which hinders parallelization. 
As Clifford circuits stabilize the Pauli group, full parallelization of the circuit is possible using parallel measurements~\cite{Jozsa:2006}. 

A Bell basis measurement projects two qubits on $\sum_{i\in \{0,1\}} \bra{ii}P^{a,b}\otimes I$, where $P^{a,b} = Z^a X^b$ and $a,b \in \bset$ correspond to the four possible measurement outcomes.
Using one Bell-basis measurement, we can in parallel apply two sequential Clifford gates $U_1$ and $U_2$ on a quantum state $\ket{\psi}$ and a GHZ state $\tfrac{1}{\sqrt{2}}(\ket{00}+\ket{11})$, which up to normalization gives:
\begin{align*}
    & \sum_{i, j\in \{0,1\}} \big[ (\bra{ii} P^{a,b} \otimes I) \otimes I \big] \big[U_1 \otimes I \otimes U_2\big] \ket{\psi}\ket{jj} \\
    & \qquad = \sum_{i,j \in \{0,1\}} \bra{i} P^{a,b} U_1 \ket{\psi} \braket{i}{j} U_2\ket{j} \\
    & \qquad = \sum_{i \in \{0,1\}}  U_2\kb{i} P^{a,b} U_1 \ket{\psi} = U_2 P^{a,b} U_1\ket{\psi}.
\end{align*}
As $U_2$ is a Clifford gate, there exists $(\hat{a},\hat{b})$ such that $U_2 P^{a,b} U_1 = P^{\hat{a},\hat{b}} U_2  U_1$.
The same argument shows that all correction terms can be postponed to the end of the computation. 

\subsection{Clifford-ladder circuit}\label{sec:LAQCC:Clifford_ladders}
Clifford-ladder circuits form a special type of Clifford circuits:
\begin{definition}[Clifford-ladder circuit]
Given a set $\{U^{i}\}_{i=0}^{n-1}$ of two-qubit Clifford unitaries. 
A Clifford-ladder circuit $C_{ladder}$ is a circuit of depth $\bigo(n)$ and width $\bigo(n)$ of the following form:
\begin{equation*}
C_{ladder} = \prod_{i=0}^{n-1} U^{i}_{(i,i+1)}
\end{equation*}
where $U^{i}_{(i,i+1)}$ denotes that unitary $U^{i}$ is applied on qubits $i$ and $i+1$.
\end{definition}
By definition, every two-qubit Clifford unitary has a constant-depth implementation. 
The next lemma shows that any Clifford-ladder circuit has an equivalent $\LAQCC$-circuit. 
\begin{lemma}\label[lemma]{lem:clifford_ladder}
Any Clifford-ladder circuit has an equivalent $\LAQCC$-circuit of depth $\bigo(1)$ and width $\bigo(n)$.
\end{lemma}
\begin{proof}
Every Clifford-ladder circuit can be made parallel using a Bell measurement and a fresh GHZ state via similar arguments as Clifford gate teleportation. 

What remains to show is that an $\nc^1$-circuit computes the Pauli-correction terms.

The $i$-th Bell measurement result gives Pauli error $P_i = Z^{a_i}X^{b_i}$. 
A Clifford-ladder circuit of size $n$ hence has an error vector $\big(a\,b\big)$ of length $2n$. 
The correction terms that have to be applied have the same form: 
we can label every Pauli correction term by an index $j$, such that $\hat{P}_j = Z^{\hat{a}_j}X^{\hat{b}_j}$. 
This gives a correction vector $\big(\hat{a}\,\hat{b}\big)$. 
Note that Pauli matrices anti-commute, hence reordering them will only incur a global phase.
This implies a binary linear map $M:\big(a\,b\big)\mapsto\big(\hat{a}\,\hat{b}\big)$. 
As matrix vector multiplication is in $\nc^1$, this error calculation is in $\nc^1$.
Hence, every Clifford-ladder circuit has an equivalent $\LAQCC$-circuit.
\end{proof}

\begin{figure}[ht]
    \centering
    \includegraphics[width=0.7\textwidth, trim={8cm 5cm 8cm 5cm}, clip]{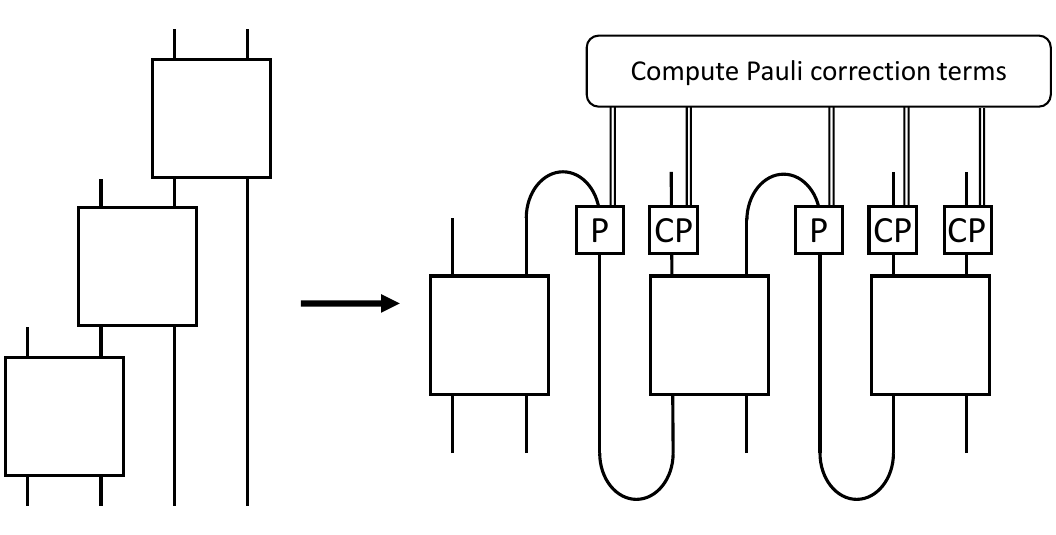}
    \caption{Graphical representation of Clifford-ladder circuit parallelization. 
    Time flows upward and lines represent qubits and boxes quantum gates. 
    Half circles represent Bell-state creation (ends pointing upwards) and Bell-state measurement (ends pointing downwards). 
    Bell-state measurements can produce Pauli errors $P = Z^a X^b$, which are corrected by the boxes CP (corrective Pauli). 
    Conventional computations determine the correction terms.}
    \label{fig:clifford_ladder}
\end{figure}

\cref{fig:clifford_ladder} shows a $\LAQCC$-circuit that implements a Clifford-ladder circuit.
Every two-qubit unitary is parallelized using gate teleportation.
Using the Clifford commutation relations, the Pauli correction terms are pushed to the end of the computation. 
The caps and cups denote Bell-state measurements and Bell-state creation, respectively. 

\begin{remark}
Constructing the binary linear map $M$ is $\Lspace$ instead of $\nc^1$. 
However, as $M$ follows from the quantum circuit and is independent from the measurement outcomes, we can compute it beforehand. 
\end{remark}

As a SWAP gate corresponds to a Clifford-ladder circuit, we see that it has an equivalent $\LAQCC$-circuit and that we can also apply any two-qubit Clifford gate on any pair of qubits in constant depth. 

\subsection{Clifford-grid circuit}\label{sec:LAQCC:Clifford_grids}
Any Clifford unitary can be mapped to a linear-depth circuit on a linear nearest-neighbor architecture~\cite{MaslovRoetteler:2018}. 
The most general representation of these circuits are so-called Clifford-grid circuits.
\begin{definition}[Clifford-grid circuit]
A Clifford-grid circuit of depth $d$ on $n$ qubits is a circuit of the form
\begin{equation*}
   C_{grid} =  \prod_{i=0}^{d}\bigotimes_{j=0}^{\frac{n}{2}}U_{i,j},
\end{equation*}
for Clifford unitaries $U_{i,j}$ and such that in the $i$-th layer, gate $U_{i,j}$ acts on qubits~$2j$ and $2j + 1$ if $i$ is even, and $2j+1$ and $2j+2$ is $i$ is odd. 
\end{definition}

The next lemma shows that Clifford-grid circuits also have an equivalent $\LAQCC$-circuit, and with that, that any Clifford unitary has an equivalent $\LAQCC$-circuit. 
\begin{lemma}\label[lemma]{lem:clifford_grid}
Any Clifford-grid circuit of depth $\bigo(n)$ has an equivalent $\LAQCC$-circuit of depth $\bigo(1)$ and width $\bigo(n^2)$.
\end{lemma}
\begin{proof}
Again, using gate teleportation we can parallelize the $\bigo(n^2)$ Clifford gates and obtain a $\LAQCC$-circuit. 
Every Clifford gate requires a Bell measurement and a GHZ state, giving a total of $\bigo(n^2)$ qubits required.

Any Bell measurement in the circuit can incur a Pauli error, which has to be dealt with at the end of the circuit. 
Similar to the Clifford-ladder circuits, a vector $(a\,b)$ exists containing information on the correction terms, this time of length $\bigo(n^2)$.
This vector induces a vector of correction terms $(\hat{a}\,\hat{b})$ of length $\bigo(n)$.

As Pauli errors anti-commute, there exists a binary linear map $M: (a\,b)\mapsto (\hat{a}\,\hat{b})$. 
The corresponding matrix is rectangular and the error-correction calculations are in $\nc^1$, completing the proof.
\end{proof}

Determining $M$ is more difficult for Clifford-grid circuits than for Clifford ladder circuits. 
This difficulty stems from errors having multiple possible paths that can contribute to an error in a single output qubit. 
The final error depends on the parity of all these paths. 
The computation still only depends on the measurement outcomes and not on the actual input of the circuit. 
Hence, $M$ can follow from a precomputation. 
This precomputation is in $\oplus\mathsf{L}\subseteq\nc^2$~\footnote{This is expected, as efficiently simulating Clifford circuits is a complete problem for $\oplus \mathsf{L}$.}.
\begin{corollary}\label[corollary]{cor:clifford_in_LAQCC}
    Any Clifford circuit has an equivalent $\LAQCC$-circuit, provided an input-independent precomputation. 
\end{corollary}

A special Clifford-grid circuit that we will use extensively is the fanout gate. 
\begin{equation}
    \text{Fanout}_n: \ket{x}\ket{y_1}\hdots\ket{y_{n}} \mapsto\ket{x}\ket{y_1\oplus x}\hdots\ket{y_{n}\oplus x}.
\end{equation}
This gate corresponds to $n$ CNOT-gates with $n$ different targets and with the same control qubit. 

%% file: LAQCC/model_definition/useful_gates_in_LAQCC.tex
\section{Useful gates and routines in \texorpdfstring{$\LAQCC$}{LAQCC}\label{sec:LAQCC:useful_gates_and_subroutines}}
As we saw in the previous section, the fanout gate is a Clifford ladder and hence has an equivalent $\LAQCC$-circuit using $\bigo(n)$ qubits. 

\citeauthor{GreenHomerMoorePollett:2002} and \citeauthor{HoyerSpalek:2005} used the fanout gate to show how to implement a set of pairwise-commuting gates in parallel~\cite{GreenHomerMoorePollett:2002,HoyerSpalek:2005}. 
\begin{theorem}\label{thm:parallel_unitaries}[Theorem~3.2~\cite{HoyerSpalek:2005}]
    Let $\{U_i\}_{i=1}^{n}$ be pairwise-commuting gates on $k$ qubits. 
    Let $K$ be a gate changing the basis in which all $U_i$ are diagonal. 
    Then, there exists a quantum circuit with fanout computing $U=\Pi_{i=1}^{n} U_i$ having depth $\max_{i=1}^n \rm{depth}(U_i) + 2\cdot\rm{depth}(K) + 2$, size $\sum_{i=1}^{n} \rm{size}(U_i) + 2\cdot\rm{size}(K) + 2nk$, and using $(n-1)k$ auxiliary qubits. 
\end{theorem}
A common setting where this theorem can be used is if qubits in one register control operations in another register, such as the modular multiplication step in Shor's algorithm. 
This result was already discovered by \citeauthor{MooreNilsson:2001}, though they used a $\bigo(\log n)$-deep implementation of the fanout gate based on cascading CNOT-gates~\cite{MooreNilsson:2001}. 
Note that for every set of pairwise-commuting gates, a gate $K$ changing the basis must exist~\cite[Theorem~1.3.19]{HornJohnson:1985}.

Using the fanout gate, we obtain $\LAQCC$-implementations for multiple complex quantum gates, as we outline below. 
Most gates follow from work by \citeauthor{TakahashiTani:2013}~\cite{TakahashiTani:2013}. 
We only consider the action on $n$-qubit computational basis states. 
The action of the gates on arbitrary states follows by linearity. 

From \cref{sec:LAQCC:Clifford_circuits} we know that any permutation $\pi\in S_n$ is in $\LAQCC$. 
\begin{align*}
    \text{Permutation}(\pi)_n: & \ket{y_1}\hdots\ket{y_n} \mapsto \ket{y_{\pi(1)}}\hdots\ket{y_{\pi(n)}}
\end{align*}

\citeauthor{HoyerSpalek:2005} provided the first constant-depth quantum implementation of the OR-gate with one-sided error~\cite{HoyerSpalek:2005}. 
\citeauthor{TakahashiTani:2013} later improved the construction of the OR-gate to be exact~\cite{TakahashiTani:2013}. 
From the OR-gate we also directly arrive at the AND- and Equal$_i$-gate:
\begin{align*}
    \text{OR}_n: & \ket{y_1}\hdots\ket{y_{n}} \ket{x} \mapsto\ket{y_1}\hdots\ket{y_n}\ket{\text{OR}_n(y)\oplus x} \\
    \text{AND}_n: & \ket{y_1}\hdots\ket{y_{n}} \ket{x} \mapsto\ket{y_1}\hdots\ket{y_n}\ket{\text{AND}_n(y)\oplus x} \\
    \text{Equal}_i: & \ket{j}\ket{b} \mapsto \begin{cases} 
    \ket{j}\ket{1\oplus b} & \text{if } j = i, \\
    \ket{j}\ket{b} & \text{else.}
    \end{cases}
\end{align*}
We will omit the subscript $n$ if the number of qubits on which they operate is clear. 
All three gates have width $\bigo(n\log n)$. 
The OR-gate follows from Theorem~1 in~\cite{TakahashiTani:2013}. 
The AND-gate follows by negating all inputs and outputs of the OR-gate. 
The Equal$_i$-gate follows from the AND-gate, combined with a negation of the inputs that correspond to zeros in the binary representation of $i$. 

By modifying the circuit for the OR-gate, we additionally obtain an implementation for the Exact$_t$-gate with the same circuit size (see also \cite[Theorem~4.6]{HoyerSpalek:2005}). 
This circuit outputs $1$ if precisely $t$ of the inputs are $1$, and outputs $0$ otherwise: 
\begin{equation*}
    \text{Exact}_t: \ket{x}\ket{b} \mapsto \begin{cases} \ket{x}\ket{b\oplus 1} & \text{if } |x| = t\\
    \ket{x}\ket{b} & \text{else.} \end{cases} 
\end{equation*}

As we have $\LAQCC$-circuits for the OR- and AND-gate, we have $\LAQCC$-circuits for all $\ac^0$-circuits, including circuits for modular addition and circuits that check (in)equalities. 
All three circuits have width~$\bigo(n^2)$. 
\begin{align*}
    \text{Add}_{n}: & \ket{x}\ket{y} \mapsto\ket{x}\ket{y + x \bmod 2^n} \\
    \text{Equality}: & \ket{x}\ket{y}\ket{b}  \mapsto \begin{cases}
    \ket{x}\ket{y}\ket{b\oplus 1} & \text{if } x = y, \\ 
    \ket{x}\ket{y}\ket{b} & \text{else,} 
    \end{cases} \\
    \text{Greaterthan}: & \ket{x}\ket{y}\ket{b} \mapsto \begin{cases}
    \ket{x}\ket{y}\ket{b\oplus 1} & \text{if } x > y, \\ 
    \ket{x}\ket{y}\ket{b} & \text{else.} \end{cases}
\end{align*}
The Equality-gate is obtained by subtracting the first register from the second and computing the OR of the second register in the auxillary register and negating it. 
The Greaterthan-gate follows by a similar construction:
Use a single auxiliar qubit in the $\ket{0}$-state to the second register and interpret $y$ as a $n+1$-bit integer;
Subtract the first register from the second and apply a CNOT-gate from this auxiliary qubit to the target output qubit. 
If $x$ is larger than $y$, then the most significant bit of the second register (the auxillary one) is in the one state. 

\citeauthor{HoyerSpalek:2005} also showed how fanout gates help implement quantum subroutines in constant depth, such as the quantum Fourier transform defined in \cref{eq:gate_QFT} with circuit size $\bigo(n^3\log n)$~\cite[Theorem~4.12]{HoyerSpalek:2005}.
With this gate, we can obtain a constant-depth implementation for the Hammingweight-gate.
This gate computes the binary representation of the Hamming weight of a string $x$ in a separate register using a $\bigo(n^2)$ wide circuit~\cite[Lemma~4]{TakahashiTani:2013}. 
Note that this gate is strictly stronger than the Exact$_t$-gate, as it actually computes the Hamming weight, instead of determining if it is equal to some integer. 
\begin{equation*}
    \text{Hammingweight}: \ket{x}_{n}\ket{0}_{\log(n)} \mapsto \ket{x}_n\ket{|x|}_{\log(n)}
\end{equation*}

\citeauthor{TakahashiTani:2013} combined the Hammingweight-gate with the Exact$_t$-gate to obtain a circuit for the Threshold$_t$-gate with circuit width $\bigo(n\log n)$ if $t\le \log n$ and $\bigo(n\sqrt{t\log n})$ for $\log n\le t\le \ceil{\tfrac{n}{2}}$~\cite[Theorem~2]{TakahashiTani:2013}. 
The circuit size for $t>\ceil{\frac{n}{2}}$ follows from the circuit size of the Threshold$_{n-t}$-gate:
\begin{equation*}
    \text{Threshold}_t: \ket{x}\ket{b} \mapsto \begin{cases} \ket{x}\ket{b\oplus 1} & \text{if } \sum_i x_i \ge t, \\
    \ket{x}\ket{b} & \text{else.} \end{cases}
\end{equation*}
Now, as the Threshold$_t$-gate is in $\LAQCC$, any $\tc^0$-circuit is also in $\LAQCC$. 
Additionally, the Threshold-gate can be turned into a weighted Threshold-gate by incorporating the weights in the rotation angles used in the implementation.
Note that we can also compute the Threshold$_t$-gate using the Greaterthan-gate, where the second register is the state~$\ket{t}$. 
However, this gives worse scaling of the circuit size than a direct Threshold$_t$-gate implementation. 

The final routine we discuss is a consequence of a quantum algorithm by \citeauthor{Long:2001}, which generalizes Grover's algorithm. 
Grover's algorithm searches a database and returns a marked item with high probability. 
The algorithm by \citeauthor{Long:2001} does the same, but with unit probability~\cite{Long:2001}. 
We modify the algorithm by \citeauthor{Long:2001} not to search a database, but instead to prepare quantum state: 
\begin{lemma}\label[lemma]{lem:grover_constant_fraction}
Let $c>0$ be a constant. 
Let $\mathcal{C}$ be a set of $2^n$ quantum states that forms a basis for all $n$-qubit quantum states. 
Let $\mathcal{G}$ and $\mathcal{B}$ partition $\mathcal{C}$, such that $\frac{|\mathcal{G}|}{|\mathcal{C}|}=c$. 
Suppose $U$ implements in constant depth the map 
\begin{equation*}
    U:\ket{y}\ket{b}\mapsto\begin{cases}
    \ket{y}\ket{b\oplus 1} & \qquad \text{if }y\in\mathcal{G}, \\
    \ket{y}\ket{b} & \qquad \text{if }y\in\mathcal{B}. 
    \end{cases}
\end{equation*}
Then, there exists a $\LAQCC$-circuit that prepares the state $\frac{1}{\sqrt{|\mathcal{G}|}}\sum_{y\in\mathcal{G}} \ket{y}$.
\end{lemma}
\begin{proof}
Write $\ket{\mathcal{G}}=\frac{1}{\sqrt{|\mathcal{G}|}}\sum_{y\in\mathcal{G}}\ket{y}$ and $\ket{\mathcal{B}}=\frac{1}{\sqrt{|\mathcal{B}|}}\sum_{y\in\mathcal{B}}\ket{y}$. 
Then we have $\braket{\mathcal{G}}{\mathcal{B}}=0$, as $\mathcal{B}$ and $\mathcal{G}$ partition~$\mathcal{C}$.

The circuit by \citeauthor{Long:2001} produces a quantum state that, upon measurement, returns a uniform random element from $\mathcal{G}$. 
Omitting that final measurement thus prepares the state $\ket{\mathcal{G}}$ as desired.

The corresponding circuit is indeed in $\LAQCC$: 
First, prepare a uniform superposition $\ket{s}=\sum_{i=0}^{2^n-1}\ket{i}$. 
Then, iteratively reflect over the state $\ket{\mathcal{B}}$ using $U$, and reflect over the uniform superposition state $\ket{s}$. 
Both reflections have a $\LAQCC$ implementation and we only need to apply them a constant number of iterations. 

The first reflection implements the map $I-(1-e^{i\phi})\kb{\mathcal{G}}$ and the second reflection the map $I - (1-e^{i\phi})\kb{s}$, where $\phi$ depends on the number of target states and the known constant $c$, see also~\cite{Long:2001}. 
The implementation for both reflections follows the same lines:
Mark the state $\ket{\mathcal{G}}$ and $\ket{s}$ using an auxillary qubit; Use an $R_Z$-gate to apply the phase $(1-e^{i\phi})$; Uncompute the auxiliary qubit. 
The first reflection uses the unitary $U$, which by definition has a constant-depth implementation. 
The second reflection uses an Exact$_0$-gate with negated output and all inputs conjugated by Hadamard gates. 

The circuit only uses operations from $\LAQCC$. 
Additionally, the total number of iterations is $\bigo(\sqrt{N/m})=\bigo(\sqrt{|\mathcal{C}|/|\mathcal{G}|})=\bigo(\sqrt{c})$, from which the constant depth of the whole circuit follows. 
\end{proof}

%% file: LAQCC/model_definition/complexity_for_LAQCC.tex
\section{Complexity results for \texorpdfstring{$\LAQCC$}{LAQCC}}\label{sec:LAQCC:complexity_results_standard}
$\LAQCC$ seems more powerful than conventional computing alone. 
Yet, most power seems to come from the conventional intermediate computations. 
This section first proves an upper bound on the power of $\LAQCC$, before showing that $\LAQCC$-circuits are unlikely to be efficiently simulatable.

Any $\LAQCC$-circuit $A$ can be written as a composition of quantum layers~$U_i$, measurements $M_i$ and conventional layers $C_i$:
\begin{equation}\label{eq:decomposition_LAQCC}
    A = M_k U_k C_k \dots M_i U_i C_i \dots M_1 U_1 C_1,
\end{equation}
for some constant $k$. 
Any unitary $U_i$ corresponds to a $\qnc^0$-circuit and any $C_i$ corresponds to an $\nc^1$-circuit. 
The measurements $M_i$ can measure any subset of the qubits.
By the principle of deferred measurements, we can always postpone them to the end of the circuit using CNOT-gates and fresh auxiliary qubits~\cite[Section~4.4]{NielsenChuang:2010}, which gives the following lemma. 
\begin{lemma}\label[lemma]{lem:LAQCC_QNC1}
    For any $\LAQCC$-circuit $A$ there is a $\QPoly$-circuit $B$ without intermediate measurements that outputs the same state as $A$.
\end{lemma}
\begin{proof}
The $\LAQCC$-circuit $A$ contains the layers $C_i$ that use the intermediate measurement results as input. 
Delaying the measurements until the end of the circuit by applying a CNOT-gate from the qubit to a fresh auxiliary qubit, replaces the conventional output wires by quantum wires. 

Next, we can replace any $\nc^1$-circuit by a $\qnc^1$-circuit that uses at most $\poly(n)$ auxiliary qubits:
1) Replace all OR-gates by AND- and NOT-gates;
2) Replace all AND-gates by Toffoli-gates, where the third input is a clean auxiliary qubit;
and, 3) Replace all NOT-gates by $X$-gates. 

Hence, for every conventional circuit $C_i$, we can obtain a quantum circuit $V_i$, such that $V_i$ computes the same output. 
The quantum circuit $B = U_k V_k \dots U_1 V_1$ would work and it appears to have logarithmic depth.
However, we are not guaranteed that the $V_i$ adheres to the linear nearest-neighbor topology constraints.
To correct for that, we can permute the qubits between every layers, which results in a quantum circuit of size $\bigo(n\log n)$, as any permutation can be implemented in size $\bigo(n)$. 
\end{proof}

Note that without the topology constraints, the power of $\LAQCC$-circuits is bounded by $\qnc^1$-circuits, as we have polynomial-sized quantum circuits of logarithmic depth. 
This makes one wonder if these circuits admit efficient conventional simulations.
Even in that case, $\LAQCC$-circuits can still have value as ``fast'' alternatives for state preparation routines.
By linking the class of Instantaneous Quantum Polynomial-time ($\mathsf{IQP}$)-circuits, first introduced in~\cite{ShepherdBremner:2009}, to $\LAQCC$-circuits, we can show that such an efficient simulation for \emph{all} $\LAQCC$-circuits is unlikely.
\begin{definition}[Definition 2~\cite{NakataMurao:2014}]
An $\mathsf{IQP}$-circuit on $n$ qubits is a quantum circuit such that: each gate in the circuit is diagonal in the Pauli-$Z$ basis, the input state is $\ket{+}^{\otimes n}$, and the output is the result of a measurement in the Pauli-$X$ basis on a specified set of output qubits.
\end{definition}
\begin{lemma}\label[lemma]{lem:IQP_in_LAQCC}
Any $\mathsf{IQP}$-circuit has an equivalent $\LAQCC$-circuit.
\end{lemma}
\begin{proof}
Gates diagonal in the Pauli-$Z$ basis commute in the Pauli-$X$ basis. 
Therefore, prepare $\ket{+}^{\otimes n}$ by a single layer of Hadamard gates on all qubits.
Then, parallelize the gates from the IQP-circuit using \cref{thm:parallel_unitaries}.
Measurements in the Pauli-$X$ basis correspond to a Hadamard gate followed by a measurement in the Pauli-$Z$ basis. 
This construction gives the equivalent $\LAQCC$-circuit. 
\end{proof}

\citeauthor{BremnerJozsaShepherd:2010} showed that efficient weak simulation of all possible IQP-circuits up to a small multiplicative error implies a collapse of the polynomial hierarchy~\cite{BremnerJozsaShepherd:2010}. 
As a direct corollary, $\LAQCC$-circuits are unlikely to be efficiently simulatable using conventional computers, unless the polynomial hierarchy collapses. 
A circuit family is weakly simulatable if its output distribution can be sampled by purely conventional means in polynomial time given the description of the circuit family.
\begin{lemma}[Corollary 1~\cite{BremnerJozsaShepherd:2010}]
If the output probability distributions generated by uniform families of $\mathsf{IQP}$-circuits could be weakly simulated to within multiplicative error $1 \leq c < \sqrt{2}$ then the polynomial hierarchy collapses to the third level, in particular, $\mathsf{PH} = \Delta^p_3$.
\end{lemma}

%% file: LAQCC/model_definition/Powerful_LAQCC.tex
\section{Complexity results for powerful \texorpdfstring{$\LAQCC$}{LAQCC}}\label{sec:LAQCC:complexity_results_powerful}
Based on the previous section, $\LAQCC$ seems more powerful than conventional computing alone. 
The conventional intermediate computations provide $\LAQCC$ with significant power. 
This added power raises the question what other capabilities are possible if we extend the conventional computational power beyond~$\nc^1$. 
In this section, we consider a different instance of $\LAQCC(\mathcal{Q}, \mathcal{C},d)$ where we allow for more powerful quantum and conventional computations, as well as more alternations between them, giving us the class $\LAQCC^*$. 
\begin{notation}
The class $\LAQCC^*$ refers to the instance
\begin{equation*}
    \LAQCC(\QPoly(n), \mathsf{ALL}, \text{poly}(n)).
\end{equation*}
That is, $\LAQCC^*$ refers to the class of circuits that alternate a polynomial number of times between polynomial-sized quantum circuits and arbitrary powerful conventional computations, together with feed forward of the conventional information to future quantum operations. 
The quantum computations are restricted to all single-qubit gates and the two-qubit CNOT-gate. 
\end{notation}
Note that $\LAQCC^*$ can trivially solve decision problems by simply using the unbounded conventional computational power and then loading the result in a quantum state. 
Therefore, we are mainly interested in the computational power of $\LAQCC^*$ with respect to state preparation. 
The definition of $\LAQCC^*$ directly defines $\mathsf{State}$$\LAQCC^*_{\eps}$ for $\eps\ge 0$. 
\begin{remark}
For any nonzero $\eps$, we can restrict ourselves to finite universal gate sets. 
The Solovay-Kitaev theorem~\cite{Kitaev:1997,NielsenChuang:2010} states that any multi-qubit unitary can be approximated to within precision $\delta$ by a quantum circuit with size depending only on $\delta$. 
Hence, any $\LAQCC^*$-circuit using a continuous gate set can be approximated by a $\LAQCC^*$-circuit using a universal finite gate set. 
\end{remark}

\begin{figure}[ht]
\centering
\includegraphics[width=0.9\textwidth, trim={7.5cm 7.5cm 7.5cm 7.5cm}, clip]{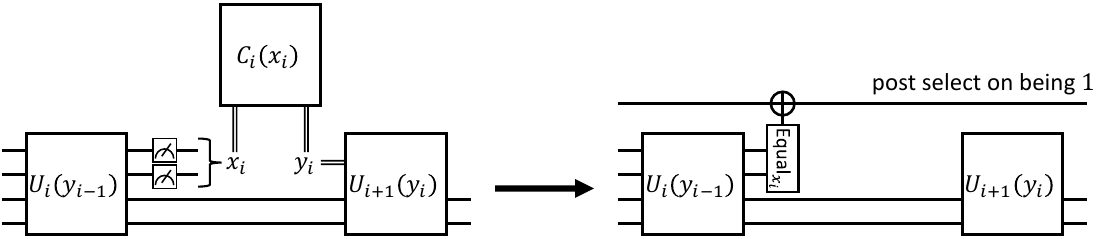}
\caption{Graphical representation of transforming a $\LAQCC^*$-circuit for generating $\ket{\psi}$ into a $\mathsf{PostQPoly}$-circuit.}
\label{fig:laqcc_to_post_bqp}
\end{figure}

We will compare $\LAQCC^*$ with the class of polynomial-sized quantum circuits equipped with an additional post-selection gate.
Even though this gate has no physical realization, as measurements outcomes cannot be chosen, it is a useful gate in analyzing different algorithms. 
\begin{definition}
The class $\mathsf{PostQPoly}$ consists of all $\QPoly$-circuits that produce a quantum state $\alpha\ket{1}\ket{\psi}+\beta\ket{0}\ket{\perp}$, where $\ket{\psi}$ is the relevant quantum state and $\ket{\perp}$ can be any arbitrary quantum state. 
The success of the quantum circuit is conditioned on the first qubit being in the $\ket{1}$-state. 
\end{definition}
We can decompose any $\LAQCC^*$-circuit as
\begin{equation}\label{eq:decomposition_LAQCC_star}
    \Pi_{i=0}^{\text{poly}(n)} M_i U_i(y_i) C_i(x_i)\ket{0}^{\otimes \text{poly}(n)}.
\end{equation}
Again, $M_i$ denotes the $i$-th measurement, and $U_i$ and $C_i$ denote the $i$-th quantum and conventional layer, respectively. 
The $x_i\in\{0,1\}^*$ denote the outcome of $M_i$ and $y_i\in \{0,1\}^*$ the bitstring outputted by $C_i$. 
Note, all $x_i$ and $y_i$ have length at most polynomial in~$n$. 
We will use this decomposition to prove a similar inclusion as in \cref{lem:LAQCC_QNC1}.
\begin{theorem}\label{thm:stateLAQCC_in_state_post_qpoly}
For every $\eps\ge 0$ it holds that 
\begin{equation*}
    \mathsf{StateLAQCC}^*_{\eps}\subseteq\mathsf{StatePostQPoly}_{\eps}
\end{equation*}
\end{theorem}
\begin{proof}
Fix $\eps\ge 0$ and a positive integer $n$ and let $\ket{\psi}\in\mathsf{State}$$\LAQCC^*_{\eps}$. 
By definition, there exists a $\LAQCC^*$-circuit $A=\Pi_{i=0}^{\text{poly}(n)} \, M_i U_i(y_i) C_i(x_i)$ that prepares a state $\ket{\phi}$ such that $|\braket{\phi}{\psi}|\ge \eps$.

Let $B=\Pi_{i=0}^{\text{poly}(n)} \,\text{Equal}_{x_i}(x_i)U_i(y_i)\ket{0}^{\otimes \text{poly}(n)}$, where the $y_i$ are hardwired inputs to the quantum circuits, based on measurement outcome $x_i$. 
This $\mathsf{PostQPoly}$-circuit prepares $\ket{\psi}$ with unit probability, conditioned on the auxillary qubit being in the $\ket{1}$-state.

The Equal$_{x_i}$-gate replaces the measurement $M_i$ by checking if the qubits that would be measured are in $\ket{x_i}$-state.
The outcomes are stored in an auxiliary qubit. 
This gives $\poly(n)$ auxiliary qubits, one for each replaced measurement layer. 
Then apply an AND-gate on these auxiliary qubits and store the result in another auxiliary qubit. 
Condition the output of the circuit is then conditioned on this final auxiliary qubit. 
\end{proof}
Note that every measurement has at least one possible output string $x_i$. 
\cref{fig:laqcc_to_post_bqp} sketches the idea behind the proof. 

%% file: LAQCC/model_definition/discussion_LAQCC_model.tex
\section{Reflections and outlook}
This chapter introduced the $\LAQCC$-model that alternates between conventional and quantum operations. 
The $\LAQCC$-model generalizes already existing models by explicitly bounding the conventional intermediate computing power. 

We focused on a specific instance of $\LAQCC$ that alternates between $\qnc^0$-circuits and $\nc^1$-circuits.
This instance generalizes the $\qnc^0[\oplus]$-circuits encountered in the first part of this work. 
The conventional computations help to implement a quantum fanout gate, which is often used as a building block for constructing more complex gates. 
Future research can focus on what other problems we can offload to the conventional computer. 
Another promising direction is to explore $\LAQCC$-instances with increased quantum or conventional computational power, or more alternations between the two. 
Finally, future research can help quantify the relation between different $\LAQCC$-instances and other known complexity classes.

At the end of the chapter, we took a first step in this direction by considering a powerful instance called $\LAQCC^*$.
This instance alternates between arbitrary conventional computations and polynomial-sized quantum circuits. 
This class naturally solves any solvable decision problem using the unbounded conventional computations. 
With respect to state preparation, we however have the inclusion
\begin{equation*}
    \mathsf{StateLAQCC}^{*}_{\eps} \subseteq \mathsf{StatePostQPoly}_{\eps}.
\end{equation*}
We conjecture that this inclusion is strict for any $\eps>0$. 

As a first step towards proving this conjecture, one might consider an oracular separation, using, for instance, the oracle \citeauthor{AaronsonKuperberg:2007} used to separate $\mathsf{QMA}$ and $\mathsf{QCMA}$~\cite{AaronsonKuperberg:2007}.
For every state $\ket{\psi}$, they define an oracle $O_{\psi}$ such that $O_{\psi}:\ket{\psi} = -\ket{\psi}$, while for every $\ket{\phi}\perp\ket{\psi}$ $O_{\psi}:\ket{\phi} = \ket{\phi}$.

Interestingly, a $\mathsf{PostQPoly}$-circuit can prepare $\ket{\psi}$ exactly using a single oracle query. 
There always exists a computational basis state $\ket{i}$ with nonzero overlap with $\ket{\psi}$.
Starting with that state and implementing $O_{\psi}$ as a bit-flip oracle gives
\begin{equation*}
    \alpha \ket{\psi^{\perp}}\ket{0} + \beta \ket{\psi}\ket{1},
\end{equation*}
for some $\alpha,\beta\in\C$ and $\beta\neq 0$. 
Conditioned on the auxiliary qubit, this circuit indeed prepares $\ket{\psi}$.

To prove that $\LAQCC^*$-circuits cannot approximate $\ket{\psi}$, we have two promising directions. 
The first uses $\eps$-nets to show that the number of possible quantum states is double exponential in~$n$~\cite[Section~4.5.4]{NielsenChuang:2010}.
However, counting the number of $\LAQCC^*$-circuits, we find that $\mathsf{StateLAQCC}^{*}_{\eps}$ contains only an exponential number of quantum states.
When counting the number of $\LAQCC^*$-circuits, we can omit the intermediate computations as their output is only used to control future quantum operations. 
We can then use that the oracle is close to the identity operator, and oracle calls only marginally help to approximate $\ket{\psi}$. 
We can omit an oracle call at the cost of a small error. 
However, if we want $\eps=1/\poly(n)$, we find that these small errors stack too much with the multiple oracle calls.

The second direction uses quantum state based on incompressible strings. 
These strings have no short program that returns them. 
Kolmogorov complexity dictates that strings exist that admit no efficient description~\cite{Li:2008}. 
Let $z$ be an incompressible string of length $|z| = N (3+\varepsilon) n$ for $N = 2^n$. 
Interpret $z$ as a concatenation of $N$ substrings of length $(3+\varepsilon)n$ each: $z = [z_0, z_1 \dots z_i \dots, z_{N-1}]$.
The following $(4+\varepsilon)n$-qubit quantum state encodes $z$
\begin{equation*}
    \ket{z} = \frac{1}{\sqrt{N}}\sum_{i=0}^{N-1} \ket{i}\ket{z_i}.
\end{equation*}
A $\LAQCC^*$-circuit that prepares $\ket{z}$ now gives an efficient description of a routine that returns the string $z$, something which we conjectured impossible due to $z$ being incompressible. 
It is again unclear how oracle calls can help approximate~$\ket{z}$.

%% file: LAQCC/LAQCC_state_preparation.tex
\chapter{State preparation by \texorpdfstring{$\LAQCC$}{LAQCC}}\label{chp:LAQCC:state_preparation}
This chapter proposes $\LAQCC$-circuits for three types of non-stabilizer states. 
First, we give a $\LAQCC$-circuit to prepare a uniform superposition over an arbitrary number of states. 
Next, we use an uncompress-compress method to prepare the $W$-state. 
We conclude with two protocols to prepare the Dicke state, uniform superpositions of all $n$-bitstrings with Hamming weight $k$. 
The first protocol prepares in constant depth Dicke states for $k=\bigo(\sqrt{n})$.
The second protocol prepares Dicke states for all $k$ in logarithmic depth. 

\input{LAQCC/state_preparation/outline_state_preparation}
\input{LAQCC/state_preparation/uniform_superposition}

\input{LAQCC/state_preparation/W_state}
\input{LAQCC/state_preparation/Dicke_small_k}

\input{LAQCC/state_preparation/Dicke_log_depth}

\input{LAQCC/state_preparation/discussion_LAQCC_state_preparation}

%% file: LAQCC/state_preparation/outline_state_preparation.tex
\section{Chapter overview}
The previous chapter introduced the $\LAQCC$-model, where quantum and conventional computations alternate to solve a problem or prepare some quantum state. 
This chapter considers the capability of this $\LAQCC$-model to prepare quantum states often used in other algorithms and for benchmarking quantum devices. 

First, \cref{sec:LAQCC:uniform_superposition_modulo_q} gives in \cref{thm:uniform_superposition_mod_q} a $\LAQCC$-circuit to prepare a uniform superposition of an arbitrary number of states.
The circuit uses \cref{lem:grover_constant_fraction}.
This state is often used as starting state by other algorithms. 

\cref{sec:LAQCC:W_state} gives a $\LAQCC$-circuit for preparing the $W$-state. 
This circuit uses the circuit for the uniform superposition. 
\cref{thm:W_state} shows how to obtain the $W$-state using an uncompress-compress method that efficiently maps between the binary number representation and the one-hot number representation. 
The first representation corresponds to the initial uniform superposition obtained from \cref{thm:uniform_superposition_mod_q}, whereas the second representation corresponds to the $W$-state. 

Next, \cref{sec:LAQCC:Dicke_small_k} uses the circuit for the $W$-state to prepare the Dicke-$(n,k)$ state for $k=\bigo(\sqrt{n})$. 
Specifically, the circuit for the Dicke state uses $k$ parallel instances of the $\LAQCC$-circuit for the $W$-state, all using the same target register. 
Additional $\LAQCC$-operations ensure that all parallel instances target different qubits in the target register. 
The result is summarized in \cref{thm:Dicke:small_k}.

\cref{sec:LAQCC:Dicke_in_LAQCC_LOG} then presents a method to prepare the Dicke state for arbitrary~$k$. 
The protocol, summarized in \cref{thm:Dicke:Log_depth}, maps between the combinatorial number representation (which uniquely labels Dicke states for fixed~$k$) and the factoradic representation. 
As this mapping requires logarithmic depth, we use a new instance called $\LAQCC$-$\mathsf{LOG}$. 

%% file: LAQCC/state_preparation/uniform_superposition.tex
\section{Uniform superposition of size \texorpdfstring{$q$}{q}}\label{sec:LAQCC:uniform_superposition_modulo_q}
Many quantum algorithms start with a uniform superposition. 
Often, a uniform superposition over $2^n$ states is used.
This state is obtained by applying a Hadamard gate on all $n$ qubits. 
In practice, uniform superpositions over fewer states might be more useful. 
However, preparing the superposition 
\begin{equation*}
    \frac{1}{\sqrt{q}}\sum_{i=0}^{q-1}\ket{i}
\end{equation*}
for arbitrary $q$ is difficult. 
As a first step, we can consider a probabilistic approach:
\begin{enumerate}
    \item Create a superposition $\frac{1}{\sqrt{2^n}}\sum_{i=0}^{2^n-1}\ket{i}$ with $n = \ceil{\log_2(q)}$ qubits;
    \item Mark the states $i< q$ using an auxiliary qubit;
    \item Measure the auxiliary qubit.
\end{enumerate}
With probability at least one half, the correct state is prepared. 
The measurement result indicates whether we have the correct state. 
The second step uses a Greaterthan-gate, using $\bigo(\ceil{\log_2(q)}^2)$ qubits, with the second register initialized in the $\ket{q}$-state. 
However, this approach is non-unitary due to the measurements. 

We can modify this probabilistic circuit into a deterministic circuit. 
The next section uses the deterministic circuit to prepare the $W$-state.
Note that if $q$ is a power of $2$, the circuit simplifies to a layer of single-qubit Hadamard gates. 
\begin{theorem}\label{thm:uniform_superposition_mod_q}
There exists a $\LAQCC$-circuit that prepares the uniform superposition on $q$ states. 
This circuit uses $\bigo(\ceil{\log_2(q)}^2)$ qubits.
\end{theorem}
\begin{proof}
Let $n = \ceil{\log_2(q)}$ and define $\mathcal{G}=\{i\mid 0\le i<q\}$ and $\mathcal{B}=\{i\mid q\le i<2^n\}$. 
Applying \cref{lem:grover_constant_fraction} with the unitary
\begin{equation*}
    U_q: \ket{y}\ket{b} \mapsto \begin{cases}
        \ket{y}\ket{b\oplus 1} & \text{if } y<q, \\
        \ket{y}\ket{b} & \text{if } y\ge q,
        \end{cases}
\end{equation*}
gives the desired state. 
This circuit is in $\LAQCC$, as the unitary corresponds to a Greaterthan-gate. 
\end{proof}

%% file: LAQCC/state_preparation/W_state.tex
\section{\texorpdfstring{$W$}{W}-state}\label{sec:LAQCC:W_state}
This section considers the $W_n$-state (or the $W$-state, if $n$ is implicit) and presents a $\LAQCC$-circuit that prepares it. 
The $W$-state is a uniform superposition over all $n$-bit strings of Hamming weight $1$:
\begin{equation*}
    \ket{W}  = \frac{1}{\sqrt{n}}\sum_i \ket{e_i},
\end{equation*}
where $\ket{e_i}$ is the state with a one on the $i$-th position and zeros elsewhere.
The main theorem of this section gives a $\LAQCC$-circuit that prepares the $W$-state:
\begin{theorem}\label{thm:W_state}
There exists a $\LAQCC$-circuit that prepares $\ket{W}$. 
This circuit uses $\bigo(n\log (n)\log\log(n))$ qubits placed in a grid of size $n\times \bigo(\log(n)\log\log(n))$.
\end{theorem}

A natural bijection exists between the states $\ket{e_i}$ and a uniform superposition over $n$ states. 
If we can find a unitary that implements this bijection, we can prepare the $W$-state. 
Hence, we wish to find a unitary that implements the map
\begin{equation}\label{eq:unitary:i_to_ei}
    \ket{i}_{\log n}\ket{0}_{n}\mapsto \ket{0}_{\log n}\ket{e_i}_{n}.
\end{equation}
This unitary maps the binary representation of $i$ (using $\log n$ qubits) to its one-hot representation (using $n$ qubits). 
We refer to registers with a binary representation as index registers and those with a one-hot representation as system registers. 
The mapping between these two representations naturally defines two operations:
\begin{align}
\text{\textbf{Uncompress}: }& \ket{i}_{\log n}\ket{0}_{n} \mapsto \ket{i}_{\log n}\ket{e_i}_{n}, \label{eq:unitary:uncompress} \\
\text{\textbf{Compress}: }& \ket{i}_{\log n}\ket{e_i}_{n} \mapsto \ket{0}_{\log n}\ket{e_i}_{n}. \label{eq:unitary:compress}
\end{align}
Combining these two operations implements the unitary given in \cref{eq:unitary:i_to_ei}. 
Both operations have a $\LAQCC$-implementation, as the next two lemmas show. 
\begin{lemma}\label[lemma]{lem:W_state_uncompress}
There exists a $\LAQCC$-circuit using $\bigo(n \log(n)\log\log(n))$ qubits that, for any $n$, implements the \textbf{Uncompress} operation:
\begin{equation}
    \frac{1}{\sqrt{n}}\sum_{i = 0}^{n-1}\ket{i}_{\log n}\ket{0}_n \mapsto \frac{1}{\sqrt{n}}\sum_{i = 0}^{n-1}\ket{i}_{\log n}\ket{e_i}_n.
\end{equation}
The qubits are placed in a grid of size $n\times \bigo(\log(n)\log\log(n))$.
\end{lemma}
\begin{proof}
Interpret the right-most column of the grid as the system qubits. 
Adjacent to each system qubit are $\log n$ index qubits. 
Additionally, every row has access to $\bigo(\log\log n)$ auxiliary qubits. 
The $\LAQCC$-circuit then consists of the following three steps (where we only show the index and system qubits): 
\begin{align*}
\frac{1}{\sqrt{n}}\sum_{i = 0}^{n-1}\ket{i}_{\log n}\ket{0}_{\log n}^{\otimes n-1}\ket{0}_n & \xrightarrow{(1)} \frac{1}{\sqrt{n}}\sum_{i = 0}^{n-1}\ket{i}_{\log n}^{\otimes n}\ket{0}_n \\
& \xrightarrow{(2)} \frac{1}{\sqrt{n}}\sum_{i = 0}^{n-1}\ket{i}_{\log n}^{\otimes n}\ket{e_i}_n \\
& \xrightarrow{(3)} \frac{1}{\sqrt{n}}\sum_{i = 0}^{n-1}\ket{i}_{\log n}\ket{0}_{\log n}^{\otimes n-1}\ket{e_i}_n 
\end{align*}
Step (1) Use fanout gates to copy the first index register;
Step (2) In parallel for all $i$, apply $\text{Equal}_i$-gates from the $i$-th index register to the corresponding system qubit. 
This creates the state $\ket{e_i}$ in the system register;
Step (3) Use fanout gates to disentangle the index registers. 
\end{proof}

\begin{figure}[ht]
\includegraphics[width=\textwidth, trim={5cm 5.9cm 5cm 5.9cm}, clip]{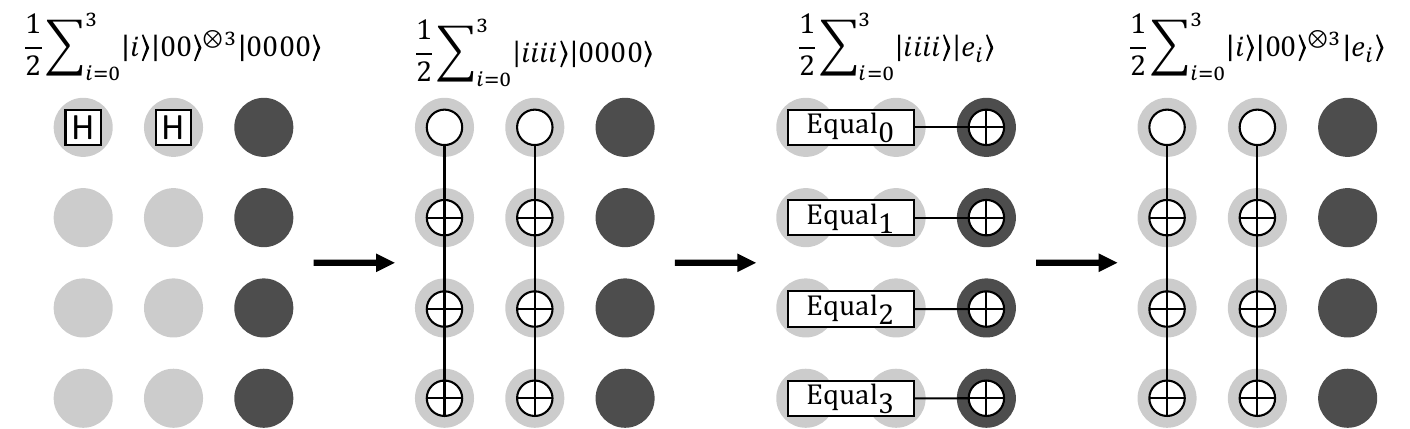}
\caption{Circuit for the \textbf{Uncompress} operation for $n=4$.
Shown is a grid of $12$ qubits: $8$ light gray index qubits, and $4$ dark gray system qubits. 
Each grid represents a single time step.}
\label{fig:W_state_uncompress}
\end{figure}

\cref{fig:W_state_uncompress} shows the steps of the \textbf{Uncompress} operation graphically for $n=4$.
We explicitly leave out auxiliary qubits in the shown circuits. 

\begin{lemma}\label[lemma]{lem:W_state_compress}
There exists a $\LAQCC$-circuit that, for any $n$, implements the \textbf{Compress} operation:
\begin{equation*}
    \frac{1}{\sqrt{n}}\sum_{i = 0}^{n-1}\ket{i}_{\log n}\ket{e_i}_{n} \mapsto \frac{1}{\sqrt{n}}\sum_{i = 0}^{n-1}\ket{0}_{\log n}\ket{e_i}_{n}.
\end{equation*}
This circuit uses $\bigo(n \log n)$ qubits placed in a grid pattern of size $n\times \log n$.
\end{lemma}
\begin{proof}
The \textbf{Compress} operation uses parallel $CNOT$-gates controlled by the system register to uncompute the index registers. 
These $CNOT$-gates commute for different indices in the system register and hence by \cref{thm:parallel_unitaries} they can be applied in parallel. 
The \textbf{Compress} operation consists of five steps: 
\begin{align*}
\frac{1}{\sqrt{n}}\sum_{i = 0}^{n}\ket{i}_{\log n}\ket{0}_{\log n}^{\otimes n - 1}\ket{e_i}_n & \xrightarrow{(1)} \frac{1}{n}\sum_{i, j = 0}^{n}(-1)^{i \cdot j}\ket{j}_{\log n}\ket{0}_{\log n}^{\otimes n - 1}\ket{e_i}_n \\
    & \xrightarrow{(2)} \frac{1}{n}\sum_{i, j = 0}^{n}(-1)^{i \cdot j}\ket{j}_{\log n}^{\otimes n}\ket{e_i}_n \\
    & \xrightarrow{(3)} \frac{1}{n}\sum_{i, j = 0}^{n}\ket{j}_{\log n}^{\otimes n}\ket{e_i}_n \\
    & \xrightarrow{(4)} \frac{1}{n}\sum_{i, j = 0}^{n}\ket{j}_{\log n}\ket{0}_{\log n}^{\otimes n-1}\ket{e_i}_n \\
    & \xrightarrow{(5)} \frac{1}{\sqrt{n}}\sum_{i =0 }^{n}\ket{0}_{\log n}^{\otimes n}\ket{e_i}_n 
\end{align*}
Step (1) Apply Hadamard gates to the first index register, changing to the Hadamard basis, in which the $NOT$-gate is diagonal;
Step~(2) Use fanout gates to copy the first index register;
Step (3) Apply controlled-$Z$-gates, controlled by system qubit $i$ and with targets those index qubits in the $i$-th index register that correspond to the ones in the binary representation of $i$;
Step (4) Use fanout gates to disentangle the index registers; 
and, Step (5) Apply Hadamard gates to clean the index register.

The controlled-$Z$-gates depend on the binary representation of $i$ and precisely cancel the phase factors $(-1)^{i\cdot j}$ and disentangle the index registers from the system register.
\end{proof}

\begin{figure}[ht]
\includegraphics[width=\textwidth, trim={1.8cm 5.8cm 1.8cm 5.8cm}, clip]{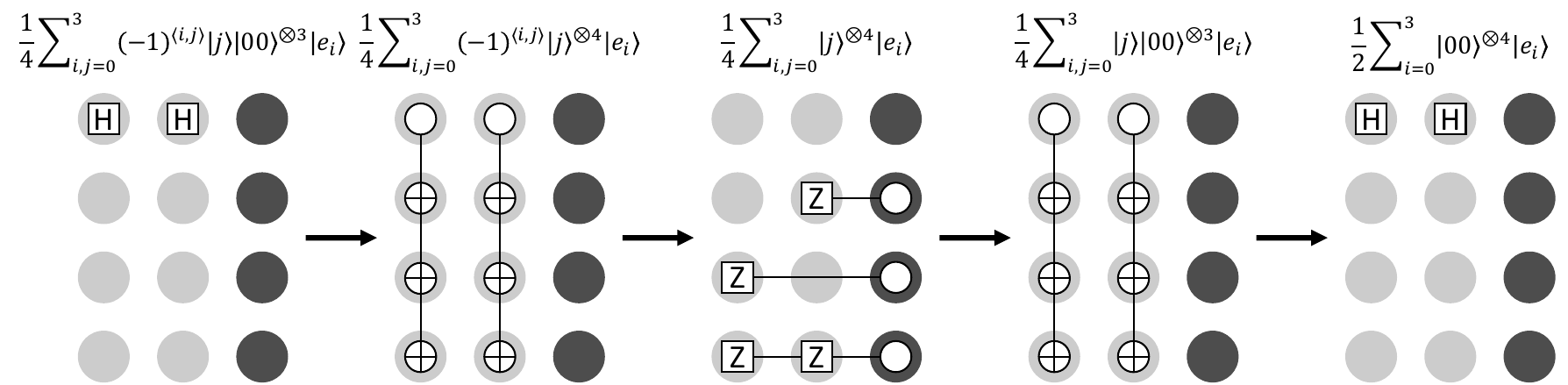}
\caption{Circuit for the \textbf{Compress} operation for $n=4$. 
Shown is a grid of $12$ qubits: $8$ light gray index qubits, and $4$ dark gray system qubits. 
Each grid represents a single time step of the \textbf{Compress} operation.}
\label{fig:W_state_compress}
\end{figure}

\cref{fig:W_state_compress} shows the steps of the \textbf{Compress} operation graphically for $n=4$.
We again omit the auxiliary qubits from the shown circuits. 

We now combine these two lemmas to prove the main result of this section. 
\begin{proof}[ of \cref{thm:W_state}]
The $\LAQCC$-circuit consists of the following three steps:
\begin{align*}
\ket{0}^{\otimes n}_{\log n}\ket{0}_n & \to \frac{1}{\sqrt{n}}\sum_{i = 0}^{n-1}\ket{i}\ket{0}^{\otimes n-1}\ket{0} && (\text{\cref{thm:uniform_superposition_mod_q}}) \\
        & \to \frac{1}{\sqrt{n}}\sum_{i = 0}^{n-1}\ket{i}\ket{0}^{\otimes n-1}\ket{e_i} && (\text{\cref{lem:W_state_uncompress}}) \\
        & \to \frac{1}{\sqrt{n}}\sum_{i =0 }^{n}\ket{0}^{\otimes n}\ket{e_i} && (\text{\cref{lem:W_state_compress}}) 
\end{align*}
%If $n=2^k$ for some integer $k$, the first step simplifies to a single layer of Hadamard gates. 
\end{proof}

%% file: LAQCC/state_preparation/Dicke_small_k.tex
\section{Dicke states for small \texorpdfstring{$k$}{k}}\label{sec:LAQCC:Dicke_small_k}
This section generalizes \cref{thm:W_state} to Dicke-$(n,k)$ states, uniform superpositions over $n$-bit strings of Hamming weight $k$:
\begin{align}
    \ket{D_k^n} = \binom{n}{k}^{-1/2}\sum_{x \in \{0,1\}^n: |x| = k} \ket{x}.
\end{align}
Other works also considered efficient state preparation routines for Dicke states. 
\citeauthor{BartschiEidenbenz:2019} for instance presented a circuit of linear depth and width in $n$ that prepares the Dicke-$(n,k)$ state, independent of $k$~\cite{BartschiEidenbenz:2019}. 
Their method uses a recursive relation for the Dicke-state:
\begin{align*}
    \ket{D_k^n} = \alpha_{k,n} \ket{D_k^{n-1}}\otimes \ket{0} + \beta_{k,n} \ket{D_{k-1}^{n-1}}\otimes \ket{1}.
\end{align*}
Our method, the main result of this section, instead extends the compress-uncompress method used for the $W$-state to small $k$:
\begin{theorem}\label{thm:Dicke:small_k}
For any $n$ and $k = \bigo(\sqrt{n})$, there exists a $\LAQCC$-circuit preparing the Dicke-$(n,k)$ state, $\ket{D^n_k}$, using $\bigo(n^3)$ qubits.
\end{theorem}
The main idea is to apply $k$ parallel \textbf{Uncompress} operations.
The parallel \textbf{Uncompress} operations might target the same system register qubit.
We thus have to filter these cases. 
Following the lines of the birthday paradox, we find that coinciding indices occur rarely for $k = \bigo(\sqrt{n})$.
\cref{lem:grover_constant_fraction} allows boosting the amplitude of the correct states to obtain the Dicke state. 

\begin{figure}[ht]
\includegraphics[width=\textwidth, trim={1.9cm 3.4cm 1.9cm 3.4cm}, clip]{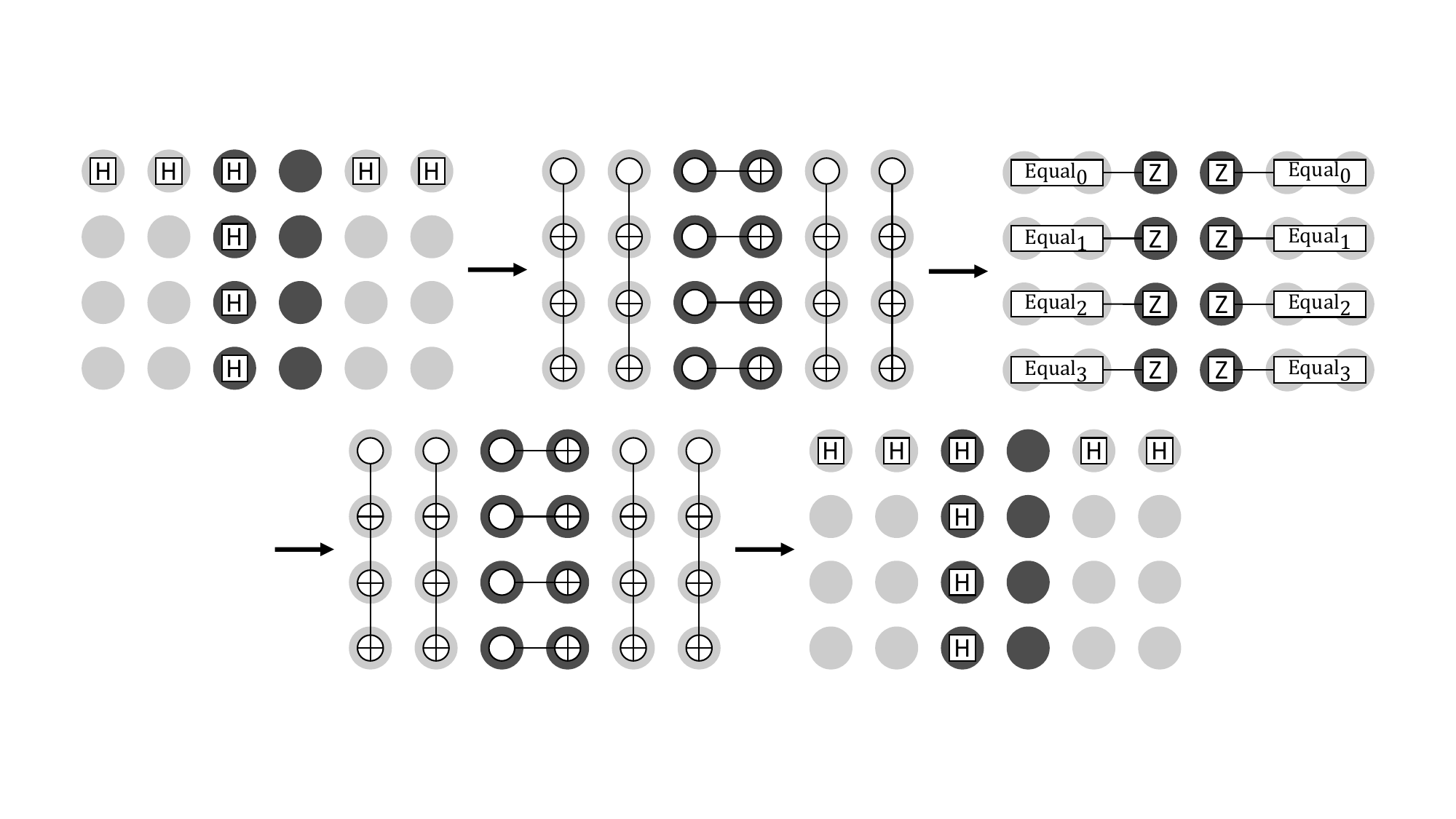}
\caption{Circuit for the \textbf{Filling} operation for $n=4$ and $k=2$.
Shown is a grid of $24$ qubits: $16$ light gray index qubits and $8$ dark gray system qubits.
Each grid represents a single step.}
\label{fig:Dicke:filling_example}
\end{figure}

We again assume the qubits to be laid out in a grid, the main difference being that we now have $k$ index registers denoted by subscripts $i\in[k]$, each consisting of $\log n$ qubits.
The first grid in \cref{fig:Dicke:filling_example} gives an example of the initial layout of the grid for $n=4$ and $k=2$. 

The algorithm consists of four operations:
\begin{align}
    & \ket{0}_{1}\dots\ket{0}_{k}\ket{0} \nonumber \\
    & \quad \to \frac{1}{n^{k/2}}\sum_{j_1, \dots, j_k = 0}^{n-1} \ket{j_1}_{1} \dots \ket{j_k}_{k} \ket{e_{j_1} \oplus \dots \oplus e_{j_k} } && (\textbf{Filling}) \label{eq:Dicke_filling} \\
    & \quad \to \sqrt{\frac{(n-k)!}{n!}}\sum_{j_1 \neq \dots \neq j_k} \ket{j_1}_{1} \dots \ket{j_k}_{k} \ket{e_{j_1} \oplus \dots \oplus e_{j_k}} && (\textbf{Filtering}) \label{eq:Dicke_filtering} \\
    & \quad \to \frac{1}{\sqrt{\binom{n}{k}}}\sum_{j_1 < \dots < j_k} \ket{j_1}_{1} \dots \ket{j_k}_{k} \ket{e_{j_1} \oplus \dots \oplus e_{j_k}} && (\textbf{Ordering}) \label{eq:Dicke_ordering} \\
    & \quad \to \frac{1}{\sqrt{\binom{n}{k}}}\sum_{j_1 < \dots < j_k} \ket{0}_{1} \dots \ket{0}_{k} \ket{e_{j_1} \oplus \dots \oplus e_{j_k}} && (\textbf{Cleaning}) \label{eq:Dicke_cleaning} 
\end{align}
The following four lemmas will give a $\LAQCC$-circuit for each of these operations. 
First, the \textbf{Filling} operation shown in \cref{eq:Dicke_filling} fills the system register. 
\begin{lemma}\label[lemma]{lem:Dicke:filling}
There exists a $\LAQCC$-circuit for the \textbf{Filling} operation using $\bigo(k n\log (n)\log\log(n))$ qubits.
\end{lemma}
\begin{proof}
The \textbf{Filling} operation uses $k$ parallel instances of the \textbf{Uncompress} operation from \cref{lem:W_state_uncompress}.
These $k$ operations commute and hence have a parallel implementation by \cref{thm:parallel_unitaries}.
This parallelization requires $k$ parallel system registers.
In total, the \textbf{Filling} operation consists of five steps: 
\begin{align*}
& \ket{0}_{1} \dots \ket{0}_{k} \ket{0}^{\otimes k} \\
& \quad \xrightarrow{(1)} \frac{1}{n^{k/2}} \sum_{j_1,\dots, j_k = 0}^{n-1} \ket{j_1}_{1} \dots \ket{j_k}_{k} \frac{1}{\sqrt{2^n}} \sum_{l=0}^{2^n - 1} \ket{l}\ket{0}^{\otimes k-1} \\
& \quad \xrightarrow{(2)} \frac{1}{n^{k/2}} \sum_{j_1,\dots, j_k = 0}^{n-1} \ket{j_1}_{1} \dots \ket{j_k}_{k} \frac{1}{\sqrt{2^n}} \sum_{l=0}^{2^n - 1} \ket{l}^{\otimes k} \\
& \quad \xrightarrow{(3)} \frac{1}{n^{k/2}} \sum_{j_1,\dots, j_k = 0}^{n-1} \ket{j_1}_{1} \dots \ket{j_k}_{k} \frac{1}{\sqrt{2^n}} \sum_{l=0}^{2^n - 1}(-1)^{(2^{j_1} + \dots + 2^{j_k}) \cdot l} \ket{l}^{\otimes k} \\
& \quad \xrightarrow{(4)} \frac{1}{n^{k/2}} \sum_{j_1,\dot,s j_k = 0}^{n-1} \ket{j_1}_{1} \dots \ket{j_k}_{k} \frac{1}{\sqrt{2^n}} \sum_{l=0}^{2^n - 1}(-1)^{(2^{j_1} + \dots + 2^{j_k}) \cdot l} \ket{l} \ket{0}^{\otimes k-1} \\ 
& \quad \xrightarrow{(5)} \frac{1}{n^{k/2}} \sum_{j_1,\dots, j_k = 0}^{n-1} \ket{j_1}_{1} \dots \ket{j_k}_{k} \ket{e_{j_1} \oplus \dots \oplus e_{j_k}} \ket{0}^{\otimes k-1}
\end{align*}
Step (1) Use \cref{thm:uniform_superposition_mod_q} to create uniform superpositions of size $n$ in all index registers and of size $2^n$ in the system register;
Step (2) Copy the registers using fanout gates;
Step (3) Apply phase-versions of Equal$_{j_i}$-gates for all $j_i$. 
These phase-versions apply a $Z$-gate instead of an $X$-gate to the target qubit;
Step~(4) Use fanout gates to clean the copies of the system register; 
Step (5) Apply Hadamard gates on the system register. 
This gives the sum of the one-hot representations of the index registers in the system register. 

We use $kn$ Equal$_i$-gates, each using $\bigo(\log(n)\log\log(n))$ qubits.
This gives a total of $\bigo(kn \log(n)\log\log(n))$ qubits required to implement the \textbf{Filling} operation.
\end{proof}

\cref{fig:Dicke:filling_example} shows these five steps graphically. 
We omitted the auxiliary qubits and included only the auxiliary system register for clarity.

The \textbf{Filling} operation can have multiple index registers with the same value, resulting in a state in the system register with Hamming weight less than $k$.
The next \textbf{Filtering} operation shown in \cref{eq:Dicke_filtering} removes these collisions, ensuring that every state in the system register has Hamming weight~$k$. 
\begin{lemma}\label[lemma]{lem:Dicke:filtering}
There exists a $\LAQCC$-circuit that implements the \textbf{Filtering} operation using $\bigo(n\log n)$ qubits.
\end{lemma}
\begin{proof}
The final state produced by the \textbf{Filling} operation splits in two substates, based on whether all indices $j_i$ are different. 
Let 
\begin{equation*}
    \ket{\psi} = \sum_{j_1 \neq \dots \neq j_k } \ket{j_1}_{1}\dots \ket{j_k}_{k} \ket{e_{j_1} \oplus \dots \oplus e_{j_k}},
\end{equation*}
then we can write the output of the \textbf{Filling} operation as
\begin{equation*}
    \frac{1}{n^{k/2}}\sum_{j_1,\dots, j_k = 0}^{n-1}\ket{j_1}_{1}\dots \ket{j_k}_{k} \ket{e_{j_1} \oplus \dots \oplus e_{j_k}} = \alpha \ket{\psi} + \beta \ket{\psi^{\perp}},
\end{equation*}
where $\ket{\psi^{\perp}}$ contains all terms with overlapping indices $j_i$. 
As a result, the states in $\ket{\psi^{\perp}}$ have Hamming weight strictly smaller than $k$.
Of all $n^k$ possible values of the index registers, only $n!/(n-k)!$ have all index register values different. 

By the birthday paradox we have that for any integers $n$ and $k < \frac{n}{2}$ it holds that
\begin{equation}
    \frac{n!}{n^k (n-k)!} > e^{-\frac{2 k^2}{n}}. \label{eq:birtday_paradox}
\end{equation}
In fact, we have the simple computation
\begin{align*}
    \frac{n!}{n^k (n-k)!} & = e^{\sum_{i = 1}^k \log(1 - \frac{i}{n})}\\
    & > e^{\sum_{i = 1}^k \frac{-i}{n-i}} \\
    & > e^{-\sum_{i = 1}^k \frac{i}{n-k}} \\
    & > e^{-\frac{k^2}{n-k}} \\
    & > e^{-\frac{2 k^2}{n}},
\end{align*}
where in the first inequality we use that $\log(1 + x)\geq \frac{x}{1+x}$ for $x > - 1$.
As a result, we obtain the lower bound $|\alpha|^2 > \exp(\frac{-2k^2}{n})$, which is constant for $k=\bigo(\sqrt{n})$.

Now note that a unitary $U_{flag}$ that flag $\ket{\psi}$ is implemented using an Exact$_t$-gate. 
The output of this gate is $1$, precisely if $t$ of the inputs are $1$, which happens precisely if every index register has a different value. 
We thus have the map 
\begin{align*}
& \frac{1}{n^{k/2}}\sum_{j_1,\dots, j_k = 0}^{n-1}\ket{j_1}_{1}\dots \ket{j_k}_{k} \ket{e_{j_1} \oplus \dots \oplus e_{j_k}} \ket{0} \\
& \quad \to \frac{1}{n^{k/2}} \sum_{j_1,\dots, j_k = 0}^{n-1}\ket{j_1}_{1}\dots \ket{j_k}_{k} \ket{e_{j_1} \oplus \dots \oplus e_{j_k}} \ket{\one_{|e_{j_1} \oplus \dots \oplus e_{j_k}|=k}} \\
& \quad = \alpha\ket{\psi}\ket{1} + \beta \ket{\psi^{\perp}}\ket{0}.
\end{align*}
Applying \cref{lem:grover_constant_fraction} with unitary $U_{flag}$ allows us to amplify $\alpha$ to $1$ in constant depth and using $\bigo(n\log n)$ qubits, to give the state
\begin{equation*}
    \sqrt{\frac{(n-k)!}{(n)!}}\sum_{j_1 \neq \dots \neq j_k} \ket{j_1}_{1} \dots \ket{j_k}_{k} \ket{e_{j_1} \oplus \dots \oplus e_{j_k}}.
\end{equation*}
\end{proof}
All states in the system register now have the correct Hamming weight. 
However, it is remains unclear which index register corresponds to which $1$ in the system register, as a permutation of the index registers gives the same state in the system register. 
The \textbf{Ordering} operation (\cref{eq:Dicke_ordering}) resolves this redundancy by imposing an order on the index registers. 
\begin{lemma}\label[lemma]{lem:Dicke:ordering}
There exists a $\LAQCC$-circuit that implements the \textbf{Ordering} operation using $\bigo(k^2 \log^2 n)$ qubits.
\end{lemma}
\begin{proof}
Start with $k-1$ auxiliary qubits per index register. 
For index register~$i$, use the first $j<i$ auxiliary qubits to store the outcome of a Greaterthan-gate evaluated on the $j$-th and $i$-th index registers. 
For the auxiliary qubits $j\ge i$, store the outcome of a Greaterthan-gate evaluated on the $j+1$-th and $i$-th index registers. 
Every Greaterthan-gate uses $\bigo(\log^2 n)$ qubits, and a total of $\bigo(k^2)$ Greaterthan-gates are evaluated. 
We use fanout gates on the index registers to parallelize the Greaterthan-gates.
Omitting a scaling factor and the system register, these parallel Greaterthan-gates implement the map
\begin{align*}
    & \sum_{j_1 \neq \dots \neq j_k} \ket{j_1}_{1}^{\otimes k-1} \ket{0}^{\otimes k-1} \dots \ket{j_k}_{k}^{\otimes k-1} \ket{0}^{\otimes k-1}\\
    & \quad \to \sum_{j_1 \neq \dots \neq j_k} 
    \big[\ket{j_1}_{1}^{\otimes k} \ket{\one_{j_1 > j_2}} \dots \ket{\one_{j_1 > j_k}}\big]
    \dots \big[ \ket{j_k}_{k}^{\otimes k} \ket{\one_{j_k > j_1}} \dots \ket{\one_{j_k > j_{k-1}}}\big].
\end{align*}
Each $\one_{j_k > j_{k'}}$ acts as an indicator variable for the event $\{j_k> j_{k'}\}$.

Next, measure the auxiliary qubits and add all measurement outcomes to determine the Hamming weights. 
These Hamming weights impose an ordering on the index registers. 
Assume without loss of generality that the imposed ordering is $j_1<\hdots <j_k$, otherwise a $\LAQCC$-circuit exists to SWAP the index registers accordingly. 
Adding the measurement results is in $\ac^0$ and sorting the measurement results is in $\tc^0$~\cite{Siu:1993}. 
As both classes are subsets of $\nc^1$, the conventional controller can impose an ordering and then determine and apply the permutation to obtain the desired state
\begin{equation*}
    \binom{n}{k}^{-1/2} \sum_{j_1 < \dots < j_k} \ket{j_1}_{1} \dots \ket{j_k}_{k}\ket{e_{j_1} \oplus \dots \oplus e_{j_k}}.
\end{equation*}
\end{proof}
Note that we can also compute the Hammingweight-gate and the permutation in superposition. 
As quantum operations are in general expensive, it is better to offload these comptuations to the conventional controller. 

Similar to the \textbf{Compress} operation in the $W$-state protocol, we now have to clean the index registers. 
The \textbf{Cleaning} operation, shown in \cref{eq:Dicke_cleaning}, cleans the index registers, taking into account the added ordering of the index registers. 
Suppose the $m$-th qubit of the system register is a $1$.
If this is the $t$-th one in the string, then the $t$-th index register has value $m$. 
Computing the Hamming weight of the first $t-1$ qubits gives precisely the required information to know which index register to uncompute. 
\begin{lemma}\label[lemma]{lem:Dicke:cleaning}
There exists a $\LAQCC$-circuit that implements the \textbf{Cleaning} operation using $\bigo(n^3)$ qubits.
\end{lemma}
\begin{proof}
Compute in parallel, the Hamming weight of the first $i$ system register qubits, for $i\in[n-1]$. 
Provided that the $j$-th system register qubit is in the $\ket{1}$-state, the Hamming weight of the first $j-1$ qubits determines which of the index registers has to be uncomputed. 
Compute the Hammingweight-gate requires $\bigo(n^2)$ qubits. 
We store the Hamming weight in an auxiliary register of size $\log k$. 

Using these auxiliary registers, we can then clean the $k$ index registers, similar to the \textbf{Compress} method of \cref{lem:W_state_compress}, combined with the information stored in the auxiliary registers. 
Cleaning index register $j$ consist of fives steps and requires a copy of the system register and a copy of each of the $n$ auxiliary Hammingweight-registers. 
\begin{enumerate}
    \item Apply Hadamard gates to bring the index register to phase space;
    \item Apply fanout gates to copy the index register;
    \item Apply $n$ parallel $Z$-gates on the index register to clean it. 
    These $Z$-gates are controlled by one system register qubit and by the Hammingweight-register, such that there have been precisely $j-1$ ones in the system register already;
    \item Apply fanout gates to clean the index register copies;
    \item Apply Hadamard gates to reset the index register qubits to the $\ket{0}$-state.
\end{enumerate}

The cost of the Hammingweight-computation dominates the requirements on the number of qubits, giving the $\bigo(n^3)$ qubit requirement for the \textbf{Cleaning} step.
\end{proof}

\begin{figure}[ht]
\includegraphics[width=\textwidth, trim={1.6cm 6.7cm 1.6cm 6.7cm}, clip]{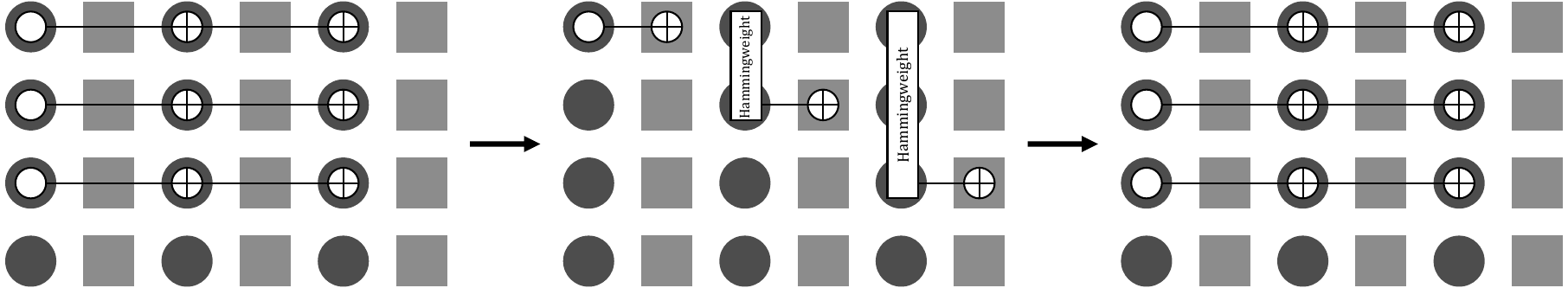}
\caption{Circuit to compute the Hamming weight of the first qubits. 
The dark gray dots represent system qubits, and the gray squares denote auxiliary registers of $\log k$ qubits each.}
\label{fig:Hammingweight}
\end{figure}

\begin{figure}[ht]
\includegraphics[width=\textwidth, trim={0cm 6.7cm 0cm 6.7cm}, clip]{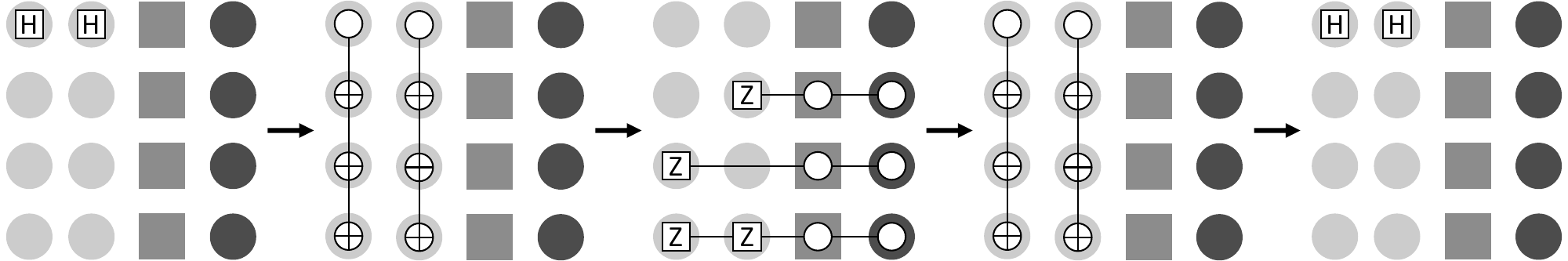}
\caption{Circuit to clean index register $j$. 
The black dots represent qubits in the system register and the light gray dots the index register and its copies.
The gray squares denote auxiliary registers of $\log k$ qubits each. 
Each grid represents a single time step.}
\label{fig:compress_dicke}
\end{figure}

\cref{fig:Hammingweight} shows the circuit to compute the Hamming weight of the system register qubits for $n=4$. 
Note that we need only three computations, as system qubit~$i$ uses the Hamming weight of the first $i-1$ system qubits. 
We omitted the index registers, and instead show the auxiliary registers that hold the Hamming-weight values. 
\cref{fig:compress_dicke} shows the steps taken to clean the $j$-th index register. 
The \textbf{Cleaning} operation uses the Hamming weight information. 

The proof of the theorem now follows by combining these four lemmas: 
\begin{proof}[ of \cref{thm:Dicke:small_k}]
The circuit that prepares the Dicke-state for $k=\bigo(\sqrt{n})$ follows directly from the circuits presented in \cref{lem:Dicke:filling,lem:Dicke:filtering,lem:Dicke:ordering,lem:Dicke:cleaning}.
These circuits use at most $\bigo(n^3)$ qubits completing the proof.
\end{proof}
\citeauthor{BartschiEidenbenz:2022} posed a conjecture on the optimal depth of quantum circuits that prepare the Dicke-$(n,k)$ state. 
They presented an algorithm for generating Dicke-$(n,k)$ states in depth $\bigo(k \log\frac{n}{k})$, given all-to-all connectivity, and conjectured that this scaling is optimal for constant $k$.
Our result shows that there is a $\LAQCC$-circuit in this regime, but does not disprove their conjecture.
However, as the circuits shown here are also accessible in $\qnc^1$ by \cref{lem:LAQCC_QNC1}, there also exists a quantum circuit of depth $\bigo(\log n)$ for $k = \nolinebreak \bigo(\sqrt{n})$ that achieves better scaling for non-constant $k$, i.e., when $k = \omega(1)$.

%% file: LAQCC/state_preparation/Dicke_log_depth.tex
\section{Dicke states for all \texorpdfstring{$k$}{k}}\label{sec:LAQCC:Dicke_in_LAQCC_LOG}
The previous section presented a $\LAQCC$-circuit that prepares the Dicke-$(n,k)$ state for $k=\bigo(\sqrt{n})$. 
That method does not scale to larger $k$, because the birthday-paradox argument fails for larger $k$. 
This section shows how to prepare Dicke-$(n,k)$ states for all $k$ using a logarithmic number of alternations. 
We let $\LAQCC$-$\mathsf{LOG}$ refer to the instance $\LAQCC(\qnc^0,\nc^1, \bigo(\log n))$. 

One way of studying the creation of Dicke states is by looking at efficient algorithms that convert numbers from one representation to another. 
An example is the \textbf{Uncompress}-\textbf{Compress} method in the $W$-state protocol that converts numbers from a binary representation to a one-hot representation. 
Dicke states generalize the $W$-state, hence the one-hot representation no longer suffices for preparing the state. 
Instead, we use a construction based on number conversion between the combinatorial and the factoradic representation.
The main theorem of this section uses an efficient conversion between these two representations. 
\begin{theorem}\label{thm:Dicke:Log_depth}
For $k\le n$ positive integers, there exists a $\LAQCC$-$\mathsf{LOG}$-circuit for preparing the Dicke-$(n,k)$ state using $\bigo(\poly(n))$ qubits. 
\end{theorem}
The next sections introduce the two representations, as well as quantum circuits that convert one representation into the other. 

\subsection{Combinatorial number system}
\citeauthor{Beckenbach:1964} showed that any integer $m\ge 0$ can be written as a sum of $k$ binomial coefficients~\cite{Beckenbach:1964}. 
This decomposition is even unique for fixed $k$, as the next lemma shows:
\begin{lemma}[\cite{Beckenbach:1964}]\label[lemma]{lem:combinatorial_numbers}
For all integers $m\geq 0$ and $k \geq 1$, there exists a unique decreasing sequence of integers $c_k, c_{k-1},\dots, c_1$ with $c_j > c_{j-1}$ and $c_1 \geq 0$ such that 
\begin{equation*}
    m = \binom{c_k}{k}  + \binom{c_{k-1}}{k-1} \dots \binom{c_1}{1} = \sum_{i=1}^k \binom{c_i}{i}.
\end{equation*}
\end{lemma}

From this lemma, we naturally arrive at a definition for the combinatorial number representation: 
\begin{definition}
Let $k \in \mathbb{N}$ be a constant. 
Any integer $m \in \mathbb{N}$ can be represent by a unique string of integers $(c_k, c_{k-1} \dots, c_1)$, according to \cref{lem:combinatorial_numbers}. 
We call this string the \emph{index representation} of $m$ and denote it by $m^{indx(k)}$.

The bit string containing $k$ ones at indices $(c_k,\dots, c_1)$ is the $m$-th bit string with~$k$ ones in the lexicographical order. 
This bit string is called the \emph{combinatatorial representation}.
We denote the $m$-th bit string with $k$ ones as $m^{comb(k)}$.
\end{definition}
The $W$-state protocol used the conversion between the binary representation of a number $m$ and its combinatorial representation $m^{comb(1)}$.

Preparing the Dicke-$(n,k)$ state requires the following steps.
Given positive integers $k$ and $n$, prepare the superposition
\begin{equation*}
    {\binom{n}{k}}^{-\frac{1}{2}}\sum_{i=0}^{\binom{n}{k} - 1}\ket{i}\ket{0};
\end{equation*}
Map between the representation $i$ and $i^{comb(k)}$ to obtain the state
\begin{equation*}
    {\binom{n}{k}}^{-\frac{1}{2}}\sum_{i=0}^{\binom{n}{k} - 1}\ket{i}\ket{i^{comb(k)}};
\end{equation*}
Map between the representation $i^{comb(k)}$ and $i$ to clean the label register
\begin{equation*}
    {\binom{n}{k}}^{-\frac{1}{2}}\sum_{i=0}^{\binom{n}{k} - 1}\ket{0}\ket{i^{comb(k)}} = \ket{D^n_k}.
\end{equation*}
The second mapping uses \cref{lem:combinatorial_numbers}. 
This calculation requires iterative multiplication and addition, both of which are in $\tc^0$~\cite{Vollmer:1999}, hence this calculation is in $\tc^0$. 

The first mapping, from binary to combinatorial representation for given $k$, has a simple greedy iterative algorithm:
On input $m$, find the biggest $c_k$ such that $m \geq \binom{c_k}{k}$ and subtract this from $m$: $\tilde{m} = m - \binom{c_k}{k}$. 
This gives $c_k$ and a residual~$\tilde{m}$. 
Repeat this process for $\tilde{m}$: 
Find the largest $c_{j}$ such that $\tilde{m}\geq \binom{c_j}{j}$ and update residual $\tilde{m} = \tilde{m} - \binom{c_j}{j}$, until all $c_j$ are found. 

This greedy algorithm is inherently linear in $k$ as it requires all previously found $\{c_i\}_{i=j}^k$ to find $c_{j-1}$. 
Hence, it is not immediately obvious if and how to achieve this mapping in constant or even logarithmic depth. 

\subsection{Factoradic representation}
A number representation closely related to the combinatorial number representation is the \textit{factoradic representation}. 
This number system uses factorials instead of binomials to represent integers. 
\begin{definition} \label{def:factoradics}
A sequence $y = (y_{n-1}, y_{n-2}, \dots, y_0)$ of integers, such that $j \geq y_j \geq 0$ is called an \textit{$n$-factoradic} (or simply a factoradic). 
The elements of an $n$-factoradic are called \textit{$n$-digits}.
An $n$-factoradic $y$ can represent a number $m$ between $0$ and $n!-1$, in the following way
\begin{equation}\label{eq:factoradic_to_integer}
    m = \sum_{j=0}^{n-1} y_j \cdot j!.
\end{equation}
We call the $n$-factoradic $y$ the \emph{factoradic representation} of an integer $m\le n!-1$.
Denote $\text{Fact}(n)$ as the set of all $n$-factoradics.
\end{definition}

A counting argument combined with the following lemma shows that every integer $m\in\{0,\hdots,n!-1\}$ has a unique factoradic representation given by \cref{eq:factoradic_to_integer}. 
\begin{lemma}\label[lemma]{factoradic_summation}
For $k \geq 0$ it holds that:
$$\sum_{i=0}^k i \cdot i!= (k + 1)! - 1.$$ 
\end{lemma}
\begin{proof}
We prove the lemma by induction. 
The base case $k=0$ follows naturally, as $0\cdot 0! = 1!-1$. 

Now assume the lemma holds for some integer $j$, then
\begin{equation*}
    \sum_{i=0}^{j+1} i \cdot i! = (j+1) \cdot (j+1)! + \sum_{i = 0}^{j} i \cdot i = (j+1) \cdot (j+1)! + (j+1)! - 1 = (j+2)! - 1,
\end{equation*}
which completes the proof.
\end{proof}
As the factoradic representation represents a unique integer, we can use it as a number system. 
The next section shows how to convert between the factoradic representation and the combinatorial number representation.

\subsection{Mapping between representations}
The next lemma gives a logspace algorithm to convert a factoradic representation to its combinatorial representation. 
\begin{lemma}\label[lemma]{lem:factoradic_to_combinatorial}
There is a logspace algorithm $\mathcal{A}$ that, given $k\in\{0,\hdots,n\}$, and a uniformly random $n$-factoradic, outputs a uniformly random $n$-bit string of Hamming weight $k$.
\end{lemma}
\begin{proof}
Let $y = (y_{n-1}, \dots, y_0)$ be an $n$-factoradic.
The logspace algorithm $\mathcal{A}$ will then output an $n$-bit string $y^{comb(k)} = y^{comb(k)}_{n-1} \dots y^{comb(k)}_0 \in \{0,1\}^n$ of Hamming weight $k$, one bit at a time, from left to right, according to the following rule:
Let $H_{>n-j} = \sum_{i=n-j+1}^{n-1} y^{comb(k)}_i$ be the Hamming weight of the bits produced before we reach bit $n-j$. 
Then $y^{comb(k)}_{n-j}$ is given by:
\begin{align}
\label{eqn:fac_to_comb}
    (\mathcal{A} (y))_{n-j} = y^{comb(k)}_{n-j} = \begin{cases} 
    1 & \text{if } y_{n-j} < k - H_{>n-j}, \\
    0 & \text{otherwise}.
    \end{cases}
\end{align}
This conversion requires comparing an $n$-digit with a constant and the Hamming weight of a bit string.
The only information that $\mathcal{A}$ needs to remember, as it goes from bit $n-j+1$ to bit $n-j$, is the Hamming weight $H_{>n-j}$ of the bits it produced so far, which requires logarithmic storage space. 

Note that the number of factoradic $n$-digit strings that map to the same combinatorial bit string is always $k!(n-k)!$:
Let $y^{comb(k)} \in \{0,1\}^n$ have Hamming weight $k$. 
For any bit position $y^{comb(k)}_{n-j}$, there are $n - j + 1 - (k - H_{>n-j})$ possible choices for the $n$-digit $y_{n-j}$ such that $y^{comb(k)}_{n-j} = 0$. 
For the leftmost index $n-j$ such that $y^{comb(k)}_{n-j} = 0$, it holds that $H_{>n-j} = j-1$. 
There then are $n-k$ possible $n$-digits $y_{n-j}$ such that $y^{comb(k)}_{n-j} = 0$. 
Then, for the second leftmost index $n-j$ such that $y^{comb(k)}_{n-j} = 0$ it holds that $H_{>n-j} = j - 2$, hence there are $n - k - 1$ possible $n$-digits $y_{n-j}$ such that $y^{comb(k)}_{n-j} = 0$. 
This argument repeats for all $k$ indices.
This results in $(n-k)!$ different possible choices for the $(n-k)$-many $n$-digits such that $y^{comb(k)}=0$. 

Similarly, for the leftmost position $n-j$ where $y^{comb(k)}_{n-j} = 1$, there are $k$ possible choices for the $n$-digit $y_{n-j}$ that given $y^{comb(k)}_{n-j} = 1$. 
The second leftmost position $n-j$ gives $k-1$ possible choices, and so forth, for a total of $k!$ possible choices of the $k$-many $n$-digits where $y^{comb(k)}=1$.

We conclude that, for every $n$-bit string $y^{comb(k)} \in \{0,1\}^n$ of Hamming weight $k$, there are exactly $k!(n-k)!$ $n$-factoradics $y$ such that $\mathcal{A}(y) = y^{comb(k)}$.
\end{proof}

This lemma gave a logspace algorithm to convert a uniformly random $n$-factoradic to a uniformly random $n$-bit string of Hamming weight $k$, for any $k$.
As logspace is contained in $\tc^1$~\cite{Johnson:1990}, the algorithm $\mathcal{A}$ has a parallel log-depth implementation, provided one has access to threshold gates.
As $\LAQCC$-$\mathsf{LOG}$ contains threshold gates, any $\tc^1$-circuit has an equivalent $\LAQCC$-$\mathsf{LOG}$-circuit:
\begin{corollary}\label[corollary]{cor:factoradic_to_combinatorial}
The following map can be implemented in $\LAQCC$-$\mathsf{LOG}$.
\begin{equation*}
\frac{1}{\sqrt{n!}} \sum_{y \in \text{Fact}(n)} \ket{y}\ket{0} \xrightarrow{} \frac{1}{\sqrt{n!}} \sum_{y \in \text{Fact}(n)} \ket{y}\ket{\mathcal{A} (y)}.
\end{equation*}
\end{corollary}
 
The next lemma gives a $\tc^0$-circuit that implements the inverse of $\mathcal{A}$. 
\begin{lemma}\label[lemma]{lem:combinatorial_to_factoradic}
Given as input an $n$-bit string $y^{comb(k)}$ of Hamming weight $k$, a uniformly-random $k$-factoradic, and a uniformly-random $(n-k)$-factoradic. 
There exists a $\tc^0$-circuit that on this input, when given this input, outputs a uniformly random $n$-factoradic $y$ such that $\mathcal{A}(y) = \nolinebreak y^{comb(k)}$.
\end{lemma}
\begin{proof}
The conversion can be done in parallel, generating an $n$-digit for every bit in $y^{comb(k)} = (y_{n-1} \dots y_0)\in\{0,1\}^n$. 
Recall that we are given as input a uniformly-random $k$-factoradic $X_{k-1}, \dots, X_0$ and a uniformly-random $(n-k)$-factoradic $Z_{n-k-1}, \dots, Z_0$.

For every bit position $n-j$, for $j\in[n]$, calculate the Hamming weight of the bits from $n-j+1$ to $n-1$: $H_{>n-j} = \sum_{i=j+1}^{n-1} y^{comb(k)}_i$. 
Recall that iterated addition is in $\tc^0$~\cite{Vollmer:1999}.

If $y^{comb(k)}_{n-j} = 1$, set $y_{n-j} = X_{k - H_{> n-j}}$.
This gives a uniform random $n$-digit between~$0$ and $k - H_{> n-j} - 1$. 
If $y^{comb(k)}_{n-j} = 0$, set $y_{n-j} = k - H_{> n-j} + Z_{n-k-H_{>n-j}}$. 
Note that this gives a uniform random $n$-digit between $k - H_{>n-j}$ and $n - j$. 
By construction, it now follows that $\mathcal{A}(y) = y^{comb(k)}$. 
Computing each $n$-digit in this way requires summation and indexing, both of which are in $\ac^0$.
\end{proof}

\begin{remark}\label[remark]{rem:combinatorial_to_factoradic}
The above algorithm establishes a bijection $(y^{comb(k)}, Z, X) \leftrightarrow \nolinebreak y$ between triples $(y^{comb(k)}, Z, X)$, where $y^{comb(k)} \in \{0,1\}^n$ has Hamming weight $k$, $Z \in \text{Fact}(n-k)$ and $X\in\text{Fact}(k)$, and an $n$-factoradic $y \in \text{Fact}(n)$. 
Let $(\mathcal{A}(y), \mathcal{Z}(y), \mathcal{X}(y))$ be the image of an $n$-factoradic $y$ under this bijection. 
The previous lemma shows that one can compute $y$ from $(y^{comb(k)}, Z, X)$ in $\tc^0$. 
Then the map $(\mathcal{A}(y), y) \mapsto (\mathcal{A}(y), y, \mathcal{Z}(y), \mathcal{X}(y))$ is also in $\tc^0$. 
Indeed, to find $\mathcal Z(y)$ and $\mathcal X(y)$, we need only invert the two defining equalities $y_{n-j} = X_{k - H_{> n-j}}$ and $y_{n-j} = k - H_{> n-j} + Z_{n-k-H_{>n-j}}$.
\end{remark}

\begin{corollary}\label[corollary]{cor:combinatorial_to_factoradic}
There exists a $\LAQCC$-circuit for the map
\begin{equation*}
    \binom{n}{k}^{-\frac{1}{2}}\sum_{y^{comb(k)}} \ket{0} \ket{y^{comb(k)}} \xrightarrow{} \frac{1}{\sqrt{n!}} \sum_{y \in \text{Fact}(n)} \ket{y}\ket{\mathcal{A} (y)},
\end{equation*}
where $y^{comb(k)}$ ranges over all $n$-bit strings of Hamming weight $k$.
\end{corollary}
\begin{proof}
The map consists of three steps: 
\begin{align*}
& \binom{n}{k}^{-\frac{1}{2}}\sum_{y^{comb(k)}} \ket{y^{comb(k)}} \ket{0}\ket{0}\ket{0}\\
\xrightarrow{(1)} \;\; &\binom{n}{k}^{-\frac{1}{2}}\sum_{y^{comb(k)}} \ket{y^{comb(k)}} \left(\bigotimes_{j = 0}^{n-k-1} \sum_{i = 0}^{j} \ket{i}\right)\left(\bigotimes_{j = 0}^{k-1} \sum_{i = 0}^{j} \ket{i}\right)\ket{0}\\
= \;\; & \frac{1}{\sqrt{n!}} \sum_{y^{comb(k)}} \ket{y^{comb(k)}} \left(\sum_{Z\in\text{Fact}(n-k)} \ket{Z}\right)\left(\sum_{X\in\text{Fact}(k)} \ket{X}\right)\ket{0}\\
\xrightarrow{(2)}\;\; & \frac{1}{\sqrt{n!}} \sum_{y\in\text{Fact}(n)} \ket{\mathcal{A}(y)} \ket{\hat Z(y)}\ket{\hat X(y)}\ket{y}\\
\xrightarrow{(3)}\;\; & \frac{1}{\sqrt{n!}} \sum_{y\in\text{Fact}(n)} \ket{\mathcal{A}(y)} \ket{0}\ket{0}\ket{y}
\end{align*}
The first step uses \cref{thm:uniform_superposition_mod_q} to prepare a uniform superposition over all $n$-factoradics.
The second step follows directly from \cref{lem:combinatorial_to_factoradic}.
The third step follows directly from \cref{rem:combinatorial_to_factoradic}.
The above steps implicitly use that the inverse of a $\LAQCC$-operation is also in $\LAQCC$.
Even though it is unclear if this inverse-property holds in general, it does hold for the used $\LAQCC$-operations. 
The only non-reversible operations used are measurements in the fanout gate construction. 
The fanout gate itself is reversible. 
\end{proof}

We now have all necessary tools to prove the main theorem of this section, a $\LAQCC$-$\mathsf{LOG}$-circuit that prepares the Dicke-$(n,k)$ state for arbitrary $k\le n$. 
\begin{proof}[ of \cref{thm:Dicke:Log_depth}]
The $\LAQCC$-$\mathsf{LOG}$-circuit combines the circuits resulting from \cref{lem:factoradic_to_combinatorial} and \cref{lem:combinatorial_to_factoradic} and consists of three steps:
\begin{align*}
\ket{0}^{\otimes n \log n}\ket{0}^{\otimes n} & \xrightarrow{(1)} \frac{1}{\sqrt{n!}}\left(\bigotimes_{j = 0}^{n-1} \sum_{i = 0}^{j} \ket{i}\right)\ket{0}^{\otimes n} = \sum_{y \in \text{Fact}(n)} \ket{y}\ket{0}^{\otimes n} \\
& \xrightarrow{(2)} \frac{1}{\sqrt{n!}} \sum_{y \in \text{Fact}(n)} \ket{y}\ket{\mathcal{A} (y)}\\
& \xrightarrow{(3)} \binom{n}{k}^{-\frac{1}{2}}\sum_{y \in \text{Fact}(n)} \ket{0}\ket{\mathcal{A} (y)} = \ket{D^n_k}.
\end{align*}
The first step uses \cref{thm:uniform_superposition_mod_q} to prepare a uniform superposition over all $n$-factoradics. 
The second step uses the circuit given in \cref{cor:factoradic_to_combinatorial}.
The third step uses the inverse of the circuit given in \cref{cor:combinatorial_to_factoradic}.
\end{proof}

%% file: LAQCC/state_preparation/discussion_LAQCC_state_preparation.tex
\section{Reflections and outlook}
This chapter introduced novel state-preparation protocols for three types on non-stabilizer states: 
The uniform superposition, the $W$-state and the Dicke state. 
For the uniform superposition, we used an exact version of Grover's search routine and $\LAQCC$-circuits given in the previous chapter. 

For the $W$-state, we used an uncompress-compress method, which efficiently maps between the binary representation and the one-hot representation of integers. 
We then extended this method to also work for Dicke-$(n,k)$ states for $k=\bigo(\sqrt{n})$. 
For arbitrary $k$, we used a mapping between two different representations, namely the combinatorial number representation and the factoradic representation. 
This second method does require a logarithmic number of alternations between the quantum and conventional circuits. 

Future research can focus on improving the circuits given in this chapter.
For the Dicke state for instance, one can try and find a constant-depth circuit that works for any $k$.
Alternatively, one might try to implement an uncompress-compress method similar to the one used for the $W$-state. 
The current approach is inspired by that method, but uses multiple intermediate steps for it. 

Another direction for future research is to focus on preparing other types of quantum states. 
These states can be close to the states we considered, such as many-body scar states~\cite[Section~4.5]{BuhrmanFolkertsmaLoffNeumann:2024} or weighted superpositions of different Dicke states~\cite{Bond:2023}, or states with certain characteristics, such as sparsity or some symmetry~\cite{Sun:2023,Ramacciotti:2024,LuoLi:2024,Mao:2024}.

Next, our protocols can improve in the circuit size. 
They are of constant depth, but require wide circuits to be implemented. 
Similarly, the constants in the protocols might be improvable. 

Finally, it is worthwhile to show that states in general can or cannot be prepared by certain $\LAQCC$-instances. 
Such a result could then indicate if our protocol for the Dicke state for arbitrary $k$ is optimal. 
Such a result can additionally imply separations between certain complexity classes.

%% file: LAQCC/LAQCC_error_analysis.tex
\chapter{Error analysis}\label{chp:LAQCC:error_analysis}
This chapter compares $\LAQCC$-circuits that prepare the GHZ state and the $W$-state with standard implementations. 
We determine the theoretical success probability for both and compare them. 

\input{LAQCC/error_analysis/outline_error_analysis}
\input{LAQCC/error_analysis/error_model}
\input{LAQCC/error_analysis/GHZ_error_analysis}

\input{LAQCC/error_analysis/GHZ_implementation_analysis}
\input{LAQCC/error_analysis/W_state_error_analysis}
\input{LAQCC/error_analysis/discussion_error}

%% file: LAQCC/error_analysis/outline_error_analysis.tex
\section{Chapter overview}
One of the main reasons to consider the class $\LAQCC$ was to offload computations to a conventional controller and assure that qubits are idle only briefly. 
As an example, when preparing a GHZ state on $n$ qubits, a standard approach using an all-to-all connectivity uses $n$ qubits, $n$ quantum gates, and has depth $\ceil{\log(n)}+1$, for a total circuit size of $\bigo(n\log n)$.
The $\LAQCC$-approach that prepares the GHZ state instead uses $2n-1$ qubits, around $4n$ quantum gates, and has constant depth, for a total circuit size of $\bigo(n)$. 

Based on the number of qubits required and the number of quantum gates applied, the standard approach seems favorable. 
On the other hand, based on the circuit size, the $\LAQCC$-circuit seems favorable. 
The $\LAQCC$-circuit is significantly more dense, meaning that qubits are idle only briefly. 

In this chapter, we explore which approach works best, based on the success probabilities of the quantum gates and the probabilities that qubits decohere while idling.
This chapter presents a first-order comparison between standard circuits and $\LAQCC$-circuits for preparing the GHZ state and the $W$-state. 

First, in \cref{sec:error:error_model}, we introduce the error models used in the remainder of this chapter.
Next, \cref{sec:error:GHZ} will provide the analysis for the GHZ state. 
For the GHZ state we consider a standard approach using an all-to-all connectivity and using a linear nearest-neighbor connectivity, and we consider a $\LAQCC$-approach. 
Additionally, we also derive expressions for the success probability of hybrid versions of a standard approach combined with a $\LAQCC$-approach. 
Next, we compare these success probabilities and, using some reductions, obtain a bound on when one protocol performs better than the other in terms of the success probabilities of the individual terms. 
\cref{thm:LAQCC:error:GHZ} summarizes the results on which protocol performs best when.

Next, \cref{sec:error:hardware_implementation} gives an implementation of both a standard approach and a $\LAQCC$-circuit on quantum hardware. 
We use these implementations to compare the performance of the two approaches with each other and with the outcomes expected by the theoretical analysis. 

Afterwards, in \cref{sec:error:W_state}, we perform the same theoretical analysis for the $W$-state as we did the the GHZ state. 
First, we derive an expression for the success probability of a standard approach and of a $\LAQCC$-approach. 
Next, we compare these approaches to determine under what circumstances, the $\LAQCC$-approach performs best. 
Due to the qubit requirement of the $\LAQCC$-approach, we have not implemented this circuit.

%% file: LAQCC/error_analysis/error_model.tex
\section{Error model}\label{sec:error:error_model}
In the next section, we analyze the protocols by comparing the success probability of these circuits in terms of the success probabilities of the individual terms in the circuits. 
Computing this overall success probability requires an error model that describes the behavior of the quantum circuit in case qubits decohere. 

We consider a worst-case error model where every error corresponds to an independent uniformly random gate being applied to the qubit(s)\footnote{Formally, the unitaries are taken uniformly at random with respect to the Haar measure.}.
Qubits can decohere both while idling and while a gate is applied to it. 
In both cases, a random unitary is applied to that qubit. 

In this error model, the probability that two errors cancel each other is $0$. 
Suppose an error $B$ occurred and suppose a gate $G$ is applied after which another error $D$ occurs. 
The probability that the two errors cancel corresponds to the probability of the event
\begin{equation*}
    D\cdot G\cdot B = G.
\end{equation*}
This means that we should have $D = G B^{\dagger} G^{\dagger}$.
As both sides of the equality correspond to independent Haar-random unitaries, the probability that they are equal is zero. 
Note that taking $G=I$, the identity gate, corresponds to the situation where no gate is applied and a qubit was instead idling. 

An error for measurement gates corresponds to a situation where the measurement outcome differs from the actually measured state. 
In the $\LAQCC$-circuits, the measurement outcomes are used to control future quantum gates.
Hence, if the incorrect control is used, the wrong gate can be applied. 
We thus `need' one specific error to cancel he measurement error and have the circuit still outputting the correct quantum state. 

In this error model, all errors are independent. 
The circuits we consider in this section have at most polynomial size. 
The set of errors that can possibly be canceled is therefore also polynomial in size. 
Yet, the group of unitary matrices from which the errors are sampled corresponds to a continuum of matrices.
That is, the group of unitary matrices is parametrized by three continuous parameters. 
Hence, the set of unitary matrices that can cancel a previous error has measure~$0$. 
As a result, in the subsequent analysis, we only have to derive expressions for the situation where all gates succeed and no errors occur at all. 

We determine the success probability of quantum circuits based on the success probability of the elementary operations used in the circuit. 
These elementary operations are single-qubit gates, two-qubit CNOT-gates, and measurements, as well as idling terms for other qubits during these operations. 
Additionally, we have the idling term during the conventional computations. 
\cref{tab:success_probabilities} summarizes these terms together with their meaning. 
\begin{table}[th]
    \centering
    \begin{tabular}{c|l}
        Success term & Probability that ... \\
        \hline
        $p_s$ & a single-qubit gate succeeds \\
        $p_d$ & a two-qubit gate succeeds \\
        $p_m$ & a measurement returns the correct value \\
        $p_c$ & the intermediate computation returns the correct value \\
        $p_{ix}$ & a qubit remains coherent, while idling during an $x$-operation
    \end{tabular}
    \caption{The considered success probabilities when analyzing state preparation protocols in the remainder of this chapter.}
    \label{tab:success_probabilities}
\end{table}

We explicitly assume that intermediate conventional computations always return the correct answer, hence $p_c=1$. 
Note that $p_d$ relates to two qubits, whereas $p_{id}$ concerns individual qubits.

In most quantum devices, multi-qubit gates have to be decomposed in single-qubit gates and CNOT-gates.
Some devices have the controlled-$Z$-gate as native gate, instead of the CNOT-gate.
The choice for one of these two gates negligibly affects the following success probabilities. 

In the remainder, we will use the term $P_{X}$ to denote the success probability of a subroutine $X$. 
Similarly, $P_{iX}$ denotes the probability that a qubit remains coherent while idling during the execution of $X$.

\begin{figure}[th]
    \centering
    \begin{quantikz}
    & \ctrl{1} & \\
    & \gate{U} & 
    \end{quantikz}=\begin{quantikz}
    & & \ctrl{1} & & \ctrl{1} & & \\
    & \gate{C} & \targ{} & \gate{B} & \targ{} & \gate{A} & 
    \end{quantikz}
    \caption{Identity for the controlled-$U$ gate.}
    \label{fig:q_circuit:controlled_U}
\end{figure}
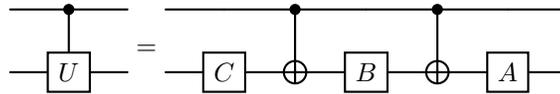

Some protocols use controlled-$U$-gates, for some single-qubit gate~$U$. 
Every such gate admits a decomposition in two CNOT-gates and three single-qubit gates, as \cref{fig:q_circuit:controlled_U} shows graphically (see also~\cite[Corollary~4.2]{NielsenChuang:2010}).
The gates $A$, $B$, and $C$ are chosen such that the product $ABC$ is the identity and $AXBXC = U$. 

Using the success probabilities from \cref{tab:success_probabilities} in the circuit shown in \cref{fig:q_circuit:controlled_U}, we obtain a success probability for the controlled-$U$-gate of
\begin{equation}\label{eq:success_controlled_U}
    P_{cU} = p_{s}^{3}p_{is}^{3}p_{d}^{2}.
\end{equation}

In the remainder of this chapter, we derive success probabilities for preparing the GHZ state and the $W$-state.
The derived expressions depend on up to seven success probability variables and the input size $n$.
Hence, for a first-order estimate of the relative theoretical performance of these protocols, we make some assumptions about the magnitude of the different terms. 

We distinguish between cheap operations, where the probability of an error is low, and expensive operations, where the probability of an error is higher. 
This distinction applies to both the operations themselves and the corresponding idling times.
In most hardware realizations, single-qubit gate errors are significantly lower than two-qubit gate errors. 
Similarly, the gate times for single-qubit gates are significantly shorter than those for two-qubit gates. 
We therefore assume that single-qubit gates and idling during single-qubit gates are cheap operations, while other operations are considered expensive.

In practice, this means that when comparing success probabilities, we assume $p_{s}\approx 1 \approx p_{is}$.
Additionally, we assume that $p_{d} \approx p_{m}$, and $p_{id} \approx p_{im} \approx p_{ic}$. 
The assumptions are supported by observed success probabilities for quantum hardware, such as the error rates reported by IBM~\cite{IBMQuantumExperience:2024}.

When comparing protocols, we use approximate inequality signs ($\gtrsim$ and $\lesssim$) to indicate that we applied these assumptions on the success probabilities. 

%% file: LAQCC/error_analysis/GHZ_error_analysis.tex
\section{Error analysis for GHZ state preparation}\label{sec:error:GHZ}
In this section, we derive an expression for the success probability of preparing a GHZ state using a standard approach using an all-to-all connectivity, a standard approach using a linear nearest-neighbor connectivity, and a $\LAQCC$-approach. 
These two hardware connectivities cover most quantum hardware realizations. 
We also derive the success probability for a hybrid version that combines a standard approach with the $\LAQCC$ approach. 

After deriving the success probabilities for the different approaches, we compare them to determine which protocol performs best in terms of the success probabilities of the individual parts of the circuit. 

Informally, the $\LAQCC$-approach performs exponentially better than the standard approach using an all-to-all connectivity if $p_{d} > p_{id}^{\Omega(\log n)}$. 
This means that the $\LAQCC$-approach performs exponentially better if the probability of a CNOT-gate introducing an error is at most the probability that an error is introduced while a qubit is idling for $\Omega(\log n)$ CNOT-gates. 
For the standard approach using a linear nearest-neighbor connectivity, we find that the $\LAQCC$-approach performs exponentially better than the standard approach if $p_{d} > p_{id}^{\Omega(n)}$.
Both results are in line with what one might expect based on the circuit sizes of the different approaches. 
\cref{thm:LAQCC:error:GHZ} provides a formal statement of the result for both comparisons.

\subsection{Success probability of GHZ state preparation}
Below we consider four possible approaches to prepare GHZ states and for each of them determine the success probability.
The four approaches considered are: 
\begin{enumerate}
    \item Standard approach using an all-to-all connectivity;
    \item Standard approach using a linear nearest-neighbor connectivity;
    \item A $\LAQCC$-approach;
    \item A hybrid version of a standard and a $\LAQCC$-approach. 
\end{enumerate}
We derive success probability expressions for preparing the GHZ state given by
\begin{equation*}
    \frac{1}{\sqrt{2}}\left(\ket{0}^{\otimes n}+\ket{1}^{\otimes n}\right).
\end{equation*}

\subsubsection{Standard approach using an all-to-all connectivity}
\begin{figure}[th]
    \centering
    \begin{quantikz}
    \lstick{\ket{0}}\slice[style=gray]{t=0} & \gate{H}\slice[style=gray]{t=1} & \ctrl{1}\slice[style=gray]{t=2} & \ctrl{2} & \slice[style=gray]{t=3} & \ctrl{4} & & & \slice[style=gray]{t=4} & \\
    \lstick{\ket{0}} & & \targ{} & & \ctrl{2} & & \ctrl{4} & & & \\
    \lstick{\ket{0}} & & & \targ{} & & & & \ctrl{4} & & \\
    \lstick{\ket{0}} & & & & \targ{} & & & & \ctrl{4} & \\
    \lstick{\ket{0}} & & & & & \targ{} & & & & \\
    \lstick{\ket{0}} & & & & & & \targ{} & & & \\
    \lstick{\ket{0}} & & & & & & & \targ{} & & \\
    \lstick{\ket{0}} & & & & & & & & \targ{} & 
    \end{quantikz}
    \caption{Circuit for preparing the GHZ state using an all-to-all connectivity for $n=8$. 
    The dotted lines indicate time steps and which gates can be applied in parallel.}
    \label{fig:q_circuit:GHZ_prep:all}
\end{figure}
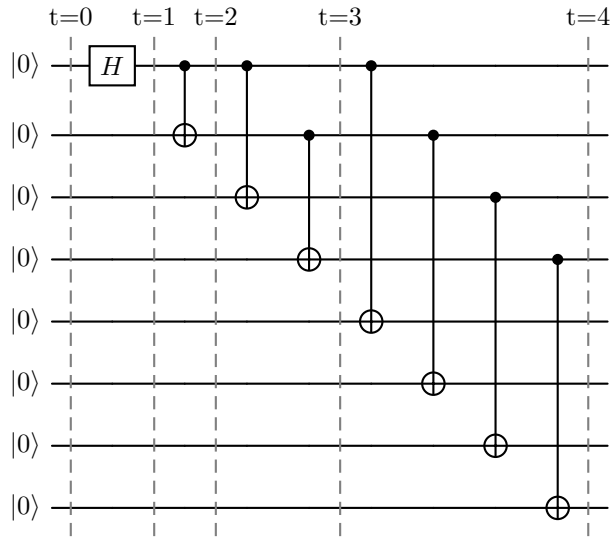

\cref{fig:q_circuit:GHZ_prep:all} shows a quantum circuit to prepare the GHZ state for $n=8$ using an all-to-all connectivity.
In every subsequent time step, twice as many qubits can be targeted.

Let $k=\floor{\log_2 n}$ and let $m = n-2^k$. 
In the first layer, only a single one-qubit gate is applied. 
Afterwards, at time step $t=i$, $2^{i-1}$ two-qubit gates are applied, while all other qubits remain idle. 
After time step $t=k$, a total of $2^{k}-1$ qubits have been targeted by a CNOT-gate.
If $n$ is a power of two, the state has now been prepared. 
Otherwise, a single extra layer is necessary with $m$ CNOT-gates. 

For the success probability $P_{GHZ_{n},all}$, we then arrive at the expression: 
\begin{align}
    P_{GHZ_{n},all} & = p_s p_{is}^{n-1} \left(\Pi_{t=1}^{k} p_d^{2^{t-1}}p_{id}^{n-2^{t}}\right)p_d^{m}p_{id}^{(n-2m)(\ceil{\log_2 n}-k)} \nonumber \\
    & = p_s p_{is}^{n-1}p_d^{n-1}p_{id}^{nk - 2^{k+1} + 2 + n(\ceil{\log_2 n}-\floor{\log_2 n}) -2m(\ceil{\log_2 n}-\floor{\log_2 n})} \nonumber \\
    & = p_s p_{is}^{n-1}p_d^{n-1}p_{id}^{n\ceil{\log_2 n} - 2(n-m) + 2 -2m(\ceil{\log_2 n}-\floor{\log_2 n})} \nonumber \\
    & = p_{s} p_{is}^{n-1} p_{d}^{n-1} p_{id}^{n(\ceil{\log_2 n}-2) + 2}. \label{eq:LAQCC:GHZ:P_all}
\end{align}
As $m(\ceil{\log_2 n}-\floor{\log_2 n}-1) = 0$, most terms in the exponent of $p_{id}$ cancel. 

To determine the probability that a qubit remains coherent while a GHZ state on $n$ qubits is prepared, we note that the circuit uses a single-qubit gate and then $\ceil{\log_2 n}$ layers of CNOT-gates. 
Combined, we have the success probability for idling qubits of 
\begin{equation*}
    P_{iGHZ_{n},all} = p_{is} p_{id}^{\ceil{\log_2 n}}.
\end{equation*}

\subsubsection{Standard approach using a linear nearest-neighbor connectivity}
\begin{figure}[th]
    \centering
    \begin{quantikz}
    \lstick{\ket{0}}\slice[style=gray]{t=0} & \slice[style=gray]{t=1} & \slice[style=gray]{t=2} & \slice[style=gray]{t=3} & \targ{}\slice[style=gray]{t=4} & \\
    \lstick{\ket{0}} & & & \targ{} & \ctrl{-1} & \\
    \lstick{\ket{0}} & \gate{H} & \ctrl{1} & \ctrl{-1} & & \\
    \lstick{\ket{0}} & & \targ{} & \ctrl{1} & & \\
    \lstick{\ket{0}} & & & \targ{} & \ctrl{1} & \\
    \lstick{\ket{0}} & & & & \targ{} &
    \end{quantikz}
    \caption{Circuit for preparing the GHZ state using a linear nearest-neighbor connectivity for $n=6$.
    The dotted lines indicate time steps and which gates can be applied in parallel.}
    \label{fig:q_circuit:GHZ_prep:linear}
\end{figure}
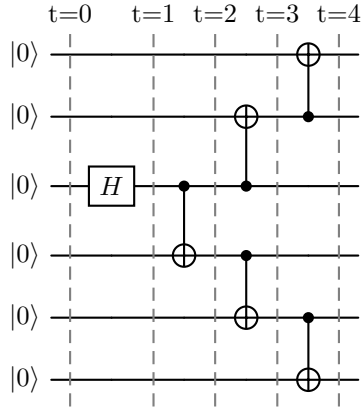

\cref{fig:q_circuit:GHZ_prep:linear} shows a quantum circuit to prepare the GHZ state for $n=6$ using a linear nearest-neighbor connectivity.
The key difference from the previous approach is that qubits can now only interact with their direct neighbors, and hence at most two CNOT-gates per layer can be applied. 

The circuit consists of the following steps: First, apply a single-qubit Hadamard gate on qubit $\floor{\frac{n+1}{2}}$ and then a CNOT-gate from that qubit to its direct neighbor with a higher index.
Let $k=\floor{\frac{n}{2}}$ and note that we can now have $k-1$ time steps consisting of $2$ CNOT-gates each. 
For odd $n$, we require a final layer consisting of a single CNOT-gate to include the last qubit in the GHZ state. 
We multiply the term in the exponent corresponding to the last layer with $n-2k$, as the term then vanishes for even $n$.

The overall probability of correctness $P_{GHZ,linear}$ is then given by
\begin{align}
    P_{GHZ_{n},linear} & = p_s p_{is}^{n-1} p_dp_{id}^{n-2}\left(p_d^{2}p_{id}^{n-4}\right)^{k-1}\left(p_d p_{id}^{n-2}\right)^{n-2k} \nonumber \\
    & = p_{s} p_{is}^{n-1} p_{d}^{n-1} p_{id}^{n(\ceil{n/2}-2)+2}. \label{eq:LAQCC:GHZ:P_linear}
\end{align}
Note that for $n\le 6$, the performance using an all-to-all connectivity and a linear nearest-neighbor connectivity is the same. 
For $n\ge 7$, the approach using an all-to-all connectivity has higher success probability as it has fewer idling qubits. 

The idling term looks similar for this approach.
Only the exponent of $p_{id}$ differs, corresponding to the extra layers of CNOT-gates applied: 
\begin{equation*}
    P_{iGHZ_{n},linear} = p_{is} p_{id}^{\ceil{n/2}}.
\end{equation*}

\subsubsection{\texorpdfstring{$\LAQCC$}{LAQCC}-approach}
\begin{figure}[th]
    \centering
    \begin{quantikz}
    \lstick{\ket{0}}\slice[style=gray]{t=0} & \gate{H}\slice[style=gray]{t=1} & \ctrl{1}\slice[style=gray]{t=2} & \slice[style=gray]{t=3} & \slice[style=gray]{t=4} & \gate[5]{\text{Correct}}\slice[style=gray]{t=5} &  \\
    \lstick{\ket{0}} & & \targ{} & \targ{} & \meter{} & \setwiretype{c} \\
    \lstick{\ket{0}} & \gate{H} & \ctrl{1} & \ctrl{-1} & & &  \\
    \lstick{\ket{0}} & & \targ{} & \targ{} & \meter{} & \setwiretype{c} \\
    \lstick{\ket{0}} & \gate{H} & & \ctrl{-1} & & & 
    \end{quantikz}
    \caption{GHZ state preparation using a $\LAQCC$ circuit.}
    \label{fig:q_circuit:GHZ_prep:LAQCC}
\end{figure}
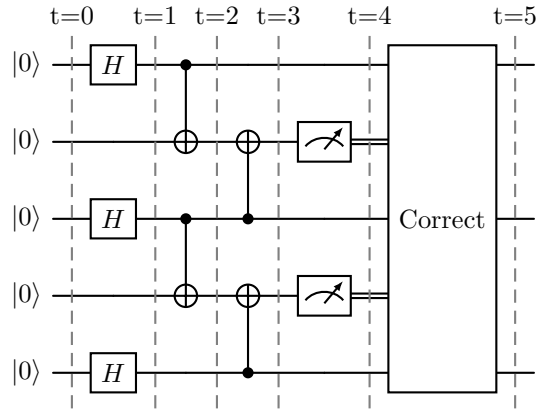

\cref{cor:clifford_in_LAQCC} implies the existence of a $\LAQCC$-circuit that prepares a GHZ state. 
In fact, we already saw such a circuit in \cref{fig:q_circuit:poor_man_cat_state} in \cref{sec:quantum:details_quantum_algorithm}.
\cref{fig:q_circuit:GHZ_prep:LAQCC} shows the same circuit as in \cref{fig:q_circuit:poor_man_cat_state} for $n=3$ with the time steps explicitly shown. 

The $\LAQCC$-circuit uses $2n-1$ qubits and the circuit depth remains constant, even with growing $n$. 
The success probability at time $t=4$ is given by 
\begin{equation*}
    p_s^n p_{is}^{n-1}p_d^{2(n-1)}p_{id}^2 p_m^{n-1}p_{im}^n.
\end{equation*}
A prefix sum computation on the measurement results indicates which qubits need correction. 
In the worst case, half of the qubits require a Pauli-$X$ correction. 
Assuming $p_s \le p_{is}$, which is valid as letting a qubit remain idle is generally better than manipulating it, we can lower bound the success probability:
\begin{equation}
    P_{GHZ_{n}, \LAQCC} \ge p_{s}^{n + \floor{n/2}} p_{is}^{n+\ceil{n/2}-1} p_{d}^{2(n-1)} p_{id}^{2} p_{m}^{n-1} p_{im}^{n} p_{ic}^{n}. \label{eq:LAQCC:GHZ:P_LAQCC}
\end{equation}

A qubit idling while a GHZ state is prepared by a $\LAQCC$-circuit has success probability
\begin{equation*}
    P_{iGHZ, \LAQCC} \ge p_{is}^{2} p_{id}^{2} p_{im} p_{ic}.
\end{equation*}

\subsubsection{Hybrid approach}
We now combine the different state preparation approaches: 
First, use a standard approach to prepare $k$ small GHZ states.
Then, use a $\LAQCC$-approach to join these small GHZ states together.

We assume that $k$ perfectly divides $n$, such that $n=kg$ for some positive integer~$g$. 
The same analysis can also be performed when we prepare GHZ states of different sizes and then combine them. 
However, the analysis will be more involved. 

When we use a $\LAQCC$-protocol to join the $k$ GHZ states, all other qubits remain idle. 
The total success probability is given by
\begin{equation*}
    P_{GHZ_{g}}^{k} P_{iGHZ_{g}}^{k} P_{GHZ_{k},\LAQCC} P_{iGHZ,\LAQCC}^{n-k}.
\end{equation*}
Note that in the worst case, we still have to correct at most half of the total number of qubits, as a correction term in the $\LAQCC$-protocol also has to be applied to all qubits in the corresponding GHZ state.
Hence, the error terms in the circuit are slightly worse than what this expression alone suggests. 

Depending on the connectivity used, we get one of two expressions for the success probability. 
For the hybrid $\LAQCC$-all-to-all approach we obtain
\begin{equation}
    P_{GHZ_{n,k},hybrid\text{-}all} \ge p_{s}^{2k + \floor{n/2}} p_{is}^{3n - k + \ceil{n/2} - 1} p_{d}^{n + k - 2} p_{id}^{(n + k)\ceil{\log_2 g} + 2} p_{m}^{k - 1} p_{im}^{n} p_{ic}^{n}. \label{eq:LAQCC:GHZ:P_hybrid_all}
\end{equation}

For the hybrid $\LAQCC$-linear nearest-neighbor approach we obtain
\begin{equation}
    P_{GHZ_{n,k},hybrid\text{-}linear} \ge p_{s}^{2k + \floor{n/2}} p_{is}^{3n - k + \ceil{n/2} - 1} p_{d}^{n + k - 2} p_{id}^{(n+k)\ceil{g/2} + 2} p_{m}^{k - 1} p_{im}^{n} p_{ic}^{n}. \label{eq:LAQCC:GHZ:P_hybrid_linear}
\end{equation}

\subsection{Comparing GHZ state preparation approaches}
In this section we compare the success probabilities for the different approaches to prepare the GHZ state. 
The main theorem proven in this section is the following.
\begin{theorem}\label{thm:LAQCC:error:GHZ}
    Let $\eps>0$ be a constant. 
    If $p_{d} \gtrsim (1+\eps)p_{id}^{\frac{n}{n-1}(\ceil{\log_2 n}/2 - 2)}$, then $P_{GHZ_n,\LAQCC}\gtrsim (1+\eps)^{2(n-1)} P_{GHZ_n,all}$. 

    If $p_{d} \gtrsim (1+\eps)p_{id}^{\frac{n}{n-1}(\ceil{n/2}/2 - 2)}$, then $P_{GHZ_n,\LAQCC}\gtrsim (1+\eps)^{2(n-1)} P_{GHZ_n,linear}$.
\end{theorem}
In each of the following subsections, we start with a comparison of the success probabilities and then derive the expression for $p_{d}$ and $p_{id}$ accordingly. 
The theorem follows directly from these comparisons.

\subsubsection{All-to-all versus \texorpdfstring{$\LAQCC$}{LAQCC}}\label{sec:LAQCC:GHZ:error_all_LAQCC}
To determine when 
\begin{equation}
    P_{GHZ_{n},\LAQCC} \ge P_{GHZ_{n},all}
\end{equation}
holds, we compare \cref{eq:LAQCC:GHZ:P_LAQCC,eq:LAQCC:GHZ:P_all} to find that the inequality holds, precisely if
\begin{equation*}
    p_{s}^{n + \floor{n/2} - 1} p_{is}^{\ceil{n/2}} p_{d}^{n-1} p_{m}^{n-1} p_{im}^{n} p_{ic}^{n} \ge p_{id}^{n(\ceil{\log_2 n}-2)}.
\end{equation*}
Applying the assumptions on the success probabilities (discussed in \cref{sec:error:error_model}), we see that this expression reduces to 
\begin{equation*}
    p_{d}^{2(n-1)} \gtrsim p_{id}^{n(\ceil{\log_2 n} - 4)}.
\end{equation*}
Hence, the $\LAQCC$-approach performs better if 
\begin{equation}\label{eq:LAQCC:error:GHZ_all_LAQCC}
    p_{d} \gtrsim p_{id}^{\frac{n}{n-1}(\ceil{\log_2 n}/2 - 2)}.
\end{equation} 
That is, the $\LAQCC$-approach outperforms the standard approach using an all-to-all connectivity if the probability of a CNOT-gate introducing an error is at most the probability of a qubit idling during $\frac{n}{n-1}(\ceil{\log_2 n}/2 - 2)$ CNOT-gates picking up an error.

Now note that we can rewrite \cref{eq:LAQCC:GHZ:P_LAQCC} in terms of \cref{eq:LAQCC:GHZ:P_all}, using the assumptions on the success probabilities, as
\begin{equation}
    P_{GHZ_n, \LAQCC} \gtrsim p_{d}^{2(n-1)} p_{id}^{n(4-\ceil{\log_2 n})} P_{GHZ_n, all}.
\end{equation}
Hence, for $p_{d} = (1+\eps)p_{id}^{\frac{n}{n-1}(\ceil{\log_2 n}/2 - 2)}$ we obtain
\begin{equation}
    P_{GHZ_n,\LAQCC} \gtrsim (1+\eps)^{2(n-1)} P_{GHZ_n,all},
\end{equation}
proving the first statement of \cref{thm:LAQCC:error:GHZ}.

Note that for large $n$, \cref{eq:LAQCC:error:GHZ_all_LAQCC} reduces to
\begin{equation} \label{eq:LAQCC:GHZ:error_informal_all}
    p_{d} \gtrsim p_{id}^{\Omega(\log_2 n)},
\end{equation}
corresponding with what we expected based on the circuit size.

\subsubsection{Linear versus \texorpdfstring{$\LAQCC$}{LAQCC}}\label{sec:LAQCC:GHZ:error_linear_LAQCC}
Similar to the previous section, to determine when
\begin{equation}
    P_{GHZ_{n},\LAQCC} \ge P_{GHZ_{n}, linear}
\end{equation}
holds, we compare \cref{eq:LAQCC:GHZ:P_LAQCC,eq:LAQCC:GHZ:P_linear}.
We then see that the inequality holds precisely if
\begin{equation*}
    p_{s}^{n + \floor{n/2} - 1} p_{is}^{\ceil{n/2}} p_{d}^{n-1} p_{m}^{n-1} p_{im}^{n} p_{ic}^{n} \ge p_{id}^{n(\ceil{n/2}-2)}.
\end{equation*}
Applying the assumptions on the success probabilities, we see that this expression reduces to 
\begin{equation*}
    p_{d}^{2(n-1)} \gtrsim p_{id}^{n(\ceil{n/2}-4)}.
\end{equation*}
Hence, the $\LAQCC$-approach performs better if 
\begin{equation}\label{eq:LAQCC:error:GHZ_linear_LAQCC}
    p_{d} \gtrsim p_{id}^{\frac{n}{n-1}(\ceil{n/2}/2 - 2)}.
\end{equation} 
That is, the $\LAQCC$-approach outperforms the standard protocol using a linear nearest-neighbor connectivity if the probability of a CNOT-gate introducing an error is at most the probability of a qubit idling during $\frac{n}{n-1}(\ceil{n/2}/2 - 2)$ CNOT-gates picking up an error.

Now note that we can rewrite \cref{eq:LAQCC:GHZ:P_LAQCC} in terms of \cref{eq:LAQCC:GHZ:P_linear}, using the assumptions on the success probabilities, as
\begin{equation}
    P_{GHZ_n, \LAQCC} \gtrsim p_{d}^{2(n-1)} p_{id}^{n(4-\ceil{n/2})} P_{GHZ_n, linear}.
\end{equation}
Hence, for $p_{d} = (1+\eps)p_{id}^{\frac{n}{n-1}(\ceil{n/2}/2 - 2)}$ we obtain
\begin{equation}
    P_{GHZ_n,\LAQCC} \gtrsim (1+\eps)^{2(n-1)} P_{GHZ_n, linear},
\end{equation}
proving the second statement of \cref{thm:LAQCC:error:GHZ}.

Note that for large $n$, \cref{eq:LAQCC:error:GHZ_linear_LAQCC} reduces to
\begin{equation} \label{eq:LAQCC:GHZ:error_informal_linear}
    p_{d} \gtrsim p_{id}^{\Omega(n)},
\end{equation}
corresponding with what we expected based on the circuit size.

\subsubsection{Comparison with the hybrid approach}
We now compare the two standard approaches with their corresponding hybrid version. 
In the hybrid approach, $k$ GHZ states of $g$ qubits each are combined using the $\LAQCC$-approach to prepare a GHZ state on $n=kg$ qubits. 
Therefore, we express the relative performance of the approaches in terms of $k$ and $g$. 

For the hybrid approach using an all-to-all connectivity, we find that it outperforms the standard approach using the same connectivity if
\begin{equation*}
    P_{GHZ_{n,k},hybrid\text{-}all} \ge P_{GHZ_{n}, all}.
\end{equation*}
Using \cref{eq:LAQCC:GHZ:P_hybrid_all,eq:LAQCC:GHZ:P_all}, this expression simplifies to
\begin{equation*}
    p_{s}^{2k + \floor{n/2} - 1} p_{is}^{2n - k + \ceil{n/2}} p_{d}^{k - 1} p_{m}^{k - 1} p_{im}^{n} p_{ic}^{n} \ge p_{id}^{n(\ceil{\log_2 n} - \ceil{\log_2 g} - 2) - k\ceil{\log_2 g}}.
\end{equation*}
Again applying the assumptions on the success probabilities, we see that this expression reduces to
\begin{equation} \label{eq:LAQCC:GHZ:error_pd_id_all}
    p_{d} \gtrsim p_{id}^{\frac{n}{2(k-1)}(\ceil{\log_2 n} - \ceil{\log_2 g} - 4) - \frac{k}{2(k-1)}\ceil{\log_2 g}}.
\end{equation}
Let $\eps>0$ and $p_{d} = (1+\eps) p_{id}^{\frac{n}{2(k-1)}(\ceil{\log_2 n} - \ceil{\log_2 g} - 4) - \frac{k}{2(k-1)}\ceil{\log_2 g}}$, then 
\begin{equation}
    P_{GHZ_{n,k},hybrid\text{-}all} \ge (1+\eps)^{2(k-1)} P_{GHZ_n, all}.
\end{equation}
Hence, the hybrid approach performs exponentially better than the standard approach. 

For the hybrid approach using a linear nearest-neighbor connectivity, we find that it outperforms the standard approach using the same connectivity if
\begin{equation*}
    P_{GHZ_{n,k},hybrid\text{-}linear} \ge P_{GHZ_{n}, linear}.
\end{equation*}
Using \cref{eq:LAQCC:GHZ:P_hybrid_linear,eq:LAQCC:GHZ:P_linear}, this expression simplifies to
\begin{equation*}
    p_{s}^{2k + \floor{n/2} - 1} p_{is}^{2n - k + \ceil{n/2}} p_{d}^{k - 1} p_{m}^{k - 1} p_{im}^{n} p_{ic}^{n} \ge p_{id}^{n(\ceil{n/2} - \ceil{g/2} - 2) - k\ceil{g/2}}.
\end{equation*}
Again applying the assumptions on the success probabilities, we see that this expression reduces to
\begin{equation} \label{eq:LAQCC:GHZ:error_pd_id_linear}
    p_{d} \gtrsim p_{id}^{\frac{n}{2(k-1)}(\ceil{n/2} - \ceil{g/2} - 4) - \frac{k}{2(k-1)}\ceil{g/2}}.
\end{equation}
Let $\eps>0$ and $p_{d} = (1+\eps) p_{id}^{\frac{n}{2(k-1)}(\ceil{n/2} - \ceil{g/2} - 4) - \frac{k}{2(k-1)}\ceil{g/2}}$, then 
\begin{equation}
    P_{GHZ_{n,k},hybrid\text{-}linear} \ge (1+\eps)^{2(k-1)} P_{GHZ_n, linear}.
\end{equation}
Hence, the hybrid approach performs exponentially better than the standard approach. 

For both hybrid approaches, we can obtain an informal estimate similar to \cref{eq:LAQCC:GHZ:error_informal_all,eq:LAQCC:GHZ:error_informal_linear}. 
Using that $n=kg$, $\ceil{x}\approx x \approx \floor{x}$ for any $x\in\R$, and $\frac{x}{x-1}\approx 1$ for large $x\in \R$, we see that \cref{eq:LAQCC:GHZ:error_pd_id_all,eq:LAQCC:GHZ:error_pd_id_linear} simplify to
\begin{equation}
    p_{d} \gtrsim p_{id}^{\Omega(g\log_2 k)}
\end{equation}
and
\begin{equation}
    p_{d} \gtrsim p_{id}^{\Omega(ng)},
\end{equation}
respectively.

Note that for $g=\bigo(1)$ (and hence $k=\Theta(n)$), the hybrid approaches perform similarly to the $\LAQCC$-approach.

%% file: LAQCC/error_analysis/GHZ_implementation_analysis.tex
\section{Implementation on quantum hardware}\label{sec:error:hardware_implementation}
In the previous sections we compared three GHZ state generation approaches.
In this section, we implement the approaches on quantum hardware and compare the results with our theoretical results. 
We first discuss the nuances and practicalities of implementing a protocol on currently available quantum hardware. 
Then, we present the results for the selected quantum device. 
Given the worst-case error model considered in the previous section, we expect to see differences when analyzing quantum hardware implementations.

\subsection{Setup and implementation details}
We expect that our theoretical bounds will give a lower bound on the actual success probabilities, as we consider a worst-case error model. 
In practice, errors may be less severe.
In some cases, errors might even cancel each other out. 

We consider the IBM Brisbane device of IBM, which is based on superconducting technology~\cite{IBMQuantumExperience:2024}. 
The device has 127 qubits and is freely available via IBM's online quantum computing platform. 
Given the device's topology and the qubit requirements of the $\LAQCC$-approach, we can prepare a GHZ state on at most $n=55$ qubits. 

We implement a standard approach and the $\LAQCC$-approach.
For each approach, we give the measurement outcomes as a quasi-probability distribution based on $4{,}096$ samples. 
We also provide the expected success probabilities based on \cref{eq:LAQCC:GHZ:P_linear,eq:LAQCC:GHZ:P_LAQCC} and using the parameters of the device.

Note that the measurements used to obtain this quasi-probability distribution can introduce errors themselves. 
Furthermore, we cannot detect phase errors, as measurements are performed in the Pauli-$Z$ basis. 
However, as both approaches apply $n$ measurements at the end of the circuit, we ignore resulting decoherence for now, expecting its impact on both outcomes to be approximately the same.

We expect differences between the hardware results and the theoretical results for multiple reasons. 
First, the worst-case error model used likely gives an upper bound on errors in practice. 
Second, quantum hardware systems typically have a limited set of native gates, requiring that some gates used in an algorithm are decomposed into native gates. 
Third, conventional pre- and post-processing techniques can help reduce the circuit depth and the errors in the circuits. 

In a noiseless situation, we expect the circuits to give a near-uniform distribution between the all-zeros and all-ones outcomes upon measurement. 
In a noisy setting, we expect to often find the two correct measurement outcomes, but we also expect to find other measurement results due to implementation imperfections.

For the theoretical success probabilities, we use the success probabilities of the individual terms in a circuit. 
Most of these terms are provided by the quantum hardware provider in terms of error probabilities.
However, the idling terms for CNOT-gates and measurements are not provided and must
be derived manually. 

We compute the idling success probability for CNOT-gate and for measurements using the relaxation time and dephasing time, given by decay constants~$T_1$ and~$T_2$, respectively. 
These constants define the probability that a qubit remains in its correct state.
Specifically, the probability that a qubit initially in the $\ket{1}$-state remains in that state after time~$t$ is given by $e^{-t/T_1}$. 
Similarly for a qubit initially in the $\ket{+}$-state, the probability is determined by the $T_2$ decay constant.
In general, $T_1$ is larger than $T_2$.
Therefore, we will compute both $p_{id}$ and $p_{im}$ with the $T_2$ decay constant.

Each qubit has its own $T_1$ and $T_2$ constants, and similarly, every measurement and every quantum gate has its own success probability.
Therefore, we take the median over all available values for every success probability term. 

The IBM Brisbane device supports the CZ-gate instead of a CNOT-gate as native two-qubit gate. 
By conjugating the target qubit of a CNOT-gate with Hadamard gates, we obtain a CZ-gate. 
We therefore compute $p_{d} = p_{d,CZ}p_{s}^{2}p_{is}^{2}$. 

We now give the code used to generate the results on quantum hardware. 
The current quantum hardware only allows control of future quantum gates by measurement outcomes. 
We cannot perform computations on these outcomes, making the current $\LAQCC$-implementation suboptimal.
\begin{python}
from qiskit import QuantumCircuit, QuantumRegister, ClassicalRegister
from qiskit_ibm_runtime import QiskitRuntimeService, SamplerV2
from qiskit.transpiler.preset_passmanagers import generate_preset_pass_manager

API_token = "<your token here>"
backend_name = "ibm_brisbane"
n_qubits = 10
service = QiskitRuntimeService(channel="ibm_quantum", token=API_token)
backend = service.backend(backend_name)
pm = generate_preset_pass_manager(backend=backend, optimization_level=1)

#%% LAQCC circuit
qrm = QuantumRegister(n_qubits)
qrx = QuantumRegister(n_qubits - 1)
crx = ClassicalRegister(n_qubits-1, name="intermediate_result")
crm = ClassicalRegister(n_qubits, name="final_result")
qcircuit_LAQCC = QuantumCircuit(qrm,qrx, crx, crm, name="GHZ")
for i in range(n_qubits):
    qcircuit_LAQCC.h(qrm[i])
for i in range(n_qubits-1):
    qcircuit_LAQCC.cx(qrm[i], qrx[i])
for i in range(1,n_qubits):
    qcircuit_LAQCC.cx(qrm[i], qrx[i-1])

for i in range(n_qubits - 1):
    qcircuit_LAQCC.measure(qrx[i], crx[i])
    
for i in range(n_qubits-1):
    with qcircuit_LAQCC.if_test((crx[i], 1)):
        for j in range(i+1, n_qubits):
            qcircuit_LAQCC.x(qrm[j])

qcircuit_LAQCC.measure(qrm, crm)

isa_circuit_LAQCC = pm.run(qcircuit_LAQCC)
sampler = Sampler(backend)
job = sampler.run([(isa_circuit_LAQCC)])
result_LAQCC = job.result()
#%% Standard circuit
qrm = QuantumRegister(n_qubits)
crm = ClassicalRegister(n_qubits)

start_qubit = (n_qubits + 1) // 2 - 1
k = n_qubits//2
qcircuit_standard = QuantumCircuit(qrm, crm, name="GHZ")

qcircuit_standard.h(start_qubit)
qcircuit_standard.cx(start_qubit, start_qubit +1)
for i in range(k-1):
    qcircuit_standard.cx(start_qubit-i, start_qubit-i-1)
    qcircuit_standard.cx(start_qubit+1+i, start_qubit+2+i)

if n_qubits - 2*k: # Check if we need a final layer
    qcircuit_standard.cx(1, 0)
    
qcircuit_standard.measure(qrm, crm)

isa_circuit_standard = pm.run(qcircuit_standard)
sampler = Sampler(backend)
job = sampler.run([(isa_circuit_standard)])
result_standard = job.result()
\end{python}

We now proceed by comparing the theoretical success probabilities from \cref{eq:LAQCC:GHZ:P_linear,eq:LAQCC:GHZ:P_LAQCC} with the implementation results.
As mentioned, we expect these success probabilities to give a lower bound on the actual success probabilities. 
Additionally, we give the expected running time of each circuit based on the obtained gate and measurement times. 
Finally, we show the measurement results for different values of $n$ for both approaches, allowing us to compare practical performance and observe how the success probabilities change as $n$ grows.

\subsection{IBM Brisbane device}
This section presents the results of the hardware experiments run on the IBM Brisbane device. 
The IBM Brisbane device has $127$ qubits, allowing for the preparation of a GHZ state on at most $55$ qubits. 
\cref{tab:success_probabilities_IBM_Brisbane} summarizes the success probabilities of different gates of this device, using a $T_2$ value of $131.71\mu$s.
\begin{table}[th]
    \centering
    \begin{tabular}{c|l|l}
        Success term & Value & Obtained via \\
        \hline
        $p_{s}$ & $1 - 2.530\cdot 10^{-4}$ & Provided by IBM \\
        $p_{is}$ & $1 - 2.530\cdot 10^{-4}$ & Provided by IBM \\
        $p_{d}$ & $1 - 9.442\cdot 10^{-3}$ & Provided by IBM and computed with $p_s$ \\
        $p_{id}$ & $1 - 4.998\cdot 10^{-3}$ & Based on a gate time of $660$ ns \\
        $p_{m}$ & $1-1.600\cdot 10^{-2}$ & Provided by IBM \\
        $p_{im}$ & $1-9.822 \cdot 10^{-3}$ & Based on a measurement time of $1300$ ns \\
        $p_{ic}$ & $1-9.822 \cdot 10^{-3}$ & Equal to $p_{im}$
    \end{tabular}
    \caption{Success probabilities of IBM Brisbane device, based on calibration on December 13, 2024 at 09.30 (UTC+2).}
    \label{tab:success_probabilities_IBM_Brisbane}
\end{table}

\cref{eq:LAQCC:GHZ:P_linear,eq:LAQCC:GHZ:P_LAQCC} give the following two success probabilities for the largest GHZ state preparable on this device
\begin{align*}
    P_{Brisbane, GHZ_{55}, standard} & = p_{s} p_{is}^{54} p_{d}^{54} p_{id}^{1432} \\
    & = 4.52 \cdot 10^{-4} \\
    P_{Brisbane, GHZ_{55}, \LAQCC} & \ge p_{s}^{82} p_{is}^{83} p_{d}^{108} p_{id}^{2} p_{m}^{54} p_{im}^{55} p_{ic}^{55} \\
    & = 4.82 \cdot 10^{-2}.
\end{align*}
We see a factor $100$ difference in the theoretical success probabilities with a worst-case error model, favoring the $\LAQCC$-approach. 

To determine the time duration of both approaches, we estimate the single-qubit gate time using $p_{s}$ and $T_2$ to be approximately $33$ ns. 
Using a two-qubit gate time of $660$ ns and a measurement time of $1300$ ns, we find running times
\begin{align*}
    T_{Brisbane, GHZ_{55}, standard} & = 33 \text{ns} + 28*660 \text{ns} = 18.51 \mu\text{s} \\
    T_{Brisbane, GHZ_{55}, \LAQCC} & = 2(33 + 660 + 1300) \text{ns} = 3.99 \mu\text{s}.
\end{align*}
Hence, we expect the $\LAQCC$-approach to have a shorter running time.

Below, we give the results of the hardware implementations for different $n$.
For small $n$, all measurement outcomes are given. 
For larger $n$, we instead give aggregated results, grouping the results of strings with the same Hamming weight.

Based on the success probabilities shown in \cref{tab:success_probabilities_IBM_Brisbane} and the theoretical relation between $p_{d}$ and $p_{id}$ shown in \cref{thm:LAQCC:error:GHZ}, we expect the $\LAQCC$-approach to outperform the standard approach for $n\ge 15$. 

\begin{figure}
    \centering
    \includegraphics[width=0.32\textwidth]{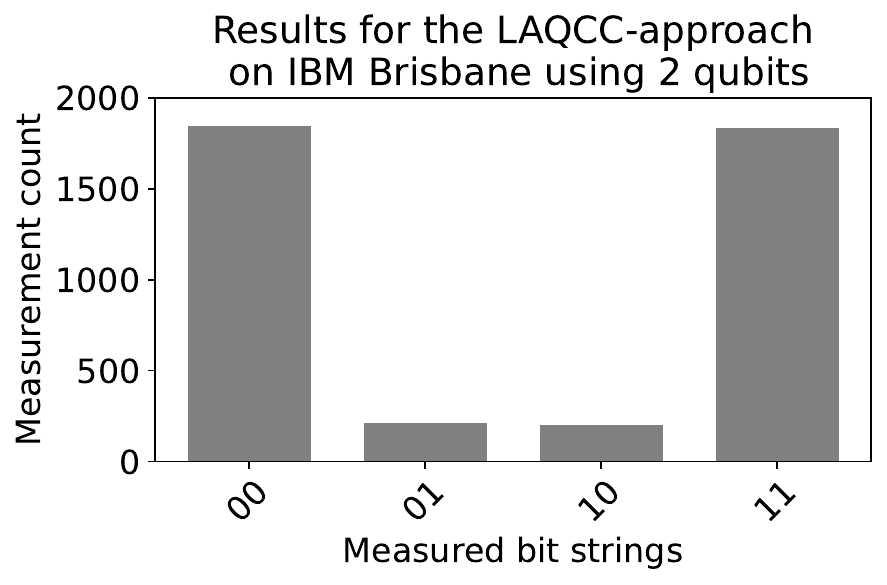}
    \includegraphics[width=0.32\textwidth]{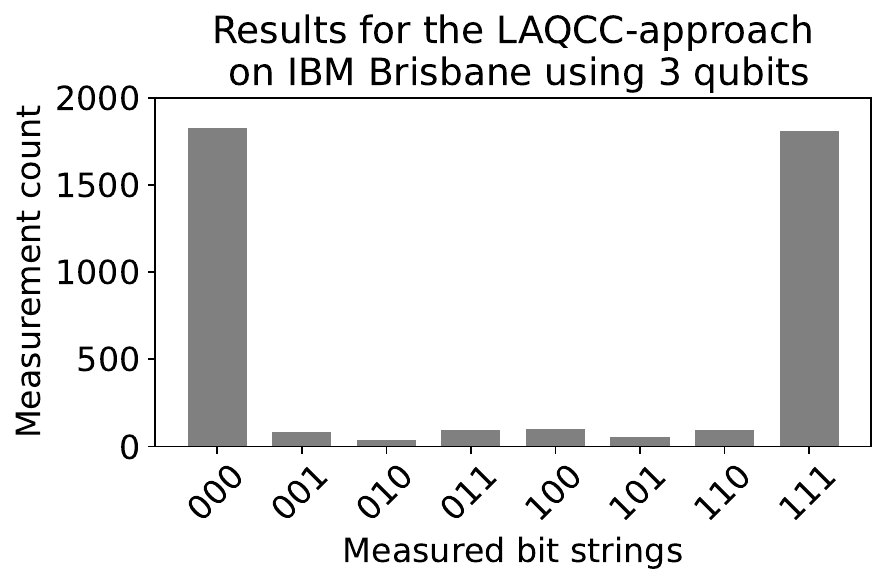}
    \includegraphics[width=0.32\textwidth]{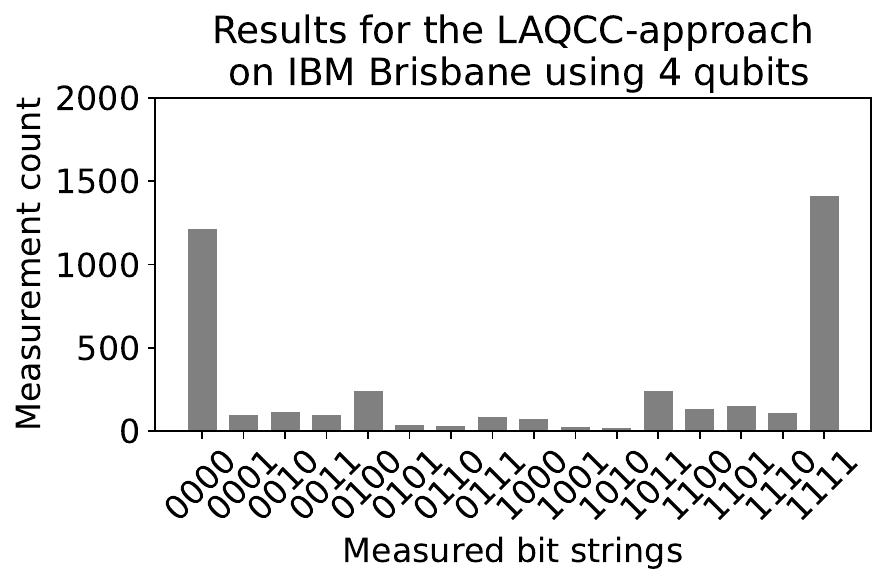}
    \includegraphics[width=0.32\textwidth]{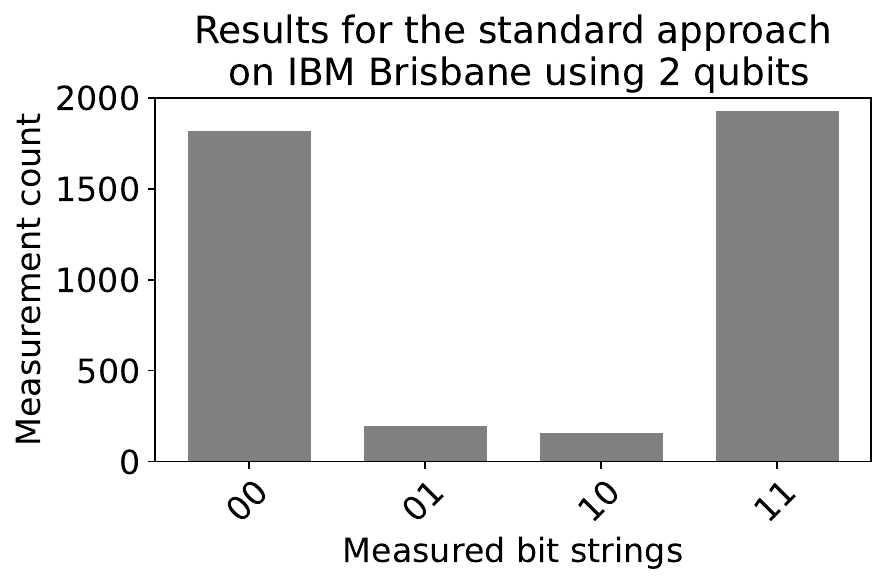}
    \includegraphics[width=0.32\textwidth]{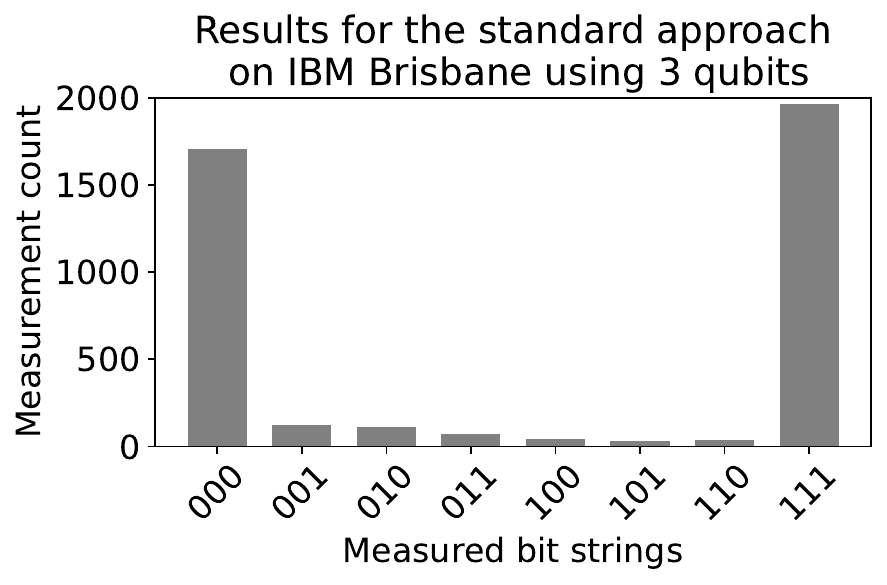}
    \includegraphics[width=0.32\textwidth]{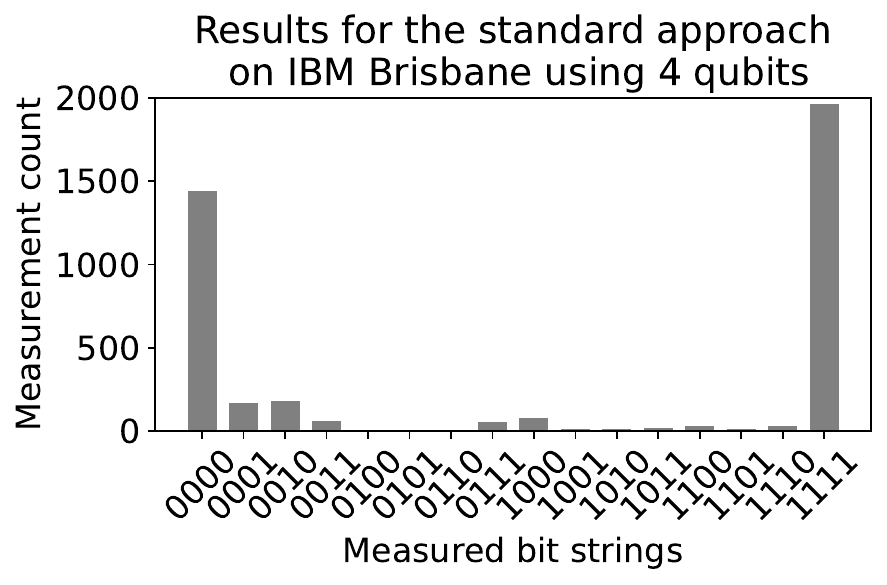}
    \caption{Measurement results for preparing a GHZ state on few qubits on the IBM Brisbane device using the $\LAQCC$-approach and the standard approach.
    Horizontally, the different measurement results are shown and the height of the bars shows how often that measurement result is found.}
    \label{fig:error:Brisbane:LAQCC_standard_small_n}
\end{figure}

\cref{fig:error:Brisbane:LAQCC_standard_small_n} shows the results for small $n$. 
The standard approach slightly outperforms the $\LAQCC$-approach, as expected based on the success probabilities. 

\begin{figure}
    \centering
    \includegraphics[width=0.45\textwidth]{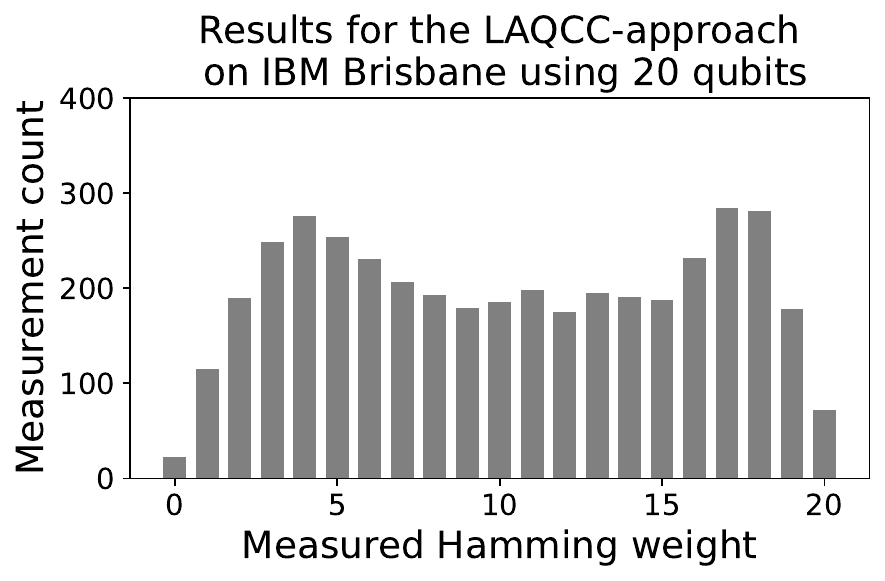}
    \includegraphics[width=0.45\textwidth]{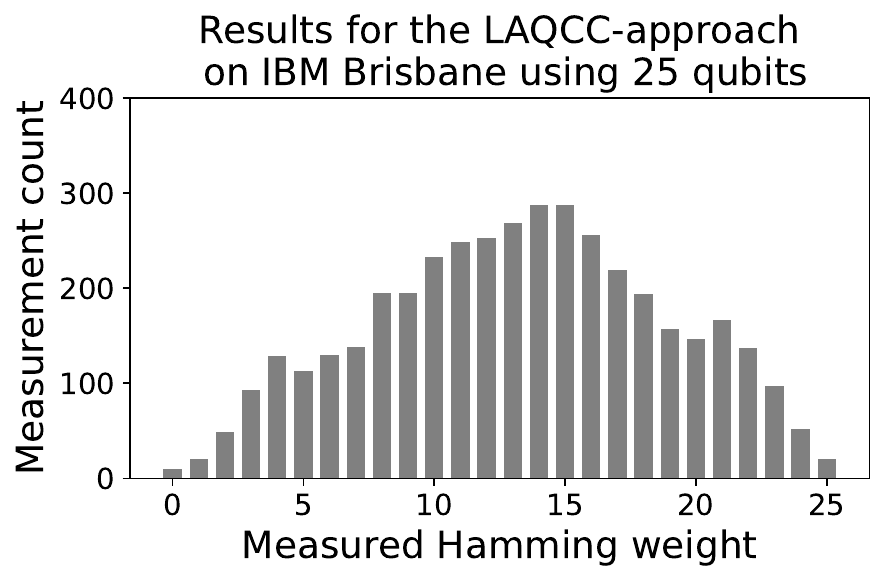}
    \includegraphics[width=0.45\textwidth]{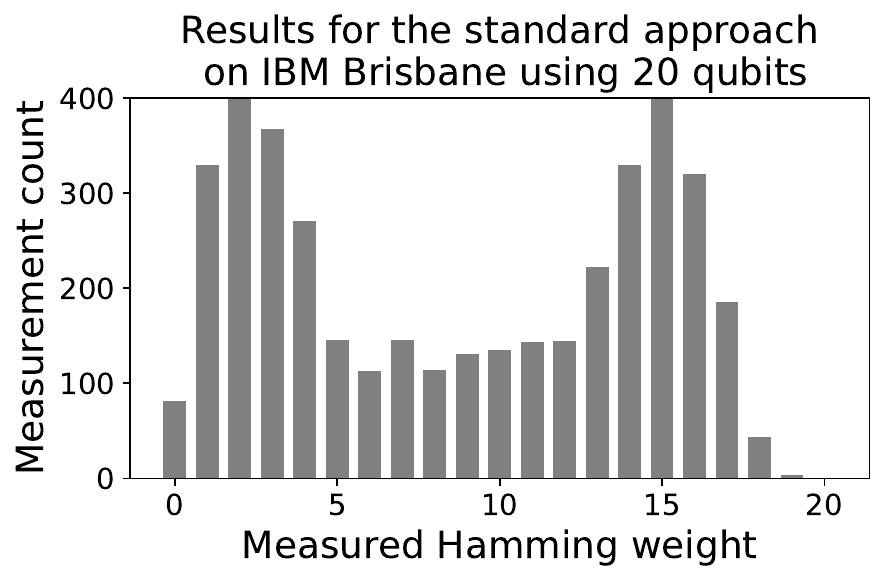}
    \includegraphics[width=0.45\textwidth]{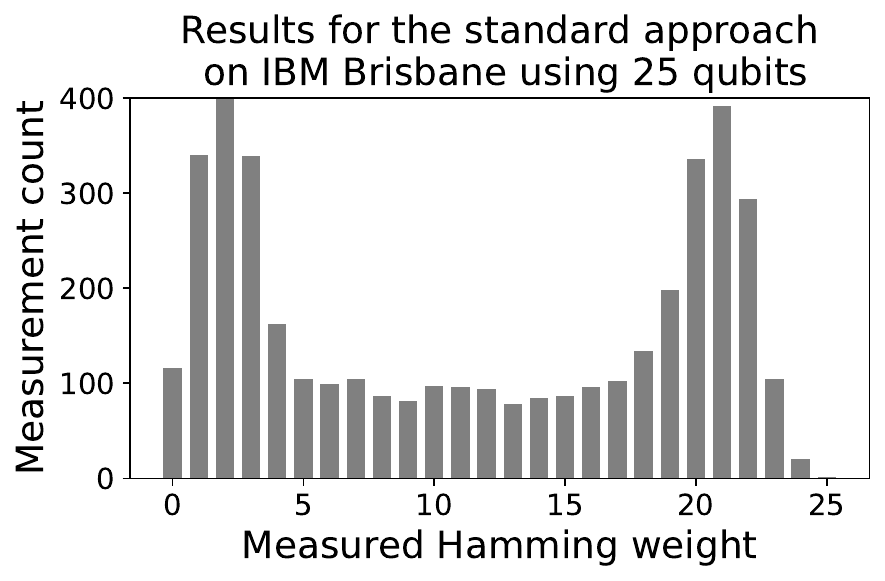}
    \caption{Measurements results for preparing a GHZ state on $n=20$ and $n=25$ qubits on the IBM Brisbane device using the $\LAQCC$-approach and the standard approach. 
    Horizontally, the different measured Hamming weights are shown and the height of the bars shows how often that Hamming weight was found.}
    \label{fig:error:Brisbane:LAQCC_standard_n20_25}
\end{figure}

\cref{fig:error:Brisbane:LAQCC_standard_n20_25} shows the aggregated results for $n=20$ and $n=25$, where measurement outcomes with the same Hamming weight are grouped. 
The aggregated results show the magnitude of the errors. 
With the $\LAQCC$-approach, both expected outcomes are measured, but the results are somewhat uniform.
In contrast, the standard approach did not return the all-ones string in any measurement, but does show two distinct peaks near the low and high Hamming weight outcomes. 

\begin{figure}
    \centering
    \includegraphics[width=0.3\textwidth]{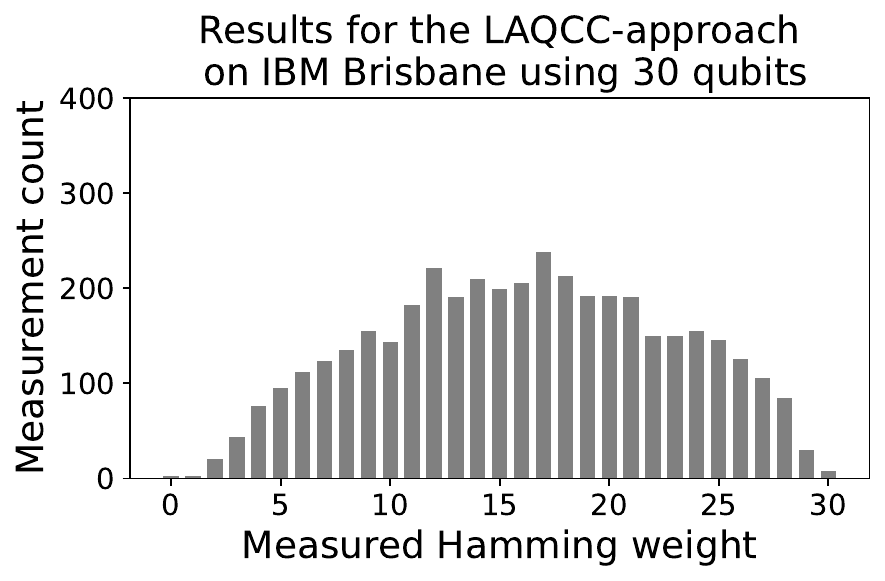}
    \includegraphics[width=0.3\textwidth]{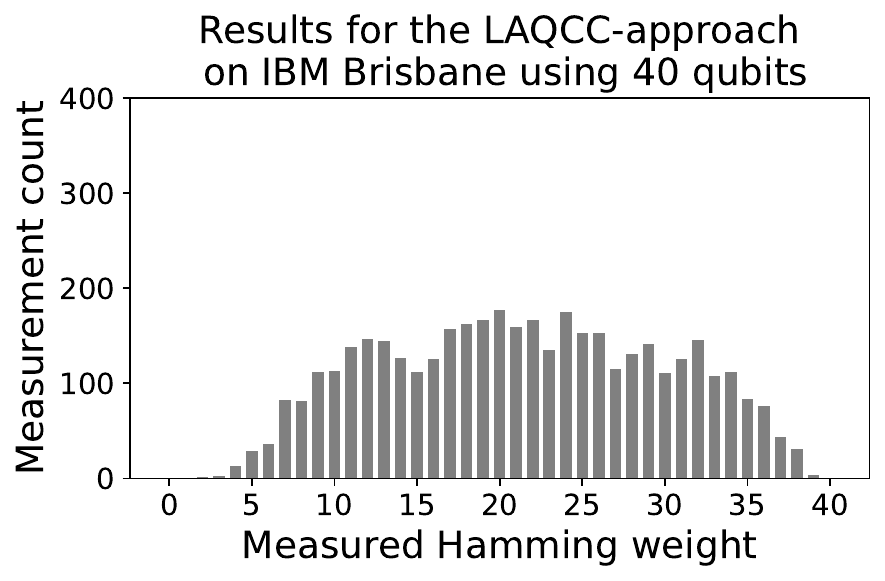}
    \includegraphics[width=0.3\textwidth]{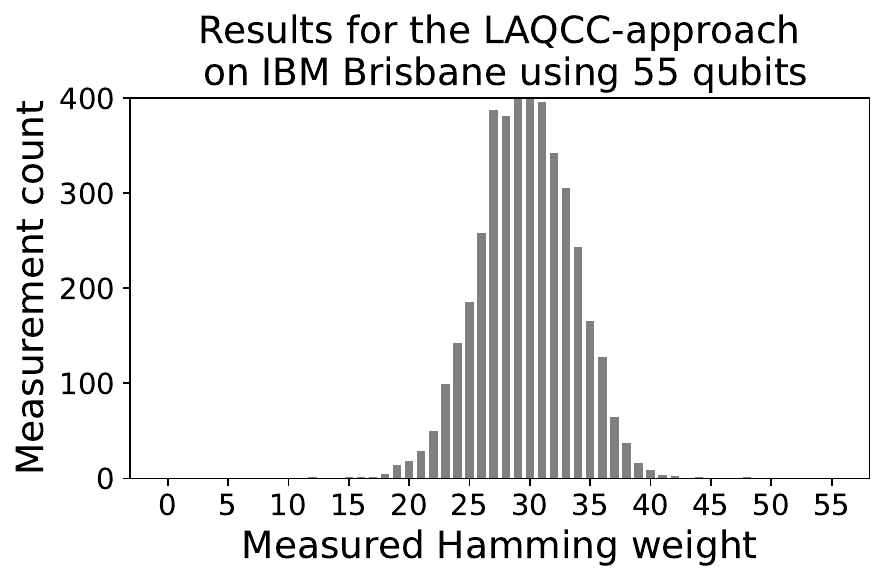}
    \includegraphics[width=0.3\textwidth]{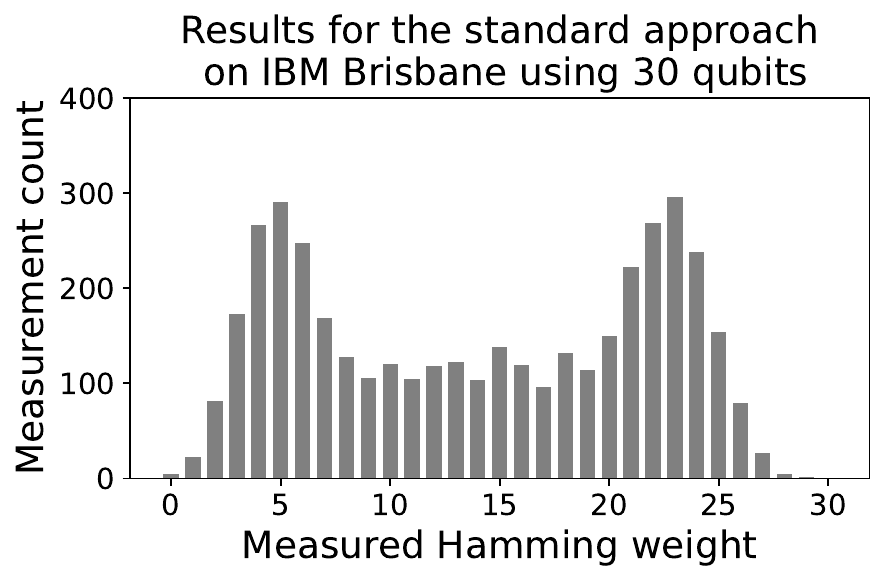}
    \includegraphics[width=0.3\textwidth]{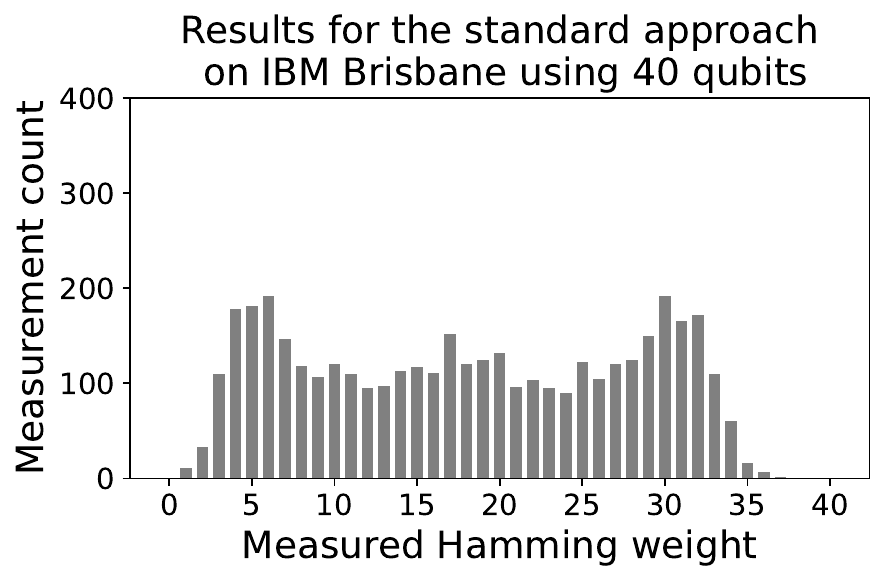}
    \includegraphics[width=0.3\textwidth]{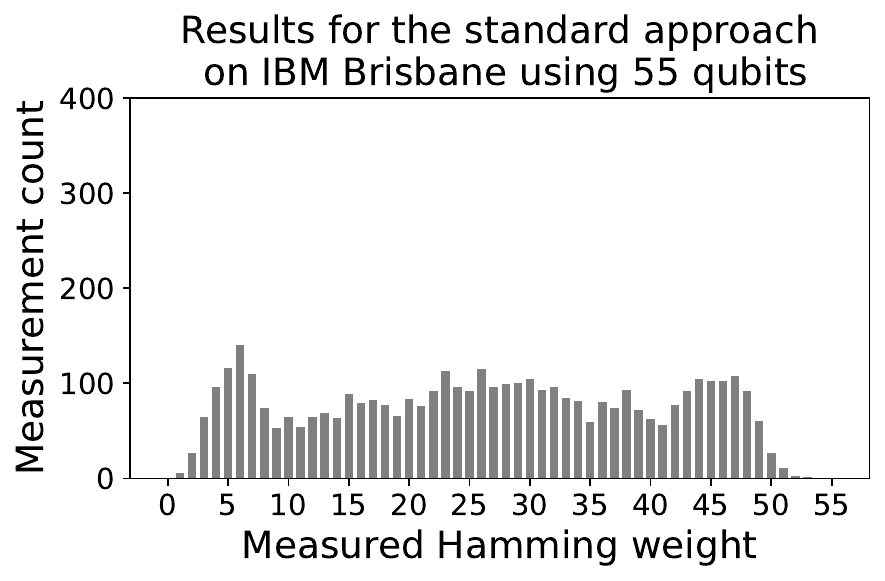}
    \caption{Measurement results for preparing a GHZ state on $n=30$, $n=40$, and $n=55$ qubits on the IBM Brisbane device using the $\LAQCC$-approach and the standard approach.
    Horizontally, the different measured Hamming weights are shown and the height of the bars shows how often that Hamming weight was found.}
    \label{fig:error:Brisbane:LAQCC_standard_large_n}
\end{figure}

\cref{fig:error:Brisbane:LAQCC_standard_large_n} shows the results for large $n$. 
The $\LAQCC$-approach appears to produce a normal distribution, as seen for $n=25$. 
This suggests the $\LAQCC$-approach samples from a uniform superposition over all bit strings. 
The results for the standard approach show similarities with the expected distribution, with a little more weight towards the two extremes of the distribution. 

We believe that to a large extend the erroneous results for the $\LAQCC$-circuits can be explained by the limitations of the intermediate computations. 
We could only control future quantum operations by single bit measurement outcomes. 
As a result, we had to manually implement the parity sum computations. 
We did so by applying an $X$-gate for every measurement result to all subsequent qubits. 
This resulted in the last qubit having up to $n-1$ $X$-gates being applied to it. 
It is quite likely that the overhead this created affected the performance of the protocols negatively. 

%% file: LAQCC/error_analysis/W_state_error_analysis.tex
\section{Error analysis for \texorpdfstring{$W$}{W}-state preparation}\label{sec:error:W_state}
In this section, we derive an expression for the success probability of preparing $W$-states with a standard approach using a linear nearest-neighbor connectivity and with the $\LAQCC$-approach from \cref{sec:LAQCC:W_state}. 
The standard approach has depth $\bigo(n)$ and uses $n$ qubits, whereas the $\LAQCC$-approach uses $\bigo(n\log (n)\log\log (n))$ qubits and has constant depth. 
Based on the circuit sizes, our intuition suggests that the $\LAQCC$-approach performs better if 
\begin{equation*}
    p_{d} \gtrsim p_{id}^{\Omega(n/(\log (n)\log\log (n)))}.
\end{equation*}
In the next sections we will derive success probabilities for both approaches and will see that our intuition is indeed correct, as summarized in the next theorem. 
\begin{theorem}\label{thm:LAQCC:error:W_state}
    Let $n=2^k$ for some integer $k$ and let $\eps >0$ be a constant. 
    If $p_d = (1+\eps)p_{id}^{3n/(59\log_2 (n) \log_2\log_2 (n))}$, then, with respect to the most significant terms, $P_{W,\LAQCC} \gtrsim (1+\eps)^{59n / (\log_2 (n)\log_2\log_2 (n))} P_{W, direct}$.
\end{theorem}
\cref{sec:success_subroutines} presents success probabilities for some $\LAQCC$-subroutines. 
\cref{sec:success_LAQCC_W_state,sec:success_standard_W_state} give the success probabilities for preparing the $W$-state using a $\LAQCC$-approach and using a standard approach. 
Finally, \cref{sec:success_compare_W_state} compares the two approaches and proves the theorem.

\subsection{Success probability for different subroutines}\label{sec:success_subroutines}
The $\LAQCC$-approach to prepare the $W$-state uses multiple subroutines, such as the fanout gate and the OR-gate. 
This section provides the success probability for these subroutines. 
Specifically, this section gives the success probability for the fanout gate and parity gate on $n$ qubits, the OR-reduction introduced by \citeauthor{HoyerSpalek:2005}~\cite{HoyerSpalek:2005} and the exact OR-gate by \citeauthor{TakahashiTani:2013}~\cite{TakahashiTani:2013}.

\subsubsection{Fanout and parity gate}
\begin{figure}[th]
    \centering
    \begin{quantikz}
    \lstick{$\ket{\phi}$}\slice[style=gray]{t=0} & \slice[style=gray]{t=1} & \slice[style=gray]{t=2} & \ctrl{3}\slice[style=gray]{t=3} & \slice[style=gray]{t=4} & \slice[style=gray]{t=5} & & \slice[style=gray]{t=6} & \slice[style=gray]{t=7} & \slice[style=gray]{t=8} & \gate{Z}\slice[style=gray]{t=9} & \\
    \lstick{$\ket{x_1}$} & & & & & & \targ{} & & & & & \\
    \lstick{$\ket{x_2}$} & & & & & & & \targ{} & & & & \\
    \lstick{$\ket{0}$} & \gate{H} & \ctrl{1} & \targ{} & \meter{} & \ctrl[vertical wire=c]{1}\setwiretype{c} \\
    \lstick{$\ket{0}$} & & \targ{} & \targ{} & \meter{} & \ctrl[vertical wire=c]{1}\setwiretype{c} \\
    \lstick{$\ket{0}$} & \gate{H} & \ctrl{1} & \ctrl{-1} & & \targ{} & \ctrl{-4} & & \gate{H} & \meter{} & \ctrl[vertical wire=c]{-5}\setwiretype{c} & \setwiretype{n} \\
    \lstick{$\ket{0}$} & & \targ{} & \targ{} & \meter{} & \ctrl[vertical wire=c]{1}\wire[u][1]{c}\setwiretype{c} & \setwiretype{n} \\
    \lstick{$\ket{0}$} & \gate{H} & & \ctrl{-1} & & \targ{} & & \ctrl{-5} & \gate{H} & \meter{} &  \ctrl[vertical wire=c]{-2}\setwiretype{c} & \setwiretype{n} 
    \end{quantikz}
    \caption{Implementation of a quantum fanout gate with the GHZ state preparation expanded.
    The time steps indicate which gates can be applied in parallel.}
    \label{fig:q_circuit:non_local_cnot_expanded}
\end{figure}
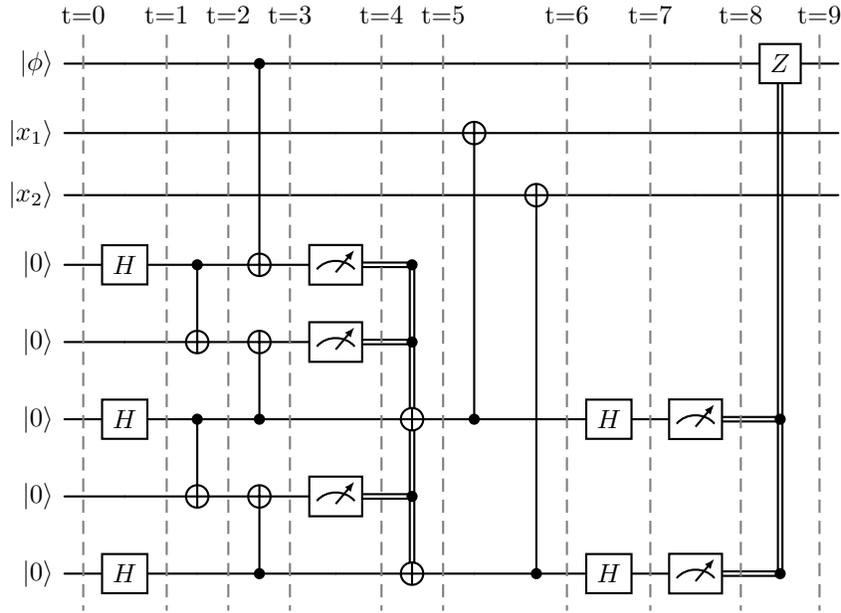

A fanout gate on $n$ qubits has one control qubit and $n-1$ target qubits. 
Our implementation requires $3n-1$ qubits and is inspired by the non-local CNOT-gate by \citeauthor{YimsiriwattanaLomonaco:2004}~\cite{YimsiriwattanaLomonaco:2004}.
The circuit first prepares a GHZ state on $n$ qubits using $2n-1$ qubits and then uses this GHZ state to apply parallel gate teleportation to implement the fanout gate. 
\cref{fig:q_circuit:non_local_cnot_expanded} gives the corresponding circuit for $n=3$, the time steps indicate which gates can be applied in parallel. 

We can extend the circuit shown in \cref{fig:q_circuit:non_local_cnot_expanded} to general~$n$ and count the gates and idling terms to obtain the following expression for the success probability:
\begin{equation}\label{eq:success:fanout}
    P_{Fanout_{n}} \ge p_{s}^{2n + \ceil{(n-1)/2}} p_{is}^{5n + \floor{(n-1)/2} - 2} p_{d}^{3n - 2} p_{id}^{2n + 1} p_{m}^{2n - 1} p_{im}^{3n - 1} p_{ic}^{3n - 1}.
\end{equation}

The fanout gate and the parity gate are closely related, as the parity gate corresponds to a fanout gate with each qubit conjugated by Hadamard gates. 
By including the Hadamard gates on the first $n$ qubits in the fanout-gate circuit, we reduce the number of idling qubits, while retaining the same circuit depth. 
This gives a success probability for the parity gate on $n$ qubits of 
\begin{equation}\label{eq:success:parity}
    P_{Parity_{n}} \ge p_{s}^{4n + \ceil{(n-1)/2} - 1} p_{is}^{3n + \floor{(n-1)/2} - 1} p_{d}^{3n - 2} p_{id}^{2n + 1} p_{m}^{2n - 1} p_{im}^{3n - 1} p_{ic}^{3n - 1}. 
\end{equation}
As the circuits for both the fanout gate and the parity gate have the same depth and apply the same type of operations, we see that a qubit idling during the execution of either of the two gates has the same success probability. 
Counting the gates gives a success probability for an idling qubit of
\begin{equation}\label{eq:success:fanout_idle}
    P_{iFanout_{n}} = P_{iParity_{n}} = p_{is}^{4} p_{id}^{3} p_{im}^{2} p_{ic}^{2}.
\end{equation}

\subsubsection{OR-reduction}
The OR-reduction introduced by \citeauthor{HoyerSpalek:2005} prepares a state on $\bigo(\log n)$ qubits, such that evaluating an OR-gate on this reduced state gives the same output as the OR-gate evaluated on the initial $n$ qubits~\cite[Lemma~5.1]{HoyerSpalek:2005}.

Let $c$ be any positive integer and $\varphi\in [0,2\pi)$, define the state
\begin{equation*}
    \ket{\mu_{\varphi}^{c}} = \frac{1+e^{i\varphi c}}{2}\ket{0} + \frac{1-e^{i\varphi c}}{2}\ket{1}.
\end{equation*}
We obtain this state by computing: $\ket{\mu_{\varphi}^{c}}= HR_Z(\varphi c) H \ket{0}$. 
The OR-reduction uses these $\ket{\mu_{\varphi}^{c}}$ states. 

\begin{figure}
    \centering
    \begin{quantikz}
    \lstick{$\ket{x_1}$}\slice[style=gray]{t=0} & \slice[style=gray]{t=1} & \slice[style=gray]{t=2} & \ctrl{4} & & \slice[style=gray]{t=3} & \slice[style=gray]{t=4} & \slice[style=gray]{t=5} & \rstick{$\ket{x_1}$} \\
    \lstick{$\ket{x_2}$} & & & & \ctrl{4} & & & & \rstick{$\ket{x_2}$} \\
    \lstick{$\vdots$} \\
    \lstick{$\ket{x_n}$} & & & & & \ctrl{4} & & & \rstick{$\ket{x_n}$} \\
    \lstick{$\ket{0}$} & \gate{H} & \ctrl{3} & \gate{R_Z(\varphi_k)} & & & \ctrl{3} & \gate{H} & \rstick{$\ket{\mu_{\varphi_k}^{|x|}}$} \\
    \lstick{$\ket{0}$} & & \targ{} & & \gate{R_Z(\varphi_k)} & & \targ{} & & \rstick{$\ket{0}$} \\
    \lstick{$\vdots$} \\
    \lstick{$\ket{0}$} & & \targ{} & & & \gate{R_Z(\varphi_k)} & \targ{} & & \rstick{$\ket{0}$}
    \end{quantikz}
    \caption{$\LAQCC$-circuit that prepares the state $\ket{\mu_{\varphi_k}^{|x|}}$. 
    These states are used to implement the OR-reduction. 
    The dotted lines indicate time steps and which gates can be applied in parallel.}
    \label{fig:q_circuit:OR_reduction}
\end{figure}
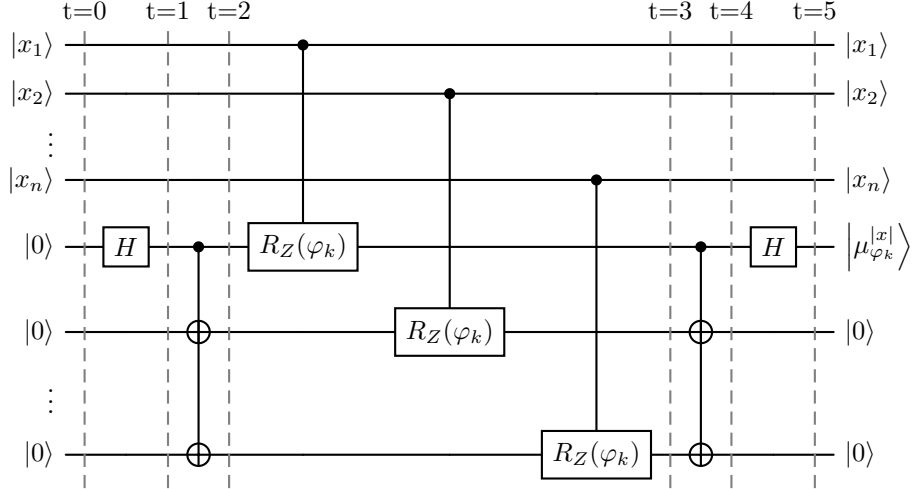

Given input $x_1,\hdots,x_n$, the OR-reduction prepares $t=\ceil{\log_2 (n+1)}$ states $\ket{\mu_{\varphi_k}^{|x|}}$ for $\varphi_k = \frac{2\pi}{2^k}$ and $k\in [t]$. 
We can prepare the states $\ket{\mu_{\varphi_k}^{|x|}}$ for each $k$ in parallel using fanout gates. 
\cref{fig:q_circuit:OR_reduction} shows the circuit for a single $k$, with the time steps again indicating which gates can be applied in parallel. 

We can parallelize this circuit for every $k$ using fanout gates. 
The $n$ fanout gates of length $t$ for copying the input qubits can be applied in parallel with the $t$ fanout gates of length $n$ for copying the auxiliary qubits. 
As a result, we have no idling terms for the fanout gates. 
Additionally, the first and last Hadamard gate in \cref{fig:q_circuit:OR_reduction} can be included in the construction of the fanout gates. 
This gives a total success probability of: 
\begin{align}
    P_{OR_{n}\text{-}reduction} & = P_{Fanout_{t}}^{2n}P_{Fanout_{n}}^{2t} \big(P_{cR_Z}^{n}\big)^{t} \nonumber \\
    & \ge p_{s}^{11nt + 2(n\ceil{(t-1)/2} + t\ceil{(n-1)/2}) + 2t} p_{is}^{23nt + 2n\floor{(t-1)/2} + 2t\floor{(n-1)/2} - 4(n+t)} \nonumber \\
    & \qquad p_{d}^{14nt - 4(n+t)} p_{id}^{8nt + 2(n+t)} p_{m}^{8nt - 2(n+t)} p_{im}^{12nt - 2(n+t)} p_{ic}^{12nt - 2(n+t)} \label{eq:success:OR_reduction}
\end{align}
We now use this expression to determine the success probability for the OR-gate.

\subsubsection{OR-gate}
\citeauthor{TakahashiTani:2013} presented an exponential-sized circuit for the OR-gate, which they applied to a state of logarithmic size~\cite{TakahashiTani:2013}. 
Combined, this gives a circuit of polynomial size.
Using the Fourier inversion formula shown in \cref{eq:Fourier_inversion_formula} we see that for a bit string $x\in\F_2^n$ we can write
\begin{equation*}
    \text{OR}_n(x) = \frac{1}{2^{n-1}}\sum_{a\in\F_2^n\setminus \{0^n\}} \text{PA}_n^a(x),
\end{equation*}
where PA$_n^a(x) = \oplus_{j=0}^{n-1} a_i x_i$ is the parity of $x$, weighted by a nonzero vector~$a$.
This weighted-parity gate is implemented by a standard parity gate on a subset of the inputs.
The exponential-sized circuit for the OR$_n$-gate consists of three steps: 
\begin{enumerate}
    \item Simultaneously, 
    \begin{enumerate}
        \item Copy the input state $2^n-1$ times and compute PA$_n^a(x)$ for every nonzero $n$-bit string $a$;
        \item Prepare a GHZ state on $2^n-1$ qubits;
    \end{enumerate}
    \item Apply $R_Z(\pi/2^{n-1})$-gates to the qubits in the GHZ state and controlled by the qubits that hold the PA$_n^a(x)$ results;
    \item Apply a fanout gate of length $2^n-1$ to the GHZ state together with a Hadamard gate to uncopy the GHZ state and obtain the OR of the input in a single auxiliary qubit. 
\end{enumerate}
We then need to uncompute the auxiliary registers by running the first two steps.
This introduces an extra factor two in the exponents in the overall success probability. 
Furthermore, in this protocol, we can choose to prepare the GHZ states only when they are needed, thereby omitting idling terms for the GHZ states. 

In total, we will have $n$ parity gates with a single target, $\binom{n}{2}$ parity gates with two targets, and in general $\binom{n}{k}$ parity gates with $k$ targets, where $k$ corresponds to the Hamming weight of $a$. 
A lower bound on the success probability follows by replacing every parity gate by the success probability for the parity gate on all inputs, resulting in an easier expression. 

Let $n$ be the length of the input and $t=\ceil{\log_2(n+1)}$. 
The success probability for the OR-gate is then given by
\begin{align}
    P_{OR_{n}} & = P_{OR_{n}\text{-}reduction}^{2} P_{Fanout_{2^{t-1}}}^{2t} \left(\Pi_{k=1}^{t} \binom{t}{k} P_{Parity_{k}}\right)^{2} P_{GHZ_{2^t-1}, \LAQCC}^{2} P_{cR_Z}^{2^{t}-1} p_{s}^{1} \nonumber \\
    & \ge P_{OR_{n}\text{-}reduction}^{2} P_{Fanout_{2^{t-1}}}^{2t} P_{Parity_{t}}^{2(2^{t}-1)} P_{GHZ_{2^t-1}, \LAQCC}^{2} P_{cR_Z}^{2^{t}-1} p_{s}^{1} \nonumber \\
    & \ge p_{s}^{22nt + 2(2n - 2^{t} - 1)\ceil{(t-1)/2} + 4t\ceil{(n-1)/2} + 2t\ceil{(2^{t-1}-1)/2} + 2\ceil{(2^t-1)/2} + 10t\cdot 2^{t} + 3\cdot 2^{t} - 4t - 2} \nonumber \\
    & \qquad p_{is}^{2(2n + 2^{t} - 1)\floor{(t-1)/2} + 4t\floor{(n-1)/2} + 2t\floor{(2^{t-1}-1)/2} + 2\floor{(2^t-1)/2}} \nonumber \\
    & \qquad p_{is}^{46nt - 8n - 18t + 11t\cdot 2^{t} + 3\cdot 2^{t} - 5} p_{d}^{28nt - 8n - 18t + 9t\cdot 2^{t} + 2\cdot 2^{t} - 6} p_{id}^{16nt + 4n + 2t + 6t\cdot 2^{t} + 2\cdot 2^{t} + 2} \nonumber \\
    & \qquad p_{m}^{16nt - 4n - 10t + 6t\cdot 2^{t} - 2} p_{im}^{24nt - 4n - 12t + 9t\cdot 2^{t}} p_{ic}^{24nt - 4n - 12t + 9t\cdot 2^{t}}. \label{eq:success_OR_gate_exact}
\end{align}
Note that this expression holds for arbitrary $n$.
For $n=2^k$, the expression simplifies, as in that case $t=\ceil{\log_2(n+1)} = k+1$, which gives
\begin{align}
    P_{OR_{n}} & \ge p_{s}^{42nk + 50n - 4k + 2(2k - 2n + 1)\ceil{k/2} + 6(k+1)\ceil{(n-1)/2} - 6} \nonumber \\
    & \qquad p_{is}^{68nk + 68n - 18k + 2(4n - 1)\floor{k/2} + 6(k+1)\floor{(n-1)/2} - 25} p_{d}^{46nk + 42n - 18k - 24} \nonumber \\
    & \qquad p_{id}^{28nk + 36n + 2k + 4} p_{m}^{28nk + 24n - 10k - 12} p_{im}^{42nk + 38n - 12k - 12} p_{ic}^{42nk + 38n - 12k - 12}. \label{eq:success_OR_gate_simplified}
\end{align}

\subsection{\texorpdfstring{$\LAQCC$}{LAQCC}-approach}\label{sec:success_LAQCC_W_state}
This section gives the success probability for preparing the $W$-state on $n=2^k$ qubits, for some integer $k$, using the $\LAQCC$-approach presented in \cref{sec:LAQCC:W_state}.
We obtain the success probability for the $W$-state by determining them for the \textbf{Uncompress} and \textbf{Compress} methods and then taking the product of the two. 
Following the steps outlined in \cref{lem:W_state_uncompress,lem:W_state_compress} gives: 
\begin{align}
    P_{\textbf{Uncompress}_n} & = p_{s}^{k} p_{is}^{nk + n - k} P_{Fanout_{n}}^{2k} P_{iFanout}^{2n} \big(\Pi_{i=0}^{n-1} P_{Equal_{i}}\big). \label{eq:success_uncompress} \\
    P_{\textbf{Compress}_n} & = p_{s}^{2k} p_{is}^{2(nk + n - k)} P_{Fanout_{n}}^{2k} P_{iFanout}^{2n} \big(\Pi_{i=0}^{n-1} P_{cZ, target_{i}}\big). \label{eq:success_compress}
\end{align}

The success probability for the Equal$_i$-gate is lower bounded by the success probability of the OR$_k$-gate with all $k$ input qubits conjugated with $X$-gates. 
We can incorporate these $X$-gates in the circuit for the OR$_k$-gate. 
The controlled-$Z$-gates with target $i$ correspond to a fanout gate with targets on the qubits corresponding to the ones in the binary representation of $i$, and with the targets conjugated by Hadamard gates. 
Hence, we can lower bound the success probability of these controlled-$Z$-gates with target $i$ by the success probability of a fanout gate of length $k+1$ with the target qubits conjugated by Hadamard gates. 
Summarizing, we have that for every $i\in\F_{2}^{k}$
\begin{align}
    P_{Equal_{i}} & \ge P_{OR_{k}}p_{s}^{2k}p_{is}^{2}, \label{eq:success_equal_i} \\
    P_{cZ, target_{i}} & \ge P_{Fanout_{k+1}}p_{s}^{2k}p_{is}^{2}. \label{eq:success_cZ_target}
\end{align}

We can now obtain a lower bound on the success probability of preparing the $W$-state by multiplying the expressions of \cref{eq:success_uncompress,eq:success_compress}, applying the lower bounds described in \cref{eq:success_equal_i,eq:success_cZ_target} and using the expression for the OR-gate given in \cref{eq:success_OR_gate_exact}.
We use $t=\ceil{\log_2(k+1)}$ and obtain
\begin{align}
    & P_{W,\LAQCC} = P_{\textbf{Uncompress}_n} P_{\textbf{Compress}_n} \nonumber \\
    & \enspace \ge p_{s}^{4nk + 3k} p_{is}^{3nk + 7n - 3k} P_{Fanout_{n}}^{4k} P_{Fanout_{k+1}}^{n} P_{iFanout}^{4n} P_{OR_{k}}^{n} \nonumber \\
    & \enspace \ge p_{s}^{22nkt + 14nk + 2n\ceil{(2^t-1)/2} + n(3\cdot 2^{t} + \ceil{k/2}) + 2nt(5\cdot 2^{t} - 2) + 3k + 4k\ceil{(n-1)/2} + 2n(2k - 2^{t} - 1)\ceil{(t-1)/2}} \nonumber \\
    & \qquad p_{s}^{4nt\ceil{(k-1)/2} + 2nt\ceil{(2^{t-1}-1)/2} + 2n\ceil{(2^t-1)/2}} p_{is}^{46nkt + 20nk + 3n\cdot 2^{t} - 18nt + 21n - 11k + n\floor{k/2}} \nonumber \\
    & \qquad p_{is}^{2n(2k + 2^{t} - 1)\floor{(t-1)/2} + 4nt\floor{(k-1)/2} + 2nt\floor{(2^{t-1}-1)/2} + 2n\floor{(2^t-1)/2} + 11nt\cdot 2^{t} + 4k\floor{(n-1)/2}} \nonumber \\
    & \qquad p_{d}^{28nkt + 7nk + 9nt(2^{t} - 2) + 2n\cdot 2^{t} - 5n - 8k} p_{id}^{16nkt + 14nk + 2nt(3\cdot 2^{t} + 1) + 2n\cdot 2^{t} + 17n + 4k} \nonumber \\
    & \qquad p_{m}^{16nkt + 6nk + 2nt (3\cdot 2^{t} - 5) - n - 4k} (p_{im}p_{ic})^{24nkt + 11nk + 3nt(3\cdot 2^{t} - 4) + 10n - 4k}. \label{eq:success_W_LAQCC_exact} 
\end{align}
Note that this expression is quite involved with dependencies on both $n$, $k=\log_2(n)$ and $t=\ceil{\log_2(k+1)}$. 
We furthermore use ceil- and floor-functions. 
We can approximate \cref{eq:success_W_LAQCC_exact} using $\ceil{x}\approx x\approx \floor{x}$ for $x\in \R$. 
In this case, we see that $2^{t}\approx k$. 
We then obtain
\begin{align}
    P_{W,\LAQCC} & \gtrsim p_{s}^{71nkt/2 + 37nk/2 + 3nk - 8nt - n + k} p_{is}^{125nkt/2 + 47nk/2 - 22nt + 21n - 13k} \nonumber \\
    & \qquad p_{d}^{37nkt + 9nk - 18nt - 5n - 8k} p_{id}^{22nkt + 16nk + 2nt + 17n + 4k} \nonumber \\
    & \qquad p_{m}^{22nkt + 6nk - 10nt - n - 4k} (p_{im}p_{ic})^{33nkt + 11nk - 12nt + 10n - 4k}. \label{eq:success_W_LAQCC_approximate} 
\end{align}

\subsection{Direct method}\label{sec:success_standard_W_state}
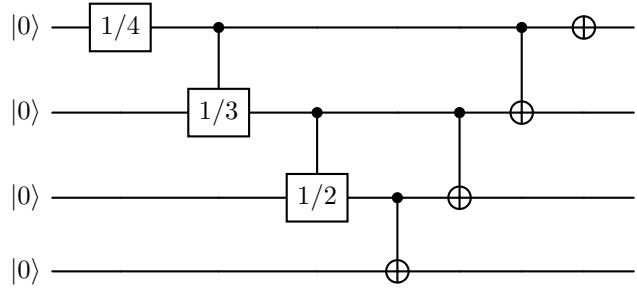
\begin{figure}
    \centering
    \begin{quantikz}
    \lstick{\ket{0}} & \gate{1/4} & \ctrl{1} & & & & \ctrl{1} & \targ{} & \\
    \lstick{\ket{0}} & & \gate{1/3} & \ctrl{1} & & \ctrl{1} & \targ{} & & \\
    \lstick{\ket{0}} & & & \gate{1/2} & \ctrl{1} & \targ{} & & & \\
    \lstick{\ket{0}} & & & & \targ{} & & & &
    \end{quantikz}
    \caption{Exact circuit for preparing the $W$-state for $n=4$.
    Every gate parametrized by $1/n$ denotes a controlled-$R_Y$-gate with argument $\theta = -2\arccos{\sqrt{1/n}}$.}
    \label{fig:q_circuit:W_prep:exact}
\end{figure}

The most direct method of preparing a $W$-state is by applying successive controlled rotations, using controlled-$R_Y$-gates (\cref{eq:R_Y_gate}), where the angle of the gates depends on the qubit index. 
\cref{fig:q_circuit:W_prep:exact} shows the circuit for $n=4$. 
Each square $1/n$ denotes an $R_Y(\theta)$-gate with argument $\theta = -2\arccos{\sqrt{1/n}}$.
The controlled-$R_Y$-gate reduce to a CNOT-gate for $n=1$.

The quantum circuit shown in \cref{fig:q_circuit:W_prep:exact}, when extended to arbitrary $n$, indeed prepares the $W$-state on $n$ qubits. 
Just before the first CNOT-gate, the quantum circuit corresponds to the state 
\begin{equation*}
    \frac{1}{\sqrt{n}} \sum_{i=0}^{n-1} \ket{1}^{i}\ket{0}^{n-i}.
\end{equation*}
Each CNOT-gate will then correctly set one additional qubit, until we have the desired $W$-state. 
The idea behind this circuit is to iteratively ``pass on'' part of the amplitude to the remaining unset qubits and thereby correctly set all qubits. 

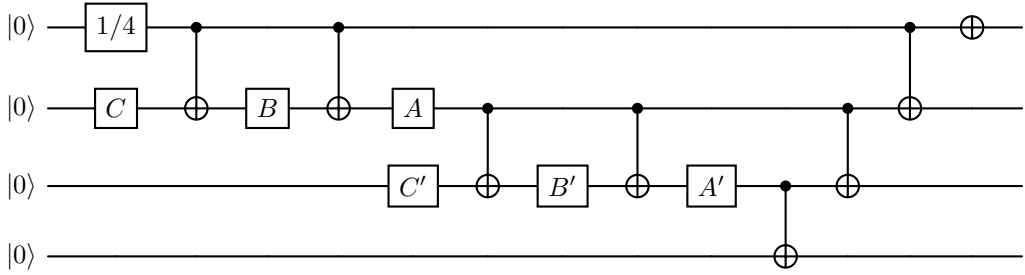
\begin{figure}
    \centering
    \begin{quantikz}
    \lstick{\ket{0}} & \gate{1/4} & \ctrl{1} & & \ctrl{1} & & & & & & & & \ctrl{1} & \targ{} & \\
    \lstick{\ket{0}} & \gate{C} & \targ{} & \gate{B} & \targ{} & \gate{A} & \ctrl{1} & & \ctrl{1} & & & \ctrl{1} & \targ{} & &  \\
    \lstick{\ket{0}} & & & & & \gate{C'} & \targ{} & \gate{B'} & \targ{} & \gate{A'} & \ctrl{1} & \targ{} & & & \\
    \lstick{\ket{0}} & & & & & & & & & & \targ{} & & & & 
    \end{quantikz}
    \caption{Exact decomposed circuit for preparing the $W$-state for $n=4$, where every controlled $R_Y$-gate is replaced by single-qubit gates and CNOT-gates.}
    \label{fig:q_circuit:W_prep:exact_decomposed}
\end{figure}

The circuit shown in \cref{fig:q_circuit:W_prep:exact} uses controlled-$R_Y$-gates, which most quantum hardware devices do not directly support. 
We can decompose the controlled-$R_Y$-gates using the decomposition shown in \cref{fig:q_circuit:controlled_U}.
In our case, the exact single-qubit gates used in the decomposition are irrelevant, as we assume the same success probability for every single-qubit gate. 
\cref{fig:q_circuit:W_prep:exact_decomposed} shows the resulting decomposed quantum circuit to prepare the $W$-state for $n=4$. 
This figure shows that some gates can be applied in parallel. 
Counting the depth gives $n-2$ groups of four layers each, with two single-qubit gates, a CNOT-gate, one single-qubit gate and again a CNOT-gate. 
Next, there is one single-qubit gate, $n-1$ layers of CNOT-gates and a final single-qubit gate, for a total depth of $5n-7$ for $n\ge 2$. 
The total success probability is given by 
\begin{equation} \label{eq:success_W_direct}
    P_{W,direct} = p_{s}^{3n - 4} p_{is}^{n(2n - 5) + 4} p_{d}^{3n - 5} p_{id}^{n(3n - 11) + 10}.
\end{equation}

\subsection{Comparison success probability}\label{sec:success_compare_W_state}
To determine when the $\LAQCC$-approach outperforms the standard approach, we have to determine when 
\begin{equation}
    P_{W,\LAQCC} \ge P_{W, direct}. 
\end{equation}
Comparing the success probabilities given in \cref{eq:success_W_LAQCC_exact,eq:success_W_direct} shows that this inequality holds approximately if the following inequality holds
\begin{align*}
    & p_{s}^{71nkt/2 + 37nk/2 + 3nk - 8nt - 4n + k + 4} p_{d}^{37nkt + 9nk - 18nt - 8n - 8k + 5} \\
    & p_{m}^{22nkt + 6nk - 10nt - n - 4k} (p_{im}p_{ic})^{33nkt + 11nk - 12nt + 10n - 4k} \\
    & \enspace \ge p_{is}^{2n^2 - 125nkt/2 - 47nk/2 + 22nt - 26n + 13k + 4} p_{id}^{3n^2 - 22nkt - 16nk - 2nt - 28n - 4k + 10}.
\end{align*}
Applying the assumptions on the success probabilities discussed in \cref{sec:error:error_model}, shows that this inequality reduces to 
\begin{equation*}
    p_{d}^{59nkt + 15nk - 28nt - 9n - 12k + 5} \gtrsim p_{id}^{3n^2 - 88nkt - 38nk + 22nt - 48n + 4k + 10}.
\end{equation*}
For a first estimate on when the $\LAQCC$-approach outperforms the standard direct approach, we only keep the most significant terms.
Using that $k=\log_2 n$ and $t\approx \log_2 \log_2 n$ shows that the $\LAQCC$-approach performs best if
\begin{equation*}
    p_{d} \gtrsim p_{id}^{3n/(59\log_2 n \log_2\log_2 n)}.
\end{equation*}
Let $\eps>0$, now if $p_d = (1+\eps)p_{id}^{3n/(59\log_2 n \log_2\log_2 n)}$, then we see that $P_{W,\LAQCC} \gtrsim (1+\eps)^{59n\log_2 n\log_2\log_2 n} P_{W, direct}$, finishing the proof of \cref{thm:LAQCC:error:W_state} and also showing our intuition was indeed correct.

%% file: LAQCC/error_analysis/discussion_error.tex
\section{Reflections and outlook}\label{sec:error:discussion}
In this chapter, we considered the success probability of different approaches to prepare the GHZ state and the $W$-state.
We used a worst-case error model, where errors correspond to Haar random unitary matrices being applied to the gates. 
A single error causes the task of perfectly preparing the quantum states to fail, as multiple errors cancel with probability zero. 
As such, we derived expressions for the probability that none of the qubits decoheres, either while idling or while being manipulated by a quantum gate.

For the GHZ state, we derived success probabilities for two standard protocols, using either an all-to-all connectivity or a linear nearest-neighbor connectivity, as well as for a $\LAQCC$-approach. 
We also did this for a hybrid version of a standard approach and a $\LAQCC$-approach. 
Next, we compared the derived success probabilities to determine which approach performs best when.
Conditional on assumptions on the magnitude of the success probabilities of the individual terms in the quantum circuit, we find that the $\LAQCC$-approach performs exponentially better than the standard approach using an all-to-all connectivity if $p_{d} \gtrsim (1+\eps)p_{id}^{\Omega(\log n)}$, and similarly for the linear nearest-neighbor connectivity if $p_{d} \gtrsim (1+\eps)p_{id}^{\Omega(n)}$. 
\cref{thm:LAQCC:error:GHZ} gives the exact constants. 
The theorem says that the $\LAQCC$-approach outperforms the standard approaches if the probability that a qubit decoheres while idling for $\Omega(\log n)$, respectively, $\Omega(n)$ two-qubit gate durations is larger than the probability that a single two-qubit gate is erroneous. 
Both results are in line with what our intuition tells us based on the circuit sizes. 

For the GHZ state we implemented both the $\LAQCC$-approach and the standard approach using a linear nearest-neighbor connectivity. 
We found that for small problem instances, both approaches perform similarly. 
For larger $n$, the $\LAQCC$-approach returns an approximately normal distribution (when aggregating the results based on the Hamming weight). 
The standard approach gives results that show some similarities with the expected output distribution: two large groups of outputs having low or high Hamming weight and some samples in between. 
Based on the results, we saw that the used error model was suboptimal and differs from practical error models.
Note that we could only use the intermediate measurement outcomes to control future quantum operations, and not to perform computations with them before controlling quantum operations. 

Next, we performed a similar analysis for the $W$-state.
The $\LAQCC$-approach has constant depth and circuit size $\bigo(n\log n\log\log n)$.
The standard direct approach uses $n$ qubits and has depth $\bigo(n)$.
Comparing the derived success probabilities shows that the $\LAQCC$-approach exponentially outperforms the standard approach if $p_{d} \gtrsim (1+\eps)p_{id}^{\Omega(n/(\log (n)\log\log (n)))}$.
\cref{thm:LAQCC:error:W_state} gives the constants for the most significant term in the exponent.

Multiple directions for future research exist. 
First, the same analysis can be used to compare different quantum approaches to prepare other quantum states. 
We expect similar relations between $p_{d}$ and $p_{id}$, depending on the circuit sizes and the number of CNOT-gates.
Ideally, one would prove such a relation on a higher level, obtaining a result for multiple quantum state preparation routines. 

Second, we used a worst-case error model in our analysis. 
The hardware implementation demonstrated that the error model is suboptimal and differs from practical error models. 
Specifically, the used error model does not allow for errors, whereas in practice, some errors can be tolerated or might even cancel. 
Furthermore, conventional post-processing techniques exist to mitigate the effect of errors~\cite{TemmeBravyiGambetta:2017,Kandala:2019,Cai:2023}.

A similar analysis with a more realistic error model is an interesting extension of this chapter. 
The depolarizing and dephasing channels applied to the quantum gates are an example of a more realistic error model. 
With these error models, gates are replaced by standard probability distributions over different gates, and two errors might cancel. 
These error models give a more realistic picture of the performance of a quantum circuit, but complicate the analysis. 
We then need both upper and lower bounds on the success probabilities to compare them. 

Another aspect in which we can extend the error model is by dropping the constraint of independent errors. 
\citeauthor{Kalai:2016} argues that the noise models underlying quantum computing are complex and highly entangled~\cite{Kalai:2016}. 
He states that error models with independent errors are oversimplified and cannot approximate the behavior for large systems well. 
Correlated noise is also seen with metronomes initialized randomly but placed on the same surface. 
Over time, these metronomes will synchronize~\cite{PikovskyRosenblumKurths:2001}, something that cannot be explained by (noise) models of individual metronomes alone. 
The computation of the success probability will become extremely complex with error models where errors depend on the state of other qubits or on operations applied to other qubits. 

A third direction for future research is to make the circuits themselves more realistic. 
In general, quantum gates have to be decomposed into the native gate set of a device, giving an overhead currently not addressed in the success probability expressions. 
Additionally, some quantum hardware devices limit on the number of parallel gates, for instance due to crosstalk between qubits. 
This crosstalk is magnified when two qubits close to each other are simultaneously manipulated by two different quantum gates~\cite{Zhao:2022}. 
Such limits on the number of parallel gates significantly reduces the power of the quantum device. 
For instance, the advantage of the $\LAQCC$-approach significantly diminishes if only a constant number of gates can be applied in parallel. 
This limitation might be one of the reasons why the $\LAQCC$-approach performs worse in practice than expected. 

Fourth, determine the output states in a different way. 
We measured the states in the Pauli-$Z$ basis, which cannot detect phase errors. 
By instead applying state tomography, for instance, by measuring the expectation value of different Pauli observables, we can also detect phase errors. 
The overhead of state tomography is however large and proves infeasible for large system sizes. 

In line with this extension, we arrive at the fifth possible direction for future research: 
relax the requirements on the output of the circuits. 
Currently, the output state must match the target state exactly;
otherwise, the circuit failed. 
In practice, a high overlap with the target state suffices. 
This relaxation directly opens the way to more possible quantum circuits. 

Approximating the target state opens the way to probabilistic algorithms. 
For instance, approximately preparing the $W$-state is possible by applying $n$ single-qubit $R_Y$-gates with parameter $\theta=\arccos\big(\sqrt{\tfrac{n-1}{n}}\big)$ and then measuring the parity of these gates.
Upon measuring an odd parity, the superposition collapses to a superposition over all bit strings of odd Hamming weight, with those having Hamming weight~$1$ having the highest amplitudes. 
Choosing a smaller $\theta$ reduce the probability of finding an odd parity. 
However, once an odd parity is measured, the resulting state better approximates the $W$-state. 

Allowing the output state to have high overlap with the target state, gives room to simplify the assumptions on the gate set.
The current analysis assumes that all single-qubit gates are available. 
In practice, most gates have to be approximated, for instance using the Solovay-Kiteav theorem, which gives a small overhead. 
Similarly, some gates are easier to implement than others, such as the simple Pauli-gates and the harder $T$-gate. 
Hence, we can imagine that we have different success probabilities for different gates being applied. 
Note that this idea aligns with the third direction for future research mentioned.

An aspect we only briefly touched upon, but which might prove vital in future applications of quantum computing is the time aspect. 
Especially on the near-term, quantum computers will most likely only solve small tasks and have to finish their computations quickly before the qubits decohere completely. 
Furthermore, with short quantum computation times, the cost of having to rerun a quantum circuit because it failed is small. 
The $\LAQCC$-circuit provides one way to overcome this barrier by giving low-depth quantum circuits with intermediate conventional computations. 
With the uncertain development of future quantum computers, we can imagine that short quantum circuits can still prove useful, even if fault-tolerant quantum computation is out of reach.

%% file: non_content/symbols.tex
\chapter*{List of symbols}

The following list gives symbols and notation used throughout this work: 
\begin{itemize}
\item $[n]=\{1,\hdots, n\}$
\item $\omega_p = e^{i2\pi/p}$
\item $\Delta_h f(x) = f(x+h)-f(x)$ the additive derivative of $f:\F_p^n\to\C$
\item $\Delta_h f(x) = f(x+h)\overline{f(x)}$ the multiplicative derivative of $f:\F_p^n\to\C$
\item $\ip{x,y}=\sum_{i=1}^{n} x_i y_i$ the inner product of vectors $x,y\in\F_p^n$
\item $\oplus$ represents addition modulo $2$
\item $\mathbb{D}$ represents the complex unit disc
\end{itemize}

The following definitions concern the behavior of functions $f,g:\N\to\R$:
\begin{itemize}
    \item $f=\bigo(g)$ means there exist constants $c,n_0>0$ such that $|f(n)|\le cg(n)$ for all $n\ge n_0$;
    \item $f=\Omega(g)$ means there exist constants $c,n_0>0$ such that $|f(n)|\ge cg(n)$ for all $n\ge n_0$;
    \item $f=o(g)$ means for every constant $c>0$, there exists a constant $n_0>0$ such that $|f(n)|<c g(n)$ for all $n\ge n_0$;
    \item $f=\omega(g)$ means for every constant $c>0$, there exists a constant $n_0>0$ such that $|f(n)|>c g(n)$ for all $n\ge n_0$;
    \item $f=\Theta(g)$ means $f=\bigo(g)$ and $f=\Omega(g)$;
    \item $\bigo_{\eps}(f)$ denotes a depends of $\bigo(f)$ on a constant $\eps$, similarly for $\Omega$, $o$, $\omega$ and $\Theta$;
    \item $\widetilde{\bigo}$ hides a dependency on logarithmic factors of the argument, similarly for $\Omega$, $o$, $\omega$ and $\Theta$;
    \item $\poly(f)$ denotes some unspecified polynomial in $f$;
    \item $\exp(f)$ denotes some unspecified exponential in $f$.
\end{itemize}

Finally, 
\begin{itemize}
    \item $\Pol_{\leq d}(\F_p^n, \F_p^k)$ denotes the space of all polynomial maps $\phi:\F_p^n\to \F_p^k$ of degree at most $d$;
    \item $\arank_d(\phi)$ denotes the analytic rank of polynomial maps $\phi\in\Pol_{\leq d}(\F_p^n, \F_p^k)$ (\cref{def:analytic_rank});
    \item $\arank(T)$ denotes the analytic rank of a tensor $T$ (\cref{def:tensor_analytic_rank});
    \item $\prank(T)$ denotes the partition rank of a tensor $T$ (\cref{def:partition_rank}).
\end{itemize}